\documentclass[11pt,a4paper]{article}

\usepackage{fontspec}
\usepackage[titletoc,title,header]{appendix}

\usepackage{amsmath}
\usepackage{amssymb}
\usepackage{amsthm}
\usepackage{mathtools}
\usepackage{xparse}    

\usepackage[margin=1in]{geometry}

\usepackage{graphicx}
\usepackage{xcolor}

\usepackage{hyperref}
\hypersetup{
  colorlinks=true,
  linkcolor=blue,
  citecolor=blue,
  urlcolor=blue,
  pdfauthor={Yoshitsugu Sekine},
  pdftitle={Notes on the Optical-Lattice Hard-Core Bose Gas of Aizenman--Lieb--Seiringer--Solovej--Yngvason}
}

\usepackage[numbers]{natbib}
\theoremstyle{plain}
\newtheorem{thm}{Theorem}[section]
\AtEndEnvironment{thm}{\qed}
\newtheorem{lem}[thm]{Lemma}
\AtEndEnvironment{lem}{\qed}
\newtheorem{prop}[thm]{Proposition}
\AtEndEnvironment{prop}{\qed}
\newtheorem{cor}[thm]{Corollary}
\AtEndEnvironment{cor}{\qed}

\AtEndEnvironment{conj}{\qed}

\AtEndEnvironment{fact}{\qed}

\theoremstyle{definition}
\newtheorem{defn}[thm]{Definition}
\AtEndEnvironment{defn}{\qed}

\AtEndEnvironment{ex}{\qed}

\theoremstyle{remark}
\newtheorem{rem}[thm]{Remark}
\AtEndEnvironment{rem}{\qed}

\makeatletter
\providecommand*{\dashv}{\mathrel{\mathpalette\@Dashv\vDash}}
\newcommand*{\@dashv}[2]{\reflectbox{$\m@th#1#2$}}
\makeatother
\newcommand{\abscard}[1]{\abs{#1}} 
\newcommand{\absvol}[1]{\abs{#1}} 
\newcommand{\abs}[1]{\left| #1 \right|} 
\NewDocumentCommand{\weaknorm}{O{\dbk} m}{#1{#2}} 
\newcommand{\norm}[1]{\left\Vert #1 \right\Vert} 
\newcommand{\opnorm}[1]{\left\Vert #1 \right\Vert_{\txtoperator}} 

\newcommand{\bkt}[2]{\left\langle #1,\,#2 \right\rangle} 
\newcommand{\rbkt}[2]{\left( #1,\,#2 \right)} 
\newcommand{\slim}{\mathrm{s} \hyphen \lim} 
\newcommand{\wlim}{\mathrm{w} \hyphen \lim} 
\newcommand{\isomto}{\mathrel{\rightarrowtail\kern-1.9ex\twoheadrightarrow}} 
\newcommand{\cbk}[1]{\left\{ #1 \right\}} 
\newcommand{\dbk}[1]{\left\langle #1 \right\rangle} 
\newcommand{\pairbk}[1]{\rbk{#1}} 
\newcommand{\rbk}[1]{\left( #1 \right)} 
\newcommand{\sqbk}[1]{\left[ #1 \right]} 
\newcommand{\vecbk}[1]{\rbk{#1}} 
\newcommand{\fun}[2]{#1 \rbk{#2}} 
\newcommand{\sqfun}[2]{#1 \sqbk{#2}} 
\newcommand{\closedinterval}[2]{\sqbk{#1,\,#2}} 
\newcommand{\leftopeninterval}[2]{\left(#1, \, #2 \right]} 
\newcommand{\nonneginterval}{\rightopeninterval{0}{\infty}} 
\newcommand{\openinterval}[2]{\rbk{#1,\,#2}} 
\newcommand{\rightopeninterval}[2]{\left[#1, \, #2 \right)} 
\newcommand{\anticommutator}[2]{\cbk{#1,\,#2}} 
\newcommand{\commutator}[2]{\sqbk{#1,\,#2}} 

\DeclareMathOperator{\arctanh}{arctanh} 
\NewDocumentCommand{\imunit}{O{\mathsf{i}}}{#1} 
\NewDocumentCommand{\placeholder}{O{\bullet}}{#1} 
\NewDocumentCommand{\trace}{O{\operatorname{Tr}}}{#1} 
\newcommand{\Ad}{\operatorname{Ad}} 
\newcommand{\Ker}{\operatorname{Ker}} 
\newcommand{\Ran}{\operatorname{Ran}} 
\newcommand{\bigmiddleslash}[2]{\left. #1 \middle/ #2 \right.} 
\newcommand{\cmpconj}[1]{\overline{#1}} 
\newcommand{\dom}{\operatorname{dom}} 
\newcommand{\dual}[1]{#1^{\ast}} 
\newcommand{\eqcsq}[1]{\sqbk{#1}} 
\newcommand{\eqisom}{\cong} 
\newcommand{\hyphen}{\hbox{-}} 
\newcommand{\kroneckerdelta}{\delta} 
\newcommand{\barmean}[1]{\overline{#1}} 
\newcommand{\napiernum}{\mathsf{e}} 
\newcommand{\net}[2]{\rbk{#1}_{#2}} 
\newcommand{\od}[2]{\frac{d #1}{d #2}} 
\newcommand{\onehalf}{\frac{1}{2}} 
\newcommand{\opchern}[1]{\operatorname{ch}} 
\newcommand{\opimag}{\operatorname{Im}} 
\newcommand{\opod}[1]{\frac{d}{d #1}} 
\newcommand{\oppd}[1]{\frac{\partial}{\partial #1}} 
\newcommand{\opreal}{\operatorname{Re}} 
\newcommand{\setSymbolDownLeft}[2]{{\vphantom{#2}}_{#1}{#2}} 
\newcommand{\setSymbolUpLeft}[2]{{\vphantom{#2}}^{#1}{#2}} 
\NewDocumentCommand{\agvariety}{O{\mathcal}}{#1} 
\NewDocumentCommand{\cmdrel}{O{\omega}}{#1} 
\NewDocumentCommand{\dfsp}{O{A}}{#1} 
\NewDocumentCommand{\eqcpointed}{O{\eqcsq} m}{#1{#2}_{\ast}} 
\NewDocumentCommand{\fnheaviside}{O{H}}{#1} 
\NewDocumentCommand{\fthol}{O{\mathcal{O}}}{#1} 
\NewDocumentCommand{\ftmero}{O{\mathcal{M}}}{#1} 
\NewDocumentCommand{\grcentralizer}{O{Z}}{#1} 
\NewDocumentCommand{\grmetform}{O{2} m m}{\grmet[#1] \! \rbkt{#2}{#3}} 
\NewDocumentCommand{\grmet}{O{2}}{\setSymbolDownLeft{#1}{g}} 
\NewDocumentCommand{\grnormalizer}{O{N}}{#1} 
\NewDocumentCommand{\gropasym}{O{A}}{#1} 
\NewDocumentCommand{\gropsym}{O{S}}{#1} 
\NewDocumentCommand{\grpermorderedpair}{O{\mathcal{P}}}{#1} 
\NewDocumentCommand{\grsym}{O{\mathfrak{S}} m}{#1_{#2}} 
\NewDocumentCommand{\gtbase}{O{\mathcal}}{#1} 
\NewDocumentCommand{\gtfilter}{O{\mathcal}}{#1} 
\NewDocumentCommand{\gtfmlclosed}{O{\mathcal}}{#1} 
\NewDocumentCommand{\gtfmlopen}{O{\mathcal}}{#1} 
\NewDocumentCommand{\gtopenball}{O{U}}{#1} 
\NewDocumentCommand{\gtopencover}{O{\mathcal}}{#1} 
\NewDocumentCommand{\gtopennbh}{O{\mathcal}}{#1} 
\NewDocumentCommand{\gtpreopencover}{O{\mathcal}}{#1} 
\NewDocumentCommand{\gtsubbase}{O{\mathcal}}{#1} 
\NewDocumentCommand{\gtvicinity}{O{\mathcal}}{#1} 
\NewDocumentCommand{\lasp}{O{\mathcal}}{#1} 
\NewDocumentCommand{\latprightrbk}{O{\top} m}{\rbk{#2}^{#1}} 
\NewDocumentCommand{\latpright}{O{\top} m}{#2^{#1}} 
\NewDocumentCommand{\latp}{O{t} m}{\setSymbolUpLeft{#1}{#2}} 
\NewDocumentCommand{\lpdistribution}{O{\mu} m}{#2_{\ast,#1}} 
\NewDocumentCommand{\lpmollifier}{O{\rho}}{#1} 
\NewDocumentCommand{\lpofpositive}{O{\chi}}{#1} 
\NewDocumentCommand{\manliederiv}{O{L}}{#1} 
\NewDocumentCommand{\mansmoothnbh}{O{\mathcal}}{#1} 
\NewDocumentCommand{\mblfmldsysgenerated}{O{d} m}{\fun{#1}{#2}} 
\NewDocumentCommand{\mblfmlgenerated}{O{\sigma} m}{\fun{#1}{#2}} 
\NewDocumentCommand{\oacorrfn}{O{\Gamma}}{#1} 
\NewDocumentCommand{\oagnsvector}{O{\Omega}}{#1} 
\NewDocumentCommand{\oaideal}{O{\mathcal}}{#1} 
\NewDocumentCommand{\oanumberoperator}{O{A}}{#1} 
\NewDocumentCommand{\oaposcone}{O{\mathcal{P}}}{#1} 
\NewDocumentCommand{\oapressure}{O{P}}{#1} 
\NewDocumentCommand{\oarepn}{O{\pi}}{#1} 
\NewDocumentCommand{\oaspnormalstate}{O{N}}{#1} 
\NewDocumentCommand{\oasppurestate}{O{P}}{#1} 
\NewDocumentCommand{\oaspstate}{O{E}}{#1} 
\NewDocumentCommand{\oastatevector}{O{\Omega}}{#1} 
\NewDocumentCommand{\oastate}{O{\omega}}{#1} 
\NewDocumentCommand{\opdilation}{O{\delta}}{#1} 
\NewDocumentCommand{\opdmat}{O{\rho}}{#1} 
\NewDocumentCommand{\opfockan}{O{a}}{#1} 
\NewDocumentCommand{\opfockcran}{O{a}}{#1^{\#}} 
\NewDocumentCommand{\opfockcrdagger}{O{a}}{#1^{\dagger}} 
\NewDocumentCommand{\opfockcr}{O{a}}{#1^{\ast}} 
\NewDocumentCommand{\opfocknumber}{O{N}}{#1} 
\NewDocumentCommand{\opfocksegalconj}{O{\pi}}{#1} 
\NewDocumentCommand{\opfocksegal}{O{\phi}}{#1} 
\NewDocumentCommand{\opspecmeas}{O{E}}{#1} 
\NewDocumentCommand{\opspec}{O{} m}{\fun{\sigma_{#1}}{#2}} 
\NewDocumentCommand{\optransl}{O{\tau}}{#1} 
\NewDocumentCommand{\opvarspec}{O{} m}{\operatorname{spec}_{#1} #2} 
\NewDocumentCommand{\physaction}{O{\mathcal{A}}}{#1} 
\NewDocumentCommand{\physcharge}{O{e}}{\mathrm{#1}} 
\NewDocumentCommand{\physcplconst}{O{\mathsf{g}}}{#1} 
\NewDocumentCommand{\physelectrostaticcapasity}{O{\mathrm{Cap}}}{#1} 
\NewDocumentCommand{\physenergy}{O{E}}{#1} 
\NewDocumentCommand{\physgse}{O{E}}{#1_{0}} 
\NewDocumentCommand{\physham}{O{H}}{#1} 
\NewDocumentCommand{\physlagdensity}{O{\mathcal{L}}}{#1} 
\NewDocumentCommand{\physlag}{O{L}}{#1} 
\NewDocumentCommand{\physliouvilean}{O{L}}{#1} 
\NewDocumentCommand{\physmass}{O{m}}{#1} 
\NewDocumentCommand{\prbbrownmv}{O{B}}{#1} 
\NewDocumentCommand{\prbcharfun}{O{\chi}}{#1} 
\NewDocumentCommand{\prbdist}{O{\mathcal{P}}}{#1} 
\NewDocumentCommand{\prbgaussianmeasure}{O{\msrcal{N}}}{#1} 
\NewDocumentCommand{\prbnormaldist}{O{N}}{#1} 
\NewDocumentCommand{\prbpoissonprocess}{O{N}}{#1} 
\NewDocumentCommand{\prbprocess}{O{X}}{#1} 
\NewDocumentCommand{\prbqspace}{O{\mathcal{Q}}}{#1} 
\NewDocumentCommand{\prbspsample}{O{\Omega}}{#1} 
\NewDocumentCommand{\psh}{O{\mathfrak}}{#1} 
\NewDocumentCommand{\qtquantumchannel}{O{\mathcal{L}}}{#1} 
\NewDocumentCommand{\repn}{O{\pi}}{#1} 
\NewDocumentCommand{\schattencls}{O{\mathbb{K}}}{#1} 
\NewDocumentCommand{\setfmlcylinder}{O{\mathcal{C}}}{#1} 
\NewDocumentCommand{\setfml}{O{\mathcal}}{#1} 
\NewDocumentCommand{\setindex}{O{\mathcal} m}{#1{#2}} 
\NewDocumentCommand{\setlattice}{O{\Gamma}}{#1} 
\NewDocumentCommand{\setspecial}{O{\mathcal} m}{#1{#2}} 
\NewDocumentCommand{\shdiffform}{O{\sheaf{A}}}{#1} 
\NewDocumentCommand{\sheaf}{O{\mathfrak}}{#1} 
\NewDocumentCommand{\smchemicalpotential}{O{\mu}}{#1} 
\NewDocumentCommand{\smenergydensity}{O{\varrho}}{#1} 
\NewDocumentCommand{\smfluctuationwithdmat}{O{\beta} m}{\smuncertaintywithdmat[#1]{#2}^2} 
\NewDocumentCommand{\sminvtemperature}{O{\beta}}{#1} 
\NewDocumentCommand{\smlocaldensityoperator}{O{\rho}}{#1} 
\NewDocumentCommand{\smmicrocanonicalstate}{O{\beta} m}{\physmean{#2}_{#1}} 
\NewDocumentCommand{\smnumberdensity}{O{\rho}}{#1} 
\NewDocumentCommand{\smooth}{O{\mathcal{E}}}{#1} 
\NewDocumentCommand{\smparticlenumber}{O{N}}{#1} 
\NewDocumentCommand{\smpressure}{O{p}}{#1} 
\NewDocumentCommand{\smspecificfreeenergy}{O{\bar{f}}}{#1} 
\NewDocumentCommand{\smthermalvac}{O{\beta}}{\Omega_{#1}} 
\NewDocumentCommand{\smuncertaintywithdmat}{O{\beta} m}{\rbk{\triangle #2}_{#1}} 
\NewDocumentCommand{\sphilbfrak}{O{\mathfrak}}{#1} 
\NewDocumentCommand{\sphilb}{O{\mathcal}}{#1} 
\NewDocumentCommand{\splowerhalf}{O{\mathbb{H}}}{#1_{\txtneg}} 
\NewDocumentCommand{\spupperhalf}{O{\mathbb{H}}}{#1_{\txtnonneg}} 
\NewDocumentCommand{\topmetric}{O{d}}{#1} 
\NewDocumentCommand{\vaoutnormal}{O{\widehat}}{#1} 
\newcommand{\category}[1]{\mathop{\mathsf{#1}}} 

\newcommand{\catpresheaf}[1]{\category{PSh}} 
\newcommand{\conti}{C} 
\newcommand{\faadj}[1]{#1^{\ast}} 
\newcommand{\faftr}[1]{\widehat{#1}} 
\newcommand{\fldcmp}{\fld{C}} 
\newcommand{\fldrat}{\fld{Q}} 
\newcommand{\fldreal}{\fld{R}} 
\newcommand{\fld}[1]{\mathbb{#1}} 
\newcommand{\fndef}[1]{\boldsymbol{1}_{#1}} 
\newcommand{\fnexp}[1]{\fun{\exp}{#1}} 
\newcommand{\fngamma}[1]{\fun{\Gamma}{#1}} 
\newcommand{\fnrestr}[2]{\left. #1 \right|_{#2}} 
\newcommand{\gtclos}[1]{\overline{#1}} 
\newcommand{\latprbk}[1]{\latp{\rbk{#1}}} 
\newcommand{\liealg}[1]{\mathfrak{#1}} 
\newcommand{\liegr}[1]{\mathrm{#1}} 
\newcommand{\lpseq}{\ell} 
\newcommand{\lp}{L} 
\newcommand{\monnat}{\mathbb{N}} 
\newcommand{\msrbb}[1]{\mathbb{#1}} 
\newcommand{\msrcal}[1]{\mathcal{#1}} 
\newcommand{\msrprb}{\mathrm{Pr}} 
\newcommand{\msr}[1]{#1} 

\newcommand{\oacommutant}[1]{#1^{\prime}} 
\newcommand{\oacstar}{C^{\ast}} 
\newcommand{\oaderiv}{\delta} 
\newcommand{\oadoublecommutant}[1]{#1^{\prime \prime}} 
\newcommand{\oa}[1]{\mathcal{#1}} 
\newcommand{\opdmsr}[1]{\mathop{d #1}} 
\newcommand{\opprojto}[1]{\sqbk{#1}} 
\newcommand{\opspecint}[1]{\mathcal{E}} 
\newcommand{\physenergyfunc}{\mathcal{E}} 
\newcommand{\physmean}[1]{\dbk{#1}} 
\newcommand{\prbexp}{\mathbb{E}} 
\newcommand{\prbvar}{\operatorname{Var}} 
\newcommand{\ringratint}{\mathbb{Z}} 
\newcommand{\semigrposint}{\monnat_1} 

\newcommand{\seq}[2]{\if\relax\detokenize{#1}\relax \rbk{#1} \else \rbk{#1}_{#2} \fi} 
\newcommand{\setcardopsharp}[1]{\operatorname{\#} #1} 
\newcommand{\setextremal}{\operatorname{ex}} 
\newcommand{\setisomorphism}[1]{\operatorname{Iso}} 
\newcommand{\setone}[1]{\cbk{#1}}
\newcommand{\setquot}[2]{\bigmiddleslash{#1}{#2}} 
\newcommand{\set}[2]{\left\{#1 \, \middle| \, #2\right\}}
\newcommand{\spfock}{\mathcal{F}} 
\newcommand{\spmat}[2]{\fun{M_{#1}}{{#2}}} 
\newcommand{\topdist}{\operatorname{dist}} 
\newcommand{\txtbsn}{\mathrm{b}} 
\newcommand{\txtgs}{\mathrm{gs}} 
\newcommand{\txtloc}{\mathrm{loc}} 
\newcommand{\txtmax}{\mathrm{max}} 
\newcommand{\txtneg}{\mathrm{-}} 
\newcommand{\txtnonneg}{\mathrm{+}} 
\newcommand{\txtoperator}{\mathrm{op}} 
\newcommand{\txttot}{\mathrm{tot}} 

\usepackage[inkscape=false,inkscapelatex=false,inkscapepath=svgpath]{svg}

\title{Notes on the Optical-Lattice Hard-Core Bose Gas of Aizenman--Lieb--Seiringer--Solovej--Yngvason\\\vspace{0.5em}{\large Reflection Positivity, Infrared Bounds, and Equilibrium-State Decompositions}}

\author{%
Yoshitsugu Sekine\\{\small\texttt{4429sekine@gmail.com}}%
}

\date{\today}

\begin{document}

\maketitle

\begin{abstract}
These are detailed expository notes on the staggered optical-lattice hard-core Bose gas of Aizenman, Lieb, Seiringer, Solovej and Yngvason. They restate and expand the arguments and their principal inputs in a common notation, and describe the central and orbitwise decompositions of thermodynamic-limit equilibrium states. The purpose is to make compressed or delegated steps explicit, not to introduce a new method, strengthen the estimates, or otherwise improve the cited results.

\end{abstract}

\setcounter{tocdepth}{3}
\tableofcontents

\section{Introduction}\label{introduction}

The optical-lattice hard-core Bose gas introduced in \cite{AizenmanLiebSeiringerSolovejYngvason001} is a rigorous model with Bose--Einstein condensed and Mott regimes. Hard-core bosons occupy the cubic lattice and are subject to a staggered one-body potential of strength \(\lambda
\geq
0\). Particle--hole symmetry centers the model at half-filling. In dimension \(d
\geq
3\), reflection positivity and an infrared bound produce off-diagonal long-range order at low temperature and small \(\lambda\). A loop representation and chessboard estimates produce exponential decay at high temperature or large \(\lambda\). Under the respective sufficient conditions, the ground-state energy has a chemical-potential gap in the Mott regime and no cusp at half-filling in the condensed regime.

These notes restate and expand the arguments of \cite{AizenmanLiebSeiringerSolovejYngvason001} and the principal references used there in a common notation. Their purpose is to make steps compressed or delegated in the original paper explicit. They do not introduce a new method, strengthen the estimates, or otherwise improve the cited results. The quasi-local formulation keeps the finite-volume and infinite-volume arguments in one setting: the UHF algebra over \(\ringratint^{d}\), its finite-range interaction, and the thermodynamic-limit dynamics. Periodic Gibbs-state limit points are then KMS states of one fixed \(\oacstar\)-dynamical system.

The equilibrium analysis separates zero-mode occupation, non-factoriality of the symmetric KMS state, and the structure of its central measure. When the explicit condensation bound \(\kappa\) is positive, the averaged annihilation operators converge in the GNS representation to a central order parameter. Its rotation-invariant law is determined by the limiting zero-mode moments and assigns mass at least \(4\kappa\) to nonzero values. The central measure disintegrates fiberwise into normalized Haar measures on gauge orbits; almost every nonzero orbit consists of mutually disjoint extremal KMS states that break gauge symmetry. The short-range estimates do not prove concentration on one orbit, nonzero order parameter in every component, or uniqueness of the KMS state in the Mott regime. The mean-field model supplies a comparison in which the equilibrium state is the Haar average of one circle of mutually disjoint factor states.

At zero temperature, the ground-state condensation hypothesis is verified at zero staggered field for \(d\geq2\) and in an explicit parameter region for \(d\geq3\). Under this hypothesis, the Koma--Tasaki estimate produces an increasing family of orthogonal charge sectors whose energies collapse onto the finite-volume ground-state energy. A sufficiently broad but slowly growing superposition of the associated ladder states converges to a symmetry-breaking infinite-volume ground state with a sharp condensate phase. Ground states with different sharp phases have non-quasi-equivalent GNS representations. This collective tower is not a spin-wave dispersion. The infrared quantity \(E_{p}\) is the lattice-Laplacian symbol in a static correlation bound rather than an excitation eigenvalue. The present estimates do not establish a momentum-resolved Nambu--Goldstone branch.

The technical development gives detailed derivations of Gaussian domination, the infrared and Falk--Bruch bounds, the loop expansion, and the chessboard estimate from the cited sources. It also records the KMS and central-decomposition results needed to pass from finite-volume estimates to equilibrium-state decompositions.

The main results are stated in Section \ref{sec:main-results}. Sections \ref{sec:algebra} and \ref{sec:dynamics} construct the quasi-local model, its dynamics, and its symmetries. Sections \ref{sec:rp}--\ref{sec:bec} derive the condensation bound, while Sections \ref{sec:loop} and \ref{sec:decay} derive exponential decay and the Mott gap. Sections \ref{sec:gap} and \ref{sec:density} treat the condensed ground-state regime and the staggered density. Section \ref{sec:decomposition} develops the central, gauge-orbit, even-translation, and condensate-density decompositions. Section \ref{sec:comparison-models} compares the non-interacting and mean-field models with the short-range system. Section \ref{sec:improvements} records the estimates and uniqueness inputs needed to sharpen the short-range conclusions. The appendices supply the finite-dimensional calculations and equilibrium-state theory used in the proofs.

Throughout in this paper, a state on an operator algebra is denoted by \(\oastate[\psi]\), and its GNS triple is written \(\rbk{\sphilb{H}_{\psi}, \oarepn_{\psi}, \oagnsvector[\Psi]}\).

\section{Main Results}\label{sec:main-results}

The results collected in these notes are stated after a common finite-volume and quasi-local setup.

\subsection{Common optical-lattice setting}\label{common-optical-lattice-setting}

The lattice geometry, observable algebras, Hamiltonians, states, and Fourier observables below are used throughout.

\subsubsection{Lattice, local algebras, and spin realization}\label{lattice-local-algebras-and-spin-realization}

For \(d
\geq
1\) and an even integer \(L
>
0\), let \begin{equation}\label{eq:main-local-algebras}
\begin{aligned}
\Lambda_{L}
&=
\set{x \in \ringratint^{d}}
{-\frac{L}{2}
<
x_{i}
\leq
\frac{L}{2},
\quad
i
=
1,
\ldots,
d},
\\ 
\oa{A}_{\Lambda}
&=
\bigotimes_{x \in \Lambda}
\spmat{2}{\fldcmp},
\quad
\Lambda
\Subset
\ringratint^{d},
\\ 
\oa{A}_{x}
&=
\oa{A}_{\setone{x}},
\quad
x
\in
\ringratint^{d},
\\ 
\oa{A}
&=
\gtclos{\oa{A}_{\txtloc}}^{\norm{\cdot}},
\quad
\oa{A}_{\txtloc}
=
\bigcup_{\Lambda \Subset \ringratint^{d}}
\oa{A}_{\Lambda}.
\end{aligned}
\end{equation} The boxes carry periodic boundary conditions. The algebra \(\oa{A}_{\txtloc}\) is called the dense algebra of local observables. For \(x
\in
\ringratint^{d}\), the algebra \(\oa{A}_{x}
\subset
\oa{A}_{\txtloc}\) is the one-site algebra at \(x\). Its elements are denoted by \(A_{x}
\in
\oa{A}_{x}\). For \(y
\in
\ringratint^{d}\), the symbol \(A_{y}
\in
\oa{A}_{y}\) denotes the image of \(A_{x}\) under the canonical identification of the tensor factors at \(x\) and \(y\). For a lattice vector \(x
=
\rbk{x_{1},\ldots,x_{d}}
\in
\ringratint^{d}\), the lattice parity is defined by \begin{equation}\label{eq:lattice-parity}
\rbk{-1}^{x}
=
\rbk{-1}^{x_{1} + \dotsb + x_{d}}.
\end{equation} The exponent \(x\) in \(\rbk{-1}^{x}\) is shorthand for the ordinary scalar sum \(x_{1} + \dotsb + x_{d}\), not for an absolute value or norm. For every finite \(\Lambda \Subset \ringratint^{d}\), the checkerboard sublattices are defined by \begin{equation}\label{eq:checkerboard-sublattices}
\Lambda_{\mathrm{A}}
=
\set{x \in \Lambda}{\rbk{-1}^{x} = 1},
\quad
\Lambda_{\mathrm{B}}
=
\set{x \in \Lambda}{\rbk{-1}^{x} = -1}.
\end{equation}

\subsubsection{One-site spin normalization and hard-core dictionary}\label{one-site-spin-normalization-and-hard-core-dictionary}

At every site, the spin operators are the copies in the corresponding tensor factor of the matrices \begin{equation}\label{eq:spin-matrices}
\begin{aligned}
S^{1}
&=
\frac{1}{2}
\begin{pmatrix}
0 & 1 \\
1 & 0
\end{pmatrix},
\quad
S^{2}
&=
\frac{1}{2}
\begin{pmatrix}
0 & -\imunit \\
\imunit & 0
\end{pmatrix},
\quad
S^{3}
&=
\frac{1}{2}
\begin{pmatrix}
1 & 0 \\
0 & -1
\end{pmatrix}.
\end{aligned}
\end{equation} Thus \(S^{i}\) is one half of the corresponding Pauli matrix; in particular, the factor \(\frac{1}{2}\) is part of the definition. For \(i,j\in\setone{1,2,3}\) and \(x,y\in\ringratint^{d}\), the commutation relations and the same-site anticommutation relations are \begin{equation}\label{eq:main-spin-relations}
\begin{aligned}
\commutator{S^{i}_{x}}{S^{j}_{y}}
&=
\imunit\delta_{xy}
\sum_{k=1}^{3}\epsilon_{ijk}S^{k}_{x},
\quad
\anticommutator{S^{i}_{x}}{S^{j}_{x}}
=
\frac{1}{2}\delta_{ij}.
\end{aligned}
\end{equation} Here \(\delta_{xy}\) is the Kronecker delta, \(\epsilon_{ijk}\) is the totally antisymmetric symbol with \(\epsilon_{123}=1\). The raising and lowering operators are \(S^{\pm}_{x}=S^{1}_{x}\pm\imunit S^{2}_{x}\). The matrices \(S^{1}\) and \(S^{3}\) are real symmetric, while \(S^{2}\) is purely imaginary and self-adjoint.

The hard-core annihilation and creation matrices are represented by the spin-\(\frac{1}{2}\) lowering and raising operators. For their copies at \(x\) we write \begin{equation}\label{eq:boson-spin-dictionary}
\begin{aligned}
a_{x}
=
S^{-}_{x},
\quad
\faadj{a_{x}}
=
S^{+}_{x},
\quad
n_{x}
=
\faadj{a_{x}} a_{x}
=
S^{3}_{x}
+
\frac{1}{2}.
\end{aligned}
\end{equation} The projection \(n_{x}\) is the local hard-core particle-number operator: its eigenvalue \(0\) means that the site is empty, and its eigenvalue \(1\) means that the site is occupied by one hard-core boson. The total particle-number operator and the total third spin component are \begin{equation}\label{eq:main-particle-number}
N_{\Lambda}
=
\sum_{x \in \Lambda} n_{x},
\quad
S^{3}_{\txttot,\Lambda}
=
\sum_{x \in \Lambda} S^{3}_{x}
=
N_{\Lambda}
-
\frac{\abscard{\Lambda}}{2}.
\end{equation} If \(N\) is an eigenvalue of \(N_{\Lambda}\), the corresponding particle density is \begin{equation}\label{eq:main-particle-density}
\varrho
=
\frac{N}{\abscard{\Lambda}}.
\end{equation}

\subsubsection{Finite-volume Fourier spin fields}\label{finite-volume-fourier-spin-fields}

For every periodic box \(\Lambda=\Lambda_{L}\), the momentum lattice and the transverse Fourier spin fields used below are \begin{equation}\label{eq:spin-wave}
\begin{aligned}
\dual{\Lambda}
&=
\set{
p\in\frac{2\pi}{L}\ringratint^{d}
}{
-\pi<p_{i}\leq\pi
},
\\ 
\widetilde{S}^{i}_{p}
&=
\frac{1}{\sqrt{\abscard{\Lambda}}}
\sum_{x\in\Lambda}
S^{i}_{x}
\napiernum^{\imunit p\cdot x},
\quad
p\in\dual{\Lambda},
\quad
i\in\setone{1,2}.
\end{aligned}
\end{equation}

\subsubsection{Finite-volume Hamiltonian, sectors, and symmetries}\label{finite-volume-hamiltonian-sectors-and-symmetries}

\begin{defn}[optical-lattice data for the main results]\label{def:main-optical-lattice-setting}
Let
$\lambda \geq 0$.
The finite-volume Hamiltonian is
\begin{equation}\label{eq:main-hamiltonian}
\begin{aligned}
\physham_{\Lambda}
&=
-\frac{1}{2}
\sum_{\dbk{xy}}
\rbk{\faadj{a_{x}} a_{y}
+a_{x} \faadj{a_{y}}}
+\lambda
\sum_{x \in \Lambda}
\rbk{\frac{1}{2}
+\rbk{-1}^{x}
\rbk{\faadj{a_{x}} a_{x}
-\frac{1}{2}}},
\end{aligned}
\end{equation}
where
$\dbk{xy}$
runs over unordered nearest-neighbor pairs.
The Hamiltonian conserves the particle number,
$\commutator{\physham_{\Lambda}}{N_{\Lambda}} = 0$.
For
$k
\in
\ringratint$
with
$0
\leq
\frac{\abscard{\Lambda}}{2} + k
\leq
\abscard{\Lambda}$,
the sector projection is
\begin{equation}\label{eq:main-sector-projection}
P_{k}
=
\opprojto{\fun{\Ker}{N_{\Lambda} - \rbk{\frac{\abscard{\Lambda}}{2} + k}}}
=
\opprojto{\fun{\Ker}{S_{\txttot,\Lambda}^{3} - k}}
\end{equation}
and the sector ground-state energy is defined by
\begin{equation}\label{eq:main-sector-energy}
\physenergyfunc_{\Lambda,k}
=
\min
\opvarspec{\fnrestr{\physham_{\Lambda}}
{\Ran P_{k}}}.
\end{equation}
The global finite-volume ground-state energy and a normalized ground-state
vector are determined by
\begin{equation}\label{eq:main-finite-volume-ground-state}
\begin{aligned}
\fun{\physenergyfunc_{\Lambda}}{\lambda}
=
\min\opvarspec{\physham_{\Lambda}},
\quad
\physham_{\Lambda}\Psi_{\txtgs,\Lambda}
=
\fun{\physenergyfunc_{\Lambda}}{\lambda}\Psi_{\txtgs,\Lambda},
\quad
\norm{\Psi_{\txtgs,\Lambda}}
=
1.
\end{aligned}
\end{equation}
For $A \in \oa{A}_{\Lambda}$, its vector state is
\begin{equation}\label{eq:main-finite-volume-ground-vector-state}
\fun{\oastate[\psi_{\txtgs,\Lambda}]}{A}
=
\bkt{\Psi_{\txtgs,\Lambda}}{A\Psi_{\txtgs,\Lambda}}.
\end{equation}
The global finite-volume ground-state energy density is defined by
\begin{equation}\label{eq:main-finite-volume-ground-state-energy-density}
\fun{e_{\Lambda}}{\lambda}
=
\frac{\fun{\physenergyfunc_{\Lambda}}{\lambda}}{\abscard{\Lambda}}.
\end{equation}
For $\varrho \in \closedinterval{0}{1}$,
the thermodynamic ground-state energy per site at particle density $\varrho$
is defined, whenever the limit exists independently of the approximating
particle numbers, by
\begin{equation}\label{eq:thermodynamic-ground-state-energy-density}
\fun{e_{\infty}}{\varrho}
=
\lim_{\Lambda \nearrow \ringratint^{d}}
\frac{
\physenergyfunc_{\Lambda,M_{\Lambda} - \frac{\abscard{\Lambda}}{2}}
}{
\abscard{\Lambda}
},
\end{equation}
where
$\net{M_{\Lambda}}{\Lambda \nearrow \ringratint^{d}}$
is any net of integers satisfying
$0 \leq M_{\Lambda} \leq \abscard{\Lambda}$
and
$M_{\Lambda} / \abscard{\Lambda} \to \varrho$
as
$\Lambda \nearrow \ringratint^{d}$.
For each box
$\Lambda
=
\Lambda_{L}$
with periodic boundary conditions
and inverse temperature
$\beta
>
0$,
the finite-volume Gibbs state on
$\oa{A}_{\Lambda}$
is
\begin{equation}\label{eq:main-periodic-gibbs-state}
\fun{\oastate[\psi_{\beta,\Lambda}]}{A}
=
\frac{\sqfun{\trace}{\napiernum^{-\beta \physham_{\Lambda}} A}}
{\sqfun{\trace}{\napiernum^{-\beta \physham_{\Lambda}}}},
\quad
A
\in
\oa{A}_{\Lambda}.
\end{equation}
The even translation group is the kernel of the lattice parity
\eqref{eq:lattice-parity}:
\begin{equation}\label{eq:even-translation-group}
\ringratint^{d}_{\mathrm{even}}
=
\set{v
\in
\ringratint^{d}}{
\rbk{-1}^{v}
=
1}.
\end{equation}
The gauge automorphisms and lattice translations are determined by
\begin{equation}\label{eq:main-symmetry-actions}
\begin{aligned}
\fun{\gamma_{\theta}}{a_{x}}
&=
\napiernum^{-\imunit \theta} a_{x},
\\ 
\fun{\eta_{v}}{A_{x}}
&=
A_{x + v},
\quad
A_{x}
\in
\oa{A}_{x},
\quad
x
\in
\ringratint^{d},
\quad
v
\in
\ringratint^{d}.
\end{aligned}
\end{equation}
The gauge automorphisms and the even translations commute with the dynamics.
When
$\lambda
\neq
0$,
odd translations interchange the checkerboard sublattices and are not
symmetries of the staggered interaction.
\end{defn}

Fix a unit lattice vector \(e_{1}\). The particle--hole automorphism \(\Theta\) of \(\oa{A}\) is determined on the local generators by \begin{equation}\label{eq:main-particle-hole-automorphism}
\begin{aligned}
\fun{\Theta}{S_x^1}
&=
S_{x+e_1}^1,
\quad
\fun{\Theta}{S_x^2}
=
-S_{x+e_1}^2,
\quad
\fun{\Theta}{S_x^3}
=
-S_{x+e_1}^3,
\\ 
\fun{\Theta}{a_x}
&=
\faadj{a_{x+e_1}},
\quad
\fun{\Theta}{n_x}
=
1-n_{x+e_1},
\quad
x\in\ringratint^d.
\end{aligned}
\end{equation} It is the composition of the one-step translation by \(e_1\) and the sitewise rotation by \(\pi\) about the first spin axis.

Let \(u_{e_{1},\Lambda}\) be the tensor-factor permutation of the periodic box characterized by \begin{equation}\label{eq:main-one-step-translation-unitary}
u_{e_{1},\Lambda} A_{x}\faadj{u_{e_{1},\Lambda}}
=
A_{x+e_{1}},
\quad
A_{x}\in\oa{A}_{x},
\quad
x\in\Lambda.
\end{equation} The finite-volume particle-hole unitary is \begin{equation}\label{eq:main-particle-hole-unitary}
u_{\mathrm{ph},\Lambda}
=
u_{e_{1},\Lambda}
\prod_{x\in\Lambda}\napiernum^{\imunit\pi S^{1}_{x}}.
\end{equation} It sends \(a_{x}\) to \(\faadj{a_{x+e_{1}}}\) and satisfies \begin{equation}\label{eq:main-particle-hole-identities}
\begin{aligned}
u_{\mathrm{ph},\Lambda}\physham_{\Lambda}\faadj{u_{\mathrm{ph},\Lambda}}
&=
\physham_{\Lambda},
\quad
u_{\mathrm{ph},\Lambda}N_{\Lambda}\faadj{u_{\mathrm{ph},\Lambda}}
=
\abscard{\Lambda}-N_{\Lambda}.
\end{aligned}
\end{equation} Thus it maps \(\Ran P_{k}\) unitarily onto \(\Ran P_{-k}\), and for every admissible \(k\), \begin{equation}\label{eq:particle-hole-sector-energy}
\physenergyfunc_{\Lambda,k}
=
\physenergyfunc_{\Lambda,-k}.
\end{equation}

\subsubsection{Equilibrium states, density matrices, and infrared constants}\label{sec:main-equilibrium-data}

The interaction defined by \eqref{eq:main-hamiltonian} generates the automorphism group \(\tau\) of Theorem \ref{thm:dynamics-existence}. Let \(K_{\beta}\) denote the compact convex set of \(\rbk{\tau,\beta}\)-KMS states. For a state \(\oastate[\psi]\), its one-particle density matrix is \begin{equation}\label{eq:main-density-matrix}
\fun{\gamma_{\psi}}{x,y}
=
\fun{\oastate[\psi]}{\faadj{a_{x}} a_{y}}.
\end{equation} For a state \(\oastate[\psi]\) on \(\oa{A}_{\Lambda}\) and \(p,q
\in
\dual{\Lambda}\), the Fourier representation of \eqref{eq:main-density-matrix} is defined by \begin{equation}\label{eq:main-fourier-density-matrix}
\fun{\faftr{\gamma_{\psi}}}{p,q}
=
\frac{1}{\abscard{\Lambda}}
\sum_{x,y
\in
\Lambda}
\napiernum^{\imunit p \cdot x-\imunit q \cdot y}
\fun{\gamma_{\psi}}{x,y}.
\end{equation} The spin-wave energy, the infrared integral, and the resulting lower bound are \begin{equation}\label{eq:main-condensation-constant}
\begin{aligned}
E_{p}
&=
\sum_{i = 1}^{d}
\rbk{1 - \cos p_{i}},
\\ 
c_{d}
&=
\frac{1}
{\rbk{2 \pi}^{d}}
\int_{\closedinterval{-\pi}{\pi}^{d}}
\frac{1}
{E_{p}}
\opdmsr{p},
\\ 
\kappa
&=
\frac{1}{2}
-\frac{1}{2}
\rbk{\frac{1}{2}
\rbk{d \rbk{d + 1}
+4 \lambda^{2}}^{\onehalf}
c_{d}}^{\onehalf}
-\frac{c_{d}}{\beta}.
\end{aligned}
\end{equation} The integral \(c_{d}\) is finite precisely in the dimensions relevant to the positive-temperature condensation theorem, namely \(d \geq 3\). The zero-temperature value of the infrared lower bound is \begin{equation}\label{eq:main-ground-state-condensation-constant}
\kappa_{\mathrm{gs}}
=
\frac{1}{2}
-
\frac{1}{2}
\rbk{
\frac{1}{2}
\rbk{
d \rbk{d + 1}
+
4 \lambda^{2}
}^{\onehalf}
c_{d}
}^{\onehalf}.
\end{equation} For \(d \geq 3\), the strict positivity of \(\kappa_{\mathrm{gs}}\) is equivalent to the explicit condition \begin{equation}\label{eq:main-ground-state-condensation-region}
\lambda^{2}
<
\frac{1}{c_{d}^{2}}
-
\frac{d \rbk{d + 1}}{4}.
\end{equation}

\subsubsection{Ground-state order-parameter input}\label{ground-state-order-parameter-input}

The ground-state vector and its vector state are defined in \eqref{eq:main-finite-volume-ground-state} and \eqref{eq:main-finite-volume-ground-vector-state}. For a constant \(\kappa_{0}
>
0\) independent of \(\Lambda\), the ground-state condensation hypothesis is \begin{equation}\label{eq:main-ground-state-condensation-hypothesis}
\fun{\oastate[\psi_{\txtgs,\Lambda}]}{
S^{-}_{\txttot,\Lambda} S^{+}_{\txttot,\Lambda}
}
\geq
\kappa_{0}
\abscard{\Lambda}^{2}.
\end{equation} Whenever \(\Psi_{\txtgs,\Lambda}\) satisfies \eqref{eq:main-ground-state-condensation-hypothesis}, define the normalized ladder states by \begin{equation}\label{eq:main-ground-state-ladder-states}
\Gamma_{\Lambda,k}
=
\begin{cases}
\displaystyle
\frac{
\rbk{S^{+}_{\txttot,\Lambda}}^{k}
\Psi_{\txtgs,\Lambda}
}{
\norm{
\rbk{S^{+}_{\txttot,\Lambda}}^{k}
\Psi_{\txtgs,\Lambda}
}
},
&
k
>
0,
\\[2ex]
\Psi_{\txtgs,\Lambda},
&
k
=
0,
\\[1ex]
\displaystyle
\frac{
\rbk{S^{-}_{\txttot,\Lambda}}^{\abs{k}}
\Psi_{\txtgs,\Lambda}
}{
\norm{
\rbk{S^{-}_{\txttot,\Lambda}}^{\abs{k}}
\Psi_{\txtgs,\Lambda}
}
},
&
k
<
0.
\end{cases}
\end{equation} The denominator is nonzero in the Koma--Tasaki box, used below. Choose a subsequence \(\seq{\Lambda_{j}}{j \in \semigrposint}\) along which every even moment appearing below converges, and define the maximal ground-state order parameter by \begin{equation}\label{eq:main-maximal-ground-state-order-parameter}
m_{\ast}
=
\lim_{r \to \infty}
\left[
\lim_{j \to \infty}
\fun{\oastate[\psi_{\txtgs,\Lambda_{j}}]}{
\rbk{
\frac{S^{1}_{\txttot,\Lambda_{j}}}{\abscard{\Lambda_{j}}}
}^{2r}
}
\right]^{1 / \rbk{2r}}.
\end{equation} For a sufficiently slowly diverging integer sequence \(M_{j}\) with \(M_{j}
\leq
c_{\mathrm{KT}}\sqrt{\abscard{\Lambda_{j}}}\), the phase-localized vector and its vector state are \begin{equation}\label{eq:main-phase-localized-ground-state}
\begin{aligned}
\Xi_{\Lambda_{j},0}
&=
\frac{1}{\sqrt{2}}
\left[
\frac{
\rbk{S^{1}_{\txttot,\Lambda_{j}}}^{M_{j}}
\Psi_{\txtgs,\Lambda_{j}}
}{
\norm{
\rbk{S^{1}_{\txttot,\Lambda_{j}}}^{M_{j}}
\Psi_{\txtgs,\Lambda_{j}}
}
}
+
\frac{
\rbk{S^{1}_{\txttot,\Lambda_{j}}}^{M_{j}+1}
\Psi_{\txtgs,\Lambda_{j}}
}{
\norm{
\rbk{S^{1}_{\txttot,\Lambda_{j}}}^{M_{j}+1}
\Psi_{\txtgs,\Lambda_{j}}
}
}
\right],
\\ 
\Xi_{\Lambda_{j},\theta}
&=
\napiernum^{-\imunit\theta S^{3}_{\txttot,\Lambda_{j}}}\Xi_{\Lambda_{j},0},
\\ 
\fun{\oastate[\xi_{\Lambda_{j},\theta}]}{A}
&=
\bkt{\Xi_{\Lambda_{j},\theta}}{
A \Xi_{\Lambda_{j},\theta}
},
\quad
A
\in
\oa{A}_{\Lambda_{j}}.
\end{aligned}
\end{equation}

\subsection{Condensed and Mott phase bounds}\label{condensed-and-mott-phase-bounds}

The eight theorem statements below collect the logically distinct condensed- and Mott-regime conclusions established in \cite{AizenmanLiebSeiringerSolovejYngvason001} and the references cited with the individual statements.

\begin{thm}[positive-temperature condensation and condensate mode]\label{thm:main-bec-condensate-mode}
Let the model be as in Definition \ref{def:main-optical-lattice-setting}.
Suppose
$d
\geq
3$
and
$\kappa
>
0$,
with
$\kappa$
defined by
\eqref{eq:main-condensation-constant}.
The finite-volume Gibbs states
$\oastate[\psi_{\beta,\Lambda}]$
on the boxes
$\Lambda
=
\Lambda_{L}$
with periodic boundary conditions,
defined by
\eqref{eq:main-periodic-gibbs-state},
satisfy
$$\liminf_{\Lambda \nearrow \ringratint^{d}}
\frac{1}{\abscard{\Lambda}^{2}}
\sum_{x,y
\in
\Lambda}
\fun{\gamma_{\psi_{\beta,\Lambda}}}{x,y}
\geq
\kappa.
$$
The one-particle density matrix has exactly one eigenvalue of order
$\abscard{\Lambda}$.
Every other eigenvalue is
$\fun{O}{\abscard{\Lambda}^{2 / d}}$,
and its normalized eigenfunction has the constant-mode overlap specified by
\eqref{eq:condensate-eigenfunction-constant-mode}.
After a volume-dependent phase choice, the eigenfunction approaches the
normalized constant vector in norm at the rate stated in
\eqref{eq:condensate-eigenfunction-norm-convergence}.
\end{thm}

The infrared lower bound is proved in Theorem \ref{thm:bec}, and the eigenvalue and overlap assertions are proved in Theorem \ref{thm:condensate-constant}. Corollary \ref{cor:condensate-asymptotically-constant} proves the norm statement after the phase choice.

\begin{thm}[zero-field ground-state condensation in two and higher dimensions; \cite{KennedyLiebShastry003}]\label{thm:main-kls-ground-state-condensation}
Let
$d
\geq
2$
and
$\lambda
=
0$.
There is a constant
$\kappa_{\mathrm{KLS}}
>
0$,
depending only on
$d$,
such that the finite-volume ground states satisfy
$$
\liminf_{\Lambda \nearrow \ringratint^{d}}
\frac{1}{\abscard{\Lambda}^{2}}
\fun{\oastate[\psi_{\txtgs,\Lambda}]}{
S^{-}_{\txttot,\Lambda} S^{+}_{\txttot,\Lambda}
}
\geq
\kappa_{\mathrm{KLS}}.
$$
Thus the ground-state condensation hypothesis
\eqref{eq:main-ground-state-condensation-hypothesis}
holds for every
$0
<
\kappa_{0}
<
\kappa_{\mathrm{KLS}}$
and all sufficiently large periodic boxes.
\end{thm}

Theorem \ref{thm:kls-ground-state-bec} proves this ground-state condensation bound.

\begin{thm}[ground-state tower]\label{thm:main-ground-state-tower}
Let the model be as in Definition \ref{def:main-optical-lattice-setting},
and suppose that the ground-state condensation hypothesis
\eqref{eq:main-ground-state-condensation-hypothesis} holds.
For
$d \geq 3$,
this hypothesis is verified for every
$\kappa_{0}
<
\kappa_{\mathrm{gs}}$
and all sufficiently large
$\Lambda$
whenever
\eqref{eq:main-ground-state-condensation-region}
holds.
For every fixed
$k
\geq
1$,
the lowest sector energies \eqref{eq:main-sector-energy} satisfy
$$\physenergyfunc_{\Lambda,k}
-
\physenergyfunc_{\Lambda,0}
\leq
\frac{c_{k}}{\abscard{\Lambda}},$$
where $c_{k} > 0$ is independent of $\Lambda$.
More precisely, there are constants
$c_{\mathrm{KT}},C_{\mathrm{KT}} > 0$,
independent of
$\Lambda$
and depending only on
$d,\lambda,\kappa_{0}$,
such that
$$\physenergyfunc_{\Lambda,k}
-
\physenergyfunc_{\Lambda,0}
\leq
C_{\mathrm{KT}} \frac{k^{2}}{\abscard{\Lambda}}
\quad
\text{for }
\abs{k}
\leq
c_{\mathrm{KT}} \sqrt{\abscard{\Lambda}}.
$$
Consequently, for every integer sequence
$K_{\Lambda}
\to
\infty$
with
$K_{\Lambda}
=
\fun{o}{\sqrt{\abscard{\Lambda}}}$,
the sectors
$\abs{k}
\leq
K_{\Lambda}$
contain
$2 K_{\Lambda} + 1$
mutually orthogonal eigenstates whose excitation energies converge uniformly
to zero.
These states form the Koma--Tasaki tower of states.
\end{thm}

The microscopic sector-energy estimate and the tower construction are proved in Theorems \ref{thm:no-gap-micro} and \ref{thm:koma-tasaki-tower}, respectively.

\begin{thm}[phase-localized ground states]\label{thm:main-phase-localized-ground-states}
Let the model be as in Definition \ref{def:main-optical-lattice-setting},
and suppose that the ground-state condensation hypothesis
\eqref{eq:main-ground-state-condensation-hypothesis} holds.
For the ladder states
\eqref{eq:main-ground-state-ladder-states},
one can choose
$M_{j} \to \infty$
sufficiently slowly in
\eqref{eq:main-phase-localized-ground-state}
so that every weak-$\ast$ limit point of
$\oastate[\xi_{\Lambda_{j},\theta}]$
is an infinite-volume ground state with a sharp nonzero condensate phase
$\napiernum^{-\imunit\theta}m_{\ast}$.
Moreover
$m_{\ast}
\geq
\sqrt{\kappa_{0}}$.
Limit states belonging to two different phases
$\theta$
and
$\theta'$
modulo
$2\pi$
have non-quasi-equivalent GNS representations.
The general construction does not prove that these ground states are
factorial;
without that additional property,
non-quasi-equivalence must not be strengthened to disjointness.
\end{thm}

Theorem \ref{thm:phase-localized-ground-states} proves the phase-localized ground-state construction.

\begin{thm}[absence of a cusp in the condensed ground-state regime]\label{thm:main-no-cusp}
Let the model be as in Definition \ref{def:main-optical-lattice-setting},
and suppose that the ground-state condensation hypothesis
\eqref{eq:main-ground-state-condensation-hypothesis} holds.
The thermodynamic ground-state energy density
$\fun{e_{\infty}}{\varrho}$,
defined by \eqref{eq:thermodynamic-ground-state-energy-density},
obeys
$$0
\leq
\fun{e_{\infty}}{\varrho}
-
\fun{e_{\infty}}{\frac{1}{2}}
\leq
C
\rbk{
\varrho
-
\frac{1}{2}
}^{2}
$$
for
$\varrho$
near
$\frac{1}{2}$.
The one-sided addition and removal chemical potentials at half-filling are
the corresponding right and left slopes of the energy density:
$$\begin{aligned}
\mu_{+}
&=
\lim_{\epsilon \downarrow 0}
\frac{
\fun{e_{\infty}}{\frac{1}{2} + \epsilon}
-
\fun{e_{\infty}}{\frac{1}{2}}
}{
\epsilon
}
=
0,
\\ 
\mu_{-}
&=
\lim_{\epsilon \downarrow 0}
\frac{
\fun{e_{\infty}}{\frac{1}{2}}
-
\fun{e_{\infty}}{\frac{1}{2} - \epsilon}
}{
\epsilon
}
=
0.
\end{aligned}$$
The energy density consequently has no cusp,
and the chemical-potential interval at half-filling collapses to
the single value $0$ in the condensed ground-state regime.
\end{thm}

Theorem \ref{thm:no-cusp} proves the energy-density estimate and its chemical-potential consequence.

\begin{thm}[exponential decay in the Mott regime]\label{thm:main-exponential-decay}
Let the model be as in Definition \ref{def:main-optical-lattice-setting}.
Let
$f
=
-\rbk{\beta \abscard{\Lambda}}^{-1}
\log
\sqfun{\trace}{
\napiernum^{-\beta \physham_{\Lambda}}
}$.
If
\begin{equation}\label{eq:main-decay-condition}
\napiernum^{-\nu}
=
\frac{d}{
\lambda
-
f
}
\rbk{
1
-
\napiernum^{
-\beta
\rbk{
\lambda
-
f
}
}
}
<
1,
\end{equation}
then every
$0
<
\xi
<
\nu$
admits a volume-independent constant
$C_{\xi}$
such that
$$\fun{\gamma_{\psi_{\beta,\Lambda}}}{x,y}
\leq
C_{\xi}
\napiernum^{
-\xi
\abs{x-y}
}.
$$
Condition
\eqref{eq:main-decay-condition}
holds in particular when
$\beta d
<
2 \log 2$
and in the large-potential region specified in Theorem
\ref{thm:exponential-decay}.
No off-diagonal long-range order occurs in this regime.
\end{thm}

Theorem \ref{thm:exponential-decay} proves the exponential-decay bound.

The particle-hole identity \eqref{eq:particle-hole-sector-energy} lets the Mott estimate be stated using the energy cost in one particle-number sector.

\begin{thm}[chemical-potential gap in the Mott regime]\label{thm:main-mott-gap}
Let the model be as in Definition \ref{def:main-optical-lattice-setting}.
With the sector energy
$\physenergyfunc_{\Lambda,k}$
defined by \eqref{eq:main-sector-energy}
and the finite-volume ground-state energy density
$\fun{e_{\Lambda}}{\lambda}$
defined by \eqref{eq:main-finite-volume-ground-state-energy-density},
the large-potential estimate is
\begin{equation}\label{eq:main-mott-sector-gap}
\physenergyfunc_{\Lambda,k}
-
\physenergyfunc_{\Lambda,0}
\geq
\rbk{
\lambda
+
\abs{
\fun{e_{\Lambda}}{\lambda}
}
-
d
}
\abs{k}.
\end{equation}
If
$\lambda
+
\abs{\fun{e_{\Lambda}}{\lambda}}
>
d$,
then the lower bound in \eqref{eq:main-mott-sector-gap} is strictly positive
for every nonzero $k$.
For sufficiently large
$\lambda$,
the lower bound stays positive uniformly in the volume
and gives the chemical-potential gap of the Mott regime.
\end{thm}

Theorem \ref{thm:mott-gap} proves this sector-energy lower bound. Proposition \ref{prop:thermodynamic-mott-energy-bound} gives its thermodynamic energy bound. Corollary \ref{cor:thermodynamic-mott-cusp} gives the Mott cusp and the chemical-potential plateau.

\begin{thm}[half-filling and staggered density]\label{thm:main-half-filling-density}
Let the model be as in Definition \ref{def:main-optical-lattice-setting}.
The finite-volume Hamiltonian has a unique ground state at half-filling, and
its global and half-filled sector energies agree:
\begin{equation}\label{eq:main-half-filling-global-sector-energy}
\fun{\physenergyfunc_{\Lambda}}{\lambda}
=
\physenergyfunc_{\Lambda,0}.
\end{equation}
For
$\lambda
\neq
0$,
the particle density is not spatially constant at positive temperature
or in the ground state.
Its non-constant part has the period of the staggered optical lattice.
\end{thm}

The unique half-filled ground state is obtained in Theorem \ref{thm:half-filling}, and Theorem \ref{thm:density-staggered} proves the staggered-density statement.

These bounds do not identify the boundary between the two parameter regions. The existing correlation inequalities do not provide the monotonicity needed to prove that the regions meet at one critical curve.

\subsection{Equilibrium states and the central decomposition}\label{equilibrium-states-and-the-central-decomposition}

The infinite-volume results pass from periodic Gibbs states to KMS limit points and then resolve their long-range order through the central decomposition and the spatial averages. For \(n \in \semigrposint\), define \begin{equation}\label{eq:periodic-box-exhaustion}
\Lambda_{n}
=
\Lambda_{2^{n}},
\quad
\Lambda_{n}
\subset
\Lambda_{n + 1},
\quad
\bigcup_{n \in \semigrposint} \Lambda_{n}
=
\ringratint^{d}.
\end{equation} Each \(\Lambda_{n}\) carries periodic boundary conditions. Because its side length \(2^{n}\) is even, the periodic identification preserves the lattice parity \eqref{eq:lattice-parity}; the even and odd sublattices of \(\Lambda_{n}\) are therefore the restrictions of those of \(\ringratint^{d}\). Fix the exhaustion \eqref{eq:periodic-box-exhaustion}. For every \(n\in\semigrposint\), choose a state \(\oastate[\widetilde{\psi}_{\beta,n}]\) on \(\oa{A}\) extending the finite-volume Gibbs state: \begin{equation}\label{eq:periodic-gibbs-state-extension}
\fun{\oastate[\widetilde{\psi}_{\beta,n}]}{A}
=
\fun{\oastate[\psi_{\beta,\Lambda_n}]}{A},
\quad
A\in\oa{A}_{\Lambda_n}.
\end{equation} Every weak-\(\ast\) limit point \(\oastate[\psi_{\beta}]\) of the extensions \eqref{eq:periodic-gibbs-state-extension} along this exhaustion belongs to \(K_{\beta}\) and is invariant under the gauge automorphisms and even translations. The same subsets \(\seq{\Lambda_{n}}{n \in \semigrposint}\) of \(\ringratint^d\), before their periodic identifications are imposed, are used for all spatial averages below. For every \(v\in\ringratint^d_{\mathrm{even}}\), they satisfy \[\lim_{n\to\infty}
\frac{\abscard{\rbk{\Lambda_n+v}\mathbin{\triangle}\Lambda_n}}
{\abscard{\Lambda_n}}
=
0.\] Thus, by Definition \ref{defn:folner-net}, they form a Følner sequence for the even translation group. Their even side lengths also allow an exact pairing in the \(e_{1}\) direction. The spatially averaged annihilation operator and spatially averaged occupation-number operator are defined by \begin{equation}\label{eq:main-spatial-averages}
\begin{aligned}
\barmean{a}_{\Lambda_n}
=
\frac{1}
{\abscard{\Lambda_n}}
\sum_{x \in \Lambda_n}
a_{x},
\quad
\barmean{n}_{\Lambda_n}
=
\frac{1}{\abscard{\Lambda_n}}
\sum_{x \in \Lambda_n}
n_{x}.
\end{aligned}
\end{equation} Thus \(\barmean{a}_{\Lambda_n}\) is the average of the local annihilation operators \(a_x\) over \(\Lambda_n\), and \(\barmean{n}_{\Lambda_n}\) is the corresponding average of the local occupation-number operators \(n_x\). For \(A
\in
\oa{A}_{\txtloc}\), the spatial average of \(A\) is \begin{equation}\label{eq:main-spatial-average}
\barmean{A}_{\Lambda_n}
=
\frac{1}{
\abscard{\Lambda_n}
}
\sum_{u
\in
\Lambda_n}
\fun{\eta_u}{A}.
\end{equation}

The decomposition results below have four levels. The central decomposition first resolves the symmetric state into extremal KMS states. The gauge action then disintegrates the central measure into Haar measures on gauge orbits. Independently, the even translations disintegrate the same central measure into ergodic invariant measures. Finally, measurable central observables such as the condensate density and the particle density give coarser push-forward decompositions. The orbit and even-translation decompositions are not generally comparable.

\begin{thm}[non-factoriality of the condensed equilibrium state]\label{thm:main-symmetric-kms-decomposition}
Suppose
$d \geq 3$
and
$\kappa > 0$.
Let
$\oastate[\psi_{\beta}]$
be any weak-$\ast$ limit point of the extensions
\eqref{eq:periodic-gibbs-state-extension}
along the exhaustion
\eqref{eq:periodic-box-exhaustion}.
This state is invariant under every gauge automorphism and every even translation:
$$\oastate[\psi_{\beta}] \circ \gamma_{\theta}
=
\oastate[\psi_{\beta}],
\quad
\oastate[\psi_{\beta}] \circ \eta_{v}
=
\oastate[\psi_{\beta}],
\quad
\theta \in \fldreal,
\quad
v \in \ringratint^{d}_{\mathrm{even}}.$$
It has off-diagonal long-range order, i.e.,
$$\liminf_{n \to \infty}
\fun{\oastate[\psi_{\beta}]}
{\faadj{\barmean{a}_{\Lambda_n}} \barmean{a}_{\Lambda_n}}
\geq
\kappa.$$
The GNS von Neumann algebra
$\oa{M}_{\psi_{\beta}}$
is not a factor,
and
$\oastate[\psi_{\beta}]$
is not extremal in
$K_{\beta}$.
Its central measure
$\mu_{\psi_{\beta}}$
is not a point mass and gives the nontrivial decomposition
$$\oastate[\psi_{\beta}]
=
\int_{\setextremal K_{\sminvtemperature}}
\oastate[\psi']
\opdmsr{
\mu_{\psi_{\beta}}(\oastate[\psi'])}.$$
Equivalently, on a standard central base $\rbk{X,\mu}$,
\begin{equation}\label{eq:main-central-direct-integral}
\begin{aligned}
\sphilb{H}_{\psi_\beta}
=
\int_X^{\oplus}
\sphilb{H}_x
\opdmsr{\mu(x)},
\quad
\oarepn_{\psi_\beta}
=
\int_X^{\oplus}
\oarepn_x
\opdmsr{\mu(x)},
\quad
\oa{Z}_{\psi_\beta}
\cong
\fun{\lp^\infty}{X,\mu}.
\end{aligned}
\end{equation}
The measure
$\mu_{\psi_{\beta}}$
is invariant under the push-forwards of the gauge automorphisms
and the even translations.
\end{thm}

Theorem \ref{thm:main-decomposition} proves the non-factoriality and the nontrivial central decomposition.

The following definition fixes the terminology used for the central decomposition. Its existence condition is established by the next theorem.

\begin{defn}[central and componentwise condensate order parameters]\label{defn:main-condensate-order-parameter}
Let $\oastate[\psi_\beta]$ have a central decomposition
$$\oastate[\psi_\beta]
=
\int_X\oastate[\psi_x]\opdmsr{\mu(x)}.$$
If the weak-operator limit
$$\wlim_{n\to\infty}
\fun{\oarepn_{\psi_\beta}}{\barmean{a}_{\Lambda_n}}
=
\barmean{a}
\in
\oa{Z}_{\psi_\beta}$$
exists, then $\barmean{a}$ is the central condensate
order-parameter operator.
Under the identification
$\oa{Z}_{\psi_\beta}\cong\fun{\lp^\infty}{X,\mu}$,
the function $\barmean{a}$ is the componentwise condensate order parameter.
Its order-parameter distribution is
$$\nu
=
\barmean{a}_{\ast}\mu.$$
\end{defn}

\begin{thm}[central order parameter distribution]\label{thm:main-componentwise-structure}
Assume the hypotheses of Theorem
\ref{thm:main-symmetric-kms-decomposition}.
Write the central decomposition over a standard probability space as
$$\oastate[\psi_{\beta}]
=
\int_{X}
\oastate[\psi_{x}]
\opdmsr{\mu(x)}.
$$
The represented averages have the weak-operator limit
$$\wlim_{n \to \infty}
\fun{\oarepn_{\psi_{\beta}}}{\barmean{a}_{\Lambda_n}}
=
\barmean{a}
\in
\oa{Z}_{\psi_{\beta}}.$$
Under the identification
$\oa{Z}_{\psi_{\beta}}
\cong
\fun{\lp^{\infty}}{X,\mu}$,
the function $\barmean{a}$ and the measure
$\nu$ are those of Definition
\ref{defn:main-condensate-order-parameter}.
The gauge action satisfies
$$V_{\theta}\barmean{a}\faadj{V_{\theta}}
=
\napiernum^{-\imunit \theta}\barmean{a},
\quad
\int_{X}
\abs{\barmean{a}(x)}^{2}
\opdmsr{\mu(x)}
\geq
\kappa.$$

The measure
$\nu$
is a rotation-invariant probability measure
on the closed disc of radius
$\frac{1}{2}$.
It is uniquely determined by the limits of all words
in
$\barmean{a}_{\Lambda_n}$
and
$\faadj{\barmean{a}_{\Lambda_n}}$.
Its second moment and its possible atom at the origin satisfy
$$\kappa
\leq
\int_{\fldcmp}
\abs{z}^{2}
\opdmsr{\nu(z)}
\leq
\frac{1}{4},
\quad
\fun{\nu}{\setone{0}}
\leq
1
-
4 \kappa.
$$
Consequently the set
$$E
=
\set{x
\in
X}{
\barmean{a}(x)
\neq
0
}
$$
has measure at least
$4 \kappa$.
For almost every
$x
\in
E$,
the extremal KMS state
$\oastate[\psi_{x}]$
breaks the gauge symmetry with trivial stabilizer.
\end{thm}

Theorem \ref{thm:componentwise-breaking} constructs the central order parameter, and Theorem \ref{thm:order-parameter-distribution} proves its distributional properties.

\begin{thm}[orbitwise phase decomposition]\label{thm:main-orbitwise-phase-decomposition}
Assume the hypotheses of Theorem
\ref{thm:main-symmetric-kms-decomposition},
and use the central decomposition and the componentwise condensate order
parameter
$\barmean{a}$
from Theorem \ref{thm:main-componentwise-structure}.
The central measure disintegrates over the gauge-orbit space
$q
\colon
X
\to
Y$
through an orbit-space probability measure
$\msr{\mu_{\mathrm{orb}}}$
as
$$\mu
=
\int_{Y}
\mu_{y}
\opdmsr{\mu_{\mathrm{orb}}(y)}.
$$
For
$\msr{\mu_{\mathrm{orb}}}$-almost every
$y$
with nonzero radial order parameter,
$\mu_{y}$
is the normalized Haar measure on a circle contained in $\setextremal K_\beta$
whose states are mutually disjoint and break the gauge symmetry.
The barycenter on such an orbit has the form
$$\oastate[\psi^{\rbk{y}}]
=
\frac{1}{
2 \pi
}
\int_{0}^{2 \pi}
\oastate[\psi'_{y}]
\circ
\gamma_{\theta}
\opdmsr{\theta}.
$$
If
$$
\sphilb{H}_{\mathrm{orb},y}
=
\int_{\fun{q^{-1}}{y}}^{\oplus}
\sphilb{H}_x
\opdmsr{\mu_y(x)},
\quad
\oarepn_{\mathrm{orb},y}
=
\int_{\fun{q^{-1}}{y}}^{\oplus}
\oarepn_x
\opdmsr{\mu_y(x)},
$$
then the central direct integral
\eqref{eq:main-central-direct-integral}
has the orbitwise form
\begin{equation}\label{eq:main-orbitwise-direct-integral}
\begin{aligned}
\sphilb{H}_{\psi_\beta}
\cong
\int_Y^{\oplus}
\sphilb{H}_{\mathrm{orb},y}
\opdmsr{\mu_{\mathrm{orb}}(y)},
\quad
\oarepn_{\psi_\beta}
\cong
\int_Y^{\oplus}
\oarepn_{\mathrm{orb},y}
\opdmsr{\mu_{\mathrm{orb}}(y)}.
\end{aligned}
\end{equation}
\end{thm}

Theorem \ref{thm:orbit-decomposition} proves this orbitwise disintegration. Equation \eqref{eq:main-orbitwise-direct-integral} follows from \eqref{eq:orbit-disintegration-family} by Fubini's theorem for direct integrals.

\begin{thm}[even-translation ergodic decomposition]\label{thm:main-even-translation-decomposition}
Assume the hypotheses of Theorem
\ref{thm:main-symmetric-kms-decomposition}.
The action
$$
\fun{T_v}{\oastate[\psi']}
=
\oastate[\psi']\circ\eta_v,
\quad
v\in\ringratint^d_{\mathrm{even}},
$$
on the central measure has an ergodic disintegration
\begin{equation}\label{eq:main-even-translation-measure-decomposition}
\mu_{\psi_\beta}
=
\int_{Y_{\mathrm{even}}}
\mu_{\mathrm{even},y}
\opdmsr{\msr{\nu}_{\mathrm{even}}(y)}.
\end{equation}
The corresponding conditional KMS states are
\begin{equation}\label{eq:main-even-translation-state-decomposition}
\begin{aligned}
\oastate[\psi_{\mathrm{even},y}]
=
\int_{\setextremal K_\beta}
\oastate[\psi']
\opdmsr{\mu_{\mathrm{even},y}(\oastate[\psi'])},
\quad
\oastate[\psi_\beta]
=
\int_{Y_{\mathrm{even}}}
\oastate[\psi_{\mathrm{even},y}]
\opdmsr{\msr{\nu}_{\mathrm{even}}(y)}.
\end{aligned}
\end{equation}
For almost every $y$, the state
$\oastate[\psi_{\mathrm{even},y}]$
is extremal among the even-translation-invariant KMS states and satisfies
mean clustering for the even translations.
The central direct integral has the iterated form
\begin{equation}\label{eq:main-even-translation-direct-integral}
\begin{aligned}
\sphilb{H}_{\psi_\beta}
\cong
\int_{Y_{\mathrm{even}}}^{\oplus}
\sphilb{H}_{\mathrm{even},y}
\opdmsr{\msr{\nu}_{\mathrm{even}}(y)},
\quad
\oarepn_{\psi_\beta}
\cong
\int_{Y_{\mathrm{even}}}^{\oplus}
\oarepn_{\mathrm{even},y}
\opdmsr{\msr{\nu}_{\mathrm{even}}(y)}.
\end{aligned}
\end{equation}
There is a measurable function
$z_{\mathrm{even}}$
such that
$$
\barmean{a}
=
z_{\mathrm{even}}\circ q_{\mathrm{even}},
\quad
\fun{\msr{\nu}_{\mathrm{even}}}{
\set{y\in Y_{\mathrm{even}}}{
\fun{z_{\mathrm{even}}}{y}\neq0
}
}
\geq
4\kappa.
$$
Every component with
$\fun{z_{\mathrm{even}}}{y}\neq0$
breaks the gauge symmetry.
\end{thm}

Proposition \ref{prop:even-translation-ergodic-decomposition} proves \eqref{eq:main-even-translation-measure-decomposition} and \eqref{eq:main-even-translation-state-decomposition}. Corollary \ref{cor:even-translation-iterated-direct-integral} proves \eqref{eq:main-even-translation-direct-integral} and mean clustering. Proposition \ref{prop:even-translation-order-parameter} proves the factorization of \(\barmean{a}\) and the weight bound. Unlike the gauge-orbit conditional measures, the measures \(\mu_{\mathrm{even},y}\) need not be supported on single translation orbits and need not be point masses.

\begin{thm}[central spatial averages]\label{thm:main-central-spatial-averages}
Assume the hypotheses of Theorem
\ref{thm:main-symmetric-kms-decomposition},
and use the central decomposition from Theorem
\ref{thm:main-componentwise-structure}.
For
$A
\in
\oa{A}_{\txtloc}$,
let
$\barmean{A}_{\Lambda_n}$
be the spatial average
\eqref{eq:main-spatial-average}.
There is a unique central element
$\widehat{A}
\in
\oa{Z}_{\psi_{\beta}}$
such that the represented averages have the weak-operator limit
$$\wlim_{n
\to
\infty}
\fun{\oarepn_{\psi_{\beta}}}{\barmean{A}_{\Lambda_n}}
=
\widehat{A}.$$
Denote the function representing
$\widehat{A}$
by
$\widehat{a}
\in
\fun{\lp^{\infty}}{X,\mu}$.

For any prescribed countable family of local observables,
there is a strictly increasing sequence
$\seq{n_{j}}{j
\in
\semigrposint}$
of positive integers
such that,
outside a single
$\mu$-null set,
the following limits hold for every observable in that family:
$$\begin{aligned}
\lim_{j \to \infty}
\fun{\oastate[\psi_{x}]}{\barmean{A}_{\Lambda_{n_j}}}
=
\fun{\widehat{a}}{x},
\quad
\lim_{j \to \infty}
\rbk{\fun{\oastate[\psi_{x}]}{\faadj{\barmean{A}_{\Lambda_{n_j}}} \barmean{A}_{\Lambda_{n_j}}}
-\abs{\fun{\oastate[\psi_{x}]}{\barmean{A}_{\Lambda_{n_j}}}}^{2}}
=
0.
\end{aligned}$$
\end{thm}

Theorem \ref{thm:macroscopic-sharpness} proves the central-limit and componentwise-sharpness statements for spatial averages.

\begin{thm}[componentwise uniform-mode condensate and density]\label{thm:main-componentwise-condensate-density}
Assume the hypotheses of Theorem
\ref{thm:main-symmetric-kms-decomposition}.
Apply Theorem \ref{thm:main-central-spatial-averages} to the family consisting of
$a_{0}$
and
$n_{0}$,
and fix the resulting common sequence.
For
$A
=
a_{0}$,
one has
$$\barmean{A}_{\Lambda_n}
=
\barmean{a}_{\Lambda_n},
\quad
\widehat{A}
=
\barmean{a},
\quad
\widehat{a}
=
\barmean{a}.$$
For
$\mu$-almost every
$x$,
the order-parameter and uniform-mode condensate-density limits are
$$\begin{aligned}
\lim_{j
\to
\infty}
\fun{\oastate[\psi_{x}]}{\barmean{a}_{\Lambda_{n_j}}}
&=
\barmean{a}(x),
\\ 
\lim_{j
\to
\infty}
\fun{\oastate[\psi_{x}]}{
\faadj{\barmean{a}_{\Lambda_{n_j}}} \barmean{a}_{\Lambda_{n_j}}
}
&=
\lim_{j
\to
\infty}
\frac{1}{
\abscard{\Lambda_{n_j}}^{2}
}
\sum_{u,v
\in
\Lambda_{n_j}}
\fun{\oastate[\psi_{x}]}{
\faadj{a_{u}} a_{v}
}
&=
\abs{\barmean{a}(x)}^{2}.
\end{aligned}$$

For
$A
=
n_{0}$,
one has
$\barmean{A}_{\Lambda_n}
=
\barmean{n}_{\Lambda_n}$.
The corresponding central element is denoted by
$\barmean{n}$.
Under the same central identification, its representing function is
$\barmean{n}$.
Along the same sequence,
the density limits for
$\mu$-almost every
$x$
are
$$\begin{aligned}
\lim_{j \to \infty}
\fun{\oastate[\psi_{x}]}{\barmean{n}_{\Lambda_{n_j}}}
=
\barmean{n}(x),
\quad
\lim_{j \to \infty}
\rbk{\fun{\oastate[\psi_{x}]}{\barmean{n}_{\Lambda_{n_j}}^{2}}
-\fun{\oastate[\psi_{x}]}{\barmean{n}_{\Lambda_{n_j}}}^{2}}
=
0.
\end{aligned}$$
\end{thm}

Corollary \ref{cor:componentwise-odlro}, derived from Theorem \ref{thm:macroscopic-sharpness}, proves these condensate and density limits.

Theorem \ref{thm:main-componentwise-condensate-density} identifies \(\abs{\barmean{a}(x)}^{2}\) as the uniform-mode condensate density of \(\oastate[\psi_{x}]\). It also identifies \(\barmean{n}(x)\) as its particle density. Its second density limit states that the density fluctuations vanish.

\begin{thm}[condensate-density disintegration]\label{thm:main-condensate-density-disintegration}
Assume the hypotheses of Theorem
\ref{thm:main-symmetric-kms-decomposition}.
Define the componentwise uniform-mode condensate density and its law by
\begin{equation}\label{eq:main-condensate-density-law}
\fun{c_{\mathrm{BEC}}}{x}
=
\abs{\barmean{a}(x)}^{2},
\quad
\msr{\nu}_{\mathrm{BEC}}
=
\rbk{c_{\mathrm{BEC}}}_*\mu.
\end{equation}
The measure $\msr{\nu}_{\mathrm{BEC}}$ is supported on
$\closedinterval{0}{\frac{1}{4}}$ and satisfies
$$
\int_{\closedinterval{0}{\frac{1}{4}}}
s
\opdmsr{\msr{\nu}_{\mathrm{BEC}}(s)}
\geq
\kappa,
\quad
\fun{\msr{\nu}_{\mathrm{BEC}}}{
\leftopeninterval{0}{\frac{1}{4}}
}
\geq
4\kappa.
$$
There are conditional central measures $\mu_s$ and conditional KMS states
\begin{equation}\label{eq:main-condensate-density-state-decomposition}
\begin{aligned}
\oastate[\psi_{\mathrm{BEC},s}]
=
\int_X
\oastate[\psi_x]
\opdmsr{\mu_s(x)},
\quad
\oastate[\psi_\beta]
=
\int_{\closedinterval{0}{\frac{1}{4}}}
\oastate[\psi_{\mathrm{BEC},s}]
\opdmsr{\msr{\nu}_{\mathrm{BEC}}(s)}.
\end{aligned}
\end{equation}
The central direct integral becomes
\begin{equation}\label{eq:main-condensate-density-direct-integral}
\begin{aligned}
\sphilb{H}_{\psi_\beta}
\cong
\int_{\closedinterval{0}{\frac{1}{4}}}^{\oplus}
\sphilb{H}_{\mathrm{BEC},s}
\opdmsr{\msr{\nu}_{\mathrm{BEC}}(s)},
\quad
\oarepn_{\psi_\beta}
\cong
\int_{\closedinterval{0}{\frac{1}{4}}}^{\oplus}
\oarepn_{\mathrm{BEC},s}
\opdmsr{\msr{\nu}_{\mathrm{BEC}}(s)}.
\end{aligned}
\end{equation}
\end{thm}

Proposition \ref{prop:condensate-density-disintegration} proves \eqref{eq:main-condensate-density-law} and \eqref{eq:main-condensate-density-state-decomposition}. Corollary \ref{cor:condensate-density-iterated-direct-integral} proves \eqref{eq:main-condensate-density-direct-integral}. The law \(\msr{\nu}_{\mathrm{BEC}}\) and the conditional KMS states are not known explicitly.

The relations among the three refinements of the central decomposition are summarized by their quotient maps. Using the same central base \(X\) for all three, they are \[
q_{\mathrm{orb}}
\colon
X
\to
Y_{\mathrm{orb}},
\quad
q_{\mathrm{even}}
\colon
X
\to
Y_{\mathrm{even}},
\quad
c_{\mathrm{BEC}}
\colon
X
\to
\closedinterval{0}{\frac{1}{4}}.
\] The central observables factor as \begin{equation}\label{eq:main-decomposition-factorizations}
\begin{aligned}
\abs{\barmean{a}}
&=
\varrho\circ q_{\mathrm{orb}},
\quad
\barmean{a}
=
z_{\mathrm{even}}\circ q_{\mathrm{even}},
\\
c_{\mathrm{BEC}}
&=
\varrho^{2}\circ q_{\mathrm{orb}}
=
\abs{z_{\mathrm{even}}}^{2}\circ q_{\mathrm{even}},
\\
\barmean{n}
&=
r_{\mathrm{orb}}\circ q_{\mathrm{orb}}
=
r_{\mathrm{even}}\circ q_{\mathrm{even}}
\end{aligned}
\end{equation} for suitable measurable functions \(\varrho\), \(z_{\mathrm{even}}\), \(r_{\mathrm{orb}}\), and \(r_{\mathrm{even}}\). The gauge-orbit factorizations follow from \eqref{eq:gauge-invariant-orbit-factorization} and \eqref{eq:condensate-density-orbit-factorization}. The even-translation factorization is \eqref{eq:even-translation-order-parameter-factorization}. The density factorizations use the two invariance statements in Proposition \ref{prop:component-density}. The first line records that gauge orbits rotate the phase of \(\barmean{a}\), whereas even translations fix it. The second line shows that the condensate-density decomposition is a common coarsening of both decompositions. The third line follows because the particle density is invariant under both actions. No factorization of \(q_{\mathrm{orb}}\) through \(q_{\mathrm{even}}\), or of \(q_{\mathrm{even}}\) through \(q_{\mathrm{orb}}\), is proved.

The short-range analysis proves the angular Haar structure orbit by orbit. It does not prove that the orbit-space measure is a point mass, that \(\fun{\mu}{E}
=
1\), or that \(\barmean{n}(x)
=
\frac{1}{2}\) almost everywhere. These assertions require radial concentration, ergodicity of the gauge action on the central measure, and decay of the truncated density correlations, respectively.

\subsection{Mean-field calibration}\label{mean-field-calibration}

The mean-field model shows that the distinctions in the preceding theorem are not artifacts of the decomposition formalism.

\begin{thm}[mean-field Haar-orbit decomposition]\label{thm:main-mean-field-calibration}
For inverse temperature
$\beta
>
2$,
the mean-field Gibbs states converge without passing to a subnet.
Their limit is
$$\oastate[\psi^{\mathrm{mf}}]
=
\frac{1}{
2 \pi
}
\int_{0}^{2 \pi}
\oastate[\psi_{\theta}]
\opdmsr{\theta},
\quad
\oastate[\psi_{\theta}]
=
\rho_{\theta}^{\otimes \infty},
$$
where
$$\rho_{\theta}
=
\frac{1}{2}
+
2 \bar{\sigma}
\rbk{
S^{1} \cos \theta
+
S^{2} \sin \theta
},
\quad
\fun{\oastate[\psi_{\theta}]}{a_{x}}
=
\bar{\sigma}
\napiernum^{-\imunit \theta}.
$$
The states
$\oastate[\psi_{\theta}]$
are mutually disjoint factor states,
their circle is one gauge orbit,
and the Haar integral defining
$\oastate[\psi^{\mathrm{mf}}]$
is its central decomposition.
The central order-parameter law is the uniform probability measure
on the circle
$\abs{z}
=
\bar{\sigma}$.
The orbit-space measure is a point mass,
the gauge action on the central measure is ergodic,
and every component has density
$\frac{1}{2}$.
As
$\beta
\uparrow
\infty$,
the radius
$\bar{\sigma}$
increases to
$\frac{1}{2}$.
\end{thm}

Lemma \ref{lem:mf-disjoint-phases} proves disjointness of the factor phases, and Theorem \ref{thm:mf-central} proves the mean-field central decomposition. Proposition \ref{prop:mf-order-parameter-law} and Corollary \ref{cor:mf-component-structure} identify its Haar-orbit law and component structure. Corollary \ref{cor:mf-optimal-constants} proves that the radius \(\frac{1}{2}\) and the second-moment bound \(\frac{1}{4}\) in Theorem \ref{thm:main-componentwise-structure} are optimal. Proposition \ref{prop:mf-staggered-ground-states} and Corollary \ref{cor:mf-ground-purity-mott-cusp} give the zero-temperature phase purity, the staggered mean-field transition, and its Mott cusp.

\section{The Quasi-Local Algebra and the Model}\label{sec:algebra}

The definitions in Main Results fix the observable algebra, finite-volume model, and basic observables. The present section records the algebraic consequences specific to the hard-core realization that are used in the subsequent arguments.

\subsection{Local and quasi-local algebras}\label{local-and-quasi-local-algebras}

The local algebras \(\oa{A}_{\Lambda}\), the dense algebra \(\oa{A}_{\txtloc}\), and the quasi-local algebra \(\oa{A}\) are defined in \eqref{eq:main-local-algebras}. For \(\Lambda
\subset
\Lambda'\), the inclusion \(\oa{A}_{\Lambda}
\subset
\oa{A}_{\Lambda'}\) is obtained by tensoring with the identity on \(\Lambda'
\setminus
\Lambda\).

The algebra \(\oa{A}\) is the UHF algebra of type \(2^{\infty}\). It is simple, unital, separable, and has a unique tracial state \cite[Section 2.6]{BratteliRobinson003}. These facts are not used quantitatively below. Simplicity does imply that every representation used later is faithful. Two local observables with disjoint supports commute: if \(\Lambda \cap \Lambda' = \emptyset\), \(A \in \oa{A}_{\Lambda}\), \(B \in \oa{A}_{\Lambda'}\), then \(\commutator{A}{B} = 0\). This trivial form of locality already implies that \(\oa{A}\) is asymptotically abelian in norm for the translation group, which is the property used for the clustering arguments of Section \ref{sec:decomposition}.

\subsection{Hard-core bosons and spin operators}\label{hard-core-bosons-and-spin-operators}

The hard-core realization and the spin normalization are fixed by \eqref{eq:spin-matrices}, \eqref{eq:main-spin-relations}, and \eqref{eq:boson-spin-dictionary}. Appendix \ref{app:spin-calculations} collects the corresponding one-site, two-site, and Fourier calculations. For distinct sites the tensor-product embedding gives \(\commutator{a_{x}}{a_{y}} = \commutator{a_{x}}{\faadj{a_{y}}} = 0\). The model therefore uses the hard-core bosonic convention rather than the fermionic CAR convention. The CAR-type relation at one site does not change this convention.

\subsection{Finite volumes and the Hamiltonian}\label{finite-volumes-and-the-hamiltonian}

The periodic boxes, their nearest-neighbor bonds, and the finite-volume Hamiltonian are defined in \eqref{eq:main-local-algebras} and \eqref{eq:main-hamiltonian}. The constant \(\lambda \frac{\abscard{\Lambda}}{2}\) built into \eqref{eq:main-hamiltonian} normalizes the staggered potential \(\lambda \rbk{-1}^{x} n_{x}\) to be nonnegative; it shifts all energies but no expectation values. Equation (1) of \cite{AizenmanLiebSeiringerSolovejYngvason001} includes the on-site repulsion \(U \sum_{x} \faadj{a_{x}} a_{x} \rbk{\faadj{a_{x}} a_{x} - 1}\). The present hard-core model corresponds to \(U = \infty\). Its local space is truncated as in \eqref{eq:hardcore-matrices}, and the repulsion term vanishes identically on that space. The dictionary \eqref{eq:boson-spin-dictionary} converts \eqref{eq:main-hamiltonian} into the spin form used for all proofs.

\begin{prop}[spin form of the Hamiltonian]\label{prop:hamiltonian-spin}
In terms of the spin operators \eqref{eq:spin-matrices},
\begin{equation}\label{eq:hamiltonian-spin}
\physham_{\Lambda} = -\sum_{\dbk{xy}} \rbk{S^{1}_{x} S^{1}_{y} + S^{2}_{x} S^{2}_{y}} + \lambda \sum_{x \in \Lambda} \sqbk{\frac{1}{2} + \rbk{-1}^{x} S^{3}_{x}}.
\end{equation}
\end{prop}

\begin{proof}
The exchange and number-operator identities
\eqref{eq:app-two-site-spin-exchange} and
\eqref{eq:app-hardcore-number-spin} convert
\eqref{eq:main-hamiltonian} into
\eqref{eq:hamiltonian-spin}.
\end{proof}

The spin form \eqref{eq:hamiltonian-spin} identifies the model with the spin-\(1/2\) XY model in a staggered magnetic field of strength \(\lambda\) \cite{MatsubaraMatsuda001}. At \(\lambda = 0\), long-range order was proved in \cite{KennedyLiebShastry003}. The periodic finite-volume derivation and its automorphism group are \begin{equation}\label{eq:periodic-finite-volume-dynamics}
\begin{aligned}
\fun{\oaderiv_{\Lambda}}{A}
&=
\lim_{t \to 0}
\frac{\fun{\tau_{\Lambda,t}}{A}-A}{t}
=
\imunit \commutator{\physham_{\Lambda}}{A},
\quad
A
\in
\oa{A}_{\Lambda},
\\ 
\tau_{\Lambda,t}
&=
\napiernum^{t \oaderiv_{\Lambda}},
\quad
t
\in
\fldreal.
\end{aligned}
\end{equation}

\subsection{Symmetries}\label{symmetries}

The gauge action, the even translations, and the particle-hole unitary are defined in \eqref{eq:main-symmetry-actions}, \eqref{eq:main-particle-hole-unitary}, and \eqref{eq:main-particle-hole-identities}. Their commutation with the Hamiltonian is the finite-volume fact used below.

The finite-volume gauge unitary and its induced automorphism are \begin{equation}\label{eq:finite-volume-gauge-unitary}
\begin{aligned}
u_{\Lambda,\theta}
&=
\napiernum^{\imunit \theta S^{3}_{\txttot,\Lambda}},
\\
\fun{\Ad_{u_{\Lambda,\theta}}}{A}
&=
u_{\Lambda,\theta}
A
\faadj{u_{\Lambda,\theta}},
\quad
A
\in
\oa{A}_{\Lambda}.
\end{aligned}
\end{equation} Here \(S^{3}_{\txttot,\Lambda}\) is the total-spin operator in \eqref{eq:main-particle-number}. The automorphism \(\Ad_{u_{\Lambda,\theta}}\) is the restriction of \(\gamma_\theta\) to \(\oa{A}_{\Lambda}\).

\begin{prop}[symmetries of the finite-volume Hamiltonian]\label{prop:finite-symmetries}
Let $\physham_{\Lambda}$ be the Hamiltonian \eqref{eq:hamiltonian-spin} on the periodic box $\Lambda$.
The following unitaries commute with $\physham_{\Lambda}$.
\begin{enumerate}
\item The gauge unitaries $u_{\Lambda,\theta}$ of
\eqref{eq:finite-volume-gauge-unitary}, which implement the restrictions of
$\gamma_\theta$ in \eqref{eq:main-symmetry-actions}.  Equivalently,
$\fun{\oaderiv_{\Lambda}}{S^{3}_{\txttot,\Lambda}}=0$.
\item The unitaries implementing the even translations $\eta_v$ of
\eqref{eq:main-symmetry-actions}, with site addition understood modulo the
periodic box.
\item The particle-hole unitary
$u_{\mathrm{ph},\Lambda}$ of \eqref{eq:main-particle-hole-unitary}, whose
action is given by \eqref{eq:main-particle-hole-identities}.
\end{enumerate}
\end{prop}

\begin{proof}
The gauge-rotation identity
\eqref{eq:app-spin-gauge-rotation}
shows that the phases cancel in each hopping term
$S^{+}_{x} S^{-}_{y} + S^{-}_{x} S^{+}_{y}$.
The potential is a function of the $S^{3}_{x}$ and commutes with $S^{3}_{\txttot,\Lambda}$.

For the even translations,
the hopping term is invariant under all translations of the periodic graph,
and the staggered field satisfies $\rbk{-1}^{x + v} = \rbk{-1}^{x}$ exactly when $\rbk{-1}^{v} = +1$.
The rotation identities
\eqref{eq:app-spin-conjugation-rotation}
show that the product in \eqref{eq:main-particle-hole-unitary} fixes
$S^{1}_{x}$
and changes the signs of
$S^{2}_{x}$
and
$S^{3}_{x}$.
It preserves $S^{1}_{x} S^{1}_{y}$.
It also preserves $S^{2}_{x} S^{2}_{y}$ because both factors change sign.
It maps $\rbk{-1}^{x} S^{3}_{x}$ to $-\rbk{-1}^{x} S^{3}_{x}$.
The subsequent odd translation by $e_{1}$ restores the sign of the staggered term because $\rbk{-1}^{x + e_{1}} = -\rbk{-1}^{x}$, while leaving the hopping term invariant.

Finally, the product in \eqref{eq:main-particle-hole-unitary} maps
$S^{3}_{\txttot,\Lambda}$ to $-S^{3}_{\txttot,\Lambda}$, and its translation factor preserves
$S^{3}_{\txttot,\Lambda}$.
Together with the particle-number formula \eqref{eq:main-particle-number}, this gives
\eqref{eq:main-particle-hole-identities}, and hence the sector exchange.
Because the particle-hole unitary preserves $\physham_{\Lambda}$, the two
restricted Hamiltonians are unitarily equivalent, which proves
\eqref{eq:particle-hole-sector-energy}.
\end{proof}

\subsection{Sector-wise uniqueness}\label{sector-wise-uniqueness}

Perron--Frobenius positivity makes the lowest-energy vector unique inside every fixed particle-number sector. This finite-volume fact will locate the global ground state at half-filling.

\begin{lem}[Perron--Frobenius in the sectors]\label{lem:perron-frobenius}
In each eigenspace of $S^{3}_{\txttot,\Lambda}$ the ground state of $\physham_{\Lambda}$ is unique, and can be chosen with strictly positive coefficients in the configuration basis.
\end{lem}

\begin{proof}
Choose $c>\norm{\physham_{\Lambda}}$.
In the configuration basis,
$-\physham_{\Lambda}+c$ has a strictly positive diagonal.
Its off-diagonal entries are nonnegative and equal to the hopping matrix
elements $\frac{1}{2}$.
The empty and completely filled sectors are one-dimensional.
In every remaining sector, nearest-neighbor hops connect any two configurations.
Indeed, adjacent transpositions along lattice paths generate every rearrangement of the particles, and the available vacancies permit these moves on the periodic graph.
The resulting configuration graph is connected.
Lemma \ref{lem:finite-perron-frobenius} therefore makes the largest
eigenvalue simple and gives a strictly positive eigenvector.
This eigenvalue corresponds to the ground state of $\physham_{\Lambda}$ in the sector.
\end{proof}

\subsection{Finite-volume ground and Gibbs states}\label{finite-volume-ground-and-gibbs-states}

The zero-field Hamiltonian, the staggered observable, and the finite-volume states are fixed here for later use. The zero-field Hamiltonian is \begin{equation}\label{eq:zero-field-hamiltonian}
\physham_{0,\Lambda}
=
-
\sum_{\dbk{xy}}
\rbk{S^{1}_{x} S^{1}_{y} + S^{2}_{x} S^{2}_{y}}.
\end{equation} The zero-field derivation, distinguished from the periodic derivation \(\oaderiv_{\Lambda}\) for the Hamiltonian containing the \(\lambda W'_{\Lambda}\) term, is \begin{equation}\label{eq:zero-field-derivation}
\fun{\oaderiv_{0,\Lambda}}{A}
=
\imunit\commutator{\physham_{0,\Lambda}}{A},
\quad
A
\in
\oa{A}_{\Lambda}.
\end{equation} It is the generator of the \(\lambda=0\) specialization of \(\tau_{\Lambda}\) in \eqref{eq:periodic-finite-volume-dynamics}. The splitting of the Hamiltonian is \begin{equation}\label{eq:zero-field-splitting}
\begin{aligned}
\physham_{\Lambda}
&=
\physham_{0,\Lambda}
+\lambda W'_{\Lambda},
\quad
W'_{\Lambda}
=
\frac{\abscard{\Lambda}}{2}
+ W_{\Lambda},
\quad
W_{\Lambda}
=
\sum_{x\in\Lambda}\rbk{-1}^{x}S^{3}_{x}.
\end{aligned}
\end{equation} The bond operator \begin{equation}\label{eq:staggered-rotation-generator}
C_{\Lambda}
=
\frac{1}{2}
\sum_{\dbk{xy}}
\rbk{-1}^{x}
\rbk{S^{1}_{x}S^{2}_{y}-S^{2}_{x}S^{1}_{y}
}
\end{equation} has the commutator relation required for the staggered-density estimate.

\begin{lem}[staggered-field commutator]\label{lem:staggered-field-commutator}
The operators in \eqref{eq:zero-field-splitting} and
\eqref{eq:staggered-rotation-generator} satisfy
\begin{equation}\label{eq:magic-commutator}
\commutator{C_{\Lambda}}{W_{\Lambda}}
=
\imunit\physham_{0,\Lambda}.
\end{equation}
\end{lem}

\begin{proof}
Only the two endpoints of a bond contribute to its commutator with
$W_{\Lambda}$.
For $\dbk{xy}$, the spin commutation relations following
\eqref{eq:spin-matrices} give
$$
\begin{aligned}
\commutator{S^{1}_{x}S^{2}_{y}}{W_{\Lambda}}
&=
-\imunit\rbk{-1}^{x}S^{2}_{x}S^{2}_{y}
+
\imunit\rbk{-1}^{y}S^{1}_{x}S^{1}_{y},
\\ 
\commutator{S^{2}_{x}S^{1}_{y}}{W_{\Lambda}}
&=
\imunit\rbk{-1}^{x}S^{1}_{x}S^{1}_{y}
-
\imunit\rbk{-1}^{y}S^{2}_{x}S^{2}_{y}.
\end{aligned}
$$
Subtracting the second identity from the first and multiplying by
$\frac{1}{2}\rbk{-1}^{x}$ yields
$$\frac{1}{2}\rbk{-1}^{x}
\rbk{\commutator{S^{1}_{x}S^{2}_{y}}{W_{\Lambda}}
-\commutator{S^{2}_{x}S^{1}_{y}}{W_{\Lambda}}}
=
-\imunit\rbk{S^{1}_{x}S^{1}_{y}+S^{2}_{x}S^{2}_{y}},$$
because $\rbk{-1}^{x}\rbk{-1}^{y}=-1$ on every nearest-neighbor bond.
Summing this identity over $\dbk{xy}$ and using
\eqref{eq:zero-field-splitting} at $\lambda=0$ proves
\eqref{eq:magic-commutator}.
\end{proof}

The global finite-volume ground-state data are defined in \eqref{eq:main-finite-volume-ground-state} and \eqref{eq:main-finite-volume-ground-vector-state}. Because \eqref{eq:zero-field-hamiltonian} is the \(\lambda=0\) case of \eqref{eq:main-hamiltonian} after the spin dictionary \eqref{eq:boson-spin-dictionary}, its ground-state energy density is \(\fun{e_{\Lambda}}{0}\) from \eqref{eq:main-finite-volume-ground-state-energy-density}. For \(\beta>0\) and \(A\in\oa{A}_{\Lambda}\), the zero-field Gibbs state and its energy density are \begin{equation}\label{eq:zero-field-gibbs-state}
\begin{aligned}
\fun{\oastate[\psi_{0,\beta,\Lambda}]}{A}
=
\frac{\sqfun{\trace}{\napiernum^{-\beta\physham_{0,\Lambda}}A}}
{\sqfun{\trace}{\napiernum^{-\beta\physham_{0,\Lambda}}}},
\quad
\fun{e_{0,\Lambda}}{\beta}
=
\frac{\fun{\oastate[\psi_{0,\beta,\Lambda}]}{\physham_{0,\Lambda}}}
{\abscard{\Lambda}}.
\end{aligned}
\end{equation} For every \(\beta>0\), the state \eqref{eq:zero-field-gibbs-state} is the unique \(\rbk{\tau_{\Lambda},\beta}\)-KMS state for the \(\lambda=0\) specialization of \eqref{eq:periodic-finite-volume-dynamics} by Proposition \ref{prop:gibbs-kms}. The finite-volume free energy at inverse temperature \(\beta>0\) is \begin{equation}\label{eq:finite-volume-free-energy}
\fun{F_{\Lambda}}{\lambda,\beta}
=
-\frac{1}{\sminvtemperature}
\log\sqfun{\trace}{\napiernum^{-\beta\physham_{\Lambda}}}.
\end{equation} For every \(\beta>0\), the periodic Gibbs state \eqref{eq:main-periodic-gibbs-state} is the unique \(\rbk{\tau_{\Lambda},\beta}\)-KMS state for the periodic finite-volume dynamics \eqref{eq:periodic-finite-volume-dynamics} on \(\oa{A}_{\Lambda}\) by Proposition \ref{prop:gibbs-kms}. This is a finite-dimensional KMS fact; the model-dependent information is the Hamiltonian in \eqref{eq:hamiltonian-spin}.

The rotation factor in the particle-hole unitary \eqref{eq:main-particle-hole-unitary} is the global \(\pi\)-rotation about the first spin axis, \begin{equation}\label{eq:zero-field-first-axis-rotation}
w_{\Lambda}
=
\prod_{x \in \Lambda}
\napiernum^{\imunit \pi S^{1}_{x}}.
\end{equation} It is a symmetry of the zero-field Hamiltonian and reverses the staggered spin observable \(W_{\Lambda}\) defined in \eqref{eq:zero-field-splitting}: \begin{equation}\label{eq:zero-field-rotation-symmetry}
\begin{aligned}
w_{\Lambda}\physham_{0,\Lambda}\faadj{w_{\Lambda}}
=
\physham_{0,\Lambda},
\quad
w_{\Lambda}W_{\Lambda}\faadj{w_{\Lambda}}
=
-W_{\Lambda}.
\end{aligned}
\end{equation}

\begin{lem}[zero-field Gibbs identities]\label{lem:zero-field-gibbs-identities}
For $\beta>0$ and $A\in\oa{A}_{\Lambda}$,
\begin{equation}\label{eq:zero-field-gibbs-identities}
\begin{aligned}
\fun{\oastate[\psi_{0,\beta,\Lambda}]}{\physham_{0,\Lambda}}
=
\fun{e_{0,\Lambda}}{\beta}\abscard{\Lambda},
\quad
\fun{\oastate[\psi_{0,\beta,\Lambda}]}{W_{\Lambda}}
=
0,
\quad
\fun{\oastate[\psi_{0,\beta,\Lambda}]}
{\fun{\oaderiv_{0,\Lambda}}{A}}
=
0.
\end{aligned}
\end{equation}
\end{lem}

\begin{proof}
The first identity is the definition in \eqref{eq:zero-field-gibbs-state}.
Equation \eqref{eq:zero-field-rotation-symmetry}, cyclicity of the trace,
and \eqref{eq:zero-field-gibbs-state} give
$$\fun{\oastate[\psi_{0,\beta,\Lambda}]}{W_{\Lambda}}
=
\fun{\oastate[\psi_{0,\beta,\Lambda}]}
{w_{\Lambda}W_{\Lambda}\faadj{w_{\Lambda}}}
=
-\fun{\oastate[\psi_{0,\beta,\Lambda}]}{W_{\Lambda}}.$$
The second identity follows.
The final identity is the generator-invariance statement of
Proposition \ref{prop:kms-generator-invariance} for the KMS state
\eqref{eq:zero-field-gibbs-state} and the dynamics
\eqref{eq:periodic-finite-volume-dynamics} at $\lambda=0$;
the generator is \eqref{eq:zero-field-derivation}.
\end{proof}

\begin{cor}[simple zero-field ground-state identities]\label{cor:zero-field-ground-identities}
Suppose that the zero-field ground state is simple.
Choose its normalized vector as $\Psi_{0,\txtgs,\Lambda}$ and its
vector state as $\oastate[\psi_{0,\txtgs,\Lambda}]$.
For every $A\in\oa{A}_{\Lambda}$,
\begin{equation}\label{eq:zero-field-ground-identities}
\begin{aligned}
\fun{\oastate[\psi_{0,\txtgs,\Lambda}]}{\physham_{0,\Lambda}}
=
\fun{\physenergyfunc_{\Lambda}}{0},
\quad
\fun{\oastate[\psi_{0,\txtgs,\Lambda}]}{W_{\Lambda}}
=
0,
\quad
\fun{\oastate[\psi_{0,\txtgs,\Lambda}]}
{\fun{\oaderiv_{0,\Lambda}}{A}}
=
0.
\end{aligned}
\end{equation}
\end{cor}

\begin{proof}
Equation \eqref{eq:zero-field-rotation-symmetry} maps the unique normalized
ground-state vector to a scalar multiple of itself.
The second identity follows from the second equality in
\eqref{eq:zero-field-rotation-symmetry}.
The first identity is the ground-state equation in
\eqref{eq:main-finite-volume-ground-state} at $\lambda=0$.
The same ground-state equation makes the vector state invariant under the
$\lambda=0$ specialization of $\tau_{\Lambda}$ in
\eqref{eq:periodic-finite-volume-dynamics}.
Its generator is \eqref{eq:zero-field-derivation}, so differentiating this
invariance at $t=0$ gives the final identity.
\end{proof}

The reflection-positivity argument uses a unitary transformation of \eqref{eq:hamiltonian-spin}. It removes the alternating sign from the potential and reverses the sign of the \(S^{2}S^{2}\) coupling.

\begin{lem}[sublattice rotation]\label{lem:sublattice-rotation}
The sublattice rotation is the unitary
\begin{equation}\label{eq:sublattice-rotation-unitary}
V_{\Lambda}
=
\prod_{x \in \Lambda_{\mathrm{B}}}
\napiernum^{\imunit \pi S^{1}_{x}}.
\end{equation}
The transformed Hamiltonian is
\begin{equation}\label{eq:sublattice-rotated-hamiltonian}
\begin{aligned}
V_{\Lambda} \physham_{\Lambda} \faadj{V_{\Lambda}}
&=
-\sum_{\dbk{xy}}
\rbk{S^{1}_{x} S^{1}_{y} - S^{2}_{x} S^{2}_{y}}
\\
&\mathrel{}+
\lambda
\sum_{x \in \Lambda}
\sqbk{\frac{1}{2} + S^{3}_{x}}.
\end{aligned}
\end{equation}
\end{lem}

\begin{proof}
The rotation identities
\eqref{eq:app-spin-conjugation-rotation}
fix
$S^{1}$
and change the signs of
$S^{2}$
and
$S^{3}$.
Every nearest-neighbor bond connects $\Lambda_{\mathrm{A}}$ with
$\Lambda_{\mathrm{B}}$.
Exactly one factor in $S^{2}_{x}S^{2}_{y}$ changes sign.
The product $S^{1}_{x}S^{1}_{y}$ remains unchanged.
On $\Lambda_{\mathrm{B}}$ the potential satisfies $\rbk{-1}^{x} \rbk{-S^{3}_{x}} = S^{3}_{x}$, and on $\Lambda_{\mathrm{A}}$ it is already $S^{3}_{x}$.
\end{proof}

\section{Infinite-Volume Dynamics}\label{sec:dynamics}

The Hamiltonians \eqref{eq:hamiltonian-spin} define local dynamics on the local algebras, and the finite range of the interaction makes these converge to a strongly continuous automorphism group of \(\oa{A}\). This construction is a special case of \cite[Theorem 6.2.4]{BratteliRobinson004}. We include the proof because its convergence estimate is also needed for the KMS property of thermodynamic-limit states in Appendix \ref{app:kms}.

\subsection{The interaction}\label{the-interaction}

The periodic Hamiltonian \eqref{eq:hamiltonian-spin} differs from the restriction of a translation-covariant interaction on \(\ringratint^{d}\) only by the boundary-wrapping bonds; both descriptions are recorded here.

\begin{defn}[interaction]\label{def:interaction}
The interaction $\Phi$ assigns to each finite $X \subset \ringratint^{d}$ the self-adjoint element
$$\fun{\Phi}{X} =
\begin{cases}
-\rbk{S^{1}_{x} S^{1}_{y} + S^{2}_{x} S^{2}_{y}} & X = \setone{x, y}, \ \abs{x - y} = 1, \\
\lambda \sqbk{\frac{1}{2} + \rbk{-1}^{x} S^{3}_{x}} & X = \setone{x}, \\
0 & \text{otherwise},
\end{cases}$$
of $\oa{A}_{X}$.
For a finite $\Lambda \subset \ringratint^{d}$, the free-boundary
Hamiltonian is
$$\physham_{\Lambda}^{\mathrm{free}}
=
\sum_{X \subset \Lambda}
\fun{\Phi}{X}.$$
For a periodic box $\Lambda_{L}$,
the notation $\physham_{\Lambda}$ denotes the periodic Hamiltonian \eqref{eq:hamiltonian-spin}.
\end{defn}

The interaction has range \(1\) and the uniform bound \(\norm{\fun{\Phi}{X}} \leq \frac{1}{2} + \lambda\) for all \(X\); each site \(x\) meets at most \(2d + 1\) sets \(X\) with \(\fun{\Phi}{X} \neq 0\). The periodic and free-boundary Hamiltonians differ by the \(O\rbk{L^{d - 1}}\) boundary-wrapping bonds, which is irrelevant for the norm limits below because every fixed local observable is eventually far from those bonds.

\subsection{Existence of the dynamics}\label{existence-of-the-dynamics}

Finite-range commutator estimates make the local Heisenberg evolutions Cauchy in norm on every compact time interval. Their limit is independent of the exhaustion and defines the strongly continuous infinite-volume dynamics. The periodic finite-volume derivation and dynamics are those in \eqref{eq:periodic-finite-volume-dynamics}. The free-boundary Hamiltonian \(\physham_{\Lambda}^{\mathrm{free}}\) instead defines \begin{equation}\label{eq:free-boundary-finite-volume-dynamics}
\begin{aligned}
\fun{\oaderiv_{\Lambda}^{\mathrm{free}}}{A}
&=
\lim_{t \to 0}
\frac{\fun{\tau_{\Lambda,t}^{\mathrm{free}}}{A}-A}{t}
=
\imunit \commutator{\physham_{\Lambda}^{\mathrm{free}}}{A},
\quad
A
\in
\oa{A}_{\Lambda},
\\
\fun{\tau_{\Lambda,t}^{\mathrm{free}}}{A}
&=
\napiernum^{\imunit t \physham_{\Lambda}^{\mathrm{free}}}
A
\napiernum^{-\imunit t \physham_{\Lambda}^{\mathrm{free}}},
\quad
t
\in
\fldreal.
\end{aligned}
\end{equation} For either boundary condition the corresponding finite-volume evolution has the norm-convergent expansion \[\fun{\tau_{\Lambda,t}}{A}
=
\sum_{n \geq 0}
\frac{t^{n}}{n!}
\oaderiv_{\Lambda}^{n}(A)\] or, for free boundary, \[\fun{\tau_{\Lambda,t}^{\mathrm{free}}}{A}
=
\sum_{n \geq 0}
\frac{t^{n}}{n!}
\fun{\rbk{\oaderiv_{\Lambda}^{\mathrm{free}}}^{n}}{A}.\]

\begin{lem}[derivation estimate]\label{lem:commutator-estimate}
Let $A \in \oa{A}_{\Lambda_{0}}$ with $\Lambda_{0}$ finite.
The constants used in the derivation bound are
\begin{equation}\label{eq:commutator-bound-constants}
a
=
2 \rbk{\frac{1}{2} + \lambda},
\quad
b
=
2d + 1.
\end{equation}
For every finite $\Lambda \supset \Lambda_{0}$ and every $n \geq 0$,
\begin{equation}\label{eq:commutator-bound}
\norm{\oaderiv_{\Lambda}^{n}(A)}
\leq
\norm{A} a^{n} b^{n}
\prod_{j=0}^{n-1} \rbk{\abscard{\Lambda_{0}} + j}.
\end{equation}
The same bound holds after replacing $\oaderiv_{\Lambda}^{n}$ by
$\rbk{\oaderiv_{\Lambda}^{\mathrm{free}}}^{n}$.
Moreover, if the distance from $\Lambda_{0}$ to the complement of $\Lambda$ exceeds $n$, then
$$\oaderiv_{\Lambda}^{n}(A)
=
\fun{\rbk{\oaderiv_{\Lambda}^{\mathrm{free}}}^{n}}{A},$$
and this common value does not depend on $\Lambda$.
\end{lem}

\begin{proof}
Expanding the periodic Hamiltonian into interaction terms gives
$$\oaderiv_{\Lambda}^{n}(A) = \imunit^{n} \sum_{X_{1}, \dotsc, X_{n}} \commutator{\fun{\Phi}{X_{n}}}{\commutator{\dotsb}{\commutator{\fun{\Phi}{X_{1}}}{A}}}.$$
Replacing the periodic interaction by the free-boundary interaction gives the
same expansion for
$\fun{\rbk{\oaderiv_{\Lambda}^{\mathrm{free}}}^{n}}{A}$.
A term vanishes unless $X_{1}$ intersects $\Lambda_{0}$.
At the $j$th step it also vanishes unless $X_{j}$ intersects
$\Lambda_{j-1}
=\Lambda_{0}\cup X_{1}\cup\dotsb\cup X_{j-1}$,
because operators with disjoint supports commute.
Fix $\Lambda_{j - 1}$.
Each of its sites meets at most $b$ interaction sets.
The number of admissible choices for $X_{j}$ is at most
$b \abscard{\Lambda_{j - 1}}$.
Every $X_{j}$ has at most $2$ elements,
and each union adds at most one new site.
The resulting size bound is
$\abscard{\Lambda_{j - 1}} \leq \abscard{\Lambda_{0}} + j - 1$.
The required bound is
$b \abscard{\Lambda_{j - 1}} \leq b \rbk{\abscard{\Lambda_{0}} + j - 1}$.
For every
$B
\in
\oa{A}_{\Lambda}$,
the single-commutator estimate is
$$
\norm{\commutator{\fun{\Phi}{X}}{B}}
\leq
2
\norm{\fun{\Phi}{X}}
\norm{B}
\leq
a
\norm{B}.
$$
Applying this estimate successively to the nested commutator multiplies the
norm of the inner operator by at most
$a$
at each step.
Multiplying the bounds proves \eqref{eq:commutator-bound}.
For the last statement,
every non-vanishing term connects $X_{j}$ to $\Lambda_{0}$ through overlapping range-$1$ sets.
The set $X_{j}$ lies within distance $j$ of $\Lambda_{0}$.
If $\fun{\topdist}{\Lambda_{0}, \Lambda^{c}} > n$,
every interaction term lies in the bulk.
The periodic and free-boundary interactions coincide there,
and no term meets the boundary.
\end{proof}

\begin{thm}[existence of the dynamics]\label{thm:dynamics-existence}
For every $A \in \oa{A}_{\txtloc}$ and every $t \in \fldreal$, the norm limits
\begin{equation}\label{eq:finite-volume-dynamics-limit}
\fun{\tau_{t}}{A}
=
\lim_{\Lambda \nearrow \ringratint^{d}}
\fun{\tau_{\Lambda,t}}{A}
=
\lim_{\Lambda \nearrow \ringratint^{d}}
\fun{\tau_{\Lambda,t}^{\mathrm{free}}}{A}
\end{equation}
exist along every increasing exhaustion of $\ringratint^{d}$.
The common limit is independent of the exhaustion and the boundary condition.
It extends to a strongly continuous one-parameter group $\tau$ of $\ast$-automorphisms of $\oa{A}$.
The convergence is uniform for $t$ in compact subsets of $\fldreal$.
\end{thm}

\begin{proof}
Fix $A \in \oa{A}_{\Lambda_{0}}$ and set
$T_{0} = \rbk{2 a b}^{-1}$ with $a$ and $b$ defined in
\eqref{eq:commutator-bound-constants}.
By \eqref{eq:commutator-bound} the series $\sum_{n} \frac{\abs{t}^{n}}{n!} \norm{\oaderiv_{\Lambda}^{n}(A)}$ is dominated by
$$\norm{A} \sum_{n \geq 0} \rbk{a b \abs{t}}^{n} \binom{\abscard{\Lambda_{0}} + n - 1}{n},$$
which converges for $\abs{t} \leq T_{0}$ uniformly in $\Lambda$, with a tail estimate depending only on $\abscard{\Lambda_{0}}$ and $\norm{A}$; indeed the ratio of consecutive terms tends to $a b \abs{t} \leq 1/2$.
Assume $\abs{t} \leq T_{0}$,
$\Lambda \subset \Lambda'$,
and $\fun{\topdist}{\Lambda_{0}, \Lambda^{c}} > N$.
Lemma \ref{lem:commutator-estimate} identifies the terms with $n \leq N$ in
the periodic and free-boundary series, as well as in the two periodic series
for $\fun{\tau_{\Lambda,t}}{A}$ and $\fun{\tau_{\Lambda',t}}{A}$.
The remaining tails satisfy
$$\norm{\fun{\tau_{\Lambda,t}}{A} - \fun{\tau_{\Lambda',t}}{A}} \leq 2 \norm{A} \sum_{n > N} \rbk{\frac{1}{2}}^{n} \binom{\abscard{\Lambda_{0}} + n - 1}{n},$$
where the right side is twice the tail of the convergent dominating series evaluated at $a b \abs{t} \leq 1/2$ and tends to $0$ as $N \to \infty$.
The periodic family $\rbk{\fun{\tau_{\Lambda,t}}{A}}_{\Lambda}$ is uniformly Cauchy for $\abs{t} \leq T_{0}$.
The same termwise identification shows that the free-boundary family
$\rbk{\fun{\tau_{\Lambda,t}^{\mathrm{free}}}{A}}_{\Lambda}$ has the same
limit.
Thus the limit $\fun{\tau_{t}}{A}$ is independent of the exhaustion and
boundary condition.
Isometry and multiplicativity pass from $\tau_{\Lambda,t}$ to the limit on $\oa{A}_{\txtloc}$.
Continuity extends the limit map to $\oa{A}$.

The finite-volume group law and uniform convergence imply
$\tau_{t} \tau_{s} = \tau_{t + s}$
when $\abs{t}$,
$\abs{s}$,
and $\abs{t + s}$ do not exceed $T_{0}$.
The required comparison is
$$\norm{\fun{\tau_{t} \tau_{s}}{A} - \fun{\tau_{\Lambda,t} \tau_{\Lambda,s}}{A}}
\leq \norm{\fun{\tau_{t}}{\fun{\tau_{s}}{A}} - \fun{\tau_{\Lambda,t}}{\fun{\tau_{s}}{A}}} + \norm{\fun{\tau_{s}}{A} - \fun{\tau_{\Lambda,s}}{A}} \to 0,$$
The first term is controlled by a diagonal approximation because $\fun{\tau_{s}}{A}$ is a norm limit of local elements.

For arbitrary $t$ write $t = m t'$ with $\abs{t'} \leq T_{0}$ and define $\tau_{t} = \rbk{\tau_{t'}}^{m}$.
The identity $\tau_{\Lambda,t} = \rbk{\tau_{\Lambda,t'}}^{m}$ reduces convergence at time $t$ to $m$ applications of the compact-time convergence.
Those applications are uniform on the norm-compact set of intermediate iterates.
The limit remains compatible with the local dynamics for every $t$.
On compact time intervals,
$t\mapsto\fun{\tau_{t}}{A}$ is a uniform limit of norm-continuous maps.
It is therefore norm continuous for local $A$.
An $\epsilon/3$ argument extends strong continuity from
$\oa{A}_{\txtloc}$ to $\oa{A}$.
\end{proof}

The same series argument shows that every local element is analytic for \(\tau\) in the disc \(\abs{z} < T_{0}\), with \begin{equation}\label{eq:analytic-extension}
\fun{\tau_{z}}{A}
=
\sum_{n \geq 0} \frac{z^{n}}{n!} \oaderiv^{n}(A),
\quad
\oaderiv^{n}(A)
=
\lim_{\Lambda}
\oaderiv_{\Lambda}^{n}(A),
\end{equation} and \(\fun{\tau_{\Lambda,z}}{A} \to \fun{\tau_{z}}{A}\) uniformly on compact subsets of the disc. Gaussian smoothing produces entire analytic elements from arbitrary \(A \in \oa{A}\): \[A_{\epsilon}
=
\sqrt{\frac{\epsilon}{\pi}}
\int_{\fldreal}
\napiernum^{-\epsilon t^{2}}
\fun{\tau_{t}}{A}
\opdmsr{t}.\] These elements form a norm-dense \(\tau\)-invariant \(\ast\)-subalgebra \cite[Proposition 2.5.22]{BratteliRobinson003}.

\subsection{Symmetries as automorphisms}\label{symmetries-as-automorphisms}

The finite-volume symmetries of Proposition \ref{prop:finite-symmetries} survive the thermodynamic limit. The gauge automorphisms \(\gamma_{\theta}\) are defined on the generators in \eqref{eq:main-symmetry-actions}. The finite-volume gauge unitaries \(u_{\Lambda,\theta}\) of \eqref{eq:finite-volume-gauge-unitary} give the implementation \begin{equation}\label{eq:gauge-automorphism}
\fun{\gamma_{\theta}}{A}
=
\lim_{\Lambda}
\fun{\Ad_{u_{\Lambda,\theta}}}{A},
\end{equation} where for \(A
\in \oa{A}_{\txtloc}\) the conjugation is eventually independent of \(\Lambda\) because the \(S^{3}_{x}\) with \(x\) outside the support of \(A\) commute with \(A\). The even translation group \(\ringratint^{d}_{\mathrm{even}}\) and its action \(\eta\) are defined in \eqref{eq:even-translation-group} and \eqref{eq:main-symmetry-actions}.

\begin{prop}[commutation of symmetries and dynamics]\label{prop:symmetry-dynamics}
For all $\theta$, all $v \in \ringratint^{d}_{\mathrm{even}}$, and all $t \in \fldreal$,
it holds that
$$\gamma_{\theta} \circ \tau_{t} = \tau_{t} \circ \gamma_{\theta},
\quad
\eta_{v} \circ \tau_{t} = \tau_{t} \circ \eta_{v}.$$
\end{prop}

\begin{proof}
For local $A$ and periodic boxes $\Lambda$,
the finite-volume gauge automorphism
$\Ad_{u_{\Lambda,\theta}}$ of \eqref{eq:finite-volume-gauge-unitary}
commutes with $\tau_{\Lambda,t}$ by Proposition
\ref{prop:finite-symmetries}.
The same argument applies to even translations of the periodic graph.
For a translated observable, the finite-volume exhaustion is shifted.
Theorem \ref{thm:dynamics-existence} makes the norm limit independent of
this shift.
Both sides therefore converge to the corresponding infinite-volume
expressions.
Continuity extends the identity from $\oa{A}_{\txtloc}$ to $\oa{A}$.
\end{proof}

The group \(\ringratint^{d}_{\mathrm{even}}\) has index \(2\) in \(\ringratint^{d}\) and is isomorphic to \(\ringratint^{d}\). The algebra \(\oa{A}\) is asymptotically abelian for this action. For local \(A,B\), the commutator \(\commutator{\fun{\eta_{v}}{A}}{B}\) vanishes for all but finitely many \(v \in \ringratint^{d}_{\mathrm{even}}\). The general KMS and central-decomposition facts used below are collected in Appendix \ref{app:kms}.

\subsection{Periodic Gibbs-state limit points}\label{periodic-gibbs-state-limit-points}

The periodic Gibbs states in \eqref{eq:main-periodic-gibbs-state} have weak-\(\ast\) limit points along the exhaustion \eqref{eq:periodic-box-exhaustion}. The KMS conclusion is an application of the general limit theorem in Appendix \ref{app:kms}; gauge and even-translation invariance use the symmetries established above.

\begin{prop}[the symmetric equilibrium state]\label{prop:limit-state}
Fix $\beta > 0$ and $\lambda \geq 0$,
and let
$\seq{\Lambda_{n}}{n \in \semigrposint}$
be the periodic-box exhaustion
\eqref{eq:periodic-box-exhaustion}.
Every weak-$\ast$ limit point $\oastate[\psi_{\beta}]$ of the extensions
\eqref{eq:periodic-gibbs-state-extension} along this exhaustion satisfies
$\oastate[\psi_{\beta}]\in K_{\beta}$.
For every $\theta\in\fldreal$,
$\oastate[\psi_{\beta}]\circ\gamma_\theta=\oastate[\psi_{\beta}]$.
For every $v\in\ringratint^{d}_{\mathrm{even}}$,
$\oastate[\psi_{\beta}]\circ\eta_v=\oastate[\psi_{\beta}]$.
\end{prop}

\begin{proof}
Fix $A,B\in\oa{A}_{\txtloc}$.
For all sufficiently large $n$, both observables belong to
$\oa{A}_{\Lambda_n}$.
Proposition \ref{prop:gibbs-kms} supplies the finite-volume KMS functions
$F_n$ and the strip bound
$$\abs{\fun{F_n}{z}}
\leq
\norm{A}\norm{B}.$$
For every compact interval $J\subset\fldreal$,
Theorem \ref{thm:dynamics-existence} gives
$$\sup_{t\in J}
\norm{\fun{\tau_{\Lambda_n,t}}{B}-\fun{\tau_t}{B}}
\to
0.$$
The map $t\mapsto A\fun{\tau_t}{B}$ has norm-compact image on $J$.
Weak-$\ast$ convergence is uniform on that compact set.
It follows that
$$\sup_{t\in J}
\abs{
\fun{\oastate[\widetilde{\psi}_{\beta,n}]}{
A\fun{\tau_{\Lambda_n,t}}{B}}
-
\fun{\oastate[\psi_\beta]}{
A\fun{\tau_t}{B}}
}
\to
0.$$
The reversed product has the same compactness property, and gives
$$\sup_{t\in J}
\abs{
\fun{\oastate[\widetilde{\psi}_{\beta,n}]}{
\fun{\tau_{\Lambda_n,t}}{B}A}
-
\fun{\oastate[\psi_\beta]}{
\fun{\tau_t}{B}A}
}
\to
0.$$
Theorem \ref{thm:limit-kms} therefore gives
$\oastate[\psi_\beta]\in K_\beta$.

Proposition \ref{prop:finite-symmetries} gives finite-volume gauge
invariance.
For a fixed local observable, the finite-volume conjugation agrees with
$\gamma_{\theta}$ once the box contains its support.
The identity
$$\fun{\oastate[\widetilde{\psi}_{\beta,n}]}{\fun{\gamma_{\theta}}{A}}
=
\fun{\oastate[\widetilde{\psi}_{\beta,n}]}{A}$$
therefore passes to the limit.

Each finite-volume Gibbs state is invariant under even translations of its
periodic graph.
Fix a local $A$ and $v\in\ringratint^{d}_{\mathrm{even}}$.
For all sufficiently large $n$, the periodic translate has no wrap-around
and equals $\fun{\eta_{v}}{A}$.
Translation invariance also passes to the limit.
\end{proof}

\section{Reflection Positivity and the Infrared Bound}\label{sec:rp}

Condensation follows by combining the infrared upper bound for the Duhamel two-point function of the spin waves with Gaussian domination of the partition function. In \cite{AizenmanLiebSeiringerSolovejYngvason001} both steps are delegated to \cite[Lemma 6.1 and Theorem 4.2]{DysonLiebSimon001}; complete derivations are given here. Throughout this section, \(\Lambda
=
\Lambda_{L}\) is a periodic box defined in \eqref{eq:main-local-algebras}. The parameters \(\sminvtemperature
>
0\) and \(\lambda
\geq
0\) are fixed. Lemma \ref{lem:sublattice-rotation} permits all traces to be computed with the rotated Hamiltonian, and the finite-volume Gibbs state is \(\oastate[\psi_{\sminvtemperature,\Lambda}]\) defined by \eqref{eq:main-periodic-gibbs-state}.

\subsection{The Duhamel two-point function}\label{the-duhamel-two-point-function}

The Duhamel two-point function is the positive sesquilinear form obtained by imaginary-time averaging in a finite-volume Gibbs state. Its spectral representation and covariance properties provide the differential calculus needed for Gaussian domination and the infrared bound.

\begin{defn}[Duhamel two-point function]\label{def:duhamel}
For
$A_{\Lambda}, B_{\Lambda}
\in
\oa{A}_{\Lambda}$
and a self-adjoint Hamiltonian
$\physham[K]_{\Lambda}$
on
$\sphilb{H}_{\Lambda}$,
set
$Z_{\sminvtemperature,\physham[K]_{\Lambda}}
=
\sqfun{\trace}{\napiernum^{-\sminvtemperature \physham[K]_{\Lambda}}}$.
The Duhamel two-point function is defined by
\begin{equation}\label{eq:duhamel-two-point}
\rbkt{A_{\Lambda}}{B_{\Lambda}}_{\sminvtemperature,\physham[K]_{\Lambda}}
=
\frac{1}{Z_{\sminvtemperature,\physham[K]_{\Lambda}}}
\int_{0}^{1}
\sqfun{\trace}{
A_{\Lambda}
\napiernum^{-s \sminvtemperature \physham[K]_{\Lambda}}
B_{\Lambda}
\napiernum^{-\rbk{1 - s} \sminvtemperature \physham[K]_{\Lambda}}
}
\opdmsr{s}.
\end{equation}
When $\physham[K]_{\Lambda}$ is specialized to the model Hamiltonian
$\physham_{\Lambda}$, the abbreviated subscripted notation is
\begin{equation}\label{eq:duhamel-finite-volume-abbreviation}
\rbkt{A_{\Lambda}}{B_{\Lambda}}_{\sminvtemperature,\Lambda}
=
\rbkt{A_{\Lambda}}{B_{\Lambda}}_{\sminvtemperature,\physham_{\Lambda}}.
\end{equation}
\end{defn}

The following finite-dimensional calculation gives the spectral representation, positivity, and Cauchy--Schwarz inequality for this form.

\begin{lem}[elementary properties]\label{lem:duhamel-properties}
Let $\seq{\Phi_{\Lambda,j}}{j \in J_{\Lambda}}$ be an orthonormal eigenbasis of a
self-adjoint operator $\physham[K]_{\Lambda}$, indexed by the finite set
$J_{\Lambda}$, and write
\begin{equation}\label{eq:duhamel-eigenbasis-data}
\begin{aligned}
\physham[K]_{\Lambda} \Phi_{\Lambda,j}
=
\varepsilon_{\Lambda,j}\Phi_{\Lambda,j},
\quad
A_{\Lambda,jk}
=
\bkt{\Phi_{\Lambda,j}}{A_{\Lambda}\Phi_{\Lambda,k}}.
\end{aligned}
\end{equation}
For the Duhamel form \eqref{eq:duhamel-two-point}, the following hold.
\begin{enumerate}
\item The spectral representation is
\begin{equation}\label{eq:duhamel-spectral-representation}
\rbkt{A_{\Lambda}}{B_{\Lambda}}_{\sminvtemperature,\physham[K]_{\Lambda}}
=
\frac{1}{Z_{\sminvtemperature,\physham[K]_{\Lambda}}}
\sum_{j,k\in J_{\Lambda}}
A_{\Lambda,jk}B_{\Lambda,kj}
\begin{dcases}
\frac{\napiernum^{-\sminvtemperature\varepsilon_{\Lambda,j}}
-\napiernum^{-\sminvtemperature\varepsilon_{\Lambda,k}}}
{\sminvtemperature\rbk{\varepsilon_{\Lambda,k}-\varepsilon_{\Lambda,j}}},
& \varepsilon_{\Lambda,j} \neq \varepsilon_{\Lambda,k},
\\ 
\napiernum^{-\sminvtemperature\varepsilon_{\Lambda,j}},
& \varepsilon_{\Lambda,j}=\varepsilon_{\Lambda,k}.
\end{dcases}
\end{equation}

\item The Duhamel form is symmetric and positive:
\begin{equation}\label{eq:duhamel-symmetry-positivity}
\begin{aligned}
\rbkt{A_{\Lambda}}{B_{\Lambda}}_{\sminvtemperature,\physham[K]_{\Lambda}}
&=
\rbkt{B_{\Lambda}}{A_{\Lambda}}_{\sminvtemperature,\physham[K]_{\Lambda}},
\\ 
\rbkt{\faadj{A_{\Lambda}}}{A_{\Lambda}}_{\sminvtemperature,\physham[K]_{\Lambda}}
&\geq
0.
\end{aligned}
\end{equation}

\item The map
$\rbk{A_{\Lambda},B_{\Lambda}}\mapsto\rbkt{\faadj{A_{\Lambda}}}{B_{\Lambda}}_{\sminvtemperature,\physham[K]_{\Lambda}}$
is a positive semi-definite sesquilinear form and therefore satisfies the
Cauchy--Schwarz inequality.

\item Let $A_{\Lambda}$ be self-adjoint and set
$$
\physham[K]_{\Lambda,t}
=
\physham[K]_{\Lambda}+tA_{\Lambda},
\quad
\fun{F_{\Lambda,A}}{t}
=
-\frac{1}{\sminvtemperature}
\log\sqfun{\trace}{\napiernum^{-\sminvtemperature
\physham[K]_{\Lambda,t}}}.
$$
Then the first two derivatives are
\begin{equation}\label{eq:free-energy-duhamel-derivatives}
\begin{aligned}
\fun{F_{\Lambda,A}'}{t}
&=
\frac{
\sqfun{\trace}{A_{\Lambda}\napiernum^{-\sminvtemperature
\physham[K]_{\Lambda,t}}}
}{
\sqfun{\trace}{\napiernum^{-\sminvtemperature
\physham[K]_{\Lambda,t}}}
},
\quad
\fun{F_{\Lambda,A}''}{t}
=
-\sminvtemperature
\left\{
\rbkt{A_{\Lambda}}{A_{\Lambda}}_{\sminvtemperature,
\physham[K]_{\Lambda,t}}
-
\rbk{\fun{F_{\Lambda,A}'}{t}}^{2}
\right\}
\leq
0.
\end{aligned}
\end{equation}
\end{enumerate}
\end{lem}

\begin{proof}
Insert the eigenbasis from \eqref{eq:duhamel-eigenbasis-data} into the
trace in \eqref{eq:duhamel-two-point}.
For $s\in\closedinterval{0}{1}$, this gives
$$\begin{aligned}
&\sqfun{\trace}{
A_{\Lambda}
\napiernum^{-s\sminvtemperature\physham[K]_{\Lambda}}
B_{\Lambda}
\napiernum^{-\rbk{1-s}\sminvtemperature\physham[K]_{\Lambda}}
}
\\
&=
\sum_{j\in J_{\Lambda}}
\bkt{\Phi_{\Lambda,j}}{
A_{\Lambda}
\napiernum^{-s\sminvtemperature\physham[K]_{\Lambda}}
B_{\Lambda}
\napiernum^{-\rbk{1-s}\sminvtemperature\physham[K]_{\Lambda}}
\Phi_{\Lambda,j}
}
\\
&=
\sum_{j,k\in J_{\Lambda}}
\bkt{\Phi_{\Lambda,j}}{A_{\Lambda}\Phi_{\Lambda,k}}
\bkt{\Phi_{\Lambda,k}}{B_{\Lambda}\Phi_{\Lambda,j}}
\napiernum^{-s\sminvtemperature\varepsilon_{\Lambda,k}}
\napiernum^{-\rbk{1-s}\sminvtemperature\varepsilon_{\Lambda,j}}
\\
&=
\sum_{j,k\in J_{\Lambda}}
A_{\Lambda,jk}B_{\Lambda,kj}
\napiernum^{-s\sminvtemperature\varepsilon_{\Lambda,k}}
\napiernum^{-\rbk{1-s}\sminvtemperature\varepsilon_{\Lambda,j}}.
\end{aligned}$$
The coefficient of $A_{\Lambda,jk}B_{\Lambda,kj}$ after integration is
$$\begin{aligned}
&q_{\Lambda,jk}
=
\int_{0}^{1}
\napiernum^{-s\sminvtemperature\varepsilon_{\Lambda,k}}
\napiernum^{-\rbk{1-s}\sminvtemperature\varepsilon_{\Lambda,j}}
\opdmsr{s}
=
\napiernum^{-\sminvtemperature\varepsilon_{\Lambda,j}}
\int_{0}^{1}
\napiernum^{-s\sminvtemperature
\rbk{\varepsilon_{\Lambda,k}-\varepsilon_{\Lambda,j}}}
\opdmsr{s}
\\ 
&=
\begin{cases}
\displaystyle
\frac{\napiernum^{-\sminvtemperature\varepsilon_{\Lambda,j}}
-\napiernum^{-\sminvtemperature\varepsilon_{\Lambda,k}}}
{\sminvtemperature\rbk{\varepsilon_{\Lambda,k}-\varepsilon_{\Lambda,j}}},
&
\varepsilon_{\Lambda,j}\neq\varepsilon_{\Lambda,k},
\\ 
\napiernum^{-\sminvtemperature\varepsilon_{\Lambda,j}},
&
\varepsilon_{\Lambda,j}=\varepsilon_{\Lambda,k}.
\end{cases}
\end{aligned}$$
Substitution into \eqref{eq:duhamel-two-point} proves
\eqref{eq:duhamel-spectral-representation}.

The coefficients $q_{\Lambda,jk}=q_{\Lambda,kj}$ are nonnegative.
The relation
$\bkt{\Phi_{\Lambda,j}}{\faadj{A_{\Lambda}}\Phi_{\Lambda,k}}
=
\cmpconj{A_{\Lambda,kj}}$
then gives
\eqref{eq:duhamel-symmetry-positivity} and the positive semi-definiteness
in assertion (3).
The Cauchy--Schwarz inequality follows for every positive semi-definite
sesquilinear form.

For assertion (4), the Duhamel derivative formula and cyclicity of the trace
give the first identity in \eqref{eq:free-energy-duhamel-derivatives}.
Differentiating it once more gives the second identity there.
Its expression in braces equals
$$
\rbkt{
A_{\Lambda}-\fun{F_{\Lambda,A}'}{t}
}{
A_{\Lambda}-\fun{F_{\Lambda,A}'}{t}
}_{\sminvtemperature,\physham[K]_{\Lambda,t}},
$$
which is nonnegative by \eqref{eq:duhamel-symmetry-positivity}.
\end{proof}

\subsection{Reflection positivity of the rotated Hamiltonian}\label{reflection-positivity-of-the-rotated-hamiltonian}

The sublattice rotation and a bond-bisecting reflection put the deformed Hamiltonian into conjugate-pair form. The resulting trace functional is reflection positive and satisfies the Schwarz inequality used in the source-field comparison. Choose the first coordinate direction. The pair of bond-bisecting planes are \begin{equation}\label{eq:bond-bisecting-planes}
\begin{aligned}
p
=
p_{\onehalf}
\cup
p_{\frac{L}{2} + \onehalf},
\quad
p_{t}
=
\set{u \in \fldreal^{d}}{u_{1} = t},
\end{aligned}
\end{equation} and the two halves of the periodic box are \[\begin{aligned}
\Lambda_{-}
=
\set{x \in \Lambda}{-\frac{L}{2} < x_{1} \leq 0},
\quad
\Lambda_{+}
=
\set{x \in \Lambda}{0 < x_{1} \leq \frac{L}{2}},
\quad
\abscard{\Lambda_{\mp}}
=
\frac{L^{d}}{2}.
\end{aligned}\] The bond-bisecting spatial reflection is \begin{equation}\label{eq:bond-bisecting-spatial-reflection}
\fun{\theta}{x_{1}, x_{2}, \dotsc, x_{d}}
=
\vecbk{1 - x_{1}, x_{2}, \dotsc, x_{d}}
\pmod{L}.
\end{equation} This map satisfies \(\fun{\theta}{\Lambda_{\mp}}
=
\Lambda_{\pm}\). Writing \(x \sim y\) for nearest-neighbor adjacency in the periodic graph, the crossing bonds are \begin{equation}\label{eq:reflection-crossing-bonds}
\begin{aligned}
M
=
\set{\dbk{x \, \theta(x)}}
{x \in \Lambda_{-}, \ x \sim \theta(x)},
\quad
\abscard{M}
=
\begin{cases}
L^{d - 1},& L = 2,\\
2 L^{d - 1},& L \geq 4.
\end{cases}
\end{aligned}
\end{equation} The internal and crossing bonds give the disjoint decomposition \begin{equation}\label{eq:reflection-bond-decomposition}
\begin{aligned}
\setindex{B}_{\Lambda}
&=
\set{\dbk{xy}}
{x,y \in \Lambda,\ x \sim y},
\\ 
\setindex{B}_{\mp}
&=
\set{\dbk{xy} \in \setindex{B}_{\Lambda}}
{x,y \in \Lambda_{\mp}},
\\ 
\setindex{B}_{\Lambda}
&=
\setindex{B}_{-}
\sqcup
\setindex{B}_{+}
\sqcup
M.
\end{aligned}
\end{equation} For every \(x\in\Lambda\), let \(\sphilb{H}_{x}=\fldcmp^{2}\) be a labelled copy of the one-site Hilbert space, with standard orthonormal basis \(\seq{e_{x}^{\sigma}}{\sigma\in\setone{\pm\onehalf}}\). For \(x,y\in\Lambda\), let \[\iota_{y,x}
\colon
\sphilb{H}_{y}
\to
\sphilb{H}_{x},
\quad
\fun{\iota_{y,x}}{e_{y}^{\sigma}}
=
e_{x}^{\sigma}
\quad
\rbk{\sigma\in\setone{\pm\onehalf}}\] be the canonical basis-preserving unitary. The half-lattice and full Hilbert spaces are defined by \begin{equation}\label{eq:reflection-hilbert-spaces}
\begin{aligned}
\sphilb{H}_{\mp}
&=
\bigotimes_{x \in \Lambda_{\mp}}
\sphilb{H}_{x},
\quad
\sphilb{H}_{\Lambda}
=
\sphilb{H}_{-}
\otimes
\sphilb{H}_{+}.
\end{aligned}
\end{equation} They carry their product inner products. The linear map \(U_{\theta}\colon\sphilb{H}_{+}\to\sphilb{H}_{-}\) is defined on elementary tensors, with \(v_{y}\in\sphilb{H}_{y}\) for every \(y\in\Lambda_{+}\), by \begin{equation}\label{eq:reflection-unitary}
\fun{U_{\theta}}
{\bigotimes_{y \in \Lambda_{+}} v_{y}}
=
\bigotimes_{x \in \Lambda_{-}}
\fun{\iota_{\theta(x),x}}{v_{\theta(x)}}.
\end{equation} It maps the standard product orthonormal basis of \(\sphilb{H}_{+}\) bijectively onto that of \(\sphilb{H}_{-}\) and is therefore unitary. For every \(x\in\Lambda_{-}\) and \(i\in\setone{1,2,3}\), it satisfies \begin{equation}\label{eq:reflection-relabeling-intertwining}
U_{\theta}S^{i}_{\theta(x)}\faadj{U_{\theta}}
=
S^{i}_{x}.
\end{equation} We use the resulting unitary identification \begin{equation}\label{eq:reflection-unitary-identification}
1\otimes U_{\theta}
\colon
\sphilb{H}_{\Lambda}
\to
\sphilb{H}_{-}\otimes\sphilb{H}_{-}.
\end{equation} For \(C \in \oa{A}_{\Lambda_{-}}\) and \(x \in \Lambda_{-}\), use the standard product basis of \(S^{3}\)-eigenvectors: i.e., we denote \(S^{3}_{x} \Phi_{\sigma}
=
\sigma \Phi_{\sigma}\) for \(\sigma
\in
\setone{\pm \onehalf}\). We define the conjugation operator \(J
\colon \sphilb{H}_{-}
\to \sphilb{H}_{-}\) as \[\fun{J}
{\sum_{\sigma} c_{\sigma} \Phi_{\sigma}}
=
\sum_{\sigma}
\cmpconj{c_{\sigma}}
\Phi_{\sigma}.\] This \(J\) satisfies \(\cmpconj{C}
=
J C J\), and \(\cmpconj{C}\) means entrywise complex conjugation. Its values on the one-site spin matrices are collected in \eqref{eq:app-spin-conjugation-rotation}.

The Gaussian domination argument requires the partition function with a source field. For \(h \colon \Lambda \to \fldreal\) define, following \cite[Eq. before Lemma 1]{AizenmanLiebSeiringerSolovejYngvason001}, the deformed rotated Hamiltonian \begin{equation}\label{eq:deformed-hamiltonian}
\fun{\widehat{K}_{\Lambda}}{h} = \sum_{\dbk{xy}} \sqbk{\frac{1}{2} \rbk{S^{1}_{x} - S^{1}_{y} - h_{x} + h_{y}}^{2} + S^{2}_{x} S^{2}_{y}} + \lambda \sum_{x \in \Lambda} S^{3}_{x}
\end{equation} and the deformed partition function \[\fun{Z_{\Lambda}}{h} = \sqfun{\trace}{\napiernum^{-\sminvtemperature \fun{\widehat{K}_{\Lambda}}{h}}}.\] Expanding the square in \eqref{eq:deformed-hamiltonian} and using \eqref{eq:app-one-site-spin-squares} gives \begin{equation}\label{eq:rotated-zero-source-hamiltonian}
\fun{\widehat{K}_{\Lambda}}{0}
=
V_{\Lambda}
\rbk{
\physham_{\Lambda}
-
\frac{\lambda \abscard{\Lambda}}{2}
}
\faadj{V_{\Lambda}}
+
\frac{d \abscard{\Lambda}}{4}
=
V_{\Lambda}
\physham_{\Lambda}
\faadj{V_{\Lambda}}
+
c_{0,\Lambda},
\quad
c_{0,\Lambda}
=
\frac{d \abscard{\Lambda}}{4}
-
\frac{\lambda \abscard{\Lambda}}{2},
\end{equation} The unitary \(V_{\Lambda}\) is defined by \eqref{eq:sublattice-rotation-unitary}. The additive constants cancel from all expectation values and ratios used below. At zero source the partition function is \[\fun{Z_{\Lambda}}{0}
=
\napiernum^{-\sminvtemperature c_{0,\Lambda}}
\sqfun{\trace}{\napiernum^{-\sminvtemperature \physham_{\Lambda}}}.\]

Across the reflection planes, \(\fun{\widehat{K}_{\Lambda}}{h}\) splits into a real left block, a real right block, and crossing terms. Each crossing term is a square of the difference between two reflection-conjugate operators.

\begin{lem}[reflection Schwarz inequality]\label{lem:reflection-schwarz}
Let $h \colon \Lambda \to \fldreal$ and let $p$ be a pair of bond-bisecting planes orthogonal to a coordinate direction, with reflection $\theta$ and crossing bonds $M$ defined by \eqref{eq:reflection-crossing-bonds} after the corresponding permutation of coordinate directions.
The reflected fields are defined by
$$\begin{aligned}
\fun{h^{-}}{x}
=
\begin{cases}
\fun{h}{x}
&
x \in \Lambda_{-},
\\
\fun{h}{\theta(x)}
&
x \in \Lambda_{+},
\end{cases}
\quad
\fun{h^{+}}{x}
=
\begin{cases}
\fun{h}{\theta(x)}
&
x \in \Lambda_{-},
\\
\fun{h}{x}
&
x \in \Lambda_{+}.
\end{cases}
\end{aligned}$$
The partition functions satisfy
$$\fun{Z_{\Lambda}}{h} \leq \fun{Z_{\Lambda}}{h^{-}}^{\onehalf} \fun{Z_{\Lambda}}{h^{+}}^{\onehalf}.$$
\end{lem}

\begin{proof}
Let $K_{\Lambda,\mp}$ collect the internal bonds and single-site terms of $\Lambda_{\mp}$.
Decompose \eqref{eq:deformed-hamiltonian} as
$\fun{\widehat{K}_{\Lambda}}{h} = K_{\Lambda,-} \otimes 1 + 1 \otimes K_{\Lambda,+} + \sum_{b \in M} Q_{\Lambda,b}$.
For a crossing bond $b = \dbk{x \, \theta x}$ the last term is
\begin{equation}\label{eq:reflection-crossing-term}
\begin{aligned}
Q_{\Lambda,b}
=
\frac{1}{2} \rbk{S^{1}_{x} \otimes 1 - 1 \otimes S^{1}_{x} - \delta_{\Lambda,b}}^{2}
+S^{2}_{x} \otimes S^{2}_{x},
\quad
\delta_{\Lambda,b}
=
h_{x} - h_{\theta x}.
\end{aligned}
\end{equation}
in the identification of $\sphilb{H}_{+}$ with $\sphilb{H}_{-}$;
here the operator $S^{i}_{\theta x}$ acting on the right factor
became $S^{i}_{x}$ on the second copy of $\sphilb{H}_{-}$ by
\eqref{eq:reflection-relabeling-intertwining}.
In the standard basis $K_{\Lambda,-}$ and $K_{\Lambda,+}$ are real matrices:
they are polynomials in the real matrices $S^{1}_{x}, S^{3}_{x}$
and in products $S^{2}_{x} S^{2}_{y}$ of two imaginary matrices.
For each $b \in M$, the crossing-bond contribution $Q_{\Lambda,b}$ in
\eqref{eq:reflection-crossing-term} consists of the summand containing
$S^{1}_{x}$ and the summand $S^{2}_{x} \otimes S^{2}_{x}$.
We rewrite these two summands separately in conjugate-pair form.
For the $S^{1}$ part, $S^{1}_{x}$ is real, so with $D_{\Lambda,b} = S^{1}_{x} / \sqrt{2}$ and $c_{\Lambda,b} = \delta_{\Lambda,b} / \sqrt{2}$,
the first identity in
\eqref{eq:app-reflection-spin-squares}
puts the square in conjugate-pair form.
For the $S^{2}$ part, $\cmpconj{S^{2}_{x}} = -S^{2}_{x}$ and $\rbk{S^{2}_{x}}^{2} = 1 / 4$, so with $G_{\Lambda,b} = S^{2}_{x} / \sqrt{2}$,
the second identity in
\eqref{eq:app-reflection-spin-squares}
gives the required form and the constant
$-\frac{1}{4}$.
The finite index set and the corresponding coefficients are
$$\begin{aligned}
\setindex{I}_{\Lambda}
=
M \times \setone{1, 2},
\quad
F_{\Lambda,\rbk{b, 1}}
=
D_{\Lambda,b},
\quad
c_{\Lambda,\rbk{b, 1}}
=
c_{\Lambda,b},
\quad
F_{\Lambda,\rbk{b, 2}}
=
G_{\Lambda,b},
\quad
c_{\Lambda,\rbk{b, 2}}
=
0.
\end{aligned}$$
Absorbing the constants $-\abscard{M} / 4$ into a common factor of all the partition functions involved (they cancel in the asserted inequality), $\fun{\widehat{K}_{\Lambda}}{h}$ takes the normal form
$$\fun{\widehat{K}_{\Lambda}}{h} = K_{\Lambda,-} \otimes 1 + 1 \otimes K_{\Lambda,+} + \sum_{\alpha \in \setindex{I}_{\Lambda}} \rbk{F_{\Lambda,\alpha} \otimes 1 - 1 \otimes \cmpconj{F_{\Lambda,\alpha}} - c_{\Lambda,\alpha}}^{2} - \frac{\abscard{M}}{4},$$
with real constants $c_{\Lambda,\alpha}$ and matrices $F_{\Lambda,\alpha}$ (real or imaginary, in either case $\cmpconj{\napiernum^{\imunit k F_{\Lambda,\alpha}}} = \napiernum^{-\imunit k \cmpconj{F_{\Lambda,\alpha}}}$ for real $k$).
For $n \in \semigrposint$ consider the Trotter approximant
$$Z_{\Lambda,n} = \sqfun{\trace}{\rbk{\napiernum^{-\frac{\sminvtemperature}{n} K_{\Lambda,-} \otimes 1} \napiernum^{-\frac{\sminvtemperature}{n} 1 \otimes K_{\Lambda,+}} \prod_{\alpha \in \setindex{I}_{\Lambda}} \napiernum^{-\frac{\sminvtemperature}{n} \rbk{F_{\Lambda,\alpha} \otimes 1 - 1 \otimes \cmpconj{F_{\Lambda,\alpha}} - c_{\Lambda,\alpha}}^{2}}}^{n}},$$
which converges to $\fun{Z_{\Lambda}}{h}$ as $n \to \infty$ by the Lie--Trotter product formula.
Each Gaussian factor is resolved by the identity, valid for any self-adjoint $Y$ and $t > 0$ by the spectral theorem and the scalar Gaussian integral,
$$\napiernum^{-t Y^{2}}
=
\frac{1}{\sqrt{4 \pi t}}
\int_{\fldreal}
\napiernum^{- \frac{k^{2}}{4 t}}
\napiernum^{\imunit k Y}
\opdmsr{k}.$$
With $Y = F_{\Lambda,\alpha} \otimes 1 - 1 \otimes \cmpconj{F_{\Lambda,\alpha}} - c_{\Lambda,\alpha}$,
$$\napiernum^{\imunit k Y} = \napiernum^{-\imunit k c_{\Lambda,\alpha}} \napiernum^{\imunit k F_{\Lambda,\alpha}} \otimes \napiernum^{-\imunit k \cmpconj{F_{\Lambda,\alpha}}}
= \napiernum^{-\imunit k c_{\Lambda,\alpha}} \napiernum^{\imunit k F_{\Lambda,\alpha}} \otimes \cmpconj{\napiernum^{\imunit k F_{\Lambda,\alpha}}},$$
because the two tensor legs commute.
For this $n$, let $\nu$ be the centered Gaussian measure with variance $2 \sminvtemperature / n$.
The product measure over the factors $\alpha \in \setindex{I}_{\Lambda}$ and the Trotter slots $1 \leq j \leq n$ is
$$\opdmsr{\mu_{\Lambda,n}(k)}
=
\prod_{\alpha \in \setindex{I}_{\Lambda}}
\prod_{j = 1}^{n}
\opdmsr{\nu(k_{\alpha j})}.$$
Substituting one Gaussian integral per factor and per Trotter slot, and using that the trace of a tensor product factorizes, gives
$$\begin{aligned}
Z_{\Lambda,n}
&=
\int_{\prod_{\alpha\in\setindex{I}_{\Lambda}}\prod_{j=1}^{n}\fldreal}
\napiernum^{-\imunit \sum_{\alpha \in \setindex{I}_{\Lambda}} \sum_{j = 1}^{n} k_{\alpha j} c_{\Lambda,\alpha}}
\fun{\mathcal{F}_{\Lambda,-}}{k}
\cmpconj{\fun{\mathcal{F}_{\Lambda,+}}{k}}
\opdmsr{\mu_{\Lambda,n}(k)},
\\ 
\fun{\mathcal{F}_{\Lambda,\mp}}{k}
&=
\sqfun{\trace_{\sphilb{H}_{-}}}
{\prod_{j = 1}^{n}
\rbk{\napiernum^{-\frac{\sminvtemperature}{n} K_{\Lambda,\mp}}
\prod_{\alpha \in \setindex{I}_{\Lambda}} \napiernum^{\imunit k_{\alpha j} F_{\Lambda,\alpha}}}};
\end{aligned}$$
the complex conjugate appears on the $+$ side
because $\cmpconj{\napiernum^{-\sminvtemperature K_{\Lambda,+} / n}}
=
\napiernum^{-\sminvtemperature K_{\Lambda,+} / n}$
($K_{\Lambda,+}$ real) and the conjugations of the unitary factors assemble into the entrywise conjugate of the whole product, whose trace is the conjugate trace.
Since $Z_{\Lambda,n} > 0$, the modulus bound
and the Cauchy--Schwarz inequality for the product measure $\opdmsr{\mu_{\Lambda,n}(k)}$ give
$$\begin{aligned}
&Z_{\Lambda,n}
\leq
\int_{\prod_{\alpha\in\setindex{I}_{\Lambda}}
\prod_{j=1}^{n}\fldreal}
\abs{\fun{\mathcal{F}_{\Lambda,-}}{k}}
\abs{\fun{\mathcal{F}_{\Lambda,+}}{k}}
\opdmsr{\mu_{\Lambda,n}(k)}
\\ 
&\leq
\rbk{\int_{\prod_{\alpha\in\setindex{I}_{\Lambda}}
\prod_{j=1}^{n}\fldreal}
\abs{\fun{\mathcal{F}_{\Lambda,-}}{k}}^{2}
\opdmsr{\mu_{\Lambda,n}(k)}}^{\onehalf}
\rbk{\int_{\prod_{\alpha\in\setindex{I}_{\Lambda}}
\prod_{j=1}^{n}\fldreal}
\abs{\fun{\mathcal{F}_{\Lambda,+}}{k}}^{2}
\opdmsr{\mu_{\Lambda,n}(k)}}^{\onehalf}.
\end{aligned}$$
The square
$\int_{\prod_{\alpha\in\setindex{I}_{\Lambda}}\prod_{j=1}^{n}\fldreal} \abs{\fun{\mathcal{F}_{\Lambda,-}}{k}}^{2} \opdmsr{\mu_{\Lambda,n}(k)}$
is a Trotter approximant with the right block $K_{\Lambda,+}$ replaced by a copy of $K_{\Lambda,-}$.
All crossing constants are then zero.
Replacing $h$ by $h^{-}$ has exactly these two effects.
It reproduces the left block on both sides,
and it sets every crossing difference $\delta_{\Lambda,b}$ to zero.
The square is the approximant $Z_{\Lambda,n}^{-}$ of $\fun{Z_{\Lambda}}{h^{-}}$.
Taking $n \to \infty$ on both sides of $Z_{\Lambda,n} \leq \rbk{Z_{\Lambda,n}^{-}}^{\onehalf} \rbk{Z_{\Lambda,n}^{+}}^{\onehalf}$ proves the lemma.
\end{proof}

\subsection{Gaussian domination}\label{gaussian-domination}

Repeated bond reflections turn a maximizing source field into a constant field. The argument uses the finite-dimensional source-field space, its quotient by constant fields, and the bond differences defined below.

\begin{defn}[finite-volume source data]
For a source field $h\in\fldreal^{\Lambda}$ and a bond $b=\dbk{xy}\in\setindex{B}_{\Lambda}$, set
\begin{equation}\label{eq:gaussian-domination-source-data}
\begin{aligned}
\fun{\delta_{b}}{h}
&=
h_{x}-h_{y},
\quad
\fun{\delta_{\txtmax}}{h}
=
\max_{b\in\setindex{B}_{\Lambda}}
\abs{\fun{\delta_{b}}{h}},
\quad
\fun{\mathcal{N}}{h}
&=
\abscard{\set{b\in\setindex{B}_{\Lambda}}{\fun{\delta_{b}}{h}=0}}.
\end{aligned}
\end{equation}
Two source fields differing by a constant have the same bond differences.
\end{defn}
\begin{lem}[existence of a maximizer]\label{lem:gaussian-domination-maximizer}
The function $h\mapsto\fun{Z_{\Lambda}}{h}$ attains its supremum on $\fldreal^{\Lambda}$.
\end{lem}

\begin{proof}
The bond set $\setindex{B}_{\Lambda}$ defined by
\eqref{eq:reflection-bond-decomposition} is finite.
The orientation chosen for $b$ changes only the sign of $\fun{\delta_{b}}{h}$,
and does not affect its absolute value, its vanishing, or the deformed Hamiltonian.
Equation \eqref{eq:deformed-hamiltonian} depends on $h$ only through the bond differences in the first equality of \eqref{eq:gaussian-domination-source-data}.
For every $c \in \fldreal$ it follows that
$$\fun{\widehat{K}_{\Lambda}}{h + c}
=
\fun{\widehat{K}_{\Lambda}}{h},
\quad
\fun{Z_{\Lambda}}{h + c}
=
\fun{Z_{\Lambda}}{h}.$$

The function $\fun{Z_{\Lambda}}{h}$ attains its supremum on the normalized source
fields.
Equip $\fldreal^{\Lambda}$ with its finite-dimensional Euclidean topology.
This is equivalently the topology induced by the norm $\norm{h}_{\infty} = \max_{x \in \Lambda} \abs{h_{x}}$.
Fix one site $x_{0} \in \Lambda$ and subtract $h_{x_{0}}$ from $h$.
This normalization leaves $\fun{Z_{\Lambda}}{h}$ unchanged and gives $h_{x_{0}} = 0$.
The normalized fields form the affine subspace
$\set{h \in \fldreal^{\Lambda}}{h_{x_{0}} = 0}
\eqisom
\fldreal^{\abscard{\Lambda}-1}$.
We use the relative Euclidean topology on this affine subspace.
For two fields $h, \widetilde{h} \in \fldreal^{\Lambda}$, the first two equalities of \eqref{eq:gaussian-domination-source-data} give the Lipschitz estimate
$$\abs{\fun{\delta_{\txtmax}}{h}
-\fun{\delta_{\txtmax}}{\widetilde{h}}}
\leq
\max_{b \in \setindex{B}_{\Lambda}}
\abs{\fun{\delta_{b}}{h - \widetilde{h}}}
\leq
2
\norm{h - \widetilde{h}}_{\infty}.$$
The function $\fun{\delta_{\txtmax}}{\cdot}$ is continuous in this topology.
Every $x \in \Lambda$ can be joined to $x_{0}$ by a nearest-neighbor path of length at most $d L$.
For $R \geq 0$, define
$$\setindex{K}_{R}
=
\set{h \in \fldreal^{\Lambda}}
{h_{x_{0}} = 0,
\fun{\delta_{\txtmax}}{h} \leq R}
\subset
\fldreal^{\abscard{\Lambda}-1}.$$
If a sequence $\seq{h^{(n)}}{n}$ in $\setindex{K}_{R}$ converges to $h$ in the norm $\norm{\cdot}_{\infty}$, then $h_{x_{0}} = 0$, and the preceding Lipschitz estimate gives
$$\fun{\delta_{\txtmax}}{h}
=
\lim_{n \to \infty}
\fun{\delta_{\txtmax}}{h^{(n)}}
\leq
R.$$
This proves $h\in\setindex{K}_{R}$.
The set $\setindex{K}_{R}$ is closed in the normalized subspace.
For each $x \in \Lambda$, choose a nearest-neighbor path
$$x_{0}
=
x^{(0)},
x^{(1)},
\ldots,
x^{(m)}
=
x,
\text{ with }
m \leq d L.$$
For every $h \in \setindex{K}_{R}$, telescoping along this path gives
$$\abs{h_{x}}
=
\abs{\sum_{j = 1}^{m}
\rbk{h_{x^{(j)}}
-h_{x^{(j - 1)}}}}
\leq
\sum_{j = 1}^{m}
\abs{h_{x^{(j)}}
-h_{x^{(j - 1)}}}
\leq
m R
\leq
d L R.$$
Consequently $\norm{h}_{\infty} \leq d L R$ for every $h \in \setindex{K}_{R}$, so $\setindex{K}_{R}$ is bounded.
Finite dimensionality now makes $\setindex{K}_{R}$ compact.

The required coercivity follows because $\fun{Z_{\Lambda}}{h}$ tends to zero outside
these compact sets.
Choose $b_{h} = \dbk{xy} \in \setindex{B}_{\Lambda}$ with
$$\abs{\fun{\delta_{b_{h}}}{h}}
=
\fun{\delta_{\txtmax}}{h},$$
and set $A_{b_{h}} = S^{1}_{x} - S^{1}_{y}$.
The operator $A_{b_{h}}$ is self-adjoint and satisfies $\norm{A_{b_{h}}} \leq 1$.
For every real $\delta$, direct expansion gives the positive operator identity
\begin{equation}\label{eq:gaussian-domination-square-completion}
\rbk{A_{b_{h}} - \delta}^{2}
-
\rbk{\frac{\delta^{2}}{2} - A_{b_{h}}^{2}}
=
2
\rbk{A_{b_{h}} - \frac{\delta}{2}}^{2}
\geq
0.
\end{equation}
Applying \eqref{eq:gaussian-domination-square-completion} with $\delta = \fun{\delta_{b_{h}}}{h}$ yields
$$\frac{1}{2}
\rbk{A_{b_{h}} - \fun{\delta_{b_{h}}}{h}}^{2}
\geq
\frac{\fun{\delta_{\txtmax}}{h}^{2}}{4}
-
\frac{A_{b_{h}}^{2}}{2}
\geq
\frac{\fun{\delta_{\txtmax}}{h}^{2}}{4}
-
\frac{1}{2}.$$
For every bond $b = \dbk{uv} \in \setindex{B}_{\Lambda} \setminus \setone{b_{h}}$,
set
$\fun{T_{b}}{h}
=
S^{1}_{u}
-
S^{1}_{v}
-
h_{u}
+
h_{v}$.
This operator is self-adjoint,
and hence $T_{b}(h)^2$ is nonnegative.
These are precisely the square terms in \eqref{eq:deformed-hamiltonian} other than the term associated with $b_{h}$.
They may therefore be discarded when deriving a lower bound for $\fun{\widehat{K}_{\Lambda}}{h}$.
For each bond and each site it also holds that
$$S^{2}_{x} S^{2}_{y}
\geq
-
\frac{1}{4},
\quad
\lambda S^{3}_{x}
\geq
-
\frac{\lambda}{2},$$
because $\norm{S^{2}_{x} S^{2}_{y}} = 1 / 4$, $\lambda \geq 0$, and the eigenvalues of $S^{3}_{x}$ are $\pm \onehalf$.
The deformed Hamiltonian consequently has the lower bound
$$\fun{\widehat{K}_{\Lambda}}{h}
\geq
\frac{\fun{\delta_{\txtmax}}{h}^{2}}{4}
-
c_{\Lambda,\lambda},
\quad
c_{\Lambda,\lambda}
=
\frac{1}{2}
+
\frac{\abscard{\setindex{B}_{\Lambda}}}{4}
+
\frac{\lambda \abscard{\Lambda}}{2}.$$
Let $\seq{\fun{E_{\Lambda,j}}{h}}{1 \leq j \leq 2^{\abscard{\Lambda}}}$ denote the eigenvalues of $\fun{\widehat{K}_{\Lambda}}{h}$ counted with multiplicity.
The trace satisfies the coercive estimate
\begin{equation}\label{eq:gaussian-domination-coercive}
\fun{Z_{\Lambda}}{h}
=
\sum_{j = 1}^{2^{\abscard{\Lambda}}}
\napiernum^{-\sminvtemperature \fun{E_{\Lambda,j}}{h}}
\leq
2^{\abscard{\Lambda}}
\napiernum^{
\sminvtemperature c_{\Lambda,\lambda}
-
\frac{\sminvtemperature \fun{\delta_{\txtmax}}{h}^{2}}{4}
}.
\end{equation}
The matrix entries of $\fun{\widehat{K}_{\Lambda}}{h}$ depend polynomially on $h$.
Continuity of the matrix exponential and the trace makes $\fun{Z_{\Lambda}}{h}$ continuous.
Since $\napiernum^{-\sminvtemperature \fun{\widehat{K}_{\Lambda}}{0}}$ is strictly positive on the finite-dimensional Hilbert space,
we obtain $\fun{Z_{\Lambda}}{0}
=
\sqfun{\trace}{\napiernum^{-\sminvtemperature \fun{\widehat{K}_{\Lambda}}{0}}}
>
0$.
We can choose $R > 0$ such that
$$\frac{\sminvtemperature R^{2}}{4}
>
\abscard{\Lambda}
\log 2
+
\sminvtemperature c_{\Lambda,\lambda}
-
\log \fun{Z_{\Lambda}}{0}.$$
If $\fun{\delta_{\txtmax}}{h} > R$,
then \eqref{eq:gaussian-domination-coercive} and the strict monotonicity of the exponential function give
$$\begin{aligned}
&\fun{Z_{\Lambda}}{h}
\leq
\fnexp{\abscard{\Lambda} \log 2
+\sminvtemperature c_{\Lambda,\lambda}
-\frac{\sminvtemperature \fun{\delta_{\txtmax}}{h}^{2}}{4}}
<
\fnexp{\abscard{\Lambda} \log 2
+\sminvtemperature c_{\Lambda,\lambda}
-\frac{\sminvtemperature R^{2}}{4}}
\\ 
&<
\napiernum^{\log \fun{Z_{\Lambda}}{0}}
=
\fun{Z_{\Lambda}}{0}.
\end{aligned}$$
The global supremum is attained on the compact set of normalized fields with $\fun{\delta_{\txtmax}}{h} \leq R$.
\end{proof}

\begin{lem}[reflection preserves maximality]\label{lem:gaussian-domination-reflected-maximizers}
Let $h^{\ast}$ maximize $\fun{Z_{\Lambda}}{\cdot}$, and let $h^{-}$ and $h^{+}$ be the reflected fields of Lemma \ref{lem:reflection-schwarz} for a bond-bisecting reflection.
Then $h^{-}$ and $h^{+}$ also maximize $\fun{Z_{\Lambda}}{\cdot}$.
\end{lem}

\begin{proof}

Let $h^{\ast}$ be a maximizer of $\fun{Z_{\Lambda}}{\cdot}$.
Suppose that a bond $b_{0} \in \setindex{B}_{\Lambda}$ satisfies
$$\fun{\delta_{b_{0}}}{h^{\ast}}
\neq
0.$$
Because the side length $L$ is even, choose a pair of bond-bisecting planes orthogonal to $b_{0}$.
For this pair, let $M$ be the set of crossing bonds given by \eqref{eq:reflection-crossing-bonds}, after the corresponding permutation of coordinate directions.
Then $b_{0} \in M$.
Let $h^{-}$ and $h^{+}$ be the two reflected fields of Lemma \ref{lem:reflection-schwarz}.
The defining maximality of $h^{\ast}$, i.e, $\fun{Z_{\Lambda}}{h^{\ast}}
=
\sup_{h \colon \Lambda \to \fldreal}
\fun{Z_{\Lambda}}{h}$, gives
$$\fun{Z_{\Lambda}}{h^{-}}
\leq
\fun{Z_{\Lambda}}{h^{\ast}},
\quad
\fun{Z_{\Lambda}}{h^{+}}
\leq
\fun{Z_{\Lambda}}{h^{\ast}}.$$
The reflection Schwarz inequality supplies the reverse product estimate
\begin{equation}\label{eq:gaussian-domination-max-product}
\fun{Z_{\Lambda}}{h^{\ast}}^{2}
\leq
\fun{Z_{\Lambda}}{h^{-}}
\fun{Z_{\Lambda}}{h^{+}}
\leq
\fun{Z_{\Lambda}}{h^{\ast}}^{2}.
\end{equation}
The lower and upper bounds in \eqref{eq:gaussian-domination-max-product} coincide.
All partition functions are strictly positive.
Dividing \eqref{eq:gaussian-domination-max-product} by $\fun{Z_{\Lambda}}{h^{\ast}}^{2}$ gives
\begin{equation}\label{eq:gaussian-domination-max-ratios}
1
\leq
\frac{\fun{Z_{\Lambda}}{h^{-}}}{\fun{Z_{\Lambda}}{h^{\ast}}}
\frac{\fun{Z_{\Lambda}}{h^{+}}}{\fun{Z_{\Lambda}}{h^{\ast}}}
\leq
1.
\end{equation}
Each of the two ratios in \eqref{eq:gaussian-domination-max-ratios} belongs to $\leftopeninterval{0}{1}$ by maximality of $h^{\ast}$.
Their product can equal $1$ only when both ratios equal $1$.
It follows that
$$\fun{Z_{\Lambda}}{h^{-}}
=
\fun{Z_{\Lambda}}{h^{+}}
=
\fun{Z_{\Lambda}}{h^{\ast}}.$$
The reflected fields $h^{-}$ and $h^{+}$ are maximizers.
\end{proof}

\begin{lem}[strict increase of vanishing bond differences]\label{lem:gaussian-domination-count-increase}
Let $h^{\ast}$ be a maximizer for which $\fun{\mathcal{N}}{h^{\ast}}$ is largest among all maximizers.
Then $h^{\ast}$ has no nonzero bond difference.
\end{lem}

\begin{proof}
We argue by contradiction.
Suppose that there is a bond $b_{0}\in\setindex{B}_{\Lambda}$ with
$$\fun{\delta_{b_{0}}}{h^{\ast}}
\neq
0.$$
Because $L$ is even, choose a pair of bond-bisecting planes orthogonal to $b_{0}$.
Let $M$ be the crossing-bond set given by \eqref{eq:reflection-crossing-bonds}, after the corresponding permutation of coordinate directions.
Then $b_{0}\in M$.
Let $h^{-}$ and $h^{+}$ be the reflected fields of Lemma \ref{lem:reflection-schwarz} for this pair of planes.
Lemma \ref{lem:gaussian-domination-reflected-maximizers} shows that $h^{-}$ and $h^{+}$ are maximizers.
The numbers of vanishing bond differences distinguish the two reflected
maximizers.
The bond decomposition is
\eqref{eq:reflection-bond-decomposition}.
For the original maximizer define
$$\mathcal{N}_{-}
=
\abscard{
\set{b \in \setindex{B}_{-}}
{\fun{\delta_{b}}{h^{\ast}} = 0}
},
\quad
\mathcal{N}_{+}
=
\abscard{
\set{b \in \setindex{B}_{+}}
{\fun{\delta_{b}}{h^{\ast}} = 0}
},$$
and define the crossing-bond count by
$$\mathcal{N}_{M}
=
\abscard{
\set{b \in M}
{\fun{\delta_{b}}{h^{\ast}} = 0}
}.$$
The third equality of \eqref{eq:gaussian-domination-source-data} gives
$$\fun{\mathcal{N}}{h^{\ast}}
=
\mathcal{N}_{-}
+
\mathcal{N}_{+}
+
\mathcal{N}_{M}.$$
The nonzero difference on $b_{0}\in M$ gives
$$\mathcal{N}_{M}
\leq
\abscard{M}
-
1.$$

For a crossing bond $b = \dbk{x \theta(x)}$ with $x \in \Lambda_{-}$, the definition of $h^{-}$ gives the explicit cancellation
$$\fun{\delta_{b}}{h^{-}}
=
\fun{h^{-}}{x}
-
\fun{h^{-}}{\theta(x)}
=
h^{\ast}_{x}
-
h^{\ast}_{\fun{\theta}{\theta(x)}}
=
h^{\ast}_{x}
-
h^{\ast}_{x}
=
0.$$
The crossing differences of $h^{+}$ vanish by the parallel computation
$$\fun{\delta_{b}}{h^{+}}
=
\fun{h^{+}}{x}
-
\fun{h^{+}}{\theta(x)}
=
h^{\ast}_{\theta(x)}
-
h^{\ast}_{\theta(x)}
=
0.$$
On $\Lambda_{-}$ the internal differences of $h^{-}$ equal those of $h^{\ast}$.
For an internal bond $b \in \setindex{B}_{-}$, consider the reflected bond $\fun{\theta}{b} \in \setindex{B}_{+}$.
The difference of $h^{-}$ on $\fun{\theta}{b}$ agrees with $\fun{\delta_{b}}{h^{\ast}}$ up to orientation.
Each vanishing internal difference of $h^{\ast}$ in $\Lambda_{-}$ occurs once in each half of $h^{-}$.
The corresponding count is
$$\fun{\mathcal{N}}{h^{-}}
=
2 \mathcal{N}_{-}
+
\abscard{M}.$$
The same calculation with the roles of the halves exchanged gives
$$\fun{\mathcal{N}}{h^{+}}
=
2 \mathcal{N}_{+}
+
\abscard{M}.$$
Subtracting the count for $h^{\ast}$ from the larger of these two counts gives
\begin{equation}\label{eq:gaussian-domination-count-increase}
\max
\setone{
\fun{\mathcal{N}}{h^{-}},
\fun{\mathcal{N}}{h^{+}}
}
-
\fun{\mathcal{N}}{h^{\ast}}
=
\abs{\mathcal{N}_{-} - \mathcal{N}_{+}}
+
\abscard{M}
-
\mathcal{N}_{M}
\geq
1.
\end{equation}
The lower bound in \eqref{eq:gaussian-domination-count-increase} states that at least one of the maximizers $h^{-}$ and $h^{+}$ has more vanishing bond differences than $h^{\ast}$.
This contradicts the choice of $h^{\ast}$.
No bond with nonzero difference can exist for $h^{\ast}$.
\end{proof}

\begin{thm}[Gaussian domination]\label{thm:gaussian-domination}
For every $h \colon \Lambda \to \fldreal$ it holds that
\begin{equation}\label{eq:gaussian-domination}
\fun{Z_{\Lambda}}{h}
\leq
\fun{Z_{\Lambda}}{0}.
\end{equation}
\end{thm}

\begin{proof}
By Lemma \ref{lem:gaussian-domination-maximizer}, choose a maximizer $h^{\ast}$ of $\fun{Z_{\Lambda}}{\cdot}$ for which $\fun{\mathcal{N}}{h^{\ast}}$ is largest.
Lemma \ref{lem:gaussian-domination-reflected-maximizers} and Lemma \ref{lem:gaussian-domination-count-increase} show that $h^{\ast}$ has no nonzero bond difference.
Connectedness of the periodic nearest-neighbor graph makes $h^{\ast}$ constant.
The constant-shift invariance gives $\fun{Z_{\Lambda}}{h^{\ast}} = \fun{Z_{\Lambda}}{0}$.
For the arbitrary source field $h$ in the statement,
$$\fun{Z_{\Lambda}}{h}
\leq
\fun{Z_{\Lambda}}{h^{\ast}}
=
\fun{Z_{\Lambda}}{0},$$
which proves the theorem.
\end{proof}

\subsection{The infrared bound}\label{the-infrared-bound}

The infrared estimate converts the source-field comparison of Theorem \ref{thm:gaussian-domination} into a momentum-space bound for the Duhamel two-point function. Gaussian domination first bounds the second derivative of the deformed partition function at zero source. The unitary covariance of the Duhamel two-point function transfers this estimate from the rotated Hamiltonian \(\fun{\widehat{K}_{\Lambda}}{0}\) to the original Hamiltonian \(\physham_{\Lambda}\). A plane-wave source diagonalizes the discrete Laplacian and produces the factor \(E_{p}\). The resulting estimate for the first spin component extends to the second component by gauge symmetry. Throughout this subsection, \(E_{p}\) denotes the spin-wave energy defined in \eqref{eq:main-condensation-constant}.

The momentum lattice and spin-wave operators are defined in \eqref{eq:spin-wave}. Their adjoint and ladder-mode identities are collected in \eqref{eq:app-spin-fourier-adjoint-ladder}.

\begin{prop}[transfer of Duhamel two-point functions]\label{prop:duhamel-rotated-transfer}
Let $V_{\Lambda}$ be the sublattice rotation defined by \eqref{eq:sublattice-rotation-unitary}.
For every $A, B \in \oa{A}_{\Lambda}$,
\begin{equation}\label{eq:duhamel-rotated-transfer}
\rbkt{A}{B}_{\sminvtemperature,\fun{\widehat{K}_{\Lambda}}{0}}
=
\rbkt{\faadj{V_{\Lambda}} A V_{\Lambda}}{\faadj{V_{\Lambda}} B V_{\Lambda}}_{\sminvtemperature,\Lambda}.
\end{equation}
For every $p \in \dual{\Lambda}$, the spin-wave two-point functions satisfy
\begin{equation}\label{eq:duhamel-spin-component-transfer}
\rbkt{\widetilde{S}^{1}_{p}}{\widetilde{S}^{1}_{-p}}_{\sminvtemperature,\fun{\widehat{K}_{\Lambda}}{0}}
=
\rbkt{\widetilde{S}^{1}_{p}}{\widetilde{S}^{1}_{-p}}_{\sminvtemperature,\Lambda}
=
\rbkt{\widetilde{S}^{2}_{p}}{\widetilde{S}^{2}_{-p}}_{\sminvtemperature,\Lambda}.
\end{equation}
\end{prop}

\begin{proof}
Equation \eqref{eq:rotated-zero-source-hamiltonian} gives, for $0 \leq s \leq 1$,
$$\napiernum^{-s \sminvtemperature \fun{\widehat{K}_{\Lambda}}{0}}
=
\napiernum^{-s \sminvtemperature c_{0,\Lambda}}
V_{\Lambda}
\napiernum^{-s \sminvtemperature \physham_{\Lambda}}
\faadj{V_{\Lambda}},
\quad
Z_{\sminvtemperature,\fun{\widehat{K}_{\Lambda}}{0}}
=
\napiernum^{-\sminvtemperature c_{0,\Lambda}}
Z_{\sminvtemperature,\physham_{\Lambda}}.$$
For arbitrary $A, B \in \oa{A}_{\Lambda}$, substitution into \eqref{eq:duhamel-two-point} gives
$$\begin{aligned}
\rbkt{A}{B}_{\sminvtemperature,\fun{\widehat{K}_{\Lambda}}{0}}
&=
\frac{1}{Z_{\sminvtemperature,\physham_{\Lambda}}}
\int_{0}^{1}
\sqfun{\trace}{
A
V_{\Lambda}
\napiernum^{-s \sminvtemperature \physham_{\Lambda}}
\faadj{V_{\Lambda}}
B
V_{\Lambda}
\napiernum^{-\rbk{1 - s} \sminvtemperature \physham_{\Lambda}}
\faadj{V_{\Lambda}}
}
\opdmsr{s}
\\
&=
\frac{1}{Z_{\sminvtemperature,\physham_{\Lambda}}}
\int_{0}^{1}
\sqfun{\trace}{
\faadj{V_{\Lambda}}
A
V_{\Lambda}
\napiernum^{-s \sminvtemperature \physham_{\Lambda}}
\faadj{V_{\Lambda}}
B
V_{\Lambda}
\napiernum^{-\rbk{1 - s} \sminvtemperature \physham_{\Lambda}}
}
\opdmsr{s}
\\
&=
\rbkt{\faadj{V_{\Lambda}} A V_{\Lambda}}{\faadj{V_{\Lambda}} B V_{\Lambda}}_{\sminvtemperature,\Lambda}.
\end{aligned}$$
This proves \eqref{eq:duhamel-rotated-transfer}.

The sitewise identity $V_{\Lambda} S^{1}_{x} \faadj{V_{\Lambda}} = S^{1}_{x}$ and
\eqref{eq:spin-wave} give
\begin{equation}\label{eq:duhamel-rotated-spin-one-invariance}
\faadj{V_{\Lambda}}
\widetilde{S}^{1}_{p}
V_{\Lambda}
=
\widetilde{S}^{1}_{p}.
\end{equation}
Applying \eqref{eq:duhamel-rotated-transfer} gives the first equality in \eqref{eq:duhamel-spin-component-transfer}.
The gauge rotation $\gamma_{\pi / 2}$ defined in \eqref{eq:gauge-automorphism} leaves $\physham_{\Lambda}$ invariant and maps
$$\widetilde{S}^{1}_{p}
\mapsto
-
\widetilde{S}^{2}_{p},
\quad
\widetilde{S}^{1}_{-p}
\mapsto
-
\widetilde{S}^{2}_{-p}.$$
The Duhamel two-point function defined in \eqref{eq:duhamel-two-point} is invariant under this simultaneous transformation of both observables.
The two minus signs cancel by bilinearity.
This proves the second equality in \eqref{eq:duhamel-spin-component-transfer}.
\end{proof}

\begin{thm}[infrared bound]\label{thm:infrared-bound}
With the momentum-space notation of \eqref{eq:spin-wave},
for every $p \in \dual{\Lambda}$ with $p \neq 0$ and $i = 1, 2$,
it holds that
\begin{equation}\label{eq:infrared-bound}
\rbkt{\widetilde{S}^{i}_{p}}{\widetilde{S}^{i}_{-p}}_{\sminvtemperature,\Lambda}
\leq
\frac{1}{2 \sminvtemperature E_{p}}.
\end{equation}
\end{thm}

\begin{proof}
First take $h$ real.
For each bond $\dbk{xy}$, the scalar $h_{x} - h_{y}$ commutes with the spin operators.
The square in \eqref{eq:deformed-hamiltonian} has the bondwise expansion
$$\begin{aligned}
\frac{1}{2}
\rbk{
S^{1}_{x}
-S^{1}_{y}
-\epsilon
\rbk{h_{x} - h_{y}}}^{2}
=
\frac{1}{2}
\rbk{S^{1}_{x} - S^{1}_{y}}^{2}
-\epsilon
\rbk{S^{1}_{x} - S^{1}_{y}}
\rbk{h_{x} - h_{y}}
+\frac{\epsilon^{2}}{2}
\rbk{h_{x} - h_{y}}^{2}.
\end{aligned}$$
Summing this identity over all bonds defines
\begin{equation}\label{eq:infrared-source-expansion}
\begin{aligned}
\fun{\widehat{K}_{\Lambda}}{\epsilon h}
&=
\fun{\widehat{K}_{\Lambda}}{0}
-\epsilon A_{\Lambda,h}
+\epsilon^{2} c_{\Lambda,h},
\\
A_{\Lambda,h}
&=
\sum_{\dbk{xy}}
\rbk{S^{1}_{x} - S^{1}_{y}}
\rbk{h_{x} - h_{y}
},
\\
c_{\Lambda,h}
&=
\frac{1}{2}
\sum_{\dbk{xy}}
\rbk{h_{x} - h_{y}}^{2}.
\end{aligned}
\end{equation}
The first two derivatives of the Hamiltonian at zero source obtained from \eqref{eq:infrared-source-expansion} are
$$\fnrestr{\opod{\epsilon}
\fun{\widehat{K}_{\Lambda}}{\epsilon h}}
{\epsilon = 0}
=
-A_{\Lambda,h},
\quad
\fnrestr{\od{^{2}}{\epsilon^{2}}
\fun{\widehat{K}_{\Lambda}}{\epsilon h}}
{\epsilon = 0}
=
2 c_{\Lambda,h}.$$

The Duhamel derivative formula for the operator exponential gives
$$\opod{\epsilon}
\napiernum^{-\sminvtemperature
\fun{\widehat{K}_{\Lambda}}{\epsilon h}}
=
-\sminvtemperature
\int_{0}^{1}
\napiernum^{-s
\sminvtemperature
\fun{\widehat{K}}{\epsilon h}}
\od{\fun{\widehat{K}}{\epsilon h}}{\epsilon}
\napiernum^{-
\rbk{1 - s}
\sminvtemperature
\fun{\widehat{K}}{\epsilon h}}
\opdmsr{s}.$$
Taking the trace and using its cyclicity reduces the first derivative of the partition function to
$$\frac{d}{d \epsilon}
\fun{Z}{\epsilon h}
=
-\sminvtemperature
\sqfun{\trace}
{\od{\fun{\widehat{K}}{\epsilon h}}{\epsilon}
\napiernum^{-\sminvtemperature
\fun{\widehat{K}}{\epsilon h}}}.$$
Differentiating this identity once more gives
$$\begin{aligned}
\frac{d^{2}}{d \epsilon^{2}}
\fun{Z}{\epsilon h}
&=
-\sminvtemperature
\sqfun{\trace}
{\od{^{2} \fun{\widehat{K}}{\epsilon h}}{\epsilon^{2}}
\napiernum^{-\sminvtemperature
\fun{\widehat{K}}{\epsilon h}}}
\\ 
&\quad+
\sminvtemperature^{2}
\int_{0}^{1}
\sqfun{\trace}
{\od{\fun{\widehat{K}}{\epsilon h}}{\epsilon}
\napiernum^{-s
\sminvtemperature
\fun{\widehat{K}}{\epsilon h}}
\od{\fun{\widehat{K}}{\epsilon h}}{\epsilon}
\napiernum^{-\rbk{1 - s}
\sminvtemperature
\fun{\widehat{K}}{\epsilon h}}}
\opdmsr{s}.
\end{aligned}$$
Substitution of the derivatives from \eqref{eq:infrared-source-expansion} and the definition \eqref{eq:duhamel-two-point} yields
\begin{equation}\label{eq:infrared-second-variation}
\fnrestr{
\frac{d^{2}}{d \epsilon^{2}}
\fun{Z}{\epsilon h}
}{
\epsilon = 0
}
=
\sminvtemperature^{2}
\fun{Z}{0}
\rbkt{A_{h}}{A_{h}}_{\sminvtemperature,\fun{\widehat{K}}{0}}
-
2
\sminvtemperature
c_{h}
\fun{Z}{0}.
\end{equation}
Theorem \ref{thm:gaussian-domination} makes $\epsilon = 0$ a global maximum of $\epsilon \mapsto \fun{Z}{\epsilon h}$.
Its second derivative in \eqref{eq:infrared-second-variation} is therefore nonpositive.
Since $\sminvtemperature > 0$ and $\fun{Z}{0} > 0$, division by $\sminvtemperature^{2} \fun{Z}{0}$ gives
$$\rbkt{A_{h}}{A_{h}}_{\sminvtemperature,\fun{\widehat{K}}{0}}
\leq
\frac{2 c_{h}}{\sminvtemperature}
=
\frac{1}{\sminvtemperature}
\sum_{\dbk{xy}} \rbk{h_{x} - h_{y}}^{2},
\quad
h \colon \Lambda \to \fldreal.$$
For complex $h = u + \imunit v$ with real $u,v$,
define $A_{h} = A_{u} + \imunit A_{v}$.
Lemma \ref{lem:duhamel-properties} gives
$\rbkt{X}{Y}_{\sminvtemperature,\fun{\widehat{K}}{0}} = \rbkt{Y}{X}_{\sminvtemperature,\fun{\widehat{K}}{0}}$
for self-adjoint $X,Y$.
Sesquilinearity and this symmetry give
$$\begin{aligned}
&\rbkt{\faadj{A_{h}}}{A_{h}}_{\sminvtemperature,\fun{\widehat{K}}{0}}
=
\rbkt{A_{u}}{A_{u}}_{\sminvtemperature,\fun{\widehat{K}}{0}}
+\rbkt{A_{v}}{A_{v}}_{\sminvtemperature,\fun{\widehat{K}}{0}}
\\ 
&\leq
\frac{1}{\sminvtemperature}
\sum_{\dbk{xy}}
\sqbk{\rbk{u_{x} - u_{y}}^{2}
+\rbk{v_{x} - v_{y}}^{2}}
=
\frac{1}{\sminvtemperature}
\sum_{\dbk{xy}}
\abs{h_{x} - h_{y}}^{2},
\end{aligned}$$
the cross terms
$\imunit
\rbk{
\rbkt{A_{u}}{A_{v}}_{\sminvtemperature,\fun{\widehat{K}}{0}}
-
\rbkt{A_{v}}{A_{u}}_{\sminvtemperature,\fun{\widehat{K}}{0}}
}$
cancel by symmetry.
Now choose the plane wave $h_{x} = \napiernum^{\imunit p \cdot x}$, $p \neq 0$; it is well defined with the periodic boundary conditions.
Summation by parts against the discrete Laplacian gives, for each $x$,
$$\sum_{y \colon \abs{y - x} = 1} \rbk{h_{x} - h_{y}} = \rbk{2 d - \sum_{i} 2 \cos p_{i}} h_{x} = 2 E_{p} h_{x},$$
so, regrouping the bond sum by the site carrying the spin operator (each unordered bond $\dbk{xy}$ contributes $S^{1}_{x} \rbk{h_{x} - h_{y}} + S^{1}_{y} \rbk{h_{y} - h_{x}}$),
$$A_{h} = \sum_{x} S^{1}_{x} \sum_{y \colon \abs{y - x} = 1} \rbk{h_{x} - h_{y}} = 2 E_{p} \sum_{x} S^{1}_{x} h_{x} = 2 E_{p} \sqrt{\abscard{\Lambda}} \, \widetilde{S}^{1}_{p}.$$
Also $\sum_{\dbk{xy}} \abs{h_{x} - h_{y}}^{2} = \sum_{x} \sum_{i} \abs{1 - \napiernum^{\imunit p_{i}}}^{2} = 2 \abscard{\Lambda} E_{p}$, since $\abs{1 - \napiernum^{\imunit p_{i}}}^{2} = 2 \rbk{1 - \cos p_{i}}$.
The definition of $A_{h}$ and the sitewise identity $\faadj{V_{\Lambda}} S^{1}_{x} V_{\Lambda} = S^{1}_{x}$, which follows from \eqref{eq:sublattice-rotation-unitary}, give $\faadj{V_{\Lambda}} A_{h} V_{\Lambda} = A_{h}$.
The two plane-wave computations and \eqref{eq:duhamel-rotated-transfer} give
$$4 E_{p}^{2} \abscard{\Lambda}
\rbkt{\widetilde{S}^{1}_{-p}}{\widetilde{S}^{1}_{p}}_{\sminvtemperature,\Lambda}
=
\rbkt{\faadj{A_{h}}}{A_{h}}_{\sminvtemperature,\fun{\widehat{K}}{0}}
\leq
\frac{2 \abscard{\Lambda} E_{p}}{\sminvtemperature},$$
which is \eqref{eq:infrared-bound} for $i = 1$ after using the symmetry of the bracket.
The second equality in \eqref{eq:duhamel-spin-component-transfer} gives the same bound for $i = 2$.
\end{proof}

\section{Finite-Volume Half-Filling from Reflection Positivity}\label{sec:halffilling}

The bond reflection used for the infrared bound has two finite-volume consequences. Its vector implementation places the ground state at half-filling. Its trace implementation shows that the half-filled canonical partition function is maximal. The two statements are proved here before their separate uses in the ground-state and winding arguments.

\subsection{Ground state at half-filling}\label{ground-state-at-half-filling}

The bond reflection identifies ground-state vectors with operators on one half of the lattice. Replacing such an operator by its absolute value preserves the relevant energy and charge estimates and selects the half-filled sector. Divide the periodic box into congruent halves \(\Lambda_{-}\) and \(\Lambda_{+}\). Use the bond-bisecting planes, the reflection \(\theta\), and the crossing-bond set \(M\) defined by \eqref{eq:bond-bisecting-planes}--\eqref{eq:reflection-crossing-bonds}. The definitions \eqref{eq:reflection-hilbert-spaces} and \eqref{eq:reflection-unitary} fix \(\sphilb{H}_{\pm}\) and \(U_{\theta}\). The tensor identification used below is \eqref{eq:reflection-unitary-identification}.

The half-lattice Hamiltonian associated with the internal bond set \(\setindex{B}_{-}\) from \eqref{eq:reflection-bond-decomposition} is \begin{equation}\label{eq:half-lattice-hamiltonian}
\physham_{-}
=
-\sum_{\dbk{xy} \in \setindex{B}_{-}}
\rbk{S^{1}_{x}S^{1}_{y}+S^{2}_{x}S^{2}_{y}}
+\lambda\sum_{x \in \Lambda_{-}}
\sqbk{\frac{1}{2}+\rbk{-1}^{x}S^{3}_{x}}.
\end{equation} Let \(M_{-}=\set{x \in \Lambda_{-}}{\dbk{x\,\theta(x)} \in M}\) be the left endpoints of the crossing bonds. The right-half rotation from Lemma \ref{lem:sublattice-rotation}, the relabeling \eqref{eq:reflection-relabeling-intertwining}, and the identification \eqref{eq:reflection-unitary-identification} give \begin{equation}\label{eq:half-filling-transformed-hamiltonian}
\physham_{\Lambda,\theta}
=
\physham_{-} \otimes 1+1 \otimes \physham_{-}
-\frac{1}{2}\sum_{x \in M_{-}}
\rbk{S^{+}_{x} \otimes S^{+}_{x}+S^{-}_{x} \otimes S^{-}_{x}}.
\end{equation} Under the same transformation, the half-lattice and transformed charges are \begin{equation}\label{eq:half-filling-transformed-charge}
S^{3}_{\txttot,-}=\sum_{x \in \Lambda_{-}}S^{3}_{x},
\quad
S^{3}_{\Lambda,\theta}
=S^{3}_{\txttot,-} \otimes 1-1 \otimes S^{3}_{\txttot,-}.
\end{equation}

The real configuration basis identifies a vector \(\Psi\) with its vectorization \(\widehat{\Psi}\) as in \eqref{eq:vector-operator-dictionary}. For this basis, \(S^{1}\) and \(S^{3}\) are symmetric, \(\latp{\rbk{S^{\pm}}}=S^{\mp}\), and all matrices of \(\physham_{-}\) are real.

\begin{thm}[half-filling; Appendix A of \cite{AizenmanLiebSeiringerSolovejYngvason001}]\label{thm:half-filling}
The Hamiltonian $\physham_{\Lambda}$ has a unique ground state, and it lies in the sector $S^{3}_{\txttot,\Lambda} = 0$, i.e., at particle number $\frac{\abscard{\Lambda}}{2}$.
\end{thm}

\begin{proof}
We work with $\physham_{\Lambda,\theta}$ and
$S^{3}_{\Lambda,\theta}$ from
\eqref{eq:half-filling-transformed-hamiltonian} and
\eqref{eq:half-filling-transformed-charge}, respectively.
The unitary rotation transfers the conclusion back to the original operators.
Let $\Psi$ be a normalized ground-state vector of
$\physham_{\Lambda,\theta}$.
Since
$\commutator{\physham_{\Lambda,\theta}}{S^{3}_{\Lambda,\theta}}=0$,
we may choose
$S^{3}_{\Lambda,\theta}\Psi
=
m\Psi$.
We prove $m=0$.

The rotation identifies the $S^{3}_{\Lambda,\theta}$ sectors with the
$S^{3}_{\txttot,\Lambda}$ sectors.
The sector ground states are simple by Lemma \ref{lem:perron-frobenius}.
Lemma \ref{lem:finite-perron-frobenius} also applies directly to
$\physham_{\Lambda,\theta}$.
Its off-diagonal hopping and pair terms are nonpositive.
They connect all configurations in a fixed
$S^{3}_{\Lambda,\theta}$ sector.
Intra-half hops preserve the particle number in each half.
Crossing pair terms change the two half-lattice particle numbers by the same amount and preserve
$S^{3}_{\Lambda,\theta}$.
By \eqref{eq:vector-operator-dictionary},
\begin{equation}\label{eq:halffilling-energy}
\begin{aligned}
\bkt{\Psi}{\physham_{\Lambda,\theta} \Psi}
&=
\sqfun{\trace}{
\widehat{\Psi}
\faadj{\widehat{\Psi}}
\physham_{-}
}
+
\sqfun{\trace}{
\faadj{\widehat{\Psi}}
\widehat{\Psi}
\physham_{-}
}
\\ 
&\quad-
\frac{1}{2}
\sum_{x \in M_{-}}
\rbk{
\sqfun{\trace}{
\faadj{\widehat{\Psi}}
S^{+}_{x}
\widehat{\Psi}
S^{-}_{x}
}
+
\sqfun{\trace}{
\faadj{\widehat{\Psi}}
S^{-}_{x}
\widehat{\Psi}
S^{+}_{x}
}
}.
\end{aligned}
\end{equation}
using $\latp{\physham_{-}} = \physham_{-}$ (real symmetric) and $\latp{\rbk{S^{\pm}}} = S^{\mp}$.
Define
$$\begin{aligned}
\widehat{\Psi}_{1}
=
\rbk{\widehat{\Psi}\faadj{\widehat{\Psi}}}^{1/2},
\quad
\widehat{\Psi}_{2}
=
\rbk{\faadj{\widehat{\Psi}}\widehat{\Psi}}^{1/2}.
\end{aligned}$$
Both operators have unit Hilbert--Schmidt norm.
Let $\Psi_{1}$ and $\Psi_{2}$ be the corresponding vectors.
The two crossing summands in \eqref{eq:halffilling-energy} are complex conjugates.
Their sum is
$$2\opreal\sqfun{\trace}{\faadj{\widehat{\Psi}}S^{+}_{x}
\widehat{\Psi}S^{-}_{x}}.$$
Apply Lemma \ref{lem:trace-schwarz} with $A=S^{+}_{x}$ and then use the arithmetic-geometric mean inequality.
This gives
$$2 \opreal \sqfun{\trace}{\faadj{\widehat{\Psi}} S^{+}_{x} \widehat{\Psi} S^{-}_{x}}
\leq \sqfun{\trace}{\widehat{\Psi}_{1} S^{+}_{x} \widehat{\Psi}_{1} S^{-}_{x}} + \sqfun{\trace}{\widehat{\Psi}_{2} S^{+}_{x} \widehat{\Psi}_{2} S^{-}_{x}},$$
For the positive matrices $\widehat{\Psi}_{i}$, the traces on the right are real.
For $i\in\setone{1,2}$, cyclicity gives
$$\sqfun{\trace}{\widehat{\Psi}_{i}S^{+}_{x}\widehat{\Psi}_{i}S^{-}_{x}}
=
\sqfun{\trace}{\widehat{\Psi}_{i}S^{-}_{x}\widehat{\Psi}_{i}S^{+}_{x}}.$$

The diagonal terms satisfy the exact identity
$$\begin{aligned}
\sqfun{\trace}{\widehat{\Psi}\faadj{\widehat{\Psi}}\physham_{-}}
+\sqfun{\trace}{\faadj{\widehat{\Psi}}\widehat{\Psi}\physham_{-}}
=\frac{1}{2}\sum_{i=1,2}
\rbk{
\sqfun{\trace}{\widehat{\Psi}_{i}\faadj{\widehat{\Psi}_{i}}\physham_{-}}
+\sqfun{\trace}{\faadj{\widehat{\Psi}_{i}}\widehat{\Psi}_{i}\physham_{-}}
}.
\end{aligned}$$
Both sides contain the same sum of
$\widehat{\Psi}\faadj{\widehat{\Psi}}$ and
$\faadj{\widehat{\Psi}}\widehat{\Psi}$.
Combining this identity with the crossing bound in
\eqref{eq:halffilling-energy} gives
\begin{equation}\label{eq:vector-rp}
\begin{aligned}
\bkt{\Psi}{\physham_{\Lambda,\theta} \Psi}
&\geq
\frac{1}{2}
\bkt{\Psi_{1}}{\physham_{\Lambda,\theta} \Psi_{1}}
+
\frac{1}{2}
\bkt{\Psi_{2}}{\physham_{\Lambda,\theta} \Psi_{2}}.
\end{aligned}
\end{equation}
The eigenvalue equation for $S^{3}_{\Lambda,\theta}$ becomes
$$S^{3}_{\txttot,-}\widehat{\Psi}
-
\widehat{\Psi}S^{3}_{\txttot,-}
=
m\widehat{\Psi}.$$
The operator $S^{3}_{\txttot,-}$ is real and diagonal.
The operator $\widehat{\Psi}$ maps the $S^{3}_{\txttot,-}$ eigenspace with eigenvalue $n-m$
into the eigenspace with eigenvalue $n$.
Both
$\widehat{\Psi}\faadj{\widehat{\Psi}}$ and
$\faadj{\widehat{\Psi}}\widehat{\Psi}$ commute with
$S^{3}_{\txttot,-}$.
Their positive square roots also commute with $S^{3}_{\txttot,-}$.
It follows that
$$\begin{aligned}
S^{3}_{\Lambda,\theta}\Psi_{1}&=0,\\
S^{3}_{\Lambda,\theta}\Psi_{2}&=0.
\end{aligned}$$

The vector $\Psi$ is a ground-state vector.
Inequality \eqref{eq:vector-rp} forces $\Psi_{1}$ and $\Psi_{2}$ to be
ground-state vectors as well.
Both belong to the sector $S^{3}_{\Lambda,\theta}=0$.
Simplicity in this sector makes them equal up to phase.
Their representing matrices are positive, so in fact
$$\widehat{\Psi}\faadj{\widehat{\Psi}}
=\faadj{\widehat{\Psi}}\widehat{\Psi}.$$
The equality of the two positive matrices gives
$$\begin{aligned}
m
=
\bkt{\Psi}{S^{3}_{\Lambda,\theta} \Psi}
=
\sqfun{\trace}{
\widehat{\Psi}
\faadj{\widehat{\Psi}}
S^{3}_{\txttot,-}}
-\sqfun{\trace}{
\faadj{\widehat{\Psi}}
\widehat{\Psi}
S^{3}_{\txttot,-}}
=
0.
\end{aligned}$$
This proves that every ground-state vector which is an
$S^{3}_{\Lambda,\theta}$ eigenvector has eigenvalue $0$.
Let $\Phi$ be an arbitrary ground-state vector and decompose it into its
$S^{3}_{\Lambda,\theta}$-sector components.
Since
$\commutator{\physham_{\Lambda,\theta}}{S^{3}_{\Lambda,\theta}}=0$,
every nonzero component is again a ground-state vector and an
$S^{3}_{\Lambda,\theta}$ eigenvector.
The preceding conclusion eliminates every component except the one in the
sector $S^{3}_{\Lambda,\theta}=0$.
Thus every ground-state vector lies in the sector
$S^{3}_{\Lambda,\theta}=0$.
Simplicity in this sector proves uniqueness.
\end{proof}

\subsection{Maximality of the half-filled partition function}\label{maximality-of-the-half-filled-partition-function}

The trace version of the same reflected Hamiltonian compares the canonical partition functions sector by sector. It supplies the normalization estimate needed for the winding bound in Theorem \ref{thm:winding-estimate}.

\begin{thm}[maximality of the half-filled partition function]\label{thm:partition-maximal}
For all $\beta > 0$ and all $k$ for which the sector projection
\eqref{eq:main-sector-projection} is nonzero,
it follows that
$$\sqfun{\trace}{P_{k} \napiernum^{-\beta \physham_{\Lambda}}}
\leq
\sqfun{\trace}{P_{0} \napiernum^{-\beta \physham_{\Lambda}}}.$$
\end{thm}

\begin{proof}
The notation
$\physham_{-}$,
$M_{-}$,
$\physham_{\Lambda,\theta}$,
and $S^{3}_{\txttot,-}$ is fixed by
\eqref{eq:half-lattice-hamiltonian}--\eqref{eq:half-filling-transformed-charge}.
In the transformed picture, set
$$\begin{aligned}
\fun{Z}{m}
=
\sqfun{\trace}{P_{\theta,m}\napiernum^{-\beta\physham_{\Lambda,\theta}}},
\quad
P_{\theta,m}
=
\sum_{\substack{
n \in \opvarspec{S^{3}_{\txttot,-}} \\
n-m \in \opvarspec{S^{3}_{\txttot,-}}
}}
P_{n}\otimes P_{n-m}.
\end{aligned}$$
Here $P_{n}$ is the real spectral projection of
$S^{3}_{\txttot,-}$ for the eigenvalue $n$.
The claim becomes $\fun{Z}{m}\leq\fun{Z}{0}$.
By the Lie--Trotter formula,
$$\begin{aligned}
\fun{Z}{m}
&=
\lim_{k \to \infty}
\sqfun{\trace}{
P_{\theta,m}
\rbk{
\napiernum^{-\frac{\beta}{k} \physham_{-} \otimes 1}
\napiernum^{-\frac{\beta}{k} 1 \otimes \physham_{-}}
\napiernum^{
\frac{\beta}{2 k}
\sum_{x \in M_{-}}
\rbk{
S^{+}_{x} \otimes S^{+}_{x}
+
S^{-}_{x} \otimes S^{-}_{x}
}
}
}^{k}
}.
\end{aligned}$$
The nilpotence $\rbk{S^{\pm}_{x}}^{2}=0$ gives the finite-product expansion
$$\napiernum^{\frac{\beta}{2 k} \sum_{x \in M_{-}} \rbk{S^{+}_{x} \otimes S^{+}_{x} + S^{-}_{x} \otimes S^{-}_{x}}}
=
\prod_{x \in M_{-}}
\rbk{1 + \frac{\beta}{2 k} S^{+}_{x} \otimes S^{+}_{x}}
\rbk{1 + \frac{\beta}{2 k} S^{-}_{x} \otimes S^{-}_{x}}
+\fun{O}{k^{-2}}.$$
Factors associated with distinct sites commute.
The $\fun{O}{k^{-2}}$ errors are uniformly summable in the Trotter limit.

Expand every product.
Each Trotter slot receives a monomial in operators
$\rbk{\beta/2k}^{1/2}S^{\pm}_{x}$.
The same monomial occurs in both tensor factors.
After the sum over $n$, each term factorizes into two real traces:
$$\begin{aligned}
A&=\sqfun{\trace_{\sphilb{H}_{-}}}{
P_{n}\napiernum^{-\frac{\beta}{k}\physham_{-}}T_{1}
\napiernum^{-\frac{\beta}{k}\physham_{-}}T_{2}\dotsm},\\
B&=\sqfun{\trace_{\sphilb{H}_{-}}}{
P_{n-m}\napiernum^{-\frac{\beta}{k}\physham_{-}}T_{1}\dotsm}.
\end{aligned}$$
The second trace is the same functional evaluated at the shifted projection.
Write $A_{n}$ for its value at $P_{n}$.
The corresponding contribution to $\fun{Z}{m}$ is
$\sum_{n}A_{n}A_{n-m}$.
The elementary inequality $2ab\leq a^{2}+b^{2}$ gives
$$\sum_{n} A_{n} A_{n - m} \leq \frac{1}{2} \sum_{n} \rbk{A_{n}^{2} + A_{n - m}^{2}} = \sum_{n} A_{n}^{2}.$$
The last expression is the corresponding contribution to $\fun{Z}{0}$.
Summing over all monomial assignments and taking the Trotter limit gives $\fun{Z}{m} \leq \fun{Z}{0}$.
\end{proof}

\section{The Falk--Bruch Inequality}\label{sec:falkbruch}

The infrared bound \eqref{eq:infrared-bound} controls the Duhamel two-point function, while condensation is a statement about the ordinary thermal two-point function. The passage between the two is the Falk--Bruch inequality, quoted in \cite{AizenmanLiebSeiringerSolovejYngvason001} from \cite[Theorem 3.1 and Corollary 3.2]{DysonLiebSimon001}; this section proves it. Throughout, \(\physham\) is a self-adjoint matrix, \(Z_{\sminvtemperature,\physham}
=
\sqfun{\trace}{\fnexp{-\sminvtemperature \physham}}\), and its Gibbs state is denoted by \(\oastate[\psi_{\sminvtemperature,\physham}]\). The corresponding Duhamel two-point function is \(\rbkt{\cdot}{\cdot}_{\sminvtemperature,\physham}\) defined by \eqref{eq:duhamel-two-point}.

For an operator \(X\) define the three quadratic quantities \begin{equation}\label{eq:fb-quantities}
\fun{b}{X}
=
\frac{1}{2}
\fun{\oastate[\psi_{\sminvtemperature,\physham}]}{
\faadj{X} X
+
X \faadj{X}
},
\quad
\fun{g}{X}
=
\rbkt{X}{\faadj{X}}_{\sminvtemperature,\physham},
\quad
\fun{c}{X}
=
\fun{\oastate[\psi_{\sminvtemperature,\physham}]}{
\commutator{\faadj{X}}{
\commutator{\sminvtemperature \physham}{X}
}
}.
\end{equation}

\begin{lem}[spectral form]\label{lem:fb-spectral}
Let $\seq{\Phi_{j}}{j \in J}$ be an orthonormal eigenbasis of $\physham$, indexed by a finite set $J$.
The eigenvalues, matrix elements and Gibbs weights are defined by
$$\physham \Phi_{j}
=
\varepsilon_{j} \Phi_{j},
\quad
X_{j k}
=
\bkt{\Phi_{j}}{X \Phi_{k}},
\quad
w_{j}
=
\frac{\fnexp{-\sminvtemperature \varepsilon_{j}}}{Z_{\sminvtemperature,\physham}}.$$
Their spectral representations are
$$\begin{aligned}
\fun{b}{X}
&=
\frac{1}{2}
\sum_{j, k \in J}
\abs{X_{j k}}^{2} \rbk{w_{j} + w_{k}},
\\ 
\fun{g}{X}
&=
\sum_{j, k \in J}
\abs{X_{j k}}^{2}
\begin{cases}
\displaystyle
\frac{w_{j} - w_{k}}
{\sminvtemperature \rbk{\varepsilon_{k} - \varepsilon_{j}}},
&
\varepsilon_{j}
\neq
\varepsilon_{k},
\\
w_{j},
&
\varepsilon_{j}
=
\varepsilon_{k},
\end{cases}
\\ 
\fun{c}{X}
&=
\sminvtemperature
\sum_{j, k \in J}
\abs{X_{j k}}^{2} \rbk{\varepsilon_{k} - \varepsilon_{j}}
\rbk{w_{j} - w_{k}},
\end{aligned}$$
In particular it holds that $\fun{c}{X} \geq 0$ and $\fun{g}{X} \geq 0$.
\end{lem}

\begin{proof}
The adjoint matrix elements satisfy
$\rbk{\faadj{X}}_{k j}
=
\cmpconj{X_{j k}}$.
Assertion (1) of Lemma \ref{lem:duhamel-properties},
applied to $\rbk{A, B} = \rbk{X, \faadj{X}}$, gives the complete spectral calculation
$$\begin{aligned}
\fun{g}{X}
&=
\rbkt{X}{\faadj{X}}_{\sminvtemperature,\physham}
\\
&=
\frac{1}{Z_{\sminvtemperature,\physham}}
\sum_{j, k \in J}
\abs{X_{j k}}^{2}
\begin{cases}
\displaystyle
\frac{
\fnexp{-\sminvtemperature \varepsilon_{j}}
-
\fnexp{-\sminvtemperature \varepsilon_{k}}
}{
\sminvtemperature
\rbk{
\varepsilon_{k}
-
\varepsilon_{j}
}
},
&
\varepsilon_{j}
\neq
\varepsilon_{k},
\\
\fnexp{-\sminvtemperature \varepsilon_{j}},
&
\varepsilon_{j}
=
\varepsilon_{k},
\end{cases}
\\
&=
\sum_{j, k \in J}
\abs{X_{j k}}^{2}
\begin{cases}
\displaystyle
\frac{
w_{j}
-
w_{k}
}{
\sminvtemperature
\rbk{
\varepsilon_{k}
-
\varepsilon_{j}
}
},
&
\varepsilon_{j}
\neq
\varepsilon_{k},
\\
w_{j},
&
\varepsilon_{j}
=
\varepsilon_{k}.
\end{cases}
\end{aligned}$$
Each coefficient in the last sum is nonnegative because $w_{j}$ decreases as $\varepsilon_{j}$ increases.
This proves the formula for $\fun{g}{X}$ and its nonnegativity.

The two Gibbs expectations in the definition of $\fun{b}{X}$ are
$$\begin{aligned}
\fun{\oastate[\psi_{\sminvtemperature,\physham}]}{X \faadj{X}}
=
\sum_{j, k \in J}
w_{j}
\abs{X_{j k}}^{2},
\quad
\fun{\oastate[\psi_{\sminvtemperature,\physham}]}{\faadj{X} X}
=
\sum_{j, k \in J}
w_{k}
\abs{X_{j k}}^{2}.
\end{aligned}$$
Their average gives the formula for $\fun{b}{X}$ in the lemma.

The matrix elements of the inner commutator in $\fun{c}{X}$ are
$$\commutator{\sminvtemperature \physham}{X}_{k j}
=
\sminvtemperature
\rbk{
\varepsilon_{k}
-
\varepsilon_{j}
}
X_{k j}.$$
Expanding the outer commutator inside the trace and relabeling the summation indices of its second term gives
$$\fun{\oastate[\psi_{\sminvtemperature,\physham}]}
{\commutator{\faadj{X}}{\commutator{\sminvtemperature \physham}{X}}}
= \sum_{j, k \in J} \rbk{w_{j} - w_{k}} \cmpconj{X_{k j}} \commutator{\sminvtemperature \physham}{X}_{k j}
= \sminvtemperature \sum_{j, k \in J} \abs{X_{k j}}^{2} \rbk{\varepsilon_{k} - \varepsilon_{j}} \rbk{w_{j} - w_{k}}.$$
A final relabeling $j \leftrightarrow k$ gives the formula for $\fun{c}{X}$ in the lemma.
Each summand is nonnegative because $w_{j}$ decreases as $\varepsilon_{j}$ increases:
i.e., it holds that
$\rbk{\varepsilon_{k} - \varepsilon_{j}}
\rbk{w_{j} - w_{k}}
\geq
0$.
\end{proof}

The per-pair quantities obey an exact hyperbolic identity, and the global inequality follows by concavity.

\begin{lem}[concavity of the coth kernel]\label{lem:coth-concavity}
Define $f_{\coth}
\colon \nonneginterval
\to \fldreal$ as $$\fun{f_{\coth}}{u}
=
\begin{dcases}
\sqrt{u} \coth \sqrt{u}, & u > 0, \\
1, & u = 0.
\end{dcases}$$
The function $f_{\coth}$ is increasing and concave, and it admits the
expansion
$$\fun{f_{\coth}}{u} = 1 + \sum_{n = 1}^{\infty} \frac{2 u}{u + \pi^{2} n^{2}},
\quad
u \geq 0.$$
\end{lem}

\begin{proof}
The classical partial-fraction expansion of the hyperbolic cotangent,
$$\coth t = \frac{1}{t} + \sum_{n = 1}^{\infty} \frac{2 t}{t^{2} + \pi^{2} n^{2}},
\quad
t > 0,$$
follows from the Mittag--Leffler expansion of $\cot$ after the substitution
$t\mapsto\imunit t$.
It also follows directly from the Fourier series of $\cosh$ on
$\closedinterval{-\pi}{\pi}$.
Substitute $t=\sqrt{u}$ and multiply by $\sqrt{u}$ to obtain the asserted
series.
Each function
$u\mapsto 2u/\rbk{u+\pi^{2}n^{2}}$ is increasing and concave on
$\rightopeninterval{0}{\infty}$.
Its first two derivatives are
$$\begin{aligned}
\opod{u}
\frac{2u}{u+\pi^{2}n^{2}}
=\frac{2\pi^{2}n^{2}}{\rbk{u+\pi^{2}n^{2}}^{2}}
>
0,
\quad
\od{^2}{u^2}
\frac{2u}{u+\pi^{2}n^{2}}
=
-\frac{4\pi^{2}n^{2}}{\rbk{u+\pi^{2}n^{2}}^{3}}
<
0.
\end{aligned}$$
The series and its derivatives converge locally uniformly.
The function $f_{\coth}$ is therefore the locally uniform limit of
increasing concave partial sums.
\end{proof}

\begin{lem}[joint concavity]\label{lem:fb-joint-concavity}
For $g \geq 0$ and $c \geq 0$, define
\begin{equation}\label{eq:fb-joint-concavity-definition}
\fun{\mathcal{G}}{g, c}
=
\begin{cases}
\displaystyle
\frac{1}{2}
\sqrt{g c}
\coth \sqrt{\frac{c}{4 g}}
=
g \cdot \fun{f_{\coth}}{\frac{c}{4 g}},
&
g > 0,
c > 0,
\\
g,
&
g > 0,
c = 0,
\\
0,
&
g = 0,
c > 0,
\\
0,
&
g = 0,
c = 0.
\end{cases}
\end{equation}
The first case is the perspective of the function $f_{\coth}$ of Lemma \ref{lem:coth-concavity}.
The function $\mathcal{G}$ is positively homogeneous of degree one,
jointly concave, superadditive, and nondecreasing in each argument.
\end{lem}

\begin{proof}
The equality between the two expressions in the first case follows by setting $u = c / 4 g$.
The resulting computation is
$$g \fun{f_{\coth}}{u}
=
g \sqrt{u} \coth \sqrt{u}
=
\frac{1}{2}
\sqrt{g c} \coth \sqrt{c / 4 g}.$$
For fixed $g > 0$, the limit $\fun{f_{\coth}}{0} = 1$ gives
$$\lim_{c \downarrow 0}
\fun{\mathcal{G}}{g, c}
=
g.$$
For $g > 0$ and $c > 0$, the elementary inequality $\coth v \leq 1 + 1 / v$ gives
$$0
\leq
\fun{\mathcal{G}}{g, c}
\leq
g
+
\frac{1}{2}
\sqrt{g c}.$$
This bound shows that $\fun{\mathcal{G}}{g, c} \to 0$ when $g \downarrow 0$ with $c > 0$ fixed and when $\rbk{g, c} \to \rbk{0, 0}$ inside the closed quadrant.
The four cases in the definition therefore give the continuous extension of the perspective to the complete closed quadrant,
including the explicit value $\fun{\mathcal{G}}{0, 0} = 0$.
Homogeneity $\fun{\mathcal{G}}{t g, t c} = t \fun{\mathcal{G}}{g, c}$ is clear from this form.

Joint concavity is the standard property of the perspective of a concave function:
for $0 < s < 1$ and pairs $\rbk{g_{i}, c_{i}}$ with $g_{i} > 0$,
writing $g = s g_{1} + \rbk{1 - s} g_{2}$,
$$\begin{aligned}
&g
\fun{f_{\coth}}{\frac{s c_{1} + \rbk{1 - s} c_{2}}{4 g}}
=
g \fun{f_{\coth}}{\frac{s g_{1}}{g} \frac{c_{1}}{4 g_{1}}
+\frac{\rbk{1 - s} g_{2}}{g} \frac{c_{2}}{4 g_{2}}}
\\ 
&\geq
s g_{1} \fun{f_{\coth}}{\frac{c_{1}}{4 g_{1}}}
+\rbk{1 - s} g_{2} \fun{f_{\coth}}{\frac{c_{2}}{4 g_{2}}},
\end{aligned}$$
by concavity of $f_{\coth}$ applied with the convex weights $s g_{1} / g$ and $\rbk{1 - s} g_{2} / g$;
the degenerate cases $g_{i} = 0$ follow by continuity.
Superadditivity follows from concavity and homogeneity:
$$\fun{\mathcal{G}}{g_{1} + g_{2}, c_{1} + c_{2}}
=
2 \fun{\mathcal{G}}{\frac{1}{2} \rbk{g_{1}, c_{1}} + \frac{1}{2} \rbk{g_{2}, c_{2}}}
\geq
\fun{\mathcal{G}}{g_{1}, c_{1}} + \fun{\mathcal{G}}{g_{2}, c_{2}}.$$

For fixed $g$, the function $\fun{\mathcal{G}}{g,c}$ is nondecreasing in
$c$.
With $v = \sqrt{c / 4 g}$,
we obtain
$\fun{\mathcal{G}}{g, c}
=
g v \coth v$ and $v \coth v$ is increasing in $v \geq 0$:
its derivative
$$\coth v - \frac{v}{\sinh^{2} v}
=
\frac{\cosh v \sinh v - v}{\sinh^{2} v}
>
0$$
since $\sinh 2 v > 2 v$,
while $v$ is increasing in $c$.

For fixed $c$, the function $\fun{\mathcal{G}}{g,c}$ is nondecreasing in
$g$.
We write $\fun{\mathcal{G}}{g, c} = \frac{c}{4} \frac{\coth v}{v}$ with $v$ as above.
The quotient $\coth v/v$ is decreasing in $v$ (both factors are positive,
$\coth$ decreasing, $1 / v$ decreasing)
and $v$ is decreasing in $g$,
so $\mathcal{G}$ is nondecreasing in $g$.
\end{proof}

\begin{thm}[Falk--Bruch inequality]\label{thm:falk-bruch}
For every operator $X$, it follows that
\begin{equation}\label{eq:falk-bruch-inequality}
\fun{b}{X}
\leq
\fun{\mathcal{G}}{\fun{g}{X}, \fun{c}{X}}
=
\frac{1}{2}
\sqrt{\fun{g}{X} \fun{c}{X}}
\coth \sqrt{\frac{\fun{c}{X}}{4 \fun{g}{X}}}.
\end{equation}
\end{thm}

\begin{proof}
By Lemma \ref{lem:fb-spectral} all three quantities are sums over ordered pairs $\rbk{j, k}$ of nonnegative per-pair contributions $b_{j k}, g_{j k}, c_{j k}$.
Fix a pair with $\varepsilon_{j} \neq \varepsilon_{k}$ and set $u_{j k} = \sminvtemperature \rbk{\varepsilon_{k} - \varepsilon_{j}} / 2$; then,
with $m = \abs{X_{j k}}^{2}$,
it holds that
$$b_{j k} = \frac{m}{2} \rbk{w_{j} + w_{k}},
\quad
g_{j k} = m \frac{w_{j} - w_{k}}{2 u_{j k}},
\quad
c_{j k} = 2 m u_{j k} \rbk{w_{j} - w_{k}},$$
so that
$$g_{j k} c_{j k} = m^{2} \rbk{w_{j} - w_{k}}^{2},
\quad
\frac{c_{j k}}{4 g_{j k}} = u_{j k}^{2},
\quad
\frac{w_{j} + w_{k}}{w_{j} - w_{k}} = \coth u_{j k},$$
the last identity because $w_{j} / w_{k} = \fnexp{2 u_{j k}}$.
This identity gives
$$b_{j k} = \frac{m}{2} \rbk{w_{j} - w_{k}} \coth u_{j k} = \frac{1}{2} \sqrt{g_{j k} c_{j k}} \coth \sqrt{\frac{c_{j k}}{4 g_{j k}}} = \fun{\mathcal{G}}{g_{j k}, c_{j k}},$$
an exact identity; for a pair with $\varepsilon_{j} = \varepsilon_{k}$ one has $c_{j k} = 0$, $b_{j k} = m w_{j} = g_{j k} = \fun{\mathcal{G}}{g_{j k}, 0}$, again an identity.
Superadditivity of $\mathcal{G}$ (Lemma \ref{lem:fb-joint-concavity}), extended from two summands to arbitrary finite sums by iteration, gives
$$\fun{b}{X} = \sum_{j, k \in J} b_{j k} = \sum_{j, k \in J} \fun{\mathcal{G}}{g_{j k}, c_{j k}} \leq \fun{\mathcal{G}}{\sum_{j, k \in J} g_{j k}, \sum_{j, k \in J} c_{j k}} = \fun{\mathcal{G}}{\fun{g}{X}, \fun{c}{X}}.$$
\end{proof}

\begin{cor}[bound through the infrared estimate]\label{cor:fb-two-operators}
Let $X_{1}, X_{2}$ be operators
and let $\bar{g}, \bar{c}
\in \nonneginterval$ be numbers satisfying
$\fun{g}{X_{1}}
+\fun{g}{X_{2}}
\leq
\bar{g}$ and
$\fun{c}{X_{1}}
+\fun{c}{X_{2}}
\leq
\bar{c}$.
One has
$$\fun{b}{X_{1}}
+
\fun{b}{X_{2}}
\leq
\fun{\mathcal{G}}{\bar{g}, \bar{c}}.$$
If $\bar{g} > 0$ and $\bar{c} > 0$, the right side is
$$\fun{\mathcal{G}}{\bar{g}, \bar{c}}
=
\frac{1}{2}
\sqrt{\bar{g} \bar{c}}
\coth
\sqrt{\frac{\bar{c}}{4 \bar{g}}}.$$
The boundary values are defined in
\eqref{eq:fb-joint-concavity-definition}.
\end{cor}

\begin{proof}
Theorem \ref{thm:falk-bruch},
superadditivity and the monotonicity in both arguments from Lemma \ref{lem:fb-joint-concavity} give
$$\begin{aligned}
\fun{b}{X_{1}}
+
\fun{b}{X_{2}}
\leq
\fun{\mathcal{G}}
{\fun{g}{X_{1}}
+\fun{g}{X_{2}},
\fun{c}{X_{1}}
+\fun{c}{X_{2}}}
\leq
\fun{\mathcal{G}}{\bar{g}, \bar{c}}.
\end{aligned}$$
\end{proof}

\begin{rem}\label{rem:fb-simplification}
For $\bar{g} > 0$ and $\bar{c} > 0$, the elementary bound $\coth v \leq 1 + 1 / v$ converts the conclusion into the additively split form
$$\fun{b}{X_{1}} + \fun{b}{X_{2}} \leq \frac{1}{2} \sqrt{\bar{g} \bar{c}} + \bar{g},$$
since $\frac{1}{2} \sqrt{\bar{g} \bar{c}} \cdot \rbk{4 \bar{g} / \bar{c}}^{\onehalf} = \bar{g}$.
The same inequality at $\bar{g} = 0$ or $\bar{c} = 0$ follows from the boundary values in Corollary \ref{cor:fb-two-operators}.
Applied modewise, this estimate gives
\eqref{eq:mode-occupation-bound}.
The proof of Theorem \ref{thm:bec} sums that bound over
$p \neq 0$.
\end{rem}

\section{Existence of Bose--Einstein Condensation}\label{sec:bec}

The infrared bound and the Falk--Bruch inequality combine with a sum rule into a lower bound for the zero-momentum occupation, which is Theorem 1 of \cite{AizenmanLiebSeiringerSolovejYngvason001}. The finite-volume Gibbs state in this section is \(\oastate[\psi_{\sminvtemperature,\Lambda}]\) from \eqref{eq:main-periodic-gibbs-state}. Every Duhamel two-point function carries the subscript \(\rbk{\sminvtemperature,\Lambda}\), and the quantities in \eqref{eq:fb-quantities} are evaluated with \(\physham
=
\physham_{\Lambda}\). The unitary equivalence \eqref{eq:rotated-zero-source-hamiltonian} permits the same quantities to be computed with the rotated Hamiltonian.

\subsection{The double commutator}\label{the-double-commutator}

The Falk--Bruch correction is determined by an exact double-commutator calculation for the Fourier spin fields. The resulting expression relates each momentum mode to the Hamiltonian, the staggered field, and the longitudinal bond correlations.

\begin{lem}[double commutator identity]\label{lem:double-commutator}
For every $p \in \dual{\Lambda}$,
$$-\imunit \sum_{i = 1, 2} \commutator{\widetilde{S}^{i}_{p}}{\fun{\oaderiv_{\Lambda}}{\widetilde{S}^{i}_{-p}}}
= -\frac{2}{\abscard{\Lambda}} \rbk{\physham_{\Lambda} - \frac{\lambda \abscard{\Lambda}}{2} + 2 \sum_{\dbk{xy}} S^{3}_{x} S^{3}_{y} \cos p \cdot \rbk{x - y}}.$$
\end{lem}

\begin{proof}
The commutator of different sites vanishes.
For $y\sim z$, the two terms of \eqref{eq:hamiltonian-spin} which contribute to the first component give
$$\begin{aligned}
\commutator{-S^{2}_{z}S^{2}_{y}}
{S^{1}_{z}}
&=
-\commutator{S^{2}_{z}}
{S^{1}_{z}}
S^{2}_{y}
=
\imunit
S^{3}_{z}
S^{2}_{y},
\\ 
\commutator{\lambda \rbk{-1}^{z} S^{3}_{z}}
{S^{1}_{z}}
&=
\imunit
\lambda
\rbk{-1}^{z}
S^{2}_{z}.
\end{aligned}$$
The periodic finite-volume derivation
\eqref{eq:periodic-finite-volume-dynamics} therefore gives
$$\fun{\oaderiv_{\Lambda}}{S^{1}_{z}}
=
-
\sum_{\substack{y\in\Lambda\\y\sim z}}
S^{3}_{z}S^{2}_{y}
-
\lambda
\rbk{-1}^{z}
S^{2}_{z}.$$

The commutator of $S^{1}_{w}$ with $\fun{\oaderiv_{\Lambda}}{S^{1}_{z}}$ expands into three terms:
$$\begin{aligned}
-\imunit
\commutator{S^{1}_{w}}
{\fun{\oaderiv_{\Lambda}}{S^{1}_{z}}}
=
\imunit
\sum_{\substack{y\in\Lambda\\y\sim z}}
\commutator{S^{1}_{w}}
{S^{3}_{z}}
S^{2}_{y}
+\imunit
\sum_{\substack{y\in\Lambda\\y\sim z}}
S^{3}_{z}
\commutator{S^{1}_{w}}
{S^{2}_{y}}
+\imunit
\lambda
\rbk{-1}^{z}
\commutator{S^{1}_{w}}
{S^{2}_{z}}.
\end{aligned}$$
If $w=z$, only the first sum and the potential term survive.
If $w\sim z$, only the summand $y=w$ in the second sum survives.
All three terms vanish in the remaining case.
The three cases give
\begin{equation}\label{eq:double-comm-11}
-\imunit \commutator{S^{1}_{w}}{\fun{\oaderiv_{\Lambda}}{S^{1}_{z}}}
=
\begin{cases}
\displaystyle
\sum_{\substack{y\in\Lambda\\y\sim z}}
S^{2}_{z}S^{2}_{y}
-
\lambda
\rbk{-1}^{z}
S^{3}_{z},
&
w=z,
\\
-
S^{3}_{z}S^{3}_{w},
&
w\sim z,
\\
0,
&
w\neq z
\quad
w\nsim z.
\end{cases}
\end{equation}

The derivation of $S^{2}_{z}$ is
$$\fun{\oaderiv_{\Lambda}}{S^{2}_{z}}
=
\sum_{\substack{y\in\Lambda\\y\sim z}}
S^{3}_{z}S^{1}_{y}
+
\lambda
\rbk{-1}^{z}
S^{1}_{z}.$$
The relations
$\commutator{S^{2}}{S^{3}}
=
\imunit S^{1}$
and
$\commutator{S^{2}}{S^{1}}
=
-\imunit S^{3}$
give the same three cases with the first and second spin components interchanged:
\begin{equation}\label{eq:double-comm-22}
-\imunit \commutator{S^{2}_{w}}{\fun{\oaderiv_{\Lambda}}{S^{2}_{z}}}
=
\begin{cases}
\displaystyle
\sum_{\substack{y\in\Lambda\\y\sim z}}
S^{1}_{z}S^{1}_{y}
-
\lambda
\rbk{-1}^{z}
S^{3}_{z},
&
w=z,
\\
-
S^{3}_{z}S^{3}_{w},
&
w\sim z,
\\
0,
&
w\neq z
\quad
w\nsim z.
\end{cases}
\end{equation}

The Fourier definition gives the exact expansion
$$\begin{aligned}
-\imunit
\commutator{\widetilde{S}^{i}_{p}}
{\fun{\oaderiv_{\Lambda}}{\widetilde{S}^{i}_{-p}}}
=
\frac{1}{\abscard{\Lambda}}
\sum_{w,z\in\Lambda}
\napiernum^{\imunit
p\cdot
\rbk{w-z}}
\rbk{-\imunit
\commutator{S^{i}_{w}}
{\fun{\oaderiv_{\Lambda}}{S^{i}_{z}}}}.
\end{aligned}$$
Substitution of the three cases in \eqref{eq:double-comm-11} gives
$$\begin{aligned}
-\imunit
\commutator{\widetilde{S}^{1}_{p}}
{\fun{\oaderiv_{\Lambda}}
{\widetilde{S}^{1}_{-p}}}
=
\frac{1}{\abscard{\Lambda}}
\sqbk{\sum_{z\in\Lambda}
\sum_{\substack{y\in\Lambda\\y\sim z}}
S^{2}_{z}S^{2}_{y}
-
\lambda
\sum_{z\in\Lambda}
\rbk{-1}^{z}
S^{3}_{z}
-
\sum_{z\in\Lambda}
\sum_{\substack{w\in\Lambda\\w\sim z}}
\napiernum^{\imunit
p\cdot
\rbk{w-z}}
S^{3}_{z}S^{3}_{w}}.
\end{aligned}$$
The corresponding substitution of \eqref{eq:double-comm-22} gives
$$\begin{aligned}
-\imunit
\commutator{\widetilde{S}^{2}_{p}}
{\fun{\oaderiv_{\Lambda}}{\widetilde{S}^{2}_{-p}}}
=
\frac{1}{\abscard{\Lambda}}
\sqbk{
\sum_{z\in\Lambda}
\sum_{\substack{y\in\Lambda\\y\sim z}}
S^{1}_{z}S^{1}_{y}
-
\lambda
\sum_{z\in\Lambda}
\rbk{-1}^{z}
S^{3}_{z}
-
\sum_{z\in\Lambda}
\sum_{\substack{w\in\Lambda\\w\sim z}}
\napiernum^{
\imunit
p\cdot
\rbk{
w-z
}
}
S^{3}_{z}S^{3}_{w}
}.
\end{aligned}$$

The ordered nearest-neighbor sums reduce to unordered bond sums according to
$$\begin{aligned}
\sum_{z\in\Lambda}
\sum_{\substack{y\in\Lambda\\y\sim z}}
\rbk{
S^{1}_{z}S^{1}_{y}
+
S^{2}_{z}S^{2}_{y}
}
&=
2
\sum_{\dbk{xy}}
\rbk{
S^{1}_{x}S^{1}_{y}
+
S^{2}_{x}S^{2}_{y}
},
\\ 
\sum_{z\in\Lambda}
\sum_{\substack{w\in\Lambda\\w\sim z}}
\napiernum^{
\imunit
p\cdot
\rbk{
w-z
}
}
S^{3}_{z}S^{3}_{w}
&=
\sum_{\dbk{xy}}
\rbk{
\napiernum^{
\imunit
p\cdot
\rbk{
y-x
}
}
+
\napiernum^{
-
\imunit
p\cdot
\rbk{
y-x
}
}
}
S^{3}_{x}S^{3}_{y}
\\
&=
2
\sum_{\dbk{xy}}
\fun{\cos}{
p\cdot
\rbk{
x-y
}
}
S^{3}_{x}S^{3}_{y}.
\end{aligned}$$
Adding the two Fourier components now gives
$$\begin{aligned}
&-\imunit
\sum_{i = 1, 2}
\commutator{\widetilde{S}^{i}_{p}}{\fun{\oaderiv_{\Lambda}}{\widetilde{S}^{i}_{-p}}}
\\ 
&=
\frac{1}{\abscard{\Lambda}}
\sqbk{
2
\sum_{\dbk{xy}}
\rbk{
S^{1}_{x}S^{1}_{y}
+
S^{2}_{x}S^{2}_{y}
}
-
2
\lambda
\sum_{z\in\Lambda}
\rbk{-1}^{z}
S^{3}_{z}
-
4
\sum_{\dbk{xy}}
\fun{\cos}{
p\cdot
\rbk{
x-y
}
}
S^{3}_{x}S^{3}_{y}
}.
\end{aligned}$$
The first two terms in the square bracket are identified directly from \eqref{eq:hamiltonian-spin}:
$$\begin{aligned}
2
\sum_{\dbk{xy}}
\rbk{S^{1}_{x}S^{1}_{y}
+S^{2}_{x}S^{2}_{y}}
-2 \lambda
\sum_{z\in\Lambda}
\rbk{-1}^{z}
S^{3}_{z}
=
-2
\physham_{\Lambda}
+\lambda
\abscard{\Lambda}.
\end{aligned}$$
Substitution of this identity gives
$$-\imunit
\sum_{i = 1, 2}
\commutator{
\widetilde{S}^{i}_{p}
}{
\fun{\oaderiv_{\Lambda}}{
\widetilde{S}^{i}_{-p}
}
}
=
-
\frac{2}{
\abscard{\Lambda}
}
\rbk{
\physham_{\Lambda}
-
\frac{
\lambda
\abscard{\Lambda}
}{
2
}
+
2
\sum_{\dbk{xy}}
S^{3}_{x}S^{3}_{y}
\fun{\cos}{
p\cdot
\rbk{
x-y
}
}
},$$
which is the claimed identity.
\end{proof}

\begin{defn}\label{def:cp}
The finite-volume double-commutator expectation is defined by
\begin{equation}\label{eq:cp}
C_{p}
=
\fun{\oastate[\psi_{\sminvtemperature,\Lambda}]}{
-\imunit
\sum_{i = 1, 2}
\commutator{\widetilde{S}^{i}_{p}}{
\fun{\oaderiv_{\Lambda}}{\widetilde{S}^{i}_{-p}}
}
}.
\end{equation}
\end{defn}

\begin{cor}[positivity of the double commutator]\label{cor:cp-positive}
For every $p \in \dual{\Lambda}$,
the quadratic quantity $c$ defined by
\eqref{eq:fb-quantities} satisfies
$$C_{p}
=
\frac{1}{
\sminvtemperature
}
\sum_{i=1,2}
\fun{c}{
\widetilde{S}^{i}_{-p}
}
\geq
0.$$
\end{cor}

\begin{proof}
Set
$X_{i}
=
\widetilde{S}^{i}_{-p}$.
The periodic finite-volume derivation
\eqref{eq:periodic-finite-volume-dynamics} gives
$$\commutator{
\sminvtemperature
\physham_{\Lambda}
}{
X_{i}
}
=
-
\imunit
\sminvtemperature
\fun{\oaderiv_{\Lambda}}{
X_{i}
}.$$
The adjoint of $X_i$ is
$\faadj{X_{i}}
=
\widetilde{S}^{i}_{p}$.
The definition \eqref{eq:fb-quantities} therefore gives
$$\begin{aligned}
\fun{c}{X_{i}}
=
\fun{\oastate[\psi_{\sminvtemperature,\Lambda}]}{
\commutator{\faadj{X_{i}}}
{\commutator{\sminvtemperature \physham_{\Lambda}}
{X_{i}}}}
=
-\imunit
\sminvtemperature
\fun{\oastate[\psi_{\sminvtemperature,\Lambda}]}{
\commutator{\widetilde{S}^{i}_{p}}
{\fun{\oaderiv_{\Lambda}}{\widetilde{S}^{i}_{-p}}}}.
\end{aligned}$$
Summation over $i\in\setone{1,2}$ and Definition \ref{def:cp} prove the equality in the corollary.
Lemma \ref{lem:fb-spectral} implies positivity.
\end{proof}

\subsection{The lowest-eigenvalue estimate}\label{the-lowest-eigenvalue-estimate}

The sum \(\sum_{p} C_{p}\) is controlled by a lower bound on the ground-state energy, which the paper quotes from \cite[Theorem C.1]{DysonLiebSimon001}; the following block-matrix argument proves it.

\begin{lem}[single-site block estimate]\label{lem:block-estimate}
Let $y_{1}, \dotsc, y_{2 d}$ be the nearest neighbors of a site $x$, let $\mu \in \fldreal$, and set
$$h_{x} = -\frac{1}{2} S^{1}_{x} \sum_{i = 1}^{2 d} S^{1}_{y_{i}} - \frac{1}{2} S^{2}_{x} \sum_{i = 1}^{2 d} S^{2}_{y_{i}} + \mu S^{3}_{x}.$$
The lowest eigenvalue of $h_{x}$ is
$-\frac{1}{4}\rbk{d\rbk{d+1}+4\mu^{2}}^{1/2}$.
\end{lem}

\begin{proof}
Write $T^{i} = \sum_{j} S^{i}_{y_{j}}$ and $T^{\pm} = T^{1} \pm \imunit T^{2}$.
The transverse ladder identity
\eqref{eq:app-spin-transverse-ladder}
rewrites the transverse part of
$h_x$
as
$-\frac{1}{4}\rbk{S^{+}_{x}T^{-}+S^{-}_{x}T^{+}}$.
In the basis of the spin at $x$, every operator on the total space is a $2 \times 2$ block matrix with entries acting on the neighbor space, and
$$h_{x} =
\begin{pmatrix}
\mu / 2 & -T^{-} / 4 \\
-T^{+} / 4 & -\mu / 2
\end{pmatrix},$$
Here $S^{+}_{x}$ is the block matrix with the identity in its upper-right entry.
The diagonal block form follows from
$S^{3}_{x} = \frac{1}{2} \operatorname{diag} \rbk{1, -1}$.
Squaring implies
$$h_{x}^{2} =
\begin{pmatrix}
\mu^{2} / 4 + T^{-} T^{+} / 16 & 0 \\
0 & \mu^{2} / 4 + T^{+} T^{-} / 16
\end{pmatrix}.$$

The $2d$ neighboring spins combine into an integer total spin
$T \in \setone{0,1,\dotsc,d}$.
The Casimir operator has eigenvalue $T \rbk{T + 1}$.
The raising and lowering products satisfy
$$T^{\mp} T^{\pm} = \rbk{T^{1}}^{2} + \rbk{T^{2}}^{2} + \rbk{T^{3}}^{2} - T^{3} \rbk{T^{3} \pm 1}.$$
The corresponding eigenvalues are
$T \rbk{T + 1} - m \rbk{m \pm 1}$
for $m \in \setone{-T,\dotsc,T}$.
Their maximum is $T \rbk{T + 1}$,
attained at $m=0$ or $m=\mp1$.
It is bounded above by $d \rbk{d + 1}$.
The spectral bound for $h_x$ is
$$\closedinterval{
-\frac{1}{4} \rbk{d \rbk{d + 1} + 4 \mu^{2}}^{\onehalf}
}{
\frac{1}{4} \rbk{d \rbk{d + 1} + 4 \mu^{2}}^{\onehalf}
}.$$

Choose neighbor-space vectors $\phi_{0},\phi_{1}$ with total spin $T=d$ and $T^{3}$-eigenvalues $0,1$.
They can be normalized so that
$$\begin{aligned}
T^{+}\phi_{0}
&=
\sqrt{d \rbk{d+1}}\phi_{1},
\\
T^{-}\phi_{1}
&=
\sqrt{d \rbk{d+1}}\phi_{0}.
\end{aligned}$$
The vectors
$\rbk{\text{spin up}}\otimes\phi_{0}$
and
$\rbk{\text{spin down}}\otimes\phi_{1}$
span an invariant two-dimensional space.
On this space $h_x$ has diagonal entries $\pm\mu/2$ and off-diagonal entries
$-\sqrt{d\rbk{d+1}}/4$.
Its lowest eigenvalue is the lower endpoint of the interval above.
\end{proof}

\begin{cor}[ground-state energy bound]\label{cor:energy-lower-bound}
$\physham_{\Lambda} \geq \frac{\lambda \abscard{\Lambda}}{2} - \frac{\abscard{\Lambda}}{4} \rbk{d \rbk{d + 1} + 4 \lambda^{2}}^{\onehalf}$,
and consequently it holds that
$$\sum_{p \in \dual{\Lambda}} C_{p}
=
-2
\fun{\oastate[\psi_{\sminvtemperature,\Lambda}]}{
\physham_{\Lambda}
}
+
\lambda \abscard{\Lambda}
\leq
\frac{\abscard{\Lambda}}{2}
\rbk{
d \rbk{d + 1}
+
4 \lambda^{2}
}^{\onehalf}.
$$
\end{cor}

\begin{proof}
Assigning half of each bond term of \eqref{eq:hamiltonian-spin} to each of its endpoints,
$$\physham_{\Lambda} - \frac{\lambda \abscard{\Lambda}}{2} = \sum_{x \in \Lambda} \sqbk{-\frac{1}{2} \sum_{y \sim x} \rbk{S^{1}_{x} S^{1}_{y} + S^{2}_{x} S^{2}_{y}} + \lambda \rbk{-1}^{x} S^{3}_{x}},$$
and each summand is the operator of Lemma \ref{lem:block-estimate} with $\mu = \lambda \rbk{-1}^{x}$, whose lowest eigenvalue does not depend on the sign of $\mu$; summing the operator inequalities gives the first claim.
For the second identity, sum Lemma \ref{lem:double-commutator} over
$p\in\dual{\Lambda}$.
For fixed nearest neighbors $x\neq y$, orthogonality of characters gives
$$\sum_{p\in\dual{\Lambda}}
\fun{\cos}{p\cdot\rbk{x-y}}=0.$$
The $S^{3}S^{3}$ terms therefore disappear, and
$$\sum_{p} C_{p}
=
-\frac{2 \abscard{\dual{\Lambda}}}{\abscard{\Lambda}}
\rbk{
\fun{\oastate[\psi_{\sminvtemperature,\Lambda}]}{
\physham_{\Lambda}
}
-
\frac{\lambda \abscard{\Lambda}}{2}
}
=
-2
\fun{\oastate[\psi_{\sminvtemperature,\Lambda}]}{
\physham_{\Lambda}
}
+
\lambda \abscard{\Lambda},$$
using $\abscard{\dual{\Lambda}} = \abscard{\Lambda}$; the first claim bounds
$$-2
\fun{\oastate[\psi_{\sminvtemperature,\Lambda}]}{
\physham_{\Lambda}
}
\leq
\frac{\abscard{\Lambda}}{2}
\rbk{
d \rbk{d + 1}
+
4 \lambda^{2}
}^{\onehalf}
-
\lambda \abscard{\Lambda}.
$$
\end{proof}

\subsection{The sum rule and the condensation bound}\label{the-sum-rule-and-the-condensation-bound}

The spin sum rule fixes the total transverse spectral weight. Combining it with the infrared and Falk--Bruch bounds forces a positive zero-momentum contribution in the condensed parameter region.

\begin{lem}[sum rule]\label{lem:sum-rule}
It holds that
\begin{equation}\label{eq:sum-rule}
\sum_{p \in \dual{\Lambda}}
\fun{\oastate[\psi_{\sminvtemperature,\Lambda}]}
{\widetilde{S}^{1}_{p} \widetilde{S}^{1}_{-p} + \widetilde{S}^{2}_{p} \widetilde{S}^{2}_{-p}}
=
\frac{\abscard{\Lambda}}{2}.
\end{equation}
Each summand on the left equals $\fun{b}{\widetilde{S}^{1}_{-p}} + \fun{b}{\widetilde{S}^{2}_{-p}}$,
where the function $b$ is defined by \eqref{eq:fb-quantities}.
\end{lem}

\begin{proof}
The Fourier Parseval identity
\eqref{eq:app-spin-fourier-parseval}
for
$i
=
1,2$
proves the first claim.
The adjoint identity
\eqref{eq:app-spin-fourier-adjoint-ladder}
and the commutator identity
\eqref{eq:app-spin-fourier-commutator}
show that
$\widetilde{S}^{i}_{p}\widetilde{S}^{i}_{-p}$
is already symmetrized.
For
$X
=
\widetilde{S}^{i}_{-p}$,
one has
$\faadj{X}
=
\widetilde{S}^{i}_{p}$,
and the definition in
\eqref{eq:fb-quantities}
gives
$$
\fun{b}{X}
=
\fun{\oastate[\psi_{\sminvtemperature,\Lambda}]}{
\widetilde{S}^{i}_{p}
\widetilde{S}^{i}_{-p}
}.
$$
\end{proof}

\begin{thm}[existence of BEC; Theorem 1 of \cite{AizenmanLiebSeiringerSolovejYngvason001}]\label{thm:bec}
Let
$E_{p}$,
$c_{d}$,
and
$\kappa$
be defined by
\eqref{eq:main-condensation-constant},
with
$d
\geq
3$.
The claimed bound is
\begin{equation}\label{eq:bec-bound}
\liminf_{\Lambda \nearrow \ringratint^{d}} \frac{1}{\abscard{\Lambda}^{2}} \sum_{x, y \in \Lambda} \fun{\gamma_{\psi_{\sminvtemperature,\Lambda}}}{x, y}
\geq
\kappa.
\end{equation}
\end{thm}

\begin{proof}
Fix $p \neq 0$ and apply Corollary \ref{cor:fb-two-operators} with $X_{i} = \widetilde{S}^{i}_{-p}$.
The infrared bound \eqref{eq:infrared-bound}
and the equation \eqref{eq:cp} give
$$\fun{g}{X_{1}}
+
\fun{g}{X_{2}}
\leq
\bar{g}
=
\frac{1}{\sminvtemperature E_{p}},
\quad
\fun{c}{X_{1}}
+
\fun{c}{X_{2}}
=
\sminvtemperature C_{p}
=
\bar{c}.$$
Remark \ref{rem:fb-simplification} and Lemma \ref{lem:sum-rule} now give the mode estimate
\begin{equation}\label{eq:mode-occupation-bound}
\fun{\oastate[\psi_{\sminvtemperature,\Lambda}]}
{\widetilde{S}^{1}_{p} \widetilde{S}^{1}_{-p}
+\widetilde{S}^{2}_{p} \widetilde{S}^{2}_{-p}}
\leq
\frac{1}{2}
\sqrt{\frac{C_{p}}{E_{p}}}
+\frac{1}{\sminvtemperature E_{p}}.
\end{equation}
Summing \eqref{eq:mode-occupation-bound} over $p\neq0$ gives the depletion
bound.
The Cauchy--Schwarz inequality and Corollary \ref{cor:energy-lower-bound} give
\begin{equation}\label{eq:bec-cauchy-schwarz-bound}
\begin{aligned}
\sum_{\substack{p \in \dual{\Lambda}\\p \neq 0}}
\frac{1}{2}
\sqrt{\frac{C_{p}}{E_{p}}}
&\leq
\frac{1}{2}
\rbk{
\sum_{\substack{p \in \dual{\Lambda}\\p \neq 0}}
C_{p}
}^{\onehalf}
\rbk{
\sum_{\substack{p \in \dual{\Lambda}\\p \neq 0}}
\frac{1}{E_{p}}
}^{\onehalf}
\\ 
&\leq
\frac{1}{2}
\rbk{
\frac{\abscard{\Lambda}}{2}
\rbk{d \rbk{d + 1}
+4 \lambda^{2}}^{\onehalf}}^{\onehalf}
\rbk{\sum_{\substack{p \in \dual{\Lambda}\\p \neq 0}}
\frac{1}{E_{p}}}^{\onehalf}.
\end{aligned}
\end{equation}

The remaining limit is the Riemann-sum approximation to the integral
defining
$c_{d}$
in
\eqref{eq:main-condensation-constant}:
\begin{equation}\label{eq:bec-infrared-riemann-sum}
\lim_{L \to \infty}
\frac{1}{\abscard{\Lambda}}
\sum_{\substack{p \in \dual{\Lambda}\\p \neq 0}}
\frac{1}{E_{p}}
=
c_{d}.
\end{equation}
To prove
\eqref{eq:bec-infrared-riemann-sum},
fix
$\delta
\in
\openinterval{0}{\pi}$
and split both the sum and the integral into the region
$\abs{p}
\geq
\delta$
and the neighborhood
$0
<
\abs{p}
<
\delta$
of the singular point.
Because
$E_{p}^{-1}$
is continuous on the compact region
$\abs{p}
\geq
\delta$,
the normalized sum over that region converges to the corresponding part of
the integral as
$L
\to
\infty$.
It remains to show that the contribution from
$0
<
\abs{p}
<
\delta$
is uniformly small in
$L$
when
$\delta$
is small.

For
$t
\in
\closedinterval{-\pi}{\pi}$,
the inequality
$1
-
\cos t
\geq
\frac{2 t^{2}}{\pi^{2}}$
gives
\begin{equation}\label{eq:bec-dispersion-quadratic-bound}
E_{p}
\geq
\frac{2}{\pi^{2}}
\abs{p}^{2}.
\end{equation}
Write
$p
=
\frac{2 \pi}{L} n$
with
$n
\in
\ringratint^{d}$,
and group the nonzero momenta according to the integer
$m
=
\max_{1 \leq i \leq d}
\abs{n_{i}}$.
Applying the mean-value theorem to the map
$r
\mapsto
r^{d}$
on
$\closedinterval{2 m - 1}{2 m + 1}$,
there exists
$\xi_{m}
\in
\openinterval{2 m - 1}{2 m + 1}$
such that the shell count satisfies
\begin{equation}\label{eq:bec-momentum-shell-count}
\begin{aligned}
\abscard{\set{n \in \ringratint^{d}}
{\max_{1 \leq i \leq d} \abs{n_{i}} = m}}
&=
\rbk{2 m + 1}^{d}
-\rbk{2 m - 1}^{d}
=
d \xi_{m}^{d - 1}
\rbk{\rbk{2 m + 1}
-\rbk{2 m - 1}}
\\ 
&=
2 d \xi_{m}^{d - 1}
\leq
2 d
\rbk{2 m + 1}^{d - 1}
\leq
2 d 3^{d - 1}
m^{d - 1}.
\end{aligned}
\end{equation}
For a momentum in this shell it holds that
\begin{equation}\label{eq:bec-shell-pointwise-bound}
\begin{aligned}
\abs{p}^{2}
=
\rbk{\frac{2 \pi}{L}}^{2}
\sum_{i = 1}^{d}
n_{i}^{2}
\geq
\rbk{\frac{2 \pi m}{L}}^{2},
\quad
\frac{1}{E_{p}}
\leq
\frac{\pi^{2}}{2 \abs{p}^{2}}
\leq
\frac{L^{2}}{8 m^{2}}.
\end{aligned}
\end{equation}
Since
$d
\geq
3$,
the remaining power sum obeys
\begin{equation}\label{eq:bec-shell-power-sum}
\sum_{1 \leq m < \delta L / \rbk{2 \pi}}
m^{d - 3}
\leq
\rbk{
\frac{\delta L}{2 \pi}
}^{d - 2}.
\end{equation}
Note that the sum in
\eqref{eq:bec-shell-power-sum}
is empty
for
$\delta L / \rbk{2 \pi}
\leq
1$.

Using
$\abscard{\Lambda}
=
L^{d}$,
the shell count
\eqref{eq:bec-momentum-shell-count},
the pointwise estimate
\eqref{eq:bec-shell-pointwise-bound},
and the power-sum estimate
\eqref{eq:bec-shell-power-sum}
gives the complete calculation
\begin{equation}\label{eq:bec-singular-riemann-sum-bound}
\begin{aligned}
&\frac{1}{\abscard{\Lambda}}
\sum_{\substack{
p \in \dual{\Lambda}\\
0 < \abs{p} < \delta}}
\frac{1}{E_{p}}
\leq
\frac{1}{L^{d}}
\sum_{1 \leq m < \delta L / \rbk{2 \pi}}
\rbk{
2 d 3^{d - 1}
m^{d - 1}
}
\frac{L^{2}}{8 m^{2}}
=
\frac{d 3^{d - 1}}{4}
L^{2 - d}
\sum_{1 \leq m < \delta L / \rbk{2 \pi}}
m^{d - 3}
\\ 
&\leq
\frac{d 3^{d - 1}}{4}
L^{2 - d}
\rbk{
\frac{\delta L}{2 \pi}
}^{d - 2}
=
\frac{d 3^{d - 1}}{
4 \rbk{2 \pi}^{d - 2}
}
\delta^{d - 2}.
\end{aligned}
\end{equation}
Since
$d
\geq
3$,
the bound
\eqref{eq:bec-singular-riemann-sum-bound}
shows that the contribution of the singular neighborhood to the normalized
sum tends to zero as
$\delta
\downarrow
0$,
uniformly in
$L$.
The bound
\eqref{eq:bec-dispersion-quadratic-bound}
similarly controls the contribution of the singular neighborhood to the
integral:
the integral over
$0
<
\abs{p}
<
\delta$
is bounded by a constant times
$\delta^{d - 2}$.
Combining these two estimates with the Riemann-sum convergence on
$\abs{p}
\geq
\delta$,
first taking
$L
\to
\infty$,
and subsequently taking
$\delta
\downarrow
0$
proves
\eqref{eq:bec-infrared-riemann-sum}.

Lemma \ref{lem:sum-rule} and \eqref{eq:mode-occupation-bound} give the zero-mode estimate
\begin{equation}\label{eq:bec-zero-mode-lower-bound}
\begin{aligned}
&\frac{1}{\abscard{\Lambda}}
\fun{\oastate[\psi_{\sminvtemperature,\Lambda}]}
{\widetilde{S}^{1}_{0} \widetilde{S}^{1}_{0}
+\widetilde{S}^{2}_{0} \widetilde{S}^{2}_{0}}
\geq
\frac{1}{2}
-\frac{1}{\abscard{\Lambda}}
\sum_{\substack{p \in \dual{\Lambda}\\p \neq 0}}
\rbk{\frac{1}{2}
\sqrt{\frac{C_{p}}{E_{p}}}
+\frac{1}{\sminvtemperature E_{p}}}
\\ 
&\geq
\frac{1}{2}
-\frac{1}{2}
\rbk{\frac{1}{2}
\rbk{d \rbk{d + 1}
+4 \lambda^{2}}^{\onehalf}
\frac{1}{\abscard{\Lambda}}
\sum_{\substack{p \in \dual{\Lambda}\\p \neq 0}}
\frac{1}{E_{p}}}^{\onehalf}
-\frac{1}
{\sminvtemperature
\abscard{\Lambda}}
\sum_{\substack{p \in \dual{\Lambda}\\p \neq 0}}
\frac{1}{E_{p}},
\end{aligned}
\end{equation}
The right side of
\eqref{eq:bec-zero-mode-lower-bound}
converges to
$\kappa$
by
\eqref{eq:main-condensation-constant}.

The spin calculation that identifies the transverse zero mode with the
average of the one-particle density matrix is collected in
Lemma
\ref{lem:app-density-zero-mode}.
The limit of
\eqref{eq:bec-zero-mode-lower-bound},
the exact identity
\eqref{eq:bec-density-zero-mode-identity},
and the error estimate
\eqref{eq:bec-density-zero-mode-error}
prove
\eqref{eq:bec-bound}.
\end{proof}

\begin{cor}[explicit positive-temperature condensation region]\label{cor:bec-explicit-region}
Let
$d
\geq
3$,
and suppose that
\eqref{eq:main-ground-state-condensation-region}
holds.
For every inverse temperature satisfying
$$
\sminvtemperature
>
\frac{c_{d}}{\kappa_{\mathrm{gs}}},
$$
the constant
$\kappa$
defined by
\eqref{eq:main-condensation-constant}
is positive.
The one-particle density matrix defined by
\eqref{eq:main-density-matrix}
satisfies
$$
\liminf_{\Lambda \nearrow \ringratint^{d}}
\frac{1}{\abscard{\Lambda}}
\max \opvarspec{\gamma_{\psi_{\sminvtemperature,\Lambda}}}
\geq
\kappa
>
0.
$$
In dimension
$d
=
3$,
the allowed interval of
$\lambda$
is nonempty.
\end{cor}

\begin{proof}
The condition
\eqref{eq:main-ground-state-condensation-region}
is equivalent to
$\kappa_{\mathrm{gs}}
>
0$.
The definitions
\eqref{eq:main-condensation-constant}
and
\eqref{eq:main-ground-state-condensation-constant}
give
$$
\kappa
=
\kappa_{\mathrm{gs}}
-
\frac{c_{d}}{\sminvtemperature}
>
0.
$$

Let
$\varphi_{0}
\in
\fun{\lp^{2}}{\Lambda}$
be the normalized constant vector.
For every
$x
\in
\Lambda$,
its value is
$\fun{\varphi_{0}}{x}
=
\abscard{\Lambda}^{-\onehalf}.$
The variational principle and Theorem
\ref{thm:bec}
give
$$
\begin{aligned}
\liminf_{\Lambda \nearrow \ringratint^{d}}
\frac{1}{\abscard{\Lambda}}
\max \opvarspec{\gamma_{\psi_{\sminvtemperature,\Lambda}}}
&\geq
\liminf_{\Lambda \nearrow \ringratint^{d}}
\frac{1}{\abscard{\Lambda}}
\bkt{\varphi_{0}}{
\gamma_{\psi_{\sminvtemperature,\Lambda}}
\varphi_{0}
}
\\ 
&=
\liminf_{\Lambda \nearrow \ringratint^{d}}
\frac{1}{\abscard{\Lambda}^{2}}
\sum_{x,y
\in
\Lambda}
\fun{\gamma_{\psi_{\sminvtemperature,\Lambda}}}{x,y}
\geq
\kappa.
\end{aligned}
$$
For
$d
=
3$,
the Appendix evaluation
\eqref{eq:app-c3-numerical-evaluation}
gives
$$
\frac{1}{c_{3}^{2}}
-
3
>
0.914.
$$
The right side of
\eqref{eq:main-ground-state-condensation-region}
is therefore positive at
$\lambda
=
0$,
so the allowed interval of
$\lambda$
is nonempty.
\end{proof}

\subsection{Uniqueness of the large eigenvalue and constancy of the condensate}\label{uniqueness-of-the-large-eigenvalue-and-constancy-of-the-condensate}

The momentum bounds also control every nonzero Fourier mode of the one-particle density matrix. They imply that only the zero mode can be macroscopically occupied and that its eigenfunction becomes constant in the thermodynamic limit.

\begin{lem}[Fourier-mode control]\label{lem:condensate-fourier-control}
The Fourier representation
$\faftr{\gamma_{\psi_{\sminvtemperature,\Lambda}}}$
defined by
\eqref{eq:main-fourier-density-matrix}
is positive semi-definite.
Set
$$\begin{gathered}
\fun{c_{\mathrm{C}}}{d,\lambda}
=
2d+3\lambda,
\quad
\fun{c_{\mathrm{F}}}{d,\lambda,\sminvtemperature}
=
\frac{\pi^{2}}{2}
\rbk{\sqrt{\frac{d \fun{c_{\mathrm{C}}}{d,\lambda}}{2}}
+\frac{1}{\sminvtemperature}
+d}.
\end{gathered}$$
For
$p
\neq
0$,
its diagonal satisfies
\begin{equation}\label{eq:condensate-nonzero-diagonal-bound}
\fun{\faftr{\gamma_{\psi_{\sminvtemperature,\Lambda}}}}{p,p}
\leq
\fun{c_{\mathrm{F}}}{d,\lambda,\sminvtemperature}
\abs{p}^{-2}
\leq
\fun{O}{\abscard{\Lambda}^{2 / d}},
\end{equation}
where the constant is independent of
$\Lambda$.
The row sparsity is expressed by
\begin{equation}\label{eq:condensate-fourier-row-sparsity}
\fun{\faftr{\gamma_{\psi_{\sminvtemperature,\Lambda}}}}{p,q}
=
0
\quad\text{whenever}\quad
p-q
\notin
\setone{0,\pi}^{d}
\pmod{2\pi}.
\end{equation}
\end{lem}

\begin{proof}
Equation
\eqref{eq:app-density-spin-fourier-representation}
proves positive semi-definiteness.
There are
$d\abscard{\Lambda}$
nearest-neighbor bonds.
The spin norms and the on-site projection in
\eqref{eq:main-hamiltonian}
give
$$\begin{aligned}
\norm{\physham_{\Lambda}}
\leq
\sum_{\dbk{xy}}
\rbk{\norm{S^{1}_{x}S^{1}_{y}}
+\norm{S^{2}_{x}S^{2}_{y}}}
+\lambda
\sum_{x\in\Lambda}
\norm{\frac{1}{2}
+\rbk{-1}^{x}S^{3}_{x}}
\leq
\rbk{\frac{d}{2} + \lambda}
\abscard{\Lambda}.
\end{aligned}$$
The double-commutator expectation
$C_{p}$
defined by
\eqref{eq:cp}
is nonnegative by
Corollary
\ref{cor:cp-positive}.
Lemma
\ref{lem:double-commutator}
and
\eqref{eq:cp}
therefore imply
$$\begin{aligned}
C_{p}
\leq
\frac{2}{\abscard{\Lambda}}
\rbk{\norm{\physham_{\Lambda}}
+\frac{\lambda\abscard{\Lambda}}{2}
+2
\sum_{\dbk{xy}}
\norm{S^{3}_{x}S^{3}_{y}}}
\leq
2
\rbk{\frac{d}{2}
+\lambda
+\frac{\lambda}{2}
+\frac{d}{2}}
=
\fun{c_{\mathrm{C}}}{d,\lambda}.
\end{aligned}
$$
The diagonal identity
\eqref{eq:app-density-fourier-diagonal},
the mode estimate
\eqref{eq:mode-occupation-bound},
and
$\norm{S^{3}_{\txttot,\Lambda}}
\leq
\abscard{\Lambda}/2$
now give
$$
\begin{aligned}
\fun{\faftr{\gamma_{\psi_{\sminvtemperature,\Lambda}}}}{p,p}
&\leq
\frac{1}{2}
\sqrt{
\frac{
\fun{c_{\mathrm{C}}}{d,\lambda}
}{
E_{p}
}
}
+
\frac{1}{\sminvtemperature E_{p}}
+
\frac{1}{2}.
\end{aligned}
$$
For
$p
\in
\closedinterval{-\pi}{\pi}^{d}$,
the dispersion and momentum bounds are
$E_{p}
\geq
\frac{2}{\pi^{2}}\abs{p}^{2}$
and
$\abs{p}
\leq
\pi\sqrt{d}$.
Consequently it holds that
$$\begin{aligned}
\frac{1}{2}
\sqrt{\frac{\fun{c_{\mathrm{C}}}{d,\lambda}}
{E_{p}}}
&\leq
\frac{\pi}{2\sqrt{2}}
\sqrt{\fun{c_{\mathrm{C}}}{d,\lambda}}
\abs{p}^{-1}
\leq
\frac{\pi^{2}}{2}
\sqrt{\frac{d
\fun{c_{\mathrm{C}}}{d,\lambda}}
{2}}
\abs{p}^{-2},
\\ 
\frac{1}{\sminvtemperature E_{p}}
&\leq
\frac{\pi^{2}}{2\sminvtemperature}
\abs{p}^{-2},
\\ 
\frac{1}{2}
&\leq
\frac{d\pi^{2}}{2}
\abs{p}^{-2}.
\end{aligned}$$
Adding these three estimates proves
$$\fun{\faftr{\gamma_{\psi_{\sminvtemperature,\Lambda}}}}{p,p}
\leq
\fun{c_{\mathrm{F}}}{d,\lambda,\sminvtemperature}
\abs{p}^{-2}.$$
Every nonzero
$p
\in
\dual{\Lambda}$
satisfies
$\abs{p}
\geq
2\pi/L$.
Since
$\abscard{\Lambda}
=
L^{d}$,
$$\fun{\faftr{\gamma_{\psi_{\sminvtemperature,\Lambda}}}}{p,p}
\leq
\frac{\fun{c_{\mathrm{F}}}{d,\lambda,\sminvtemperature}}
{4\pi^{2}}
\abscard{\Lambda}^{2/d}.$$
This proves
\eqref{eq:condensate-nonzero-diagonal-bound}.

The even-translation symmetry gives the remaining vanishing condition.
For an even translation
$v$,
Proposition
\ref{prop:finite-symmetries}
and
\eqref{eq:app-density-spin-fourier-representation}
give
$$\begin{aligned}
\fun{\faftr{\gamma_{\psi_{\sminvtemperature,\Lambda}}}}{p,q}
&=
\napiernum^{\imunit\rbk{q-p}\cdot v}
\fun{\faftr{\gamma_{\psi_{\sminvtemperature,\Lambda}}}}{p,q}.
\end{aligned}$$
If this matrix element is nonzero,
then
$\napiernum^{\imunit\rbk{q-p}\cdot v}
=
1$
for every even
$v$.
Taking
$v
=
2e_{j}$
for
$j
\in
\setone{1,\ldots,d}$
gives
$$q_{j}-p_{j}
\in
\setone{0,\pi}
\pmod{2\pi}.$$
There are at most
$2^{d}$
such values of
$q$
for each
$p$.
\end{proof}

\begin{lem}[off-zero compression]\label{lem:condensate-off-zero-compression}
The normalized constant vector and its orthogonal-complement projection are
defined by
\begin{equation}\label{eq:condensate-constant-vector-projection}
\begin{aligned}
\fun{\varphi_{0,\Lambda}}{x}
&=
\abscard{\Lambda}^{-\onehalf},
&
x
&\in
\Lambda,
\\
P_{\perp,\Lambda}
&=
\opprojto{\setone{\varphi_{0,\Lambda}}^{\perp}},
&
\varphi_{0,\Lambda}
&\in
\fun{\lp^{2}}{\Lambda}.
\end{aligned}
\end{equation}
The one-particle density matrix satisfies
\begin{equation}\label{eq:condensate-off-zero-compression}
\norm{
P_{\perp,\Lambda}
\gamma_{\psi_{\sminvtemperature,\Lambda}}
P_{\perp,\Lambda}
}
=
\fun{O}{\abscard{\Lambda}^{2 / d}}.
\end{equation}
\end{lem}

\begin{proof}
In the Fourier representation
\eqref{eq:main-fourier-density-matrix},
the range of
$P_{\perp,\Lambda}$
is the subspace supported on
$p
\neq
0$.
Positive semi-definiteness gives
$$
\abs{\fun{\faftr{\gamma_{\psi_{\sminvtemperature,\Lambda}}}}{p,q}}
\leq
\fun{\faftr{\gamma_{\psi_{\sminvtemperature,\Lambda}}}}{p,p}^{\onehalf}
\fun{\faftr{\gamma_{\psi_{\sminvtemperature,\Lambda}}}}{q,q}^{\onehalf}.
$$
Equation
\eqref{eq:condensate-fourier-row-sparsity}
shows that each row has at most
$2^{d}$
nonzero entries.
The Fourier compression of
$P_{\perp,\Lambda}
\gamma_{\psi_{\sminvtemperature,\Lambda}}
P_{\perp,\Lambda}$
is Hermitian.
Equation
\eqref{eq:app-schur-hermitian}
shows that its operator norm is at most its maximal row sum.
Positive semi-definiteness,
the row sparsity in
\eqref{eq:condensate-fourier-row-sparsity},
and
\eqref{eq:condensate-nonzero-diagonal-bound}
give the explicit row estimate
\begin{equation}\label{eq:condensate-off-zero-block-bound}
\max_{p \neq 0}
\sum_{q \neq 0}
\abs{\fun{\faftr{\gamma_{\psi_{\sminvtemperature,\Lambda}}}}{p, q}}
\leq
2^{d}
\max_{p \neq 0}
\fun{\faftr{\gamma_{\psi_{\sminvtemperature,\Lambda}}}}{p, p}
=
\fun{O}{\abscard{\Lambda}^{2 / d}}.
\end{equation}
Equations
\eqref{eq:app-schur-hermitian}
and
\eqref{eq:condensate-off-zero-block-bound}
give
$$
\norm{
P_{\perp,\Lambda}
\gamma_{\psi_{\sminvtemperature,\Lambda}}
P_{\perp,\Lambda}
}
\leq
\max_{p\neq0}
\sum_{q\neq0}
\abs{
\fun{\faftr{\gamma_{\psi_{\sminvtemperature,\Lambda}}}}{p,q}
}
=
\fun{O}{\abscard{\Lambda}^{2/d}}.$$
\end{proof}

The two lemmas isolate the model-dependent input. The remaining argument uses the variational and min--max principles.

\begin{thm}[condensate wave function]\label{thm:condensate-constant}
Assume
$d
\geq
3$
and
$\kappa
>
0$,
where
$\kappa$
is defined by
\eqref{eq:main-condensation-constant}.
For the one-particle density matrix
$\gamma_{\psi_{\sminvtemperature,\Lambda}}$
defined by
\eqref{eq:main-density-matrix},
the largest eigenvalue satisfies
$$\max
\opvarspec{\gamma_{\psi_{\sminvtemperature,\Lambda}}}
\geq
\rbk{\kappa-\fun{o}{1}}
\abscard{\Lambda}.$$
Every other eigenvalue is
$\fun{O}{\abscard{\Lambda}^{2 / d}}$.
For all sufficiently large
$\Lambda$,
the largest eigenvalue is simple.
With
$P_{\perp,\Lambda}$
defined in \eqref{eq:condensate-constant-vector-projection},
its normalized eigenfunction
$\varphi_{\Lambda}$
satisfies
$$\norm{P_{\perp,\Lambda} \varphi_{\Lambda}}
=
\fun{O}
{\abscard{\Lambda}^{\frac{1}{d}-\onehalf}}.$$
The same eigenfunction obeys
\begin{equation}\label{eq:condensate-eigenfunction-constant-mode}
\lim_{\Lambda \nearrow \ringratint^{d}}
\frac{1}{\abscard{\Lambda}}
\abs{\sum_{x \in \Lambda}
\fun{\varphi_{\Lambda}}{x}}^{2}
=
1.
\end{equation}
\end{thm}

\begin{proof}
The finite-dimensional variational principle gives
$$\max
\opvarspec{\gamma_{\psi_{\sminvtemperature,\Lambda}}}
=
\max_{\varphi\neq0}
\frac{\bkt{\varphi}{\gamma_{\psi_{\sminvtemperature,\Lambda}} \varphi}}
{\bkt{\varphi}{\varphi}}.$$
The constant vector
$\varphi_{0,\Lambda}$
defined in \eqref{eq:condensate-constant-vector-projection}
has norm
$1$.
Evaluating the preceding maximum at this vector and using
\eqref{eq:bec-bound}
gives
\begin{equation}\label{eq:condensate-largest-eigenvalue-lower-bound}
\begin{aligned}
\nu_{1,\Lambda}
=
\max
\opvarspec{\gamma_{\psi_{\sminvtemperature,\Lambda}}}
\geq
\bkt{\varphi_{0,\Lambda}}
{\gamma_{\psi_{\sminvtemperature,\Lambda}}
\varphi_{0,\Lambda}}
=
\frac{1}{\abscard{\Lambda}}
\sum_{x,y \in \Lambda}
\fun{\gamma_{\psi_{\sminvtemperature,\Lambda}}}{x,y}
\geq
\rbk{\kappa-\fun{o}{1}}
\abscard{\Lambda}.
\end{aligned}
\end{equation}
Decompose the normalized eigenfunction for the largest eigenvalue $\nu_{1,\Lambda}$ as
$$\varphi_{\Lambda}
=
\alpha_{\Lambda}\varphi_{0,\Lambda}
+
\varphi_{\perp,\Lambda}.$$
The orthogonal component satisfies
$\varphi_{\perp,\Lambda}
\in
\setone{\varphi_{0,\Lambda}}^{\perp}$.
The hard-core bound
$0
\leq
\faadj{a_{x}}a_{x}
\leq
1$
gives
$$\norm{\gamma_{\psi_{\sminvtemperature,\Lambda}}}
\leq
\sqfun{\trace}{\gamma_{\psi_{\sminvtemperature,\Lambda}}}
=
\sum_{x \in \Lambda}
\fun{\oastate[\psi_{\sminvtemperature,\Lambda}]}
{\faadj{a_{x}}a_{x}}
\leq
\abscard{\Lambda}.$$
The compression estimate
\eqref{eq:condensate-off-zero-compression}
gives
$$\bkt{\varphi_{\perp,\Lambda}}
{\gamma_{\psi_{\sminvtemperature,\Lambda}}
\varphi_{\perp,\Lambda}}
\leq
\fun{O}{\abscard{\Lambda}^{2 / d}}
\norm{\varphi_{\perp,\Lambda}}^{2}.$$
Cauchy--Schwarz for the positive form defined by
$\gamma_{\psi_{\sminvtemperature,\Lambda}}$
and the trace bound give
$$\begin{aligned}
\abs{\bkt{\varphi_{\perp,\Lambda}}
{\gamma_{\psi_{\sminvtemperature,\Lambda}}
\varphi_{0,\Lambda}}}
\leq
\bkt{\varphi_{\perp,\Lambda}}
{\gamma_{\psi_{\sminvtemperature,\Lambda}}
\varphi_{\perp,\Lambda}}^{\onehalf}
\bkt{\varphi_{0,\Lambda}}
{\gamma_{\psi_{\sminvtemperature,\Lambda}}
\varphi_{0,\Lambda}}^{\onehalf}
\leq
\fun{O}
{\abscard{\Lambda}^{\onehalf+\frac{1}{d}}}
\norm{\varphi_{\perp,\Lambda}}.
\end{aligned}$$
The orthogonal decomposition first gives
$$\begin{aligned}
\bkt{\varphi_{\perp,\Lambda}}{\varphi_{\Lambda}}
=
\bkt{\varphi_{\perp,\Lambda}}
{\alpha_{\Lambda}\varphi_{0,\Lambda}
+\varphi_{\perp,\Lambda}}
=
\alpha_{\Lambda}
\bkt{\varphi_{\perp,\Lambda}}{\varphi_{0,\Lambda}}
+\bkt{\varphi_{\perp,\Lambda}}{\varphi_{\perp,\Lambda}}
=
\norm{\varphi_{\perp,\Lambda}}^{2}.
\end{aligned}$$
The eigenvalue equation
$\gamma_{\psi_{\sminvtemperature,\Lambda}}\varphi_{\Lambda}
=
\nu_{1,\Lambda}\varphi_{\Lambda}$
therefore gives the tested identity.
The two preceding estimates and
$\abs{\alpha_{\Lambda}}
\leq
1$
then yield constants
$C_{1},C_{2}
>
0$,
independent of
$\Lambda$,
such that
$$\begin{aligned}
&\nu_{1,\Lambda}
\norm{\varphi_{\perp,\Lambda}}^{2}
=
\nu_{1,\Lambda}
\bkt{\varphi_{\perp,\Lambda}}{\varphi_{\Lambda}}
=
\bkt{\varphi_{\perp,\Lambda}}
{\gamma_{\psi_{\sminvtemperature,\Lambda}}
\varphi_{\Lambda}}
\\ 
&=
\alpha_{\Lambda}
\bkt{\varphi_{\perp,\Lambda}}
{\gamma_{\psi_{\sminvtemperature,\Lambda}}
\varphi_{0,\Lambda}}
+\bkt{\varphi_{\perp,\Lambda}}
{\gamma_{\psi_{\sminvtemperature,\Lambda}}
\varphi_{\perp,\Lambda}}
\\ 
&\leq
C_{1}
\abscard{\Lambda}^{\onehalf+\frac{1}{d}}
\norm{\varphi_{\perp,\Lambda}}
+
C_{2}
\abscard{\Lambda}^{2 / d}
\norm{\varphi_{\perp,\Lambda}}^{2}.
\end{aligned}
$$
The lower bound
\eqref{eq:condensate-largest-eigenvalue-lower-bound}
implies
\begin{equation}\label{expedition0025008}
\nu_{1,\Lambda}
\geq
\frac{\kappa}{2}
\abscard{\Lambda}
\end{equation}
for every sufficiently large
$\Lambda$.
The assumption
$d
\geq
3$
gives
$2/d
<
1$.
After increasing
$\Lambda$
once more,
$$
C_{2}
\abscard{\Lambda}^{2/d}
\leq
\frac{\kappa}{4}
\abscard{\Lambda}.
$$
Moving the final term in the preceding eigenvalue estimate to the left
therefore gives
$$\begin{aligned}
\frac{\kappa}{4}
\abscard{\Lambda}
\norm{\varphi_{\perp,\Lambda}}^{2}
&\leq
\rbk{\nu_{1,\Lambda}
-C_{2}
\abscard{\Lambda}^{2/d}}
\norm{\varphi_{\perp,\Lambda}}^{2}
\leq
C_{1}
\abscard{\Lambda}^{1/2+1/d}
\norm{\varphi_{\perp,\Lambda}}.
\end{aligned}$$
If
$\varphi_{\perp,\Lambda}
=
0$,
the required estimate is immediate.
Otherwise,
division by
$\abscard{\Lambda}
\norm{\varphi_{\perp,\Lambda}}$
gives the explicit bound
$$\norm{\varphi_{\perp,\Lambda}}
\leq
\frac{4C_{1}}{\kappa}
\abscard{\Lambda}^{1/d-1/2}.$$
Hence we obtain
$$\norm{\varphi_{\perp,\Lambda}}
=
\fun{O}
{\abscard{\Lambda}^{\frac{1}{d}-\onehalf}}.$$
The constant-mode overlap is therefore
$$\begin{aligned}
\frac{1}{\abscard{\Lambda}}
\abs{\sum_{x \in \Lambda}
\fun{\varphi_{\Lambda}}{x}}^{2}
=
\abs{\bkt{\varphi_{0,\Lambda}}{\varphi_{\Lambda}}}^{2}
=
1
-
\norm{\varphi_{\perp,\Lambda}}^{2}
\to
1.
\end{aligned}$$
Order the eigenvalues of
$\gamma_{\psi_{\sminvtemperature,\Lambda}}$
in nonincreasing order and count them with multiplicity:
$$\nu_{1,\Lambda}
\geq
\nu_{2,\Lambda}
\geq
\dotsb
\geq
0.$$
The min--max principle and the compression estimate
\eqref{eq:condensate-off-zero-compression}
bound the second eigenvalue by
$$
\nu_{2,\Lambda}
\leq
\sup_{\substack{\xi \in \setone{\varphi_{0,\Lambda}}^{\perp}
\\
\norm{\xi} = 1}}
\bkt{\xi}
{\gamma_{\psi_{\sminvtemperature,\Lambda}} \xi}
=
\norm{P_{\perp,\Lambda}
\gamma_{\psi_{\sminvtemperature,\Lambda}}
P_{\perp,\Lambda}}
=
\fun{O}{\abscard{\Lambda}^{2 / d}}.$$
Hence there is a constant
$C_{\perp}
>
0$,
independent of
$\Lambda$,
such that
$$\nu_{j,\Lambda}
\leq
\nu_{2,\Lambda}
\leq
C_{\perp}
\abscard{\Lambda}^{2/d}$$
for every
$j
\geq
2$.
Since
$2/d-1
<
0$
and the estimate \eqref{expedition0025008} holds,
the ratio of the second and first eigenvalues satisfies
$$\frac{\nu_{2,\Lambda}}{\nu_{1,\Lambda}}
\leq
\frac{2C_{\perp}}{\kappa}
\abscard{\Lambda}^{2/d-1}
\to
0.$$
In particular,
$\nu_{2,\Lambda}
<
\nu_{1,\Lambda}$
for every sufficiently large
$\Lambda$.
If the largest eigenvalue had multiplicity at least
$2$,
then counting eigenvalues with multiplicity would give
$\nu_{2,\Lambda}
=
\nu_{1,\Lambda}$.
The strict inequality excludes this possibility,
so the largest eigenvalue is simple for every sufficiently large
$\Lambda$.
\end{proof}
\begin{cor}[asymptotic constancy of the condensate eigenfunction]\label{cor:condensate-asymptotically-constant}
Under the hypotheses of Theorem \ref{thm:condensate-constant}, choose the
normalized eigenfunction $\varphi_\Lambda$ for the largest eigenvalue of
$\gamma_{\psi_{\sminvtemperature,\Lambda}}$.
For every sufficiently large $\Lambda$, there is a phase
$\zeta_\Lambda
\in
\fun{\liegr{U}}{1}$
such that
\begin{equation}\label{eq:condensate-eigenfunction-norm-convergence}
\norm{
\zeta_\Lambda\varphi_\Lambda
-
\varphi_{0,\Lambda}
}
=
\fun{O}{\abscard{\Lambda}^{\frac{1}{d}-\onehalf}}.
\end{equation}
In particular, the normalized condensate eigenfunction is asymptotically
constant in the one-particle Hilbert-space norm, up to its arbitrary phase.
No exact finite-volume or pointwise constancy is asserted.
\end{cor}

\begin{proof}
Use the orthogonal decomposition from the proof of Theorem
\ref{thm:condensate-constant}:
$$
\varphi_\Lambda
=
\alpha_\Lambda\varphi_{0,\Lambda}
+
\varphi_{\perp,\Lambda}.
$$
The estimate in that theorem gives
$$
\norm{\varphi_{\perp,\Lambda}}
=
\fun{O}{\abscard{\Lambda}^{\frac{1}{d}-\onehalf}},
\quad
\abs{\alpha_\Lambda}^{2}
=
1
-
\norm{\varphi_{\perp,\Lambda}}^{2}.
$$
The second identity makes $\alpha_\Lambda$ nonzero for every sufficiently
large $\Lambda$.
Choose $\zeta_\Lambda$ so that
$\zeta_\Lambda\alpha_\Lambda=\abs{\alpha_\Lambda}$.
Orthogonality gives
$$\begin{aligned}
\norm{\zeta_\Lambda\varphi_\Lambda-\varphi_{0,\Lambda}}^{2}
&=
\rbk{1-\abs{\alpha_\Lambda}}^{2}
+
\norm{\varphi_{\perp,\Lambda}}^{2}
\leq
2\norm{\varphi_{\perp,\Lambda}}^{2}.
\end{aligned}$$
The estimate for $\varphi_{\perp,\Lambda}$ proves
\eqref{eq:condensate-eigenfunction-norm-convergence}.
\end{proof}

\section{Loop Representation and Chessboard Estimates}\label{sec:loop}

The complementary regime has exponential decay of the one-particle density matrix at large \(\lambda\) or high temperature. The proof represents the Gibbs trace by quasi-particle world lines, following \cite{AizenmanNachtergaele001}. It then applies the chessboard estimate of \cite{FrohlichLieb001,FrohlichSimon001}. This section constructs the representation and proves the reflection positivity and chessboard bounds; the next section derives the decay and the Mott gap. Throughout, \(\Lambda = \Lambda_{L}\) is a periodic box with even side \(L\). The chessboard estimate is proved for every even side by the maximizer argument of \cite{FrohlichLieb001,FrohlichSimon001}. The proof uses dyadic time discretizations.

\subsection{Quasi-particles and the path expansion}\label{quasi-particles-and-the-path-expansion}

The checkerboard sublattices \(\Lambda_{\mathrm{A}}\) and \(\Lambda_{\mathrm{B}}\) are defined by \eqref{eq:checkerboard-sublattices}. The occupation-number basis turns the Dyson expansion into a positive measure on quasi-particle world lines. This representation converts matrix elements and partition functions into geometric events that can be reflected and estimated. Work in the joint eigenbasis of the occupation numbers \(\setone{n_{x}}\), labeled by configurations \(\sigma \colon \Lambda \to \setone{0, 1}\). Split \eqref{eq:main-hamiltonian} as \(\physham_{\Lambda} = T + D\). The hopping part is \(T = -\frac{1}{2} \sum_{\dbk{xy}} \rbk{S^{+}_{x} S^{-}_{y} + S^{-}_{x} S^{+}_{y}}\). The diagonal part \(D\) has eigenvalue \(\lambda \fun{n_{\mathrm{qp}}}{\sigma}\) on the configuration \(\sigma\), where \[\fun{n_{\mathrm{qp}}}{\sigma} = \abscard{\set{x \in \Lambda_{\mathrm{A}}}{\sigma_{x} = 1}} + \abscard{\set{x \in \Lambda_{\mathrm{B}}}{\sigma_{x} = 0}}\] is the number of quasi-particles: an occupied crest site or an empty trough site. The configuration \(\sigma^{0}\) fills \(\Lambda_{\mathrm{B}}\) and empties \(\Lambda_{\mathrm{A}}\). It satisfies \(\fun{n_{\mathrm{qp}}}{\sigma^{0}}=0\) and minimizes the potential energy. Each particle hop creates or annihilates quasi-particles on the two sublattices.

\begin{figure}[!htbp]
\centering
\includesvg[width=0.82\linewidth]{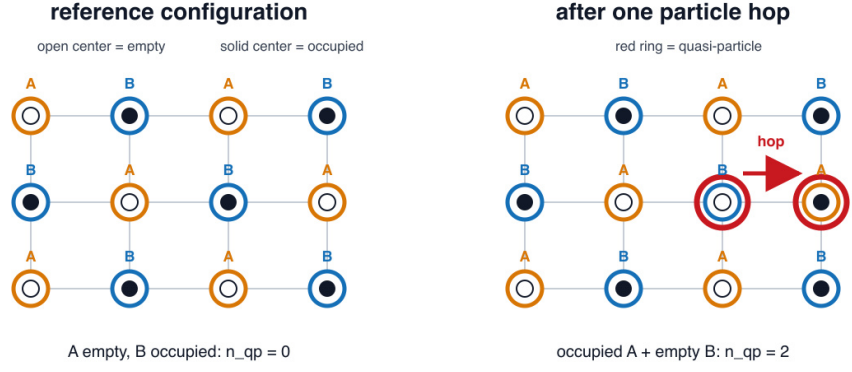}
\caption{The checkerboard quasi-particle convention.
An open center denotes an empty site and a solid center an occupied site.
The reference configuration on the left has no quasi-particles.
On the right, one hop from a trough site in $\Lambda_{\mathrm{B}}$ to a
neighboring crest site in $\Lambda_{\mathrm{A}}$ produces an empty trough and
an occupied crest, hence two quasi-particles.}
\label{fig:quasiparticle-convention}
\end{figure}

Let \(n_{\mathrm{A}}\) and \(n_{\mathrm{B}}\) count these quasi-particles. Pairwise cancellation of the remaining sites gives \[S^{3}_{\txttot,\Lambda}
=
\fun{n_{\mathrm{A}}}{\sigma}
-
\fun{n_{\mathrm{B}}}{\sigma}.\]

In this basis, the matrix of \(-T\) has entry \(\frac{1}{2}\) between two configurations that differ by a single hop, and all other entries vanish. In particular, every matrix entry is nonnegative. This entrywise nonnegativity underlies the positivity arguments below. The Dyson expansion of \(\napiernum^{-\beta \physham_{\Lambda}} = \napiernum^{-\beta \rbk{D + T}}\) in the interaction picture of \(D\), \begin{equation}\label{eq:dyson-expansion}
\napiernum^{-\beta \physham_{\Lambda}}
=
\sum_{n \geq 0}
\int_{0 < t_{1} < \dotsb < t_{n} < \beta}
\napiernum^{-\rbk{\beta - t_{n}} D}
\rbk{-T}
\napiernum^{-\rbk{t_{n} - t_{n - 1}} D}
\dotsm
\rbk{-T}
\napiernum^{-t_{1} D}
\opdmsr{t_{1}}
\dotsm
\opdmsr{t_{n}},
\end{equation} converges in norm (\(n\)-th term bounded by \(\rbk{\beta \norm{T}}^{n} / n!\)) and is verified by differentiation in \(\beta\). Insert resolutions of the identity in the configuration basis. Each term becomes a sum over discrete paths. A path consists of an initial configuration \(\sigma\), bonds \(b_{1},\dotsc,b_{n}\), and jump times \(t_{1}<\dotsb<t_{n}\). The hop along \(b_i\) must be admissible at time \(t_i\), and the path must return to \(\sigma\) at time \(\beta\). The periodic space-time and its quotient map are defined by \begin{equation}\label{eq:periodic-spacetime-quotient}
\begin{aligned}
\Gamma
&=
\Lambda
\times
\setquot{\fldreal}{\beta \ringratint},
\\
q_{\beta}
&\colon
\Lambda
\times
\fldreal
\to
\Gamma,
\\
\fun{q_{\beta}}{u,t}
&=
\rbk{u, t + \beta \ringratint},
\quad
u \in \Lambda,
\quad
t \in \fldreal.
\end{aligned}
\end{equation} The quasi-particle world lines are vertical on occupied crest sites and empty trough sites. Horizontal rungs record the jumps.

\begin{figure}[!htbp]
\centering
\includesvg[width=0.82\linewidth]{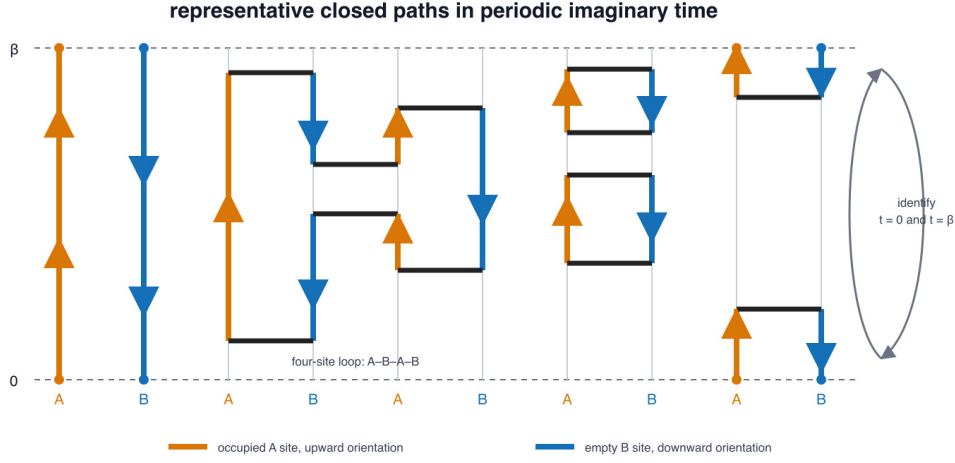}
\caption{A representative closed quasi-particle path in periodic imaginary
time.
Vertical segments on $A$ sites represent occupied crests and are oriented
upward; those on $B$ sites represent empty troughs and are oriented downward.
Horizontal rungs represent particle hops.
The examples include components with nonzero temporal winding,
one contractible component that visits four successive sites,
two components supported on the same pair of sites,
and a component crossing the periodic time boundary.
The levels $0$ and $\beta$ are identified.}
\label{fig:quasiparticle-world-lines}
\end{figure}

Figure \ref{fig:quasiparticle-convention} compares the reference configuration with the configuration produced by one particle hop. Its left panel represents \(\sigma^{0}\), and its right panel marks the two sites counted by \(\fun{n_{\mathrm{qp}}}{\sigma}\) after the hop.

Figure \ref{fig:quasiparticle-world-lines} represents the geometric path associated with the discrete data \(\rbk{\sigma;b,t}\). The two leftmost components wind through the time direction. The four-site component illustrates a path that successively visits \(A\), \(B\), \(A\), and \(B\) sites. The remaining components illustrate different rung patterns, including a loop that closes through the identification of \(t = 0\) with \(t = \beta\).

The ordered jump-time simplices are defined by \begin{equation}\label{eq:ordered-jump-time-simplex}
\fun{\Delta_{n}}{\beta}
=
\begin{cases}
\set{\seq{t_{i}}{1 \leq i \leq n} \in \openinterval{0}{\beta}^{n}}
{0 < t_{1} < \dotsb < t_{n} < \beta},
&n \geq 1,
\\
\setone{\varnothing},
&n = 0.
\end{cases}
\end{equation} The zeroth simplex is the singleton consisting of the empty time sequence. Fix a configuration \(\sigma\) and a bond sequence \(b = \seq{b_{i}}{1 \leq i \leq n}\), where each \(b_{i} = \dbk{x_{i} y_{i}}\) belongs to the nearest-neighbor bond set \(\setindex{B}_{\Lambda}\). The bond set \(\setindex{B}_{\Lambda}\) is defined by \eqref{eq:reflection-bond-decomposition}. Set \(\sigma^{\rbk{0}} = \sigma\). For \(1 \leq i \leq n\), the hop along \(b_i = \dbk{x_i y_i}\) exchanges the occupation numbers at \(x_i\) and \(y_i\) and leaves every other site unchanged. For every \(u \in \Lambda\), the resulting configuration is \[\sigma^{\rbk{i}}_{u}
=
\begin{cases}
\sigma^{\rbk{i - 1}}_{y_i},
& u = x_i,\\
\sigma^{\rbk{i - 1}}_{x_i},
& u = y_i,\\
\sigma^{\rbk{i - 1}}_{u},
& u \in \Lambda \setminus \setone{x_i,y_i}.
\end{cases}\] For \(n \geq 0\), the admissible discrete data with \(n\) hops form the set \[\oa{C}_{n}
=
\set{
\rbk{\sigma,b}
}{
\parbox{23em}{
\ensuremath{\sigma \colon \Lambda \to \setone{0,1}},
\ensuremath{b = \seq{b_{i}}{1 \leq i \leq n}},
\ensuremath{b_{i} = \dbk{x_{i}y_{i}} \in \setindex{B}_{\Lambda}}
for every \ensuremath{1 \leq i \leq n},
and
\ensuremath{\sigma^{\rbk{i - 1}}_{x_{i}}
+
\sigma^{\rbk{i - 1}}_{y_{i}}
=
1} for every \ensuremath{1 \leq i \leq n}.
}
}.\] For \(n = 0\), \(b = \varnothing\) is the unique empty bond sequence. The datum \(\rbk{\sigma,\varnothing} \in \oa{C}_{0}\) represents a path with no hops. For \(n \geq 1\), every bond \(b_i \in \setindex{B}_{\Lambda}\) has distinct endpoints. No stationary step with \(x_i = y_i\) belongs to the admissible discrete data. The closed discrete data are \[\oa{C}_{n}^{\mathrm{o}}
=
\set{\rbk{\sigma,b} \in \oa{C}_{n}}
{\sigma^{\rbk{n}} = \sigma^{\rbk{0}}}.\] For distinct sites \(x,y \in \Lambda\), the \(\rbk{x,y}\)-open discrete data are \[\oa{C}_{n}^{\rbk{x,y}}
=
\set{\rbk{\sigma,b} \in \oa{C}_{n}}
{
\parbox{23em}{
\ensuremath{\sigma^{\rbk{0}}_{x} = 1},
\ensuremath{\sigma^{\rbk{0}}_{y} = 0},
and
\ensuremath{\sigma^{\rbk{n}}_{u}
=
\sigma^{\rbk{0}}_{u}
-
\kroneckerdelta_{u x}
+
\kroneckerdelta_{u y}}
for every \ensuremath{u \in \Lambda}.
}
}.\] For \(\alpha \in \setone{\mathrm{o},\rbk{x,y}}\), the corresponding path space is

\begin{equation}\label{eq:loop-path-space}
\oa{A}^{\alpha}
=
\bigcup_{n \geq 0}
\set{
\fun{\omega}{\sigma;b,t}
}{
\rbk{\sigma,b} \in \oa{C}_{n}^{\alpha},
t \in \fun{\Delta_{n}}{\beta}
}.
\end{equation}

For \(\alpha \in \setone{\mathrm{o},\rbk{x,y}}\) and \(\omega \in \oa{A}^{\alpha}\), write \(\fun{\sigma_t}{\omega}\) for its configuration at time \(t\). Its total vertical length is

\begin{equation}\label{eq:loop-total-vertical-length}
\absvol{\omega}
=
\int_{0}^{\beta}
\fun{n_{\mathrm{qp}}}{\sigma_t}
\opdmsr{t}.
\end{equation}

The conserved winding number is

\begin{equation}\label{eq:loop-winding-number}
\fun{\nu}{\omega}
=
\fun{n_{\mathrm{A}}}{\sigma_t}
-
\fun{n_{\mathrm{B}}}{\sigma_t}.
\end{equation}

It is independent of \(t\) because \(T\) commutes with \(S^{3}_{\txttot,\Lambda}\). Geometrically, it is the net temporal winding of the oriented world lines. Lines on \(\Lambda_{\mathrm{A}}\) run upward, and lines on \(\Lambda_{\mathrm{B}}\) run downward.

\begin{figure}[!htbp]
\centering
\includesvg[width=0.82\linewidth]{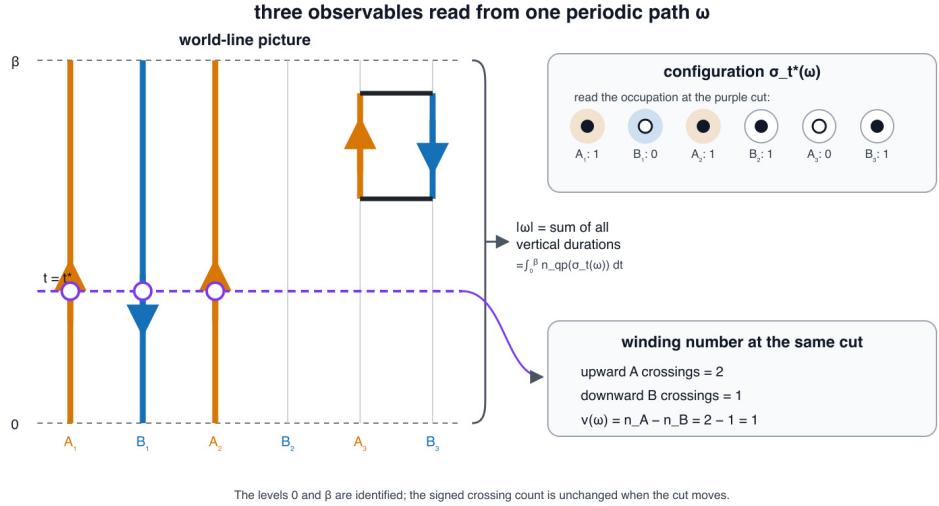}
\caption{Three quantities read from the same periodic path $\omega$.
The horizontal cut at $t=t_{*}$ determines the occupation configuration
$\fun{\sigma_{t_{*}}}{\omega}$.
The quantity $\absvol{\omega}$ is the sum of the durations of all vertical
segments, equivalently the time integral of the quasi-particle number.
The winding number is the signed number of world lines crossing any
constant-time cut: upward $A$ crossings count positively and downward $B$
crossings negatively.
In the displayed example the cut has two $A$ crossings and one $B$ crossing,
so $\fun{\nu}{\omega}=1$.}
\label{fig:path-observables}
\end{figure}

Figure \ref{fig:path-observables} shows how the three path observables are read geometrically. Moving the purple cut changes \(\fun{\sigma_t}{\omega}\) and may change \(\fun{n_{\mathrm{qp}}}{\sigma_t}\), whereas the signed difference \(\fun{n_{\mathrm{A}}}{\sigma_t}-\fun{n_{\mathrm{B}}}{\sigma_t}\) remains constant.

For \(z > 0\), the measure \(v_{z}\) on either path space is defined by the following identity for every nonnegative measurable function \(F\) on \(\oa{A}^{\alpha}\):

\begin{equation}\label{eq:loop-measure-integral}
\int_{\oa{A}^{\alpha}}
\fun{F}{\omega}
\opdmsr{v_{z}(\omega)}
=
\sum_{n \geq 0}
z^{n}
\sum_{\rbk{\sigma,b} \in \oa{C}_{n}^{\alpha}}
\int_{\fun{\Delta_{n}}{\beta}}
\fun{F}{\fun{\omega}{\sigma;b,t}}
\opdmsr{t_{1}}
\dotsm
\opdmsr{t_{n}}.
\end{equation}

The \(n = 0\) term uses the empty time integral.

\begin{defn}[loop measure]\label{def:loop-measure}
The family of path-space measures specified by
Equations \eqref{eq:loop-path-space} and \eqref{eq:loop-measure-integral}
is called the loop measure.
\end{defn}

\begin{figure}[!htbp]
\centering
\includesvg[width=0.94\linewidth]{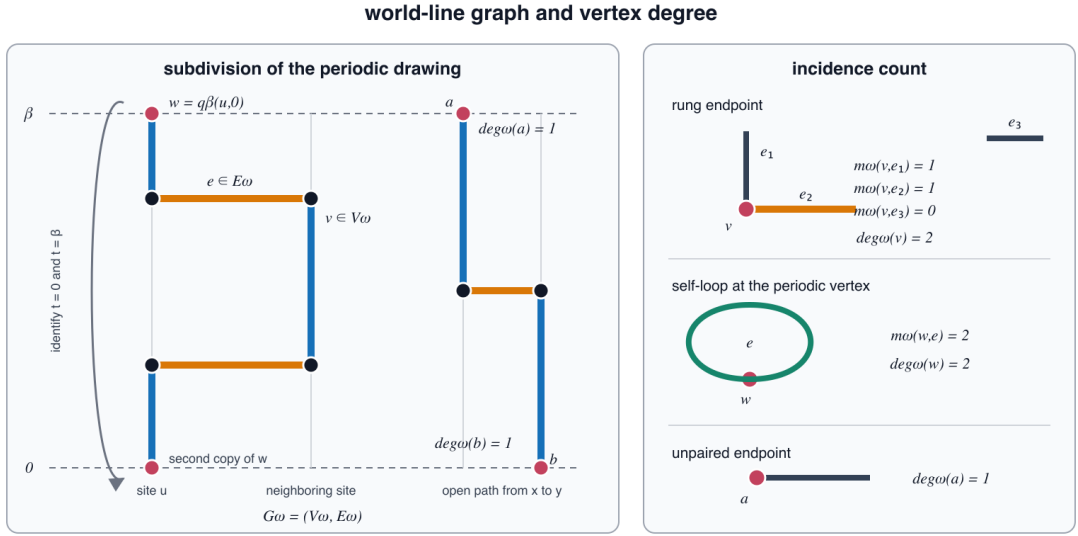}
\caption{The graph and degree in
Definition \ref{def:world-line-graph-degree}.
The left panel shows the subdivision of a periodic world line:
the filled circles form $V_{\omega}$,
and the intervening vertical segments and horizontal rungs form
$E_{\omega}$.
The upper and lower copies of $w=\fun{q_{\beta}}{u,0}$ represent one vertex
after the identification of times $0$ and $\beta$.
The right panel displays the incidence count.
A rung endpoint has one vertical and one horizontal incident edge,
while the separate edge $e_{3}$ is not incident with that vertex.
Their incidence counts are respectively $1$, $1$, and $0$,
and the vertex has degree $2$.
A self-loop has two ends incident with the same vertex and also contributes
$2$.
An unpaired endpoint of an open path has degree $1$.}
\label{fig:world-line-graph-degree}
\end{figure}

\begin{defn}[world-line graph and vertex degree]\label{def:world-line-graph-degree}
Let either $\alpha=\mathrm{o}$,
or let $x,y\in\Lambda$ be distinct and
$\alpha=\rbk{x,y}$.
The quotient map $q_{\beta}$ is defined by \eqref{eq:periodic-spacetime-quotient}.
For each
$\omega
\in
\oa{A}^{\alpha}$,
subdivide the world-line drawing of $\omega$ at every endpoint of a
horizontal rung and at every point
$\fun{q_{\beta}}{u,0}$
met by a vertical world line.
After the identification of times $0$ and $\beta$,
the resulting finite multigraph is denoted by
$$G_{\omega}
=
\rbk{V_{\omega},E_{\omega}}.$$
Its edges are the horizontal rungs and the maximal vertical world-line
segments between consecutive subdivision points.
At a rung endpoint,
the rung and the unique vertical quasi-particle segment beginning or ending
there are incident with the same vertex.
At
$\fun{q_{\beta}}{u,0}$,
the vertical half-edges at times $0$ and $\beta$ are paired whenever the
boundary condition of the path pairs them.

For
$v
\in
V_{\omega}$
and
$e
\in
E_{\omega}$,
let
$\fun{m_{\omega}}{v,e}
\in
\setone{0,1,2}$
be the number of ends of $e$ incident with $v$.
The degree of $v$ is
$$\fun{\deg_{\omega}}{v}
=
\sum_{e \in E_{\omega}}
\fun{m_{\omega}}{v,e}.$$
The construction of $G_{\omega}$,
the three possible values of $\fun{m_{\omega}}{v,e}$,
and the resulting vertex degrees are illustrated in
Figure \ref{fig:world-line-graph-degree}.
\end{defn}

An edge whose two ends are incident with the same vertex contributes \(2\) to that vertex degree. A world-line component of \(\omega\) is the geometric realization in \(\Gamma\) of a connected component of \(G_{\omega}\).

\begin{lem}[component decomposition of an open path]\label{lem:loop-component-decomposition}
Let $x,y \in \Lambda$ be distinct.
Every $\omega \in \oa{A}^{\rbk{x,y}}$ has exactly one open world-line component $\gamma$.
This component admits a parametrization satisfying
$$\begin{aligned}
\gamma
\colon
\closedinterval{0}{1}
\to
\Gamma,
\quad
\fun{\gamma}{0}
=
\fun{q_{\beta}}{x,0},
\quad
\fun{\gamma}{1}
=
\fun{q_{\beta}}{y,\beta}
=
\fun{q_{\beta}}{y,0}.
\end{aligned}$$
Every other component is closed.
Consequently,
there are closed components $\ell_{1},\dotsc,\ell_{m}$,
with $m \geq 0$,
and their disjoint decomposition is
\begin{equation}\label{eq:loop-component-decomposition}
\omega
=
\gamma
\sqcup
\ell_{1}
\sqcup
\dotsb
\sqcup
\ell_{m}.
\end{equation}
\end{lem}

\begin{figure}[!htbp]
\centering
\includesvg[width=0.82\linewidth]{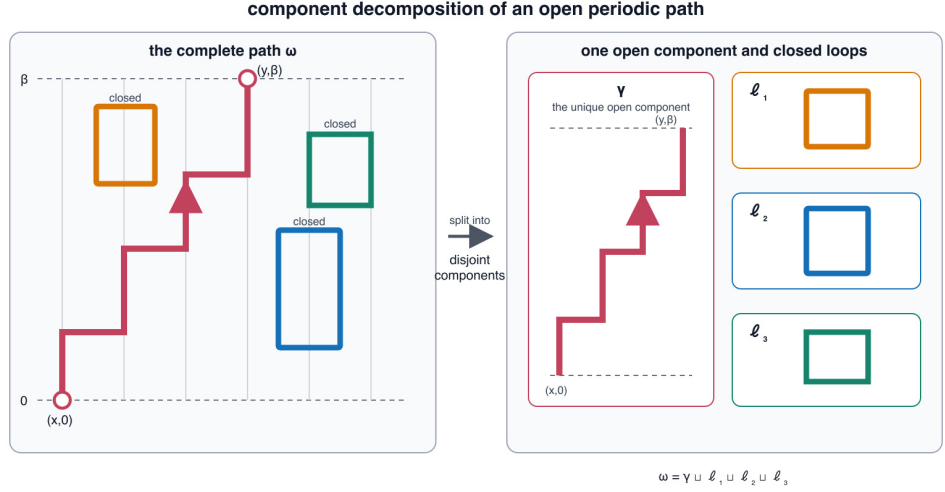}
\caption{The component decomposition in
Lemma \ref{lem:loop-component-decomposition}.
With the vertex degree of Definition \ref{def:world-line-graph-degree},
the two vertices of degree $1$ are joined by the unique open
component $\gamma$, shown in red.
Every other connected component satisfies
$\fun{\deg_{\omega}}{v}=2$
at every vertex and is
therefore a closed loop $\ell_j$.
The right panel separates the disjoint components of the path shown on the
left.}
\label{fig:loop-component-decomposition}
\end{figure}

Figure \ref{fig:loop-component-decomposition} illustrates the conclusion of Lemma \ref{lem:loop-component-decomposition}. The complete open path consists of one component connecting \(\fun{q_{\beta}}{x,0}\) to \(\fun{q_{\beta}}{y,\beta}\) and a finite collection of closed components.

\begin{proof}
Write
$\omega
=
\fun{\omega}{\sigma;b,t}$,
where
$\rbk{\sigma,b}
\in
\oa{C}_{n}^{\rbk{x,y}}$
and
$t
\in
\fun{\Delta_{n}}{\beta}$.
Let
$G_{\omega}
=
\rbk{V_{\omega},E_{\omega}}$
be the world-line graph of Definition
\ref{def:world-line-graph-degree}.
It is finite because $\Lambda$ and $n$ are finite.

Between consecutive jump times,
each world line has a unique vertical continuation.
At every admissible jump,
the quasi-particle status changes simultaneously at the two endpoints of the
jump bond.
Consequently,
the horizontal rung is incident with exactly one vertical half-edge at each
of its endpoints,
so every rung endpoint has degree $2$.

The boundary condition in the definition of
$\oa{C}_{n}^{\rbk{x,y}}$
glues the segments at times $0$ and $\beta$ at every site other than $x$ and $y$.
At $x$ and $y$ it leaves precisely the two unpaired endpoints
$\fun{q_{\beta}}{x,0}$
and
$\fun{q_{\beta}}{y,\beta}
=
\fun{q_{\beta}}{y,0}$.
With the convention that a self-loop contributes twice,
Definition \ref{def:world-line-graph-degree} therefore gives
$$\fun{\deg_{\omega}}{v}
=
\begin{cases}
1,
&
v
\in
\setone{
\fun{q_{\beta}}{x,0},
\fun{q_{\beta}}{y,0}
},
\\
2,
&
v
\in
V_{\omega}
\setminus
\setone{
\fun{q_{\beta}}{x,0},
\fun{q_{\beta}}{y,0}
}.
\end{cases}$$

For a connected component $C$ of $G_{\omega}$,
write $V_{C}$ and $E_{C}$ for its vertex and edge sets.
The handshaking identity is
$$\sum_{v \in V_{C}}
\fun{\deg_{\omega}}{v}
=
2\abscard{E_{C}}.$$
Hence $C$ contains an even number of vertices of odd degree.
The two vertices at which
$\fun{\deg_{\omega}}{v}=1$
therefore belong to the same component.
That component is a path from
$\fun{q_{\beta}}{x,0}$
to
$\fun{q_{\beta}}{y,0}$
and is the unique open component $\gamma$.
Every remaining finite connected component satisfies
$\fun{\deg_{\omega}}{v}=2$
at every vertex and is therefore a closed loop.
The finiteness of the graph gives finitely many such loops,
which proves \eqref{eq:loop-component-decomposition}.
\end{proof}

\begin{prop}[loop representation]\label{prop:loop-representation}
Let $P_{k}$ be the sector projection defined in
\eqref{eq:main-sector-projection}.
The total vertical length $\absvol{\omega}$,
winding number $\nu$,
and measure $v_{\onehalf}$ are defined by
\eqref{eq:loop-total-vertical-length},
\eqref{eq:loop-winding-number},
and \eqref{eq:loop-measure-integral}, respectively.
The loop trace is
\begin{equation}\label{eq:loop-trace}
\sqfun{\trace}{P_{k} \napiernum^{-\beta \physham_{\Lambda}}}
=
\int_{\oa{A}^{\mathrm{o}}}
\napiernum^{-\lambda \absvol{\omega}}
\fndef{\setone{\fun{\nu}{\omega} = k}}
\opdmsr{v_{\onehalf}(\omega)},
\end{equation}
For nearest neighbors or arbitrary $x \neq y$,
the inserted loop trace is
\begin{equation}\label{eq:loop-twopoint}
\sqfun{\trace}{S^{+}_{x} S^{-}_{y} \napiernum^{-\beta \physham_{\Lambda}}} = \int_{\oa{A}^{\rbk{x, y}}} \napiernum^{-\lambda \absvol{\omega}} \opdmsr{v_{\onehalf}(\omega)},
\end{equation}
so that the thermal two-point function is the ratio
\begin{equation}\label{eq:loop-ratio}
\fun{\oastate[\psi_{\beta,\Lambda}]}{S^{+}_{x} S^{-}_{y}} = \frac{\int_{\oa{A}^{\rbk{x, y}}} \napiernum^{-\lambda \absvol{\omega}} \opdmsr{v_{\onehalf}(\omega)}}{\int_{\oa{A}^{\mathrm{o}}} \napiernum^{-\lambda \absvol{\omega}} \opdmsr{v_{\onehalf}(\omega)}}.
\end{equation}
All integrands are nonnegative.
\end{prop}

\begin{proof}
Insert configuration-basis resolutions between the factors of \eqref{eq:dyson-expansion} inside the trace.
The diagonal factors contribute
$$\prod_{i}
\napiernum^{-\rbk{t_{i+1}-t_{i}}
\lambda\fun{n_{\mathrm{qp}}}{\sigma^{\rbk{i}}}}
=\napiernum^{-\lambda\absvol{\omega}}.$$
The exponent is the time integral of the quasi-particle number.
It equals the total vertical length.
Each factor $-T$ contributes its matrix element,
which is $\frac{1}{2}$ for each admissible bond and vanishes otherwise;
this yields the weight $z^{n}$ with $z = \frac{1}{2}$ and restricts to admissible sequences.
The projection $P_{k}$ restricts the initial configuration to the sector $S^{3}_{\txttot,\Lambda} = k$,
which by the conservation law is the constraint $\fun{\nu}{\omega} = k$.
This proves \eqref{eq:loop-trace}.

For \eqref{eq:loop-twopoint},
the factor $S^{+}_{x}S^{-}_{y}$ at time $0$ changes the temporal boundary condition to the one defining $\oa{A}^{\rbk{x,y}}$.
Its matrix element between the corresponding initial and final configurations is $1$.
Summing \eqref{eq:loop-trace} over $k$ removes the winding constraint and
gives
$$\sqfun{\trace}{\napiernum^{-\beta\physham_{\Lambda}}}
=\sum_{k}
\sqfun{\trace}{P_{k}\napiernum^{-\beta\physham_{\Lambda}}}.$$
Dividing by this unrestricted integral proves \eqref{eq:loop-ratio}.
\end{proof}

The component decomposition induces a restricted product decomposition of the loop measure. Let \(\oa{B}^{\rbk{x,y}}
\subset
\oa{A}^{\rbk{x,y}}\) consist of the paths for which the decomposition \eqref{eq:loop-component-decomposition} has no closed component. For \(\gamma
\in
\oa{B}^{\rbk{x,y}}\), define the compatible closed paths by \[\oa{A}^{\mathrm{o}}_{\gamma}
=
\set{\eta \in \oa{A}^{\mathrm{o}}}
{\gamma \sqcup \eta \in \oa{A}^{\rbk{x,y}}}.\]

\begin{prop}[factorization along the open component]\label{prop:loop-open-component-factorization}
Let $x,y \in \Lambda$ be distinct and let $z > 0$.
For every nonnegative measurable function
$F$
on
$\oa{A}^{\rbk{x,y}}$,
the loop measure satisfies
$$\int_{\oa{A}^{\rbk{x,y}}}
\fun{F}{\omega}
\opdmsr{v_z(\omega)}
=
\int_{\oa{B}^{\rbk{x,y}}}
\rbk{\int_{\oa{A}^{\mathrm{o}}_{\gamma}}
\fun{F}{\gamma \sqcup \eta}
\opdmsr{v_z(\eta)}}
\opdmsr{v_z(\gamma)}.$$
The weighted path integral satisfies
\begin{equation}\label{eq:loop-factorization}
\int_{\oa{A}^{\rbk{x,y}}}
\napiernum^{-\lambda\absvol{\omega}}
\opdmsr{v_z(\omega)}
=
\int_{\oa{B}^{\rbk{x,y}}}
\napiernum^{-\lambda\absvol{\gamma}}
\sqbk{
\int_{\oa{A}^{\mathrm{o}}_{\gamma}}
\napiernum^{-\lambda\absvol{\eta}}
\opdmsr{v_z(\eta)}
}
\opdmsr{v_z(\gamma)}.
\end{equation}
\end{prop}

\begin{proof}
Lemma \ref{lem:loop-component-decomposition} gives the unique decomposition
$$\begin{aligned}
\omega
=
\gamma
\sqcup
\eta,
\quad
\gamma
\in
\oa{B}^{\rbk{x,y}},
\quad
\eta
\in
\oa{A}^{\mathrm{o}}_{\gamma}.
\end{aligned}$$
Conversely,
every pair in the indicated spaces has an admissible union in
$\oa{A}^{\rbk{x,y}}$.
The union map is the bijection
$$\begin{aligned}
\set{\rbk{\gamma,\eta}}
{\gamma \in \oa{B}^{\rbk{x,y}},
\eta \in \oa{A}^{\mathrm{o}}_{\gamma}}
\to
\oa{A}^{\rbk{x,y}};
\quad
\rbk{\gamma,\eta}
\mapsto
\gamma
\sqcup
\eta.
\end{aligned}$$

Suppose that $\gamma$ has $r$ rungs and $\eta$ has $s$ rungs.
Their admissible union has $r+s$ rungs,
and its fugacity factor satisfies
$$z^{r+s}
=
z^r z^s.$$
Up to subsets of Lebesgue measure zero,
the product
$\fun{\Delta_r}{\beta}
\times
\fun{\Delta_s}{\beta}$
is the disjoint union of the possible interlacings of the two ordered time
lists.
Merging the time lists on each interlacing gives the ordered simplex
$\fun{\Delta_{r+s}}{\beta}$
in \eqref{eq:loop-measure-integral}.
The bond sequences decompose in the same way into the rungs of $\gamma$ and
the rungs of $\eta$.
Applying \eqref{eq:loop-measure-integral} to these decompositions gives
$$\int_{\oa{A}^{\rbk{x,y}}}
\fun{F}{\omega}
\opdmsr{v_z(\omega)}
=
\int_{\oa{B}^{\rbk{x,y}}}
\sqbk{
\int_{\oa{A}^{\mathrm{o}}_{\gamma}}
\fun{F}{\gamma \sqcup \eta}
\opdmsr{v_z(\eta)}
}
\opdmsr{v_z(\gamma)}.$$

The total vertical length is additive:
$$\absvol{\gamma \sqcup \eta}
=
\absvol{\gamma}
+
\absvol{\eta}.$$
The corresponding weight factorization is
$$\napiernum^{-\lambda\absvol{\gamma \sqcup \eta}}
=
\napiernum^{-\lambda\absvol{\gamma}}
\napiernum^{-\lambda\absvol{\eta}}.$$
Substitution of this weight for $F$ in the measure decomposition proves
\eqref{eq:loop-factorization}.
\end{proof}

\subsection{Reflection positivity of the loop measure}\label{reflection-positivity-of-the-loop-measure}

Space and imaginary-time reflections preserve the world-line weights and yield reflection positivity of the loop measure. This positivity is the input for the chessboard estimate. The periodic space-time \(\Gamma\) and quotient map \(q_{\beta}\) are defined by \eqref{eq:periodic-spacetime-quotient}. For \(t_{0} \in \rightopeninterval{0}{\beta/2}\), the imaginary-time reflection and the space-time lift of the spatial reflection \eqref{eq:bond-bisecting-spatial-reflection} are \begin{equation}\label{eq:loop-spacetime-reflections}
\begin{aligned}
\fun{\vartheta_{\mathrm{tm},t_{0}}}{\fun{q_{\beta}}{x,t}}
&=
\fun{q_{\beta}}{x,2t_{0}-t},
\quad
\fun{\vartheta_{\mathrm{sp}}}{\fun{q_{\beta}}{x,t}}
=
\fun{q_{\beta}}{\fun{\theta}{x},t}.
\end{aligned}
\end{equation} The corresponding negative and positive halves are \begin{equation}\label{eq:loop-reflection-halves}
\begin{aligned}
\Gamma_{\mathrm{tm},t_{0},-}
&=
\fun{q_{\beta}}{\Lambda \times \closedinterval{t_{0}}{t_{0}+\beta/2}},
\quad
\Gamma_{\mathrm{tm},t_{0},+}
=
\fun{q_{\beta}}{\Lambda \times \closedinterval{t_{0}-\beta/2}{t_{0}}},
\\
\Gamma_{\mathrm{sp},-}
&=
\fun{q_{\beta}}{\Lambda_{-} \times \fldreal},
\quad
\Gamma_{\mathrm{sp},+}
=
\fun{q_{\beta}}{\Lambda_{+} \times \fldreal}.
\end{aligned}
\end{equation} The reflection \(\vartheta_{\mathrm{tm},t_{0}}\) exchanges \(\Gamma_{\mathrm{tm},t_{0},-}\) and \(\Gamma_{\mathrm{tm},t_{0},+}\), whereas \(\vartheta_{\mathrm{sp}}\) exchanges \(\Gamma_{\mathrm{sp},-}\) and \(\Gamma_{\mathrm{sp},+}\). For either \(\vartheta=\vartheta_{\mathrm{tm},t_{0}}\) or \(\vartheta=\vartheta_{\mathrm{sp}}\), the unified half-space notation is defined by \begin{equation}\label{eq:loop-reflection-unified-halves}
\rbk{\Gamma_{\vartheta,-},\Gamma_{\vartheta,+}}
=
\begin{cases}
\rbk{\Gamma_{\mathrm{tm},t_{0},-},\Gamma_{\mathrm{tm},t_{0},+}},
&\vartheta=\vartheta_{\mathrm{tm},t_{0}},
\\
\rbk{\Gamma_{\mathrm{sp},-},\Gamma_{\mathrm{sp},+}},
&\vartheta=\vartheta_{\mathrm{sp}}.
\end{cases}
\quad
\vartheta
\in
\setone{\vartheta_{\mathrm{sp}}}
\cup
\set{\vartheta_{\mathrm{tm},t_{0}}}
{t_{0} \in \rightopeninterval{0}{\beta/2}}.
\end{equation} The path \(\fun{\vartheta}{\omega}\) is obtained by applying \(\vartheta\) to every vertical segment and rung of \(\omega\). The two half-path spaces are \begin{equation}\label{eq:loop-half-path-spaces}
\begin{aligned}
\oa{A}^{\mathrm{o}}_{\vartheta,-}
=
\set{\fnrestr{\omega}{\Gamma_{\vartheta,-}}}
{\omega \in \oa{A}^{\mathrm{o}}},
\quad
\oa{A}^{\mathrm{o}}_{\vartheta,+}
=
\set{\fnrestr{\omega}{\Gamma_{\vartheta,+}}}
{\omega \in \oa{A}^{\mathrm{o}}},
\quad
\vartheta
\in
\setone{\vartheta_{\mathrm{sp}}}
\cup
\set{\vartheta_{\mathrm{tm},t_{0}}}
{t_{0} \in \rightopeninterval{0}{\beta/2}}.
\end{aligned}
\end{equation} For a bounded measurable function \(h\colon\oa{A}^{\mathrm{o}}_{\vartheta,-}\to\fldcmp\), its reflected conjugate \(\widetilde h_{\vartheta}\colon\oa{A}^{\mathrm{o}}_{\vartheta,+}\to\fldcmp\) is defined by \begin{equation}\label{eq:loop-reflected-function}
\fun{\widetilde{h}_{\vartheta}}{\omega_{+}}
=
\cmpconj{\fun{h}{\fun{\vartheta}{\omega_{+}}}},
\quad
\omega_{+}
\in
\oa{A}^{\mathrm{o}}_{\vartheta,+},
\quad
\vartheta
\in
\setone{\vartheta_{\mathrm{sp}}}
\cup
\set{\vartheta_{\mathrm{tm},t_{0}}}
{t_{0} \in \rightopeninterval{0}{\beta/2}}.
\end{equation}

\begin{lem}[reflection positivity]\label{lem:loop-rp}
For every
$\vartheta \in \setone{\vartheta_{\mathrm{sp}}} \cup \set{\vartheta_{\mathrm{tm},t_{0}}}{t_{0} \in \rightopeninterval{0}{\beta/2}}$
and every bounded measurable $h$ as above,
\begin{equation}\label{eq:loop-reflection-positivity}
\int_{\oa{A}^{\mathrm{o}}}
\fun{h}{\fnrestr{\omega}{\Gamma_{\vartheta,-}}}
\fun{\widetilde{h}_{\vartheta}}{\fnrestr{\omega}{\Gamma_{\vartheta,+}}}
\napiernum^{-\lambda \absvol{\omega}}
\opdmsr{v_{z}(\omega)}
\geq
0.
\end{equation}
The same holds after multiplication by
$\fndef{\fun{\nu}{\omega} = 0}$.
\end{lem}

\begin{proof}
For a horizontal reflection, time-translation invariance permits us to take
$t_{0}=0$.
The two fixed time slices are then $0$ and $\beta/2$.
For
$\sigma,\eta\colon\Lambda\to\setone{0,1}$,
let
$\oa{A}^{-}_{\sigma,\eta}$
be the admissible half-paths on
$\Lambda\times\closedinterval{0}{\beta/2}$
whose configurations at the two endpoints are $\sigma$ and $\eta$.
Let
$v^{-}_{z,\sigma,\eta}$
be the half-path measure obtained from
\eqref{eq:loop-measure-integral}
by replacing $\beta$ with $\beta/2$ and fixing these endpoint configurations.
Set
$$\fun{W_{\sigma,\eta}}{h}
=
\int_{\oa{A}^{-}_{\sigma,\eta}}
\fun{h}{\omega_{-}}
\napiernum^{-\lambda\absvol{\omega_{-}}}
\opdmsr{v^{-}_{z,\sigma,\eta}(\omega_{-})}.$$
Define
$\oa{A}^{+}_{\sigma,\eta}$
and
$v^{+}_{z,\sigma,\eta}$
on the other half of the time circle in the same way.
Time reflection gives
$$\begin{aligned}
\fun{\vartheta_{\mathrm{tm},0}}{\oa{A}^{+}_{\sigma,\eta}}
=
\oa{A}^{-}_{\sigma,\eta},
\quad
\absvol{\fun{\vartheta_{\mathrm{tm},0}}{\omega_{+}}}
=
\absvol{\omega_{+}},
\quad
\rbk{\vartheta_{\mathrm{tm},0}}_{*}v^{+}_{z,\sigma,\eta}
=
v^{-}_{z,\sigma,\eta}.
\end{aligned}$$
The definition \eqref{eq:loop-reflected-function} for
$\vartheta=\vartheta_{\mathrm{tm},0}$ gives
$$\int_{\oa{A}^{+}_{\sigma,\eta}}
\fun{\widetilde h_{\vartheta_{\mathrm{tm},0}}}{\omega_{+}}
\napiernum^{-\lambda\absvol{\omega_{+}}}
\opdmsr{v^{+}_{z,\sigma,\eta}(\omega_{+})}
=
\cmpconj{\fun{W_{\sigma,\eta}}{h}}.$$
Paths with a rung on either fixed time slice have zero loop measure.
Every remaining path has a unique decomposition into the two half-paths.
The unique half-path decomposition gives
$$\begin{aligned}
&\int_{\oa{A}^{\mathrm{o}}}
\fun{h}{\fnrestr{\omega}{\Gamma_{\mathrm{tm},0,-}}}
\fun{\widetilde h_{\vartheta_{\mathrm{tm},0}}}
{\fnrestr{\omega}{\Gamma_{\mathrm{tm},0,+}}}
\napiernum^{-\lambda\absvol{\omega}}
\opdmsr{v_z(\omega)}
\\ 
&=
\sum_{\sigma,\eta\colon\Lambda\to\setone{0,1}}
\fun{W_{\sigma,\eta}}{h}
\cmpconj{\fun{W_{\sigma,\eta}}{h}}
=
\sum_{\sigma,\eta\colon\Lambda\to\setone{0,1}}
\abs{\fun{W_{\sigma,\eta}}{h}}^{2}
\geq0.
\end{aligned}$$
The winding number at the fixed slice $t=0$ is
$\fun{\nu}{\omega}
=
\fun{n_{\mathrm{A}}}{\sigma}
-\fun{n_{\mathrm{B}}}{\sigma}$.
The same decomposition yields
$$\begin{aligned}
&\int_{\oa{A}^{\mathrm{o}}}
\fun{h}{\fnrestr{\omega}{\Gamma_{\mathrm{tm},0,-}}}
\fun{\widetilde h_{\vartheta_{\mathrm{tm},0}}}
{\fnrestr{\omega}{\Gamma_{\mathrm{tm},0,+}}}
\fndef{\fun{\nu}{\omega}=0}
\napiernum^{-\lambda\absvol{\omega}}
\opdmsr{v_z(\omega)}
\\ 
&=
\sum_{\sigma,\eta\colon\Lambda\to\setone{0,1}}
\fndef{\fun{n_{\mathrm{A}}}{\sigma}
-\fun{n_{\mathrm{B}}}{\sigma}
=
0}
\abs{\fun{W_{\sigma,\eta}}{h}}^{2}
\geq0.
\end{aligned}$$

For a vertical reflection,
the crossing-bond set $M$ and ordered simplex $\fun{\Delta_m}{\beta}$ are
defined by \eqref{eq:reflection-crossing-bonds} and
\eqref{eq:ordered-jump-time-simplex}, respectively.
The crossing-data space is defined by
\begin{equation}\label{eq:loop-crossing-data-space}
\setindex{R}
=
\bigcup_{m \geq 0}
\set{\seq{\rbk{b_j,t_j}}{1 \leq j \leq m}}
{
\rbk{b_1,\dotsc,b_m}
\in
M^m,
\quad
\rbk{t_1,\dotsc,t_m}
\in
\fun{\Delta_m}{\beta}
}.
\end{equation}
Its $m=0$ member is the empty crossing datum.
Let $\rho_z$ be the positive measure on the crossing-data space
$\setindex{R}$ in \eqref{eq:loop-crossing-data-space}, defined by
$$\int_{\setindex{R}}
\fun{F}{R}
\opdmsr{\rho_z(R)}
=
\sum_{m\geq0}
z^m
\sum_{\rbk{b_1,\dotsc,b_m}\in M^m}
\int_{\fun{\Delta_m}{\beta}}
\fun{F}{\seq{\rbk{b_j,t_j}}{1\leq j\leq m}}
\opdmsr{t_1}
\dotsm
\opdmsr{t_m}.$$
For
$s\in\setone{\pm 1}^{\abscard{R}}$,
let
$\oa{A}^{-}_{R,s}$
be the negative-half paths whose sources and sinks at the crossing rungs are
specified by $s$.
The measure
$v^{-}_{z,R,s}$
includes the fugacities of the rungs contained entirely in the negative half.
Set
$$\fun{V_{R,s}}{h}
=
\int_{\oa{A}^{-}_{R,s}}
\fun{h}{\omega_{-}}
\napiernum^{-\lambda\absvol{\omega_{-}}}
\opdmsr{v^{-}_{z,R,s}(\omega_{-})}.$$
Spatial reflection interchanges the sublattices and reverses every crossing
signature.
For the corresponding positive-half path space and measure,
$$\begin{aligned}
\fun{\vartheta_{\mathrm{sp}}}{\oa{A}^{+}_{R,-s}}
=
\oa{A}^{-}_{R,s},
\quad
\absvol{\fun{\vartheta_{\mathrm{sp}}}{\omega_{+}}}
=
\absvol{\omega_{+}},
\quad
\rbk{\vartheta_{\mathrm{sp}}}_{*}v^{+}_{z,R,-s}
=
v^{-}_{z,R,s}.
\end{aligned}$$
The definition \eqref{eq:loop-reflected-function} for
$\vartheta=\vartheta_{\mathrm{sp}}$ gives
$$\int_{\oa{A}^{+}_{R,-s}}
\fun{\widetilde h_{\vartheta_{\mathrm{sp}}}}{\omega_{+}}
\napiernum^{-\lambda\absvol{\omega_{+}}}
\opdmsr{v^{+}_{z,R,-s}(\omega_{+})}
=
\cmpconj{\fun{V_{R,s}}{h}},$$
The decomposition over the crossing data and signatures gives
$$\begin{aligned}
\int_{\oa{A}^{\mathrm{o}}}
\fun{h}{\fnrestr{\omega}{\Gamma_{\mathrm{sp},-}}}
\fun{\widetilde h_{\vartheta_{\mathrm{sp}}}}
{\fnrestr{\omega}{\Gamma_{\mathrm{sp},+}}}
\napiernum^{-\lambda\absvol{\omega}}
\opdmsr{v_z(\omega)}
=
\int_{\setindex{R}}
\sum_{s\in\setone{-1,+1}^{\abscard{R}}}
\abs{\fun{V_{R,s}}{h}}^{2}
\opdmsr{\rho_z(R)}
\geq0.
\end{aligned}$$

For a decomposed path
$\omega=\omega_{-}\cup\omega_{+}$,
write
$\fun{\nu_{\mp}}{\omega_{\mp}}$
for the signed winding contribution of the indicated half.
Spatial reflection exchanges the sublattices, and therefore
$$\begin{aligned}
\fun{\nu}{\omega}
=
\fun{\nu_{-}}{\omega_{-}}
+
\fun{\nu_{+}}{\omega_{+}},
\quad
\fun{\nu_{+}}{\omega_{+}}
=
-\fun{\nu_{-}}{\fun{\vartheta_{\mathrm{sp}}}{\omega_{+}}}.
\end{aligned}$$
For $k\in\ringratint$, define
$$\fun{V_{R,s,k}}{h}
=
\int_{\oa{A}^{-}_{R,s}}
\fun{h}{\omega_{-}}
\fndef{\fun{\nu_{-}}{\omega_{-}}=k}
\napiernum^{-\lambda\absvol{\omega_{-}}}
\opdmsr{v^{-}_{z,R,s}(\omega_{-})}.$$
The identity
$\fndef{\fun{\nu}{\omega}=0}
=
\sum_{k\in\ringratint}
\fndef{\fun{\nu_{-}}{\omega_{-}}=k}
\cdot
\fndef{\fun{\nu_{+}}{\omega_{+}}=-k}$
then gives
$$\begin{aligned}
&\int_{\oa{A}^{\mathrm{o}}}
\fun{h}{\fnrestr{\omega}{\Gamma_{\mathrm{sp},-}}}
\fun{\widetilde h_{\vartheta_{\mathrm{sp}}}}
{\fnrestr{\omega}{\Gamma_{\mathrm{sp},+}}}
\fndef{\fun{\nu}{\omega}=0}
\napiernum^{-\lambda\absvol{\omega}}
\opdmsr{v_z(\omega)}
\\ 
&=
\int_{\setindex{R}}
\sum_{s\in\setone{-1,+1}^{\abscard{R}}}
\sum_{k\in\ringratint}
\abs{\fun{V_{R,s,k}}{h}}^{2}
\opdmsr{\rho_z(R)}
\geq0.
\end{aligned}$$
\end{proof}

\subsection{The chessboard estimate}\label{the-chessboard-estimate}

Repeated reflection Schwarz inequalities spread a local event over the periodic space-time cells. The resulting chessboard estimate bounds simultaneous local constraints by a fully disseminated event.

\begin{lem}[chessboard estimate]\label{lem:chessboard}
Let $\rbk{\Omega,\msrprb}$ be a probability space.
Assume that the cell-boundary reflections of the torus
$T_N=\ringratint/N\ringratint$
act on it by measure-preserving transformations.
The integer $N$ is even.
Each reflection fixes two antipodal cell boundaries and exchanges the complementary arcs of $N/2$ cells.
Assume reflection positivity with respect to every such reflection in the sense of Lemma \ref{lem:loop-rp}.
Let $F_{c}$, $c \in T_{N}$, be a covariant family of $\closedinterval{0}{1}$-valued functions, $F_{c}$ depending only on the cell $c$.
Then
\begin{equation}\label{eq:chessboard-estimate}
\sqfun{\prbexp_{\msrprb}}{\prod_{c \in S} F_{c}}
\leq
\sqfun{\prbexp_{\msrprb}}{\prod_{c \in T_{N}} F_{c}}^{\abscard{S} / N},
\quad
S \subseteq T_{N}.
\end{equation}
The same estimate holds on
$T_{\boldsymbol{N}}=\prod_{r=1}^{d}T_{N_{r}}$,
where every $N_{r}$ is even,
with reflections in each coordinate direction and
$N=\prod_{r=1}^{d}N_{r}$.
\end{lem}

\begin{figure}[!htbp]
\centering
\includesvg[width=0.98\linewidth]{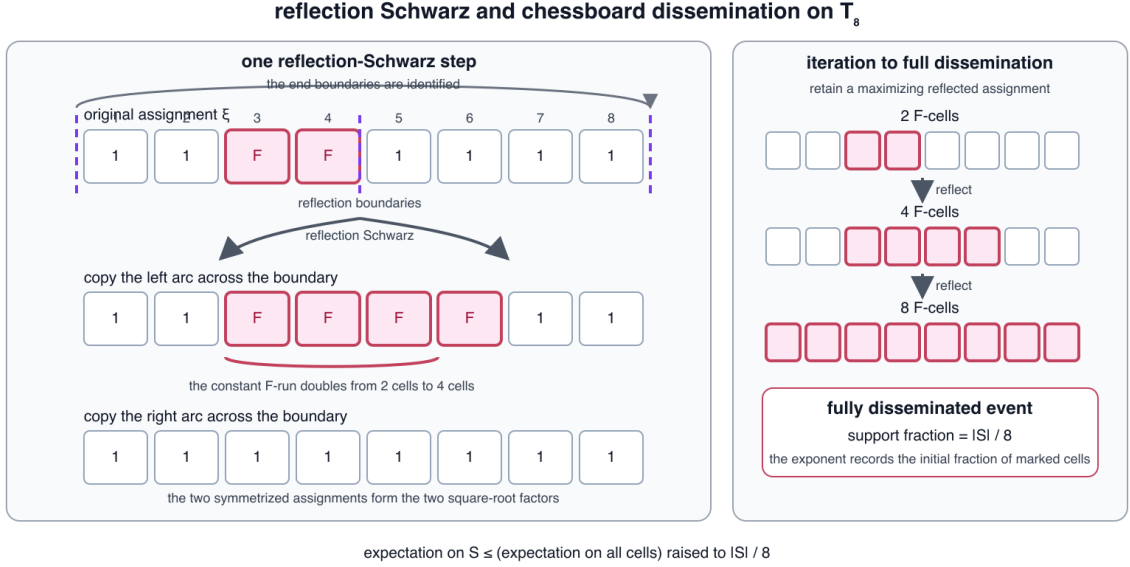}
\caption{The reflection-Schwarz mechanism behind the chessboard estimate for
$N = 8$ and $S = \setone{3,4}$.
The original assignment carries $F$ on two cells and the identity function on
the other six cells.
Reflection through the two dashed cell boundaries produces the two
symmetrized assignments entering the square-root factors.
Copying the left arc doubles the consecutive $F$-run,
while copying the right arc produces the identity assignment.
Repeating the reflection on a maximizing symmetrized assignment fills all
eight cells.
The accumulated square-root powers give the exponent
$\abscard{S} / N = 2 / 8$.}
\label{fig:chessboard-dissemination}
\end{figure}

Figure \ref{fig:chessboard-dissemination} illustrates the one-dimensional reflection step in Lemma \ref{lem:chessboard}. The example has \(N = 8\) and \(S = \setone{3,4}\). One symmetrized assignment doubles the consecutive run of cells carrying \(F\), whereas the other symmetrized assignment carries the identity function on every cell. Iteration of the first alternative disseminates \(F\) over all eight cells.

\begin{proof}
The cell-boundary reflection group acts simply transitively on $T_N$.
Indeed, the composition of two cell-boundary reflections is a rotation by
an even offset, and the group consists of $N/2$ even rotations and $N/2$
reflections.
No nonidentity rotation fixes a cell, and a cell-boundary reflection fixes
no cell.
Thus every stabilizer is trivial.
The rotations reach the cells at even offsets from $c$, whereas the
reflections $c\mapsto 2j+1-c$ reach the cells at odd offsets.
Consequently, for every $c,c'\in T_N$, there is a unique group element that
maps $c$ to $c'$.
This element transports a function $f$ on $c$ to a function
$f^{\rbk{c'}}$ on $c'$, and uniqueness makes these transports consistent
under composition.
In particular, the covariance assumption in the statement is
$F_c^{\rbk{c'}}=F_{c'}$.

For a cell-boundary reflection $\vartheta$ with complementary arcs
$H_{-}$ and $H_{+}$,
write $\vartheta(g)=g\circ\vartheta$ for its pullback action on a
bounded function $g$.
If $f$ and $g$ are supported on $H_{-}$,
the reflection Schwarz inequality is
$$
\sqfun{\prbexp_{\msrprb}}{f\vartheta(g)}
\leq
\sqfun{\prbexp_{\msrprb}}{f\vartheta(f)}^{1/2}
\sqfun{\prbexp_{\msrprb}}{g\vartheta(g)}^{1/2}.
$$

For a bounded nonnegative function $f$ supported on one cell, define
$$q(f)
=
\sqfun{\prbexp_{\msrprb}}{\prod_{c'\in T_N}f^{\rbk{c'}}}^{1/N}.$$
The lemma follows from the stronger product estimate
$$\sqfun{\prbexp_{\msrprb}}{\prod_{c\in T_N}f_c}
\leq
\prod_{c\in T_N}\fun{q}{f_c}$$
for bounded nonnegative single-cell functions $f_1,\dotsc,f_N$, where
$f_c$ is placed at cell $c$.
To recover the statement of the lemma, set
$$f_c
=
\begin{cases}
F_c,&c\in S,\\
1,&c\notin S.
\end{cases}$$
Covariance gives
$\fun{q}{F_c}
=
\sqfun{\prbexp_{\msrprb}}{\prod_{c'\in T_N}F_{c'}}^{1/N}$,
whereas $q(1)=1$.
The stronger product estimate then gives the asserted exponent
$\abscard{S}/N$.

Both sides of the stronger estimate are homogeneous of degree one in each
$f_c$.
For $\epsilon>0$,
the canonical transports fix the constant function $\epsilon$.
Since every transported copy of $f_c$ is nonnegative,
the definition of $q$ gives
$$\begin{aligned}
\fun{q}{f_c+\epsilon}^{N}
=
\sqfun{\prbexp_{\msrprb}}
{\prod_{c' \in T_N}
\rbk{f_c^{\rbk{c'}}+\epsilon}}
\geq
\sqfun{\prbexp_{\msrprb}}
{\prod_{c' \in T_N}\epsilon}
=
\sqfun{\prbexp_{\msrprb}}{\epsilon^{N}}
=
\epsilon^{N}.
\end{aligned}$$
Hence we obtain $\fun{q}{f_c+\epsilon}\geq\epsilon$.
Dominated convergence recovers the original expressions as
$\epsilon\downarrow0$.
It is therefore enough to treat the case
$\fun{q}{f_c}>0$ for every $c$, and homogeneity permits the normalization
$\fun{q}{f_c}=1$
for every $c\in T_N$.

Let $\xi$ range over the finitely many assignments
$$
\xi\colon T_N\to\setone{f_1,\dotsc,f_N}.
$$
Place each letter at its assigned cell by the canonical transport, and
define
$$
\fun{F}{\xi}
=
\sqfun{\prbexp_{\msrprb}}{\prod_{c\in T_N}\xi_c^{\rbk{c}}}.
$$
Let $K$ be the maximum of $\fun{F}{\xi}$ over these assignments.
Every constant assignment with value $f_j$ satisfies
$$
\fun{F}{\xi}
=
\fun{q}{f_j}^{N}
=
1,
$$
and hence $K\geq1$.
The left side of the stronger product estimate is
$\fun{F}{\xi^{(0)}}$ for the assignment
$\xi^{(0)}_c=f_c$.
The equality $K=1$ follows once a constant maximizing assignment is
constructed.

Let $\vartheta$ be a cell-boundary reflection with complementary arcs
$H_{-}$ and $H_{+}$.
The two symmetrized assignments are defined by
\begin{equation}\label{eq:chessboard-symmetrized-assignments}
\begin{aligned}
\fun{\xi_{-,\vartheta}}{c}
&=
\begin{cases}
\fun{\xi}{c},&c \in H_{-},
\\
\fun{\xi}{\fun{\vartheta}{c}},&c \in H_{+},
\end{cases}
\\
\fun{\xi_{+,\vartheta}}{c}
&=
\begin{cases}
\fun{\xi}{\fun{\vartheta}{c}},&c \in H_{-},
\\
\fun{\xi}{c},&c \in H_{+}.
\end{cases}
\end{aligned}
\end{equation}
Applying the reflection Schwarz inequality to the products over these arcs
gives
$$
\fun{F}{\xi}
\leq
\fun{F}{\xi_{-,\vartheta}}^{1/2}
\fun{F}{\xi_{+,\vartheta}}^{1/2}.
$$
The consistency of the transports ensures that both assignments use the
same alphabet as $\xi$.
If $\fun{F}{\xi}=K$, maximality gives
$$
K
=
\fun{F}{\xi}
\leq
\fun{F}{\xi_{-,\vartheta}}^{1/2}
\fun{F}{\xi_{+,\vartheta}}^{1/2}
\leq
K.
$$
Both factors on the middle line must therefore equal $K$, so both
reflected assignments are maximizers.

Among all maximizers, choose $\xi$ whose longest constant run has maximal
length $\ell$.
Here a run is a cyclically consecutive sequence of cells carrying the same
letter.
If $\ell=N$, then $\xi$ is constant.
Suppose that $\ell<N$, and let the cells
$i+1,\dotsc,i+\ell$ carry the letter $a$.
Choose the reflection $\vartheta$ through the boundary between $i+\ell$ and
$i+\ell+1$ and through the antipodal boundary.
Let $H_{-}$ be the arc of $N/2$ cells ending at $i+\ell$.
If $\ell\geq N/2$, then $H_{-}$ lies inside the run.
The maximizing assignment $\xi_{-,\vartheta}$ is therefore constant
with value $a$.
If $\ell<N/2$, then the last $\ell$ cells of $H_{-}$ carry $a$.
The maximizing assignment $\xi_{-,\vartheta}$ consequently carries $a$
on the $2\ell$ consecutive cells
$i+1,\dotsc,i+2\ell$.
This longer run contradicts the defining maximality of $\ell$.
Thus a constant maximizing assignment exists and $K=1$, which proves the
stronger product estimate and the lemma in one dimension.

On $T_{\boldsymbol{N}}$,
the cross-sections for a fixed coordinate direction $r$ are
$$\setone{c_{r}}
\times
\prod_{\substack{1 \leq s \leq d \\ s \neq r}}
T_{N_{s}},
\quad
c_{r}
\in
T_{N_{r}}.$$
Regard each cross-section as one cell whose internal variables are the
remaining coordinates.
The one-dimensional argument disseminates the function through coordinate
direction $r$.
Applying this argument successively for $1 \leq r \leq d$ disseminates the
function through $T_{\boldsymbol{N}}$ and proves the multidimensional
statement.
\end{proof}

\begin{lem}[contour bound; Lemma 3 of \cite{AizenmanLiebSeiringerSolovejYngvason001}]\label{lem:contour-bound}
The free energy per site is
\begin{equation}\label{eq:contour-free-energy-density}
f
=
-\frac{1}{\beta \abscard{\Lambda}}
\log \sqfun{\trace}{\napiernum^{-\beta \physham_{\Lambda}}}.
\end{equation}
Let $\setone{\gamma_{j}}$ be a family of disjoint curves in $\Gamma$.
The contour estimate is
\begin{equation}\label{eq:contour-bound}
\int_{\oa{A}^{\mathrm{o}}_{\setone{\gamma_{j}}}}
\napiernum^{-\lambda \absvol{\omega}}
\opdmsr{v_{\onehalf}(\omega)}
\leq
\napiernum^{\sum_{j} \absvol{\gamma_{j}} f}
\int_{\oa{A}^{\mathrm{o}}}
\napiernum^{-\lambda \absvol{\omega}}
\opdmsr{v_{\onehalf}(\omega)}.
\end{equation}
The same bound holds after both integrals are restricted by
$\fndef{\fun{\nu}{\omega} = 0}$.
In that case, replace $f$ by
\begin{equation}\label{eq:contour-zero-winding-free-energy-density}
f_{0}
=-\frac{1}{\beta\abscard{\Lambda}}
\log\sqfun{\trace}{P_{0}\napiernum^{-\beta\physham_{\Lambda}}}.
\end{equation}
\end{lem}

\begin{proof}
Normalize to the probability measure $\opdmsr{\msrbb{P}} = \napiernum^{-\lambda \absvol{\omega}} \opdmsr{v_{\onehalf}(\omega)} / Z$ on $\oa{A}^{\mathrm{o}}$, $Z = \sqfun{\trace}{\napiernum^{-\beta \physham_{\Lambda}}}$.
Partition $\closedinterval{0}{\beta}$ into $K=2^{s}$ intervals $I_{j}$.
Let $\setindex{C}_{K}$ be the set of
$N=\abscard{\Lambda}K$ cells
$c=\setone{x}\times I_{j}$,
where $x \in \Lambda$ and $1 \leq j \leq K$.
For each cell $c$,
let $E_{c}$ be the event that no world line of $\omega$ meets $c$.
Cell reflections and translations act covariantly on these events.
Lemma \ref{lem:loop-rp} makes $\msrbb{P}$ reflection positive for every horizontal and vertical cell-boundary reflection.
Time reflections may be placed at the dyadic slices,
and each spatial direction has an even number $L$ of cells.
For a union $D$ of cells, Lemma \ref{lem:chessboard} gives
$$\sqfun{\prbexp_{\msrbb{P}}}{\prod_{\substack{c \in \setindex{C}_{K} \\ c \subset D}} \fndef{E_{c}}} \leq \sqfun{\prbexp_{\msrbb{P}}}{\prod_{c \in \setindex{C}_{K}} \fndef{E_{c}}}^{\absvol{D} / \absvol{\Gamma}},$$
exponents measured in space-time volume, $\absvol{\Gamma} = \beta \abscard{\Lambda}$.
The full product
$\prod_{c \in \setindex{C}_{K}}\fndef{E_{c}}$
is the definition function of the empty configuration.
Every nonempty world line has positive vertical length and meets a cell.
The empty configuration contributes $1$ to the unnormalized integral,
so its $\msrbb{P}$-weight is $1/Z$.
The chessboard estimate gives
$$\sqfun{\prbexp_{\msrbb{P}}}{\prod_{\substack{c \in \setindex{C}_{K} \\ c \subset D}} \fndef{E_{c}}} \leq Z^{-\absvol{D} / \rbk{\beta \abscard{\Lambda}}} = \napiernum^{f \absvol{D}}.$$
Let $D_K$ be the union of cells contained in the vertical segments of the curves $\setone{\gamma_j}$.
A configuration in $\oa{A}^{\mathrm{o}}_{\setone{\gamma_j}}$ has no world line meeting these cells.
After normalization by $Z$,
the left side of \eqref{eq:contour-bound} is bounded by
$$\sqfun{\prbexp_{\msrbb{P}}}{\prod_{\substack{c \in \setindex{C}_{K} \\ c \subset D_K}}\fndef{E_{c}}}
\leq
\napiernum^{f\absvol{D_K}}.$$
The vertical segments have total length $\sum_j\absvol{\gamma_j}$ and finitely many endpoints.
As $K\to\infty$,
their contained cells exhaust this length:
$\absvol{D_K}\to\sum_j\absvol{\gamma_j}$.
Continuity of the exponential gives the limit
$\napiernum^{f\sum_j\absvol{\gamma_j}}$,
which proves \eqref{eq:contour-bound}.
For the winding-restricted measure,
Lemma \ref{lem:loop-rp} supplies reflection positivity.
The empty configuration has winding zero.
Its normalized weight is
$\sqfun{\trace}{P_{0} \napiernum^{-\beta \physham_{\Lambda}}}^{-1}$.
The same argument applies with $f_0$.
\end{proof}

\section{Exponential Decay and the Mott Gap}\label{sec:decay}

The loop representation expresses the two-point function as a sum over open curves. The contour bound \eqref{eq:contour-bound} penalizes each unit of curve length by the free energy per site. A random-walk estimate gives exponential decay when the effective fugacity is subcritical. This proves Theorem 2 of \cite{AizenmanLiebSeiringerSolovejYngvason001}, and the same machinery bounds the winding-number partition functions, which yields the chemical-potential gap of the Mott phase.

\subsection{Free energy bounds}\label{free-energy-bounds}

The contour estimate depends on a negative free-energy density that penalizes world-line length. The bounds below compare this density with the finite-volume ground-state energy and an elementary product-state trial pressure.

\begin{defn}[dimer covering]\label{def:dimer-covering}
Let $\setindex{B}_{\Lambda}$ be the set of unordered nearest-neighbor bonds
of the periodic graph on $\Lambda$, and call each
$b\in\setindex{B}_{\Lambda}$ a dimer.
A dimer covering of $\Lambda$ is a subset
$\setindex{M}\subset\setindex{B}_{\Lambda}$ satisfying,
for all $x \in \Lambda$,
\begin{equation}\label{eq:dimer-covering-condition}
\abscard{\set{\dbk{uv} \in \setindex{M}}
{\text{$x = u$ or $x = v$}}}
=
1.
\end{equation}
\end{defn}

\begin{lem}[canonical dimer covering of the periodic box]\label{lem:canonical-dimer-covering}
Let $\Lambda=\Lambda_{L}$ be the periodic box
\eqref{eq:main-local-algebras}, where $L$ is even, and let $e_{1}$ be the
first coordinate vector.
The bond set
\begin{equation}\label{eq:canonical-dimer-covering}
\setindex{M}_{\mathrm{dimer}}
=
\set{\dbk{x,x+e_{1}}}
{x \in \Lambda,
x_{1}
+\frac{L}{2}
\in
2 \ringratint + 1}
\end{equation}
is a dimer covering in the sense of Definition \ref{def:dimer-covering}.
It contains $\abscard{\Lambda}/2$ dimers.
\end{lem}

\begin{proof}
The possible first coordinates in \eqref{eq:main-local-algebras} are
$-\frac{L}{2}+1,
-\frac{L}{2}+2,
\ldots,
\frac{L}{2}$.
The condition in \eqref{eq:canonical-dimer-covering} selects
$-\frac{L}{2}+1,
-\frac{L}{2}+3,
\ldots,
\frac{L}{2}-1$
as the first coordinates of the left endpoints.
For every fixed choice of the remaining $d-1$ coordinates, the selected
bonds join each selected coordinate to the next coordinate in the list.
Every allowed first coordinate is therefore an endpoint of exactly one
selected bond.
Therefore equation \eqref{eq:dimer-covering-condition} follows.
There are $L/2$ selected bonds for each of the $L^{d-1}$ choices of the
remaining coordinates, and hence
$$
\abscard{\setindex{M}_{\mathrm{dimer}}}
=
\frac{L}{2}L^{d-1}
=
\frac{\abscard{\Lambda}}{2}.
$$
\end{proof}

\begin{lem}[energy of the dimer product state]\label{lem:dimer-product-state}
Let $\setindex{M}_{\mathrm{dimer}}$ be the dimer covering
\eqref{eq:canonical-dimer-covering}.
For each $b\in\setindex{M}_{\mathrm{dimer}}$, let
$x_{b}\in\Lambda_{\mathrm{B}}$ and
$y_{b}\in\Lambda_{\mathrm{A}}$ be its two endpoints.
The two-site Hamiltonian consisting of the hopping term on $b$ and the
potential terms at its endpoints is
$$
h_{\lambda,b}
=
-\frac{1}{2}
\rbk{
S^{+}_{x_{b}}S^{-}_{y_{b}}
+
S^{-}_{x_{b}}S^{+}_{y_{b}}
}
+
\lambda
\rbk{
\frac{1}{2}
-
S^{3}_{x_{b}}
}
+
\lambda
\rbk{
\frac{1}{2}
+
S^{3}_{y_{b}}
}.
$$
In the ordered one-particle basis in which the particle occupies $x_{b}$
first and $y_{b}$ second, its restriction has the matrix
\begin{equation}\label{eq:dimer-hamiltonian-matrix}
\begin{pmatrix}
0&-1/2\\
-1/2&2\lambda
\end{pmatrix}.
\end{equation}
Let $\Phi_{\lambda,b}$ be a normalized eigenvector for its lowest
eigenvalue.
Define the product vector
\begin{equation}\label{eq:dimer-product-state}
\Phi_{\mathrm{dimer}}
=
\bigotimes_{b\in\setindex{M}_{\mathrm{dimer}}}
\Phi_{\lambda,b},
\end{equation}
and its vector state is denoted by $\oastate[\psi_{\mathrm{dimer}}]$.
The vector $\Phi_{\mathrm{dimer}}$ belongs to the sector
$N_{\Lambda}=\abscard{\Lambda}/2$ and satisfies
\begin{equation}\label{eq:dimer-product-energy}
\fun{\oastate[\psi_{\mathrm{dimer}}]}{\physham_{\Lambda}}
=
\frac{\abscard{\Lambda}}{2}
\rbk{
\lambda
-
\sqrt{
\lambda^{2}
+
\frac{1}{4}
}
}.
\end{equation}
\end{lem}

\begin{proof}
Use the ordered basis in which the particle occupies $x_b$ in the first
vector and $y_b$ in the second vector.
The staggered potential energies are respectively $0$ and $2\lambda$.
The hopping matrix element between the two vectors is $-1/2$.
This gives \eqref{eq:dimer-hamiltonian-matrix}.
Its characteristic equation is
$$
\epsilon^{2}
-
2\lambda\epsilon
-
\frac{1}{4}
=
0,
$$
and its lowest eigenvalue is
$$
\epsilon_{-}
=
\lambda
-
\sqrt{
\lambda^{2}
+
\frac{1}{4}
}.
$$
Every factor $\Phi_{\lambda,b}$ has one particle.
Lemma \ref{lem:canonical-dimer-covering} gives
$\abscard{\setindex{M}_{\mathrm{dimer}}}
=
\abscard{\Lambda}/2$,
which places $\Phi_{\mathrm{dimer}}$ in the asserted sector.

Let $\dbk{xy}\notin\setindex{M}_{\mathrm{dimer}}$ be a nearest-neighbor bond.
Its endpoints belong to two distinct dimers $b_x$ and $b_y$.
The vector $S^{+}_{x}\Phi_{\lambda,b_x}$ belongs to the two-particle
sector of $b_x$,
whereas $S^{-}_{x}\Phi_{\lambda,b_x}$ belongs to its zero-particle sector.
Both sectors are orthogonal to the one-particle vector
$\Phi_{\lambda,b_x}$.
The same argument at $y$ gives
$$\bkt{\Phi_{\lambda,b_x}}
{S^{+}_{x}\Phi_{\lambda,b_x}}
=
\bkt{\Phi_{\lambda,b_x}}
{S^{-}_{x}\Phi_{\lambda,b_x}}
=
0,
\quad
\bkt{\Phi_{\lambda,b_y}}
{S^{+}_{y}\Phi_{\lambda,b_y}}
=
\bkt{\Phi_{\lambda,b_y}}
{S^{-}_{y}\Phi_{\lambda,b_y}}
=
0.$$
Since $b_x\neq b_y$,
the product structure in \eqref{eq:dimer-product-state} gives the two
factorizations
\begin{equation}\label{eq:non-dimer-bond-expectation}
\begin{aligned}
\fun{\oastate[\psi_{\mathrm{dimer}}]}{S^{+}_{x}S^{-}_{y}}
&=
\bkt{\Phi_{\lambda,b_x}}
{S^{+}_{x}\Phi_{\lambda,b_x}}
\bkt{\Phi_{\lambda,b_y}}
{S^{-}_{y}\Phi_{\lambda,b_y}}
\prod_{b \in
\setindex{M}_{\mathrm{dimer}}
\setminus
\setone{b_x,b_y}}
\bkt{\Phi_{\lambda,b}}{\Phi_{\lambda,b}}
=
0,
\\
\fun{\oastate[\psi_{\mathrm{dimer}}]}{S^{-}_{x}S^{+}_{y}}
&=
\bkt{\Phi_{\lambda,b_x}}
{S^{-}_{x}\Phi_{\lambda,b_x}}
\bkt{\Phi_{\lambda,b_y}}
{S^{+}_{y}\Phi_{\lambda,b_y}}
\prod_{b \in
\setindex{M}_{\mathrm{dimer}}
\setminus
\setone{b_x,b_y}}
\bkt{\Phi_{\lambda,b}}{\Phi_{\lambda,b}}
=
0,
\\
\fun{\oastate[\psi_{\mathrm{dimer}}]}{
S^{+}_{x}S^{-}_{y}
+
S^{-}_{x}S^{+}_{y}
}
&=
0.
\end{aligned}
\end{equation}
Every dimer bond contributes $\epsilon_{-}$, and every non-dimer bond
contributes zero by \eqref{eq:non-dimer-bond-expectation}.
The dimer count in Lemma \ref{lem:canonical-dimer-covering} now gives
\eqref{eq:dimer-product-energy}.
\end{proof}

\begin{lem}[uniform strict negativity of the energy density]\label{lem:energy-negative}
The global finite-volume ground-state energy density
$\fun{e_{\Lambda}}{\lambda}$,
defined by \eqref{eq:main-finite-volume-ground-state-energy-density},
satisfies
\begin{equation}\label{eq:uniform-negative-energy-density}
\fun{e_{\Lambda}}{\lambda}
\leq
\frac{1}{2}
\rbk{
\lambda
-
\sqrt{
\lambda^{2}
+
\frac{1}{4}
}
}
<
0.
\end{equation}
\end{lem}

\begin{proof}
Lemma \ref{lem:dimer-product-state} provides a normalized trial vector in
the half-filled sector with energy \eqref{eq:dimer-product-energy}.
The variational principle and
\eqref{eq:main-finite-volume-ground-state-energy-density}
give \eqref{eq:uniform-negative-energy-density}.
\end{proof}

\begin{lem}[free energy bounds]\label{lem:free-energy-bounds}
Let $f=\fun{f}{\beta,\lambda}$ and
$f_{0}=\fun{f_{0}}{\beta,\lambda}$ be the free-energy densities defined in
\eqref{eq:contour-free-energy-density} and
\eqref{eq:contour-zero-winding-free-energy-density}.
Let $\fun{e_{\Lambda}}{\lambda}$ be the global finite-volume ground-state
energy density defined in
\eqref{eq:main-finite-volume-ground-state-energy-density}.
The identity \eqref{eq:main-half-filling-global-sector-energy} identifies its numerator
with the sector energy $\physenergyfunc_{\Lambda,0}$ from
\eqref{eq:main-sector-energy}.
The free-energy densities and the ground-state energy density satisfy
\begin{equation}\label{eq:free-energy-bounds}
\max \setone{\abs{\fun{e_{\Lambda}}{\lambda}}, \ \frac{1}{\sminvtemperature} \log 2 - \frac{\lambda}{2}}
\leq
-f,
\quad
\abs{\fun{e_{\Lambda}}{\lambda}}
\leq
-f_{0},
\quad
\fun{e_{\Lambda}}{\lambda} < 0.
\end{equation}
\end{lem}

\begin{proof}
The strict inequality in \eqref{eq:uniform-negative-energy-density} gives
$\fun{e_{\Lambda}}{\lambda}<0$.

The global ground-state eigenvalue in
\eqref{eq:main-finite-volume-ground-state} gives
$$\sqfun{\trace}{\napiernum^{-\beta \physham_{\Lambda}}} \geq \napiernum^{-\beta \fun{\physenergyfunc_{\Lambda}}{\lambda}}.$$
Because $\fun{e_{\Lambda}}{\lambda}<0$, this gives
$$-f \geq -\fun{e_{\Lambda}}{\lambda} = \abs{\fun{e_{\Lambda}}{\lambda}}.$$
The Peierls--Bogoliubov inequality
\eqref{eq:peierls-bogoliubov}, applied with
$A=0$ and $B=-\beta\physham_{\Lambda}$, gives
$$\sqfun{\trace}{\napiernum^{-\beta \physham_{\Lambda}}}
\geq
2^{\abscard{\Lambda}}
\napiernum^{
-\frac{\beta \sqfun{\trace}{\physham_{\Lambda}}}{2^{\abscard{\Lambda}}}
}.$$
Every spin operator is traceless,
so the normalized trace of $\physham_{\Lambda}$ is $\lambda\abscard{\Lambda}/2$.
The resulting free-energy bound is
$-f \geq \beta^{-1}\log2-\lambda/2$.
The sector definition \eqref{eq:main-sector-energy} gives
$$
\sqfun{\trace}{P_{0} \napiernum^{-\beta \physham_{\Lambda}}}
\geq
\napiernum^{-\beta \physenergyfunc_{\Lambda,0}}.
$$
By \eqref{eq:main-half-filling-global-sector-energy} and \eqref{eq:main-finite-volume-ground-state-energy-density},
this exponent equals
$-\beta\abscard{\Lambda}\fun{e_{\Lambda}}{\lambda}$.
The preceding estimate gives
$-f_{0}
\geq
\abs{\fun{e_{\Lambda}}{\lambda}}$.
\end{proof}

\subsection{The random-walk bound}\label{the-random-walk-bound}

Open world lines are bounded by a renewal expansion over successive spatial jumps and holding times. Subcriticality of the resulting walk gives a finite susceptibility and exponential spatial weights. Let \(\hat{x}=\rbk{x,t_{x}}\) and \(\hat{y}=\rbk{y,t_{y}}\) be space-time points. Denote by \(\oa{B}^{\rbk{\hat{x},\hat{y}}}\) the single open curves from \(\hat{x}\) to \(\hat{y}\). This is the time-shifted version of Definition \ref{def:loop-measure}. Define \begin{equation}\label{eq:chi-definition}
\fun{\chi}{z, \lambda}
=
\sup_{\hat{x}, \hat{y}}
\int_{\oa{B}^{\rbk{\hat{x}, \hat{y}}}}
\napiernum^{-\lambda \absvol{\gamma}}
\opdmsr{v_{z}(\gamma)}.
\end{equation} For every \(r\in\fldreal\), the total weight of one holding time is \begin{equation}\label{eq:renewal-holding-time-mass}
\int_{0}^{\beta}
\napiernum^{-r t}
\opdmsr{t}
=
\begin{cases}
\displaystyle
\frac{1-\napiernum^{-\beta r}}{r},
&
r\neq0,
\\
\beta,
&
r=0.
\end{cases}
\end{equation} The renewal formulas retain the integral form because it is the total mass of the holding-time kernel in the renewal expansion leading to \eqref{eq:chi-bound}. It also shows directly the later replacements of the decay rate by \(\lambda-f\) and \(\lambda-\alpha\).

\begin{lem}[renewal bound; Lemma 4 of \cite{AizenmanLiebSeiringerSolovejYngvason001}]\label{lem:renewal}
If $2 z d \int_{0}^{\beta} \napiernum^{-\lambda t} \opdmsr{t} < 1$, then
\begin{equation}\label{eq:chi-bound}
\fun{\chi}{z, \lambda}
\leq
\frac{1}{1 - 2 z d
\int_{0}^{\beta}
\napiernum^{-\lambda t}
\opdmsr{t}}.
\end{equation}
Furthermore, for any $\hat{x}, \hat{y}$ and $\xi, \alpha > 0$,
it holds that
\begin{equation}\label{eq:chi-reweight}
\begin{aligned}
\int_{\oa{B}^{\rbk{\hat{x}, \hat{y}}}}
\napiernum^{-\lambda \absvol{\gamma}}
\opdmsr{v_{z}(\gamma)}
&\leq
\napiernum^{-\xi \abs{x - y}}
\fun{\chi}{z \napiernum^{\xi}, \lambda},
\\
\int_{\oa{B}^{\rbk{\hat{x}, \hat{y}}}}
\napiernum^{-\lambda \absvol{\gamma}}
\fndef{\absvol{\gamma} \geq t}
\opdmsr{v_{z}(\gamma)}
&\leq
\napiernum^{-\alpha t}
\fun{\chi}{z, \lambda - \alpha}.
\end{aligned}
\end{equation}
\end{lem}

\begin{proof}
A curve $\gamma \in \oa{B}^{\rbk{\hat{x}, \hat{y}}}$ starts with a vertical stretch at $x$ from time $t_{x}$ until its first rung, or has no rung at all.
Curves without rungs exist only when $x=y$.
There is one such curve for every admissible vertical connection.
Its weight is at most one, and the measure $v_{z}$ counts it once.
Their total contribution is therefore at most one.
Suppose that the first rung occurs after a vertical stretch of length
$t\in\rightopeninterval{0}{\beta}$.
The rung has $2d$ possible directions and contributes fugacity $z$.
The initial stretch contributes $\napiernum^{-\lambda t}$.
The remaining curve starts at the new space-time point and ends at $\hat y$.
Integration over the first holding time gives the mass
\eqref{eq:renewal-holding-time-mass}.
Bounding the remaining curve by \eqref{eq:chi-definition} gives
$$\fun{\chi}{z, \lambda} \leq 1 + 2 z d \fun{\chi}{z, \lambda} \int_{0}^{\beta} \napiernum^{-\lambda t} \opdmsr{t}.$$
In finite volume the total $v_z$-mass of curves is finite.
Each term of Definition \ref{def:loop-measure} is a finite sum of finite integrals,
and the full series has an exponential majorant.
The finiteness of $\chi$ permits solving the preceding inequality for $\chi$.
The result is \eqref{eq:chi-bound}.
For the first inequality in \eqref{eq:chi-reweight},
let $\setcardopsharp{\gamma}$ be the number of rungs of $\gamma$.
Each rung moves the curve by one lattice unit.
On $\oa{B}^{\rbk{\hat x,\hat y}}$ one has
$\setcardopsharp{\gamma}\geq\abs{x-y}$.
The two required identities are
$$\begin{aligned}
1
\leq
\napiernum^{\xi\setcardopsharp{\gamma}}
\napiernum^{-\xi\abs{x-y}},
\quad
\napiernum^{\xi\setcardopsharp{\gamma}}
\opdmsr{v_{z}(\gamma)}
=
\opdmsr{v_{z\napiernum^{\xi}}(\gamma)}.
\end{aligned}$$
For the second, $\fndef{\absvol{\gamma} \geq t} \leq \napiernum^{\alpha \rbk{\absvol{\gamma} - t}}$ and $\napiernum^{-\lambda \absvol{\gamma}} \napiernum^{\alpha \absvol{\gamma}} = \napiernum^{-\rbk{\lambda - \alpha} \absvol{\gamma}}$.
\end{proof}

\subsection{Exponential decay of the two-point function}\label{exponential-decay-of-the-two-point-function}

The contour and renewal bounds combine to suppress every open curve joining two distant sites. Their sum yields exponential decay of the one-particle density matrix in the Mott parameter region.

\begin{thm}[Mott regime decay; Theorem 2 of \cite{AizenmanLiebSeiringerSolovejYngvason001}]\label{thm:exponential-decay}
Let $\fun{e_{\Lambda}}{\lambda}$ be the global finite-volume ground-state
energy density defined by
\eqref{eq:main-finite-volume-ground-state-energy-density}.
The identity \eqref{eq:main-half-filling-global-sector-energy} gives its
half-filled-sector form
$\fun{e_{\Lambda}}{\lambda}
=
\frac{\physenergyfunc_{\Lambda,0}}{\abscard{\Lambda}}$.
Suppose
\begin{equation}\label{eq:decay-condition}
\napiernum^{-\nu} = \frac{d}{\lambda - f} \rbk{1 - \napiernum^{-\beta \rbk{\lambda - f}}} < 1,
\end{equation}
which defines $\nu > 0$.
For every $0<\xi<\nu$,
\begin{equation}\label{eq:exponential-decay}
\fun{\gamma_{\psi_{\beta,\Lambda}}}{x, y} = \fun{\oastate[\psi_{\beta,\Lambda}]}{\faadj{a_{x}} a_{y}} \leq C_{\xi} \napiernum^{-\xi \abs{x - y}},
\quad
C_{\xi} = \frac{1}{1 - \napiernum^{\xi - \nu}},
\end{equation}
uniformly in the volume.
The condition \eqref{eq:decay-condition} holds in particular in the two regions (i) $\beta d < 2 \log 2$ (any $\lambda \geq 0$), and (ii) $\lambda + \abs{\fun{e_{\Lambda}}{\lambda}} > d$ (any $\beta > 0$).
\end{thm}

\begin{proof}
Combining \eqref{eq:loop-ratio}, the factorization \eqref{eq:loop-factorization}, and the contour bound \eqref{eq:contour-bound} applied to the single curve $\gamma$,
$$\fun{\oastate[\psi_{\beta,\Lambda}]}{S^{+}_{x} S^{-}_{y}}
= \frac{\int_{\oa{B}^{\rbk{x, y}}} \napiernum^{-\lambda \absvol{\gamma}} \sqbk{\int_{\oa{A}^{\mathrm{o}}_{\gamma}} \napiernum^{-\lambda \absvol{\omega}} \opdmsr{v_{\onehalf}(\omega)}} \opdmsr{v_{\onehalf}(\gamma)}}{\int_{\oa{A}^{\mathrm{o}}} \napiernum^{-\lambda \absvol{\omega}} \opdmsr{v_{\onehalf}(\omega)}}
\leq \int_{\oa{B}^{\rbk{x, y}}} \napiernum^{-\absvol{\gamma} \rbk{\lambda - f}} \opdmsr{v_{\onehalf}(\gamma)}.$$
Apply the first reweighting of \eqref{eq:chi-reweight} with fugacity $z = \frac{1}{2}$ and rate $\lambda - f$, then \eqref{eq:chi-bound} with $z = \frac{1}{2} \napiernum^{\xi}$:
$$\fun{\oastate[\psi_{\beta,\Lambda}]}{S^{+}_{x} S^{-}_{y}} \leq \napiernum^{-\xi \abs{x - y}} \fun{\chi}{\frac{1}{2} \napiernum^{\xi}, \lambda - f},
\quad
2 \cdot \frac{1}{2} \napiernum^{\xi} d \int_{0}^{\beta} \napiernum^{-\rbk{\lambda - f} t} \opdmsr{t} = \napiernum^{\xi - \nu} < 1,$$
where the middle quantity was computed with \eqref{eq:decay-condition}; the geometric bound \eqref{eq:chi-bound} gives the constant $C_{\xi}$.
Finally $\fun{\oastate[\psi_{\beta,\Lambda}]}{\faadj{a_{x}} a_{y}} = \fun{\oastate[\psi_{\beta,\Lambda}]}{S^{+}_{x} S^{-}_{y}}$ by \eqref{eq:boson-spin-dictionary}.
For region (i), \eqref{eq:free-energy-bounds} gives
$$\lambda-f
\geq\frac{\lambda}{2}+\frac{\log2}{\beta}
\geq\frac{\log2}{\beta}.$$
The function
$t\mapsto\rbk{1-\napiernum^{-t}}/t$ is decreasing.
Apply this fact at
$t=\beta\rbk{\lambda-f}\geq\log2$:
$$\napiernum^{-\nu} = \beta d \, \frac{1 - \napiernum^{-\beta \rbk{\lambda - f}}}{\beta \rbk{\lambda - f}} \leq \beta d \, \frac{1 - \napiernum^{-\log 2}}{\log 2} = \frac{\beta d}{2 \log 2} < 1
\iff \beta d < 2 \log 2.$$
For region (ii):
$1-\napiernum^{-\beta\rbk{\lambda-f}}<1$
and
$\lambda-f
\geq
\lambda+\abs{\fun{e_{\Lambda}}{\lambda}}
>
d$
give
$\napiernum^{-\nu}
<
d/\rbk{\lambda-f}
<
1$.
\end{proof}

\begin{cor}[absence of BEC in the Mott region]\label{cor:mott-no-bec}
Let the thermodynamic ground-state energy density
$\fun{e_{\infty}}{\varrho}$ be defined by
\eqref{eq:thermodynamic-ground-state-energy-density}.
Suppose that the finite-volume ground-state energy density
$\fun{e_{\Lambda}}{\lambda}$ defined by
\eqref{eq:main-finite-volume-ground-state-energy-density} satisfies
$$
\lim_{n \to \infty}
\fun{e_{\Lambda_{n}}}{\lambda}
=
\fun{e_{\infty}}{\frac{1}{2}}
$$
along the periodic exhaustion
\eqref{eq:periodic-box-exhaustion}
and that
\begin{equation}\label{eq:thermodynamic-mott-decay-region}
\lambda
+
\abs{\fun{e_{\infty}}{\frac{1}{2}}}
>
d.
\end{equation}
For every $\beta>0$, there exist $\xi>0$ and
$C<\infty$, independent of all sufficiently large periodic volumes, such
that
$$
\fun{\gamma_{\psi_{\beta,\Lambda}}}{x,y}
\leq
C
\napiernum^{-\xi\abs{x-y}}.
$$
Every weak-$\ast$ limit point $\oastate[\psi_{\beta}]$ of the periodic
Gibbs states satisfies
\begin{equation}\label{eq:mott-spatial-average-vanishing}
\lim_{n \to \infty}
\fun{\oastate[\psi_{\beta}]}{\faadj{\barmean{a}_{\Lambda_n}}\barmean{a}_{\Lambda_n}}
=
0,
\end{equation}
where $\barmean{a}_{\Lambda_n}$ is defined by \eqref{eq:main-spatial-averages}.
Thus the limit state has neither off-diagonal long-range order nor
Bose--Einstein condensation in the spatial-average sense.
The sufficient condition \eqref{eq:thermodynamic-mott-decay-region} holds
for every $\lambda\geq d$.
\end{cor}

\begin{proof}
Set
$$
\delta
=
\frac{
\lambda+\abs{\fun{e_{\infty}}{\frac{1}{2}}}-d
}{2}
>
0.
$$
The convergence of the energy densities and
\eqref{eq:free-energy-bounds} imply, for all sufficiently large $\Lambda$,
$$
\lambda-f
\geq
\lambda+\abs{\fun{e_{\Lambda}}{\lambda}}
\geq
d+\delta.
$$
The decay parameter from \eqref{eq:decay-condition} consequently satisfies
$$
\napiernum^{-\nu}
\leq
\frac{d}{\lambda-f}
\leq
\frac{d}{d+\delta}
=
q
<
1.
$$
Choose $0<\xi<-\log q$.
The decay estimate \eqref{eq:exponential-decay} and
$\napiernum^{-\nu}
\leq
q$
give the volume-independent constant
$C
=
\frac{1}{1-\napiernum^{\xi}q}.$
Taking the weak-$\ast$ limit preserves the estimate for every fixed pair
$x,y$.
The definition \eqref{eq:main-spatial-averages} gives
$$\begin{aligned}
\fun{\oastate[\psi_{\beta}]}{\faadj{\barmean{a}_{\Lambda_n}}\barmean{a}_{\Lambda_n}}
=
\frac{1}{\abscard{\Lambda_n}^{2}}
\sum_{x,y\in \Lambda_n}
\fun{\gamma_{\psi_{\beta}}}{x,y}
\leq
\frac{C}{\abscard{\Lambda_n}}
\sum_{z\in\ringratint^{d}}
\napiernum^{-\xi\abs{z}}.
\end{aligned}$$
The lattice sum is finite, and the last expression tends to zero as
$R\to\infty$.
This proves \eqref{eq:mott-spatial-average-vanishing}.

Finally, \eqref{eq:uniform-negative-energy-density} gives
$$\abs{\fun{e_{\infty}}{\frac{1}{2}}}
\geq
\frac{1}{2}
\rbk{\sqrt{\lambda^{2}
+\frac{1}{4}}
-\lambda}
>
0.$$
For $\lambda\geq d$, this strict inequality implies
\eqref{eq:thermodynamic-mott-decay-region}.
\end{proof}

\subsection{Winding numbers and the gap in the chemical potential}\label{winding-numbers-and-the-gap-in-the-chemical-potential}

Temporal winding records the deviation of the conserved particle number from half-filling. Exponential suppression of nonzero winding sectors produces a positive interval of chemical potentials with fixed density. The sector energies \(\physenergyfunc_{\Lambda,k}\) are defined in \eqref{eq:main-sector-energy}, with their particle-number interpretation fixed by \eqref{eq:main-particle-number}. The particle-hole identity \eqref{eq:particle-hole-sector-energy} identifies the energy costs for adding and removing the same number of particles. Consequently, a chemical-potential gap at half-filling is equivalently a volume-uniform lower bound of the form \begin{equation}\label{eq:chemical-potential-gap-one-sided}
\physenergyfunc_{\Lambda,k}
-
\physenergyfunc_{\Lambda,0}
\geq
c\abs{k}
\end{equation} with \(c > 0\). In the loop picture, winding \(k\) requires total absolute winding at least \(\abs{k}\). Every winding line has vertical length at least \(\beta\) and carries an exponential energy cost. The winding estimate below compares this cost with the placement entropy.

\begin{thm}[winding estimate; Eq. (16) of \cite{AizenmanLiebSeiringerSolovejYngvason001}]\label{thm:winding-estimate}
Let $\alpha > 0$ satisfy
\begin{equation}\label{eq:winding-susceptibility-bound}
\fun{B}{\alpha} = d \int_{0}^{\beta} \napiernum^{-\rbk{\lambda + \abs{f_{0}} - \alpha} t} \opdmsr{t} < 1,
\quad
\fun{\widehat{\chi}}{\alpha} = \fun{\chi}{\frac{1}{2}, \lambda + \abs{f_{0}} - \alpha} \leq \frac{1}{1 - \fun{B}{\alpha}}.
\end{equation}
For every $k\neq0$,
\begin{equation}\label{eq:winding-bound}
\frac{\sqfun{\trace}{P_{k} \napiernum^{-\beta \physham_{\Lambda}}}}{\sqfun{\trace}{P_{0} \napiernum^{-\beta \physham_{\Lambda}}}}
\leq 2 \napiernum^{-\alpha \beta \abs{k}} \rbk{\napiernum^{2} \fun{\widehat{\chi}}{\alpha} \frac{\abscard{\Lambda}}{\abs{k}}}^{\abs{k}}.
\end{equation}
\end{thm}

\begin{proof}
By \eqref{eq:loop-trace} the numerator integrates over configurations with $\fun{\nu}{\omega} = k \neq 0$.
Such a configuration contains noncontractible loops $\gamma_{j}$ with
windings $\nu_{j}\neq0$.
Their signed windings sum to $k$, and hence
$$\sum_{j}\abs{\nu_{j}}\geq\abs{k}.$$
A loop of winding $\nu_{j}$ crosses the periodic time interval
$\abs{\nu_{j}}$ times.
The length bound is
$$\absvol{\gamma_{j}}\geq\beta\abs{\nu_{j}}.$$
Label each non-contractible loop by the site $x_{j}$ where it crosses the slice $t = 0$ (choosing, say, the first crossing); distinct loops have distinct labels.
Split $\omega$ into its non-contractible loops and the remaining
configuration.
The remainder has zero total winding and avoids those loops.
The factorization property and the winding-restricted contour bound
\eqref{eq:contour-bound} give
$$\sqfun{\trace}{P_{k} \napiernum^{-\beta \physham_{\Lambda}}}
\leq \sum_{\setone{x_{1}, x_{2}, \dotsc} \subset \Lambda} \sum_{\substack{m_{j} \geq 1 \\ \sum_{j} m_{j} \geq \abs{k}}} \prod_{j} \sqbk{\int_{\oa{B}^{\rbk{x_{j}, x_{j}}}} \napiernum^{-\rbk{\lambda + \abs{f_{0}}} \absvol{\gamma}} \fndef{\absvol{\gamma} \geq \beta m_{j}} \opdmsr{v_{\onehalf}(\gamma)}} \cdot \sqfun{\trace}{P_{0} \napiernum^{-\beta \physham_{\Lambda}}},$$
This sum deliberately overcounts.
It includes every label set and every collection of winding magnitudes
$m_{j}=\abs{\nu_{j}}$.
The loop through $x_{j}$ is treated as a closed curve at $x_{j}$ of length
at least $\beta m_{j}$.
The factor from the contour bound has been absorbed into the exponential
rate.
By the second reweighting in \eqref{eq:chi-reweight},
$$\int_{\oa{B}^{\rbk{x, x}}} \napiernum^{-\rbk{\lambda + \abs{f_{0}}} \absvol{\gamma}} \fndef{\absvol{\gamma} \geq \beta m} \opdmsr{v_{\onehalf}(\gamma)} \leq \napiernum^{-\alpha \beta m} \fun{\widehat{\chi}}{\alpha}.$$
If $n$ denotes the number of loops, the preceding estimate becomes
$$\frac{\sqfun{\trace}{P_{k} \napiernum^{-\beta \physham_{\Lambda}}}}{\sqfun{\trace}{P_{0} \napiernum^{-\beta \physham_{\Lambda}}}}
\leq \sum_{n \geq 1} \binom{\abscard{\Lambda}}{n} \sum_{\substack{m_{1}, \dotsc, m_{n} \geq 1 \\ \sum m_{j} \geq \abs{k}}} \prod_{j = 1}^{n} \fun{\widehat{\chi}}{\alpha} \napiernum^{-\alpha \beta m_{j}} = \Sigma,$$
say.
The generating-function device of \cite{AizenmanLiebSeiringerSolovejYngvason001} evaluates $\Sigma$:
with $\delta = \napiernum^{-\alpha \beta}$,
$$\fun{P}{w}
=
1
+\fun{\widehat{\chi}}{\alpha}
\sum_{i = 1}^{\frac{\abscard{\Lambda}}{2}} \rbk{w \delta}^{i}.$$
Define the nonnegative coefficients $c_{l}$ by
$$\fun{P}{w}^{\abscard{\Lambda}}
=
\sum_{l \geq 0} c_{l} w^{l}.$$
The coefficient $c_{l}$ is the total weight of all label-and-winding choices
with $\sum m_{j}=l$.
At each site, one chooses either no loop or a loop of winding
$i\leq\abscard{\Lambda}/2$.
The latter is the largest possible winding.
For any $1 < R < 1 / \delta$, Cauchy's coefficient formula and the geometric
series for $\rbk{1-w^{-1}}^{-1}$ give
$$\Sigma
\leq
\sum_{l \geq \abs{k}} c_{l}
=
\frac{1}{2 \pi \imunit}
\oint_{\abs{w} = R}
\frac{\fun{P}{w}^{\abscard{\Lambda}}}{
w^{\abs{k} + 1} \rbk{1 - w^{-1}}
}
\opdmsr{w}.$$
On the circle $\abs{w} = R$,
$$\abs{\fun{P}{w}}
\leq
1 + \fun{\widehat{\chi}}{\alpha} \frac{R \delta}{1 - R \delta}
\leq
\napiernum^{\fun{\widehat{\chi}}{\alpha} \frac{R \delta}{1 - R \delta}},$$
so
$$\Sigma \leq \frac{1}{1 - R^{-1}} \, R^{-\abs{k}}
\napiernum^{\abscard{\Lambda} \fun{\widehat{\chi}}{\alpha} \frac{R \delta}{1 - R \delta}}.$$
Choose
$$R=\frac{\abs{k}}{
\delta\fun{\widehat{\chi}}{\alpha}\abscard{\Lambda}}$$
and first assume $R\geq2$.
Since $\abs{k}\leq\abscard{\Lambda}/2$ and
$\fun{\widehat{\chi}}{\alpha}\geq1$, one has
$\delta R\leq\frac{1}{2}$.
It follows that
$$\begin{aligned}
\frac{1}{1-R^{-1}}
\leq2,
\quad
\frac{R\delta}{1-R\delta}
\leq
2R\delta
=
\frac{2\abs{k}}
{\fun{\widehat{\chi}}{\alpha}\abscard{\Lambda}}.
\end{aligned}$$
Substitution into the preceding estimate yields
$$\Sigma
\leq
2
\rbk{\frac{\delta \fun{\widehat{\chi}}{\alpha} \abscard{\Lambda}}
{\abs{k}}}^{\abs{k}} \napiernum^{2 \abs{k}}
=
2
\napiernum^{-\alpha \beta \abs{k}}
\rbk{\napiernum^{2}
\fun{\widehat{\chi}}{\alpha}
\frac{\abscard{\Lambda}}{\abs{k}}}^{\abs{k}},$$
which is \eqref{eq:winding-bound}.
If $R<2$, then
$$\frac{\napiernum^{2}\fun{\widehat{\chi}}{\alpha}
\abscard{\Lambda}}{\abs{k}}
>\frac{\napiernum^{2}}{2\delta}.$$
The right side of \eqref{eq:winding-bound} is then larger than
$2\rbk{\napiernum^{2}/2}^{\abs{k}}>1$.
The left side is at most $1$ by
Theorem \ref{thm:partition-maximal}.
The estimate is automatic in this case.
\end{proof}

By \eqref{eq:particle-hole-sector-energy}, the symmetric estimate in Eq. (15) of \cite{AizenmanLiebSeiringerSolovejYngvason001} is equivalent to the following one-sided bound.

\begin{thm}[Mott gap; Eq. (15) of \cite{AizenmanLiebSeiringerSolovejYngvason001}]\label{thm:mott-gap}
Let $\fun{e_{\Lambda}}{\lambda}$ be the global finite-volume ground-state
energy density defined by
\eqref{eq:main-finite-volume-ground-state-energy-density}.
The identity \eqref{eq:main-half-filling-global-sector-energy} gives its
half-filled-sector form
$\fun{e_{\Lambda}}{\lambda}
=
\frac{\physenergyfunc_{\Lambda,0}}{\abscard{\Lambda}}$.
For all $k$,
\begin{equation}\label{eq:mott-sector-gap}
\physenergyfunc_{\Lambda,k}
-
\physenergyfunc_{\Lambda,0}
\geq
\rbk{
\lambda
+
\abs{\fun{e_{\Lambda}}{\lambda}}
-
d
}
\abs{k}.
\end{equation}
If $\lambda + \abs{\fun{e_{\Lambda}}{\lambda}} > d$, this lower bound is
strictly positive for every nonzero $k$, in particular for large $\lambda$,
uniformly in the volume.
\end{thm}

\begin{proof}
The case $k=0$ is immediate.
If
$\lambda
+
\abs{\fun{e_{\Lambda}}{\lambda}}
-
d
\leq
0$,
then \eqref{eq:main-half-filling-global-sector-energy} and
\eqref{eq:main-sector-energy} give a nonnegative left side in
\eqref{eq:mott-sector-gap}.
It remains to fix $k \neq 0$ and
$0
<
\alpha
<
\lambda
+
\abs{\fun{e_{\Lambda}}{\lambda}}
-
d$.
The free-energy bound \eqref{eq:free-energy-bounds} gives
$\abs{f_{0}}
\geq
\abs{\fun{e_{\Lambda}}{\lambda}}$
for every $\beta$.
The quantities in \eqref{eq:winding-susceptibility-bound} therefore satisfy
$$\fun{B}{\alpha}
\leq
\frac{d}{\lambda + \abs{f_{0}} - \alpha}
\leq
\frac{d}{\lambda + \abs{\fun{e_{\Lambda}}{\lambda}} - \alpha}
<
1,
\quad
\fun{\widehat{\chi}}{\alpha}
\leq
\frac{1}{1 - \fun{B}{\alpha}},$$
uniformly as $\beta \to \infty$.

At fixed $\Lambda$ and for each admissible $r$, let
$\seq{E_{\Lambda,r,j}}{1 \leq j \leq D_{\Lambda,r}}$
be the eigenvalues of
$\fnrestr{\physham_{\Lambda}}{\Ran P_{r}}$
in nondecreasing order and counted with multiplicity, with $P_{r}$ defined by
\eqref{eq:main-sector-projection}.
The sector energy definition \eqref{eq:main-sector-energy} gives the
following dimension, lowest eigenvalue, and residual trace factor.
\begin{equation}\label{eq:sector-partition-ground-factor}
\begin{aligned}
D_{\Lambda,r}
&=
\dim \Ran P_{r},
\quad
E_{\Lambda,r,1}
=
\physenergyfunc_{\Lambda,r},
\\
\fun{g_{\Lambda,r}}{\beta}
&=
\sum_{j=1}^{D_{\Lambda,r}}
\napiernum^{-\beta \rbk{E_{\Lambda,r,j} - E_{\Lambda,r,1}}},
\quad
1
\leq
\fun{g_{\Lambda,r}}{\beta}
\leq
D_{\Lambda,r},
\\
\sqfun{\trace}{P_{r} \napiernum^{-\beta \physham_{\Lambda}}}
&=
\napiernum^{-\beta \physenergyfunc_{\Lambda,r}}
\fun{g_{\Lambda,r}}{\beta},
\quad
\lim_{\beta \to \infty}
\frac{1}{\beta}
\log\fun{g_{\Lambda,r}}{\beta}
=
0.
\end{aligned}
\end{equation}
Applying $-\beta^{-1}\log$ to \eqref{eq:winding-bound} and using
\eqref{eq:sector-partition-ground-factor} gives
\begin{equation}\label{eq:winding-bound-logarithmic}
\physenergyfunc_{\Lambda,k}
-
\physenergyfunc_{\Lambda,0}
\geq
\alpha \abs{k}
-
\frac{\abs{k}}{\beta}
\log\rbk{
\napiernum^{2}
\fun{\widehat{\chi}}{\alpha}
\frac{\abscard{\Lambda}}{\abs{k}}
}
-
\frac{\log 2}{\beta}
+
\frac{1}{\beta}
\log\rbk{
\frac{\fun{g_{\Lambda,k}}{\beta}}{\fun{g_{\Lambda,0}}{\beta}}
}.
\end{equation}
The uniform bound for $\fun{\widehat{\chi}}{\alpha}$ and the last line of
\eqref{eq:sector-partition-ground-factor} make all three correction terms
in \eqref{eq:winding-bound-logarithmic} vanish as $\beta \to \infty$.
The limit of \eqref{eq:winding-bound-logarithmic} gives
$\physenergyfunc_{\Lambda,k}
-
\physenergyfunc_{\Lambda,0}
\geq
\alpha \abs{k}$.
Letting $\alpha$ approach the endpoint gives
$$
\physenergyfunc_{\Lambda,k}
-
\physenergyfunc_{\Lambda,0}
\geq
\rbk{
\lambda
+
\abs{\fun{e_{\Lambda}}{\lambda}}
-
d
}
\abs{k}.
$$
This is \eqref{eq:mott-sector-gap}.
\end{proof}

\begin{prop}[thermodynamic Mott energy bound]\label{prop:thermodynamic-mott-energy-bound}
Assume that the thermodynamic ground-state energy density
$\fun{e_{\infty}}{\varrho}$ in
\eqref{eq:thermodynamic-ground-state-energy-density} exists as a finite
function on $\closedinterval{0}{1}$ independently of the approximating
particle numbers.
Define the thermodynamic Mott lower bound by
\begin{equation}\label{eq:thermodynamic-mott-gap-constant}
\Delta_{\infty}
=
\lambda
+
\abs{\fun{e_{\infty}}{\frac{1}{2}}}
-
d.
\end{equation}
The thermodynamic energy is convex and obeys
\begin{equation}\label{eq:particle-hole-thermodynamic-energy}
\fun{e_{\infty}}{\varrho}
=
\fun{e_{\infty}}{1 - \varrho},
\quad
\fun{e_{\infty}}{\varrho}
-
\fun{e_{\infty}}{\frac{1}{2}}
\geq
\Delta_{\infty}
\abs{\varrho - \frac{1}{2}}.
\end{equation}
The one-particle addition and removal energies satisfy
\begin{equation}\label{eq:thermodynamic-mott-one-particle-gap}
\liminf_{\Lambda \nearrow \ringratint^{d}}
\rbk{
\physenergyfunc_{\Lambda,\sigma}
-
\physenergyfunc_{\Lambda,0}
}
\geq
\Delta_{\infty},
\quad
\sigma
\in
\setone{-1,1}.
\end{equation}
\end{prop}

\begin{proof}
The identities
\eqref{eq:main-half-filling-global-sector-energy} and
\eqref{eq:main-finite-volume-ground-state-energy-density} give
\begin{equation}\label{eq:half-filled-energy-density-limit}
\fun{e_{\Lambda}}{\lambda}
=
\frac{\physenergyfunc_{\Lambda,0}}{\abscard{\Lambda}}
\to
\fun{e_{\infty}}{\frac{1}{2}}
\quad
\rbk{\Lambda \nearrow \ringratint^{d}}.
\end{equation}
For $\sigma = -1$, the finite-volume particle-hole identity
\eqref{eq:particle-hole-sector-energy} gives
$$
\physenergyfunc_{\Lambda,-1}
-
\physenergyfunc_{\Lambda,0}
=
\physenergyfunc_{\Lambda,1}
-
\physenergyfunc_{\Lambda,0}.
$$
Applying \eqref{eq:mott-sector-gap} with $k = 1$ and using
\eqref{eq:half-filled-energy-density-limit} proves
\eqref{eq:thermodynamic-mott-one-particle-gap}.

Fix $\varrho \in \closedinterval{0}{1}$ and choose particle numbers
$M_{\Lambda}$ such that
$M_{\Lambda}/\abscard{\Lambda} \to \varrho$.
Set
$k_{\Lambda} = M_{\Lambda} - \abscard{\Lambda}/2$.
The estimate \eqref{eq:mott-sector-gap}, divided by
$\abscard{\Lambda}$, is
\begin{equation}\label{eq:finite-volume-mott-density-bound}
\frac{
\physenergyfunc_{\Lambda,k_{\Lambda}}
-
\physenergyfunc_{\Lambda,0}
}{
\abscard{\Lambda}
}
\geq
\rbk{
\lambda
+
\abs{\fun{e_{\Lambda}}{\lambda}}
-
d
}
\abs{
\frac{M_{\Lambda}}{\abscard{\Lambda}}
-
\frac{1}{2}
}.
\end{equation}
The thermodynamic limit in
\eqref{eq:finite-volume-mott-density-bound} proves the second formula in
\eqref{eq:particle-hole-thermodynamic-energy}.
The complementary particle numbers
$\abscard{\Lambda} - M_{\Lambda}$ converge in density to $1 - \varrho$.
Equation \eqref{eq:particle-hole-sector-energy} identifies their
finite-volume energies with those of $M_{\Lambda}$.
The assumed independence of the approximating particle numbers proves the
first formula in \eqref{eq:particle-hole-thermodynamic-energy}.

The finite-range interaction also makes
$\fun{e_{\infty}}{\varrho}$ convex.
Place ground-state vectors of two disjoint boxes with densities
$\varrho_{1}$ and $\varrho_{2}$ in a larger box so that their volume
fractions converge to $t$ and $1 - t$.
The product vector has density
$t \varrho_{1} + \rbk{1 - t} \varrho_{2}$, while the interaction terms crossing
the two boxes have surface-order norm.
Division by the total volume and passage to the limit give
$$
\fun{e_{\infty}}{t \varrho_{1} + \rbk{1 - t} \varrho_{2}}
\leq
t \fun{e_{\infty}}{\varrho_{1}}
+
\rbk{1 - t} \fun{e_{\infty}}{\varrho_{2}},
\quad
t
\in
\closedinterval{0}{1}.
$$
\end{proof}

\begin{cor}[Mott cusp and chemical-potential plateau]\label{cor:thermodynamic-mott-cusp}
Assume the hypotheses of Proposition
\ref{prop:thermodynamic-mott-energy-bound}, with $\Delta_{\infty} > 0$.
The one-sided chemical potentials satisfy
\begin{equation}\label{eq:thermodynamic-mott-chemical-potentials}
\begin{aligned}
\mu_{+}
&=
\lim_{\epsilon \downarrow 0}
\frac{
\fun{e_{\infty}}{\frac{1}{2} + \epsilon}
-
\fun{e_{\infty}}{\frac{1}{2}}
}{
\epsilon
},
\\
\mu_{-}
&=
\lim_{\epsilon \downarrow 0}
\frac{
\fun{e_{\infty}}{\frac{1}{2}}
-
\fun{e_{\infty}}{\frac{1}{2} - \epsilon}
}{
\epsilon
}.
\end{aligned}
\end{equation}
The Mott cusp is
\begin{equation}\label{eq:thermodynamic-mott-cusp}
\mu_{-}
\leq
-\Delta_{\infty}
<
\Delta_{\infty}
\leq
\mu_{+}.
\end{equation}
The function $\fun{e_{\infty}}{\varrho}$ is not differentiable at
half-filling.
The interval width obeys
$\mu_{+} - \mu_{-} \geq 2 \Delta_{\infty}$.
For every $\mu \in \closedinterval{\mu_{-}}{\mu_{+}}$,
\begin{equation}\label{eq:mott-chemical-potential-plateau}
\fun{e_{\infty}}{\varrho}
-
\mu\rbk{\varrho - \frac{1}{2}}
\geq
\fun{e_{\infty}}{\frac{1}{2}},
\quad
\varrho
\in
\closedinterval{0}{1}.
\end{equation}
The minimizer is unique and equals $\varrho = \frac{1}{2}$ whenever
$\abs{\mu} < \Delta_{\infty}$.
\end{cor}

\begin{proof}
Convexity gives the one-sided limits in
\eqref{eq:thermodynamic-mott-chemical-potentials}.
The symmetry in \eqref{eq:particle-hole-thermodynamic-energy} gives
$\mu_{-} = -\mu_{+}$, and its lower bound gives
$\mu_{+} \geq \Delta_{\infty}$.
These inequalities prove \eqref{eq:thermodynamic-mott-cusp} and the stated
non-differentiability.
The subgradient inequality for a convex function is precisely
\eqref{eq:mott-chemical-potential-plateau}.
If $\abs{\mu} < \Delta_{\infty}$, the lower bound in
\eqref{eq:particle-hole-thermodynamic-energy} gives
$$
\fun{e_{\infty}}{\varrho}
-
\mu\rbk{\varrho - \frac{1}{2}}
-
\fun{e_{\infty}}{\frac{1}{2}}
\geq
\rbk{\Delta_{\infty}-\abs{\mu}}
\abs{\varrho - \frac{1}{2}},
$$
which is strictly positive away from half-filling.
\end{proof}

\begin{lem}[finite-volume ground state]\label{lem:kls-finite-volume-ground-state}
The Hamiltonian
$\physham_{\Lambda}$
has a unique ground-state vector
$\Psi_{\txtgs,\Lambda}$
up to phase, and
$$
S^{3}_{\txttot,\Lambda}\Psi_{\txtgs,\Lambda}
=
0.
$$
\end{lem}

\begin{proof}
Theorem \ref{thm:half-filling} proves both uniqueness and
$S^{3}_{\txttot,\Lambda}\Psi_{\txtgs,\Lambda}=0$.
\end{proof}

\section{Absence of a Gap in the Condensed Regime}\label{sec:gap}

The Mott gap of Theorem \ref{thm:mott-gap} disappears in the condensed regime. For a fixed number of added or removed particles, the cost per particle is \(\fun{O}{\abscard{\Lambda}^{-1}}\). For a macroscopic density change, the cost per site is \(\fun{O}{\rbk{\varrho-1/2}^{2}}\). Throughout, \(\Psi_{\txtgs,\Lambda}\) denotes the finite-volume ground state of \(\physham_{\Lambda}\), unique and lying in the sector \(S^{3}_{\txttot,\Lambda}
=
0\) by Lemma \ref{lem:kls-finite-volume-ground-state}. The sector energies \(\physenergyfunc_{\Lambda,k}\) are those of \eqref{eq:main-sector-energy}, and \(\fun{\oastate[\psi_{\txtgs,\Lambda}]}{\cdot}\) denotes the ground-state expectation. The standing assumption is condensation in the ground state, \begin{equation}\label{eq:gs-bec-assumption}
\fun{\oastate[\psi_{\txtgs,\Lambda}]}{S^{-}_{\txttot,\Lambda} S^{+}_{\txttot,\Lambda}} \geq \kappa \abscard{\Lambda}^{2},
\quad
S^{\pm}_{\txttot,\Lambda} = \sum_{x \in \Lambda} S^{\pm}_{x},
\end{equation} with \(\kappa > 0\) independent of the volume. Its proved and presently undetermined parameter ranges are distinguished next.

\subsection{Status of the ground-state condensation hypothesis}\label{status-of-the-ground-state-condensation-hypothesis}

The hypothesis \eqref{eq:gs-bec-assumption} is not an unproved input throughout the parameter space. For \(d \geq 3\) and \eqref{eq:main-ground-state-condensation-region}, it follows rigorously from the infrared estimate of Theorem \ref{thm:bec}. The finite-volume zero-mode estimate \eqref{eq:bec-zero-mode-lower-bound} holds before the thermodynamic limit. The finite-volume conclusion of Lemma \ref{lem:kls-finite-volume-ground-state} gives \[
\lim_{\beta \to \infty}
\fun{\oastate[\psi_{\beta,\Lambda}]}{A}
=
\fun{\oastate[\psi_{\txtgs,\Lambda}]}{A},
\quad
A
\in
\oa{A}_{\Lambda},
\] because the finite-dimensional Gibbs density matrix converges to the projection onto the unique ground state. Taking this limit in \eqref{eq:bec-zero-mode-lower-bound} and then taking \(\Lambda \nearrow \ringratint^{d}\) gives \begin{equation}\label{eq:ground-state-condensation-liminf}
\liminf_{\Lambda \nearrow \ringratint^{d}}
\frac{1}{\abscard{\Lambda}^{2}}
\fun{\oastate[\psi_{\txtgs,\Lambda}]}{
S^{-}_{\txttot,\Lambda} S^{+}_{\txttot,\Lambda}
}
\geq
\kappa_{\mathrm{gs}},
\end{equation} where \(\kappa_{\mathrm{gs}}\) is defined by \eqref{eq:main-ground-state-condensation-constant}. The identification of the zero mode uses the exact ground-state relation \[
\fun{\oastate[\psi_{\txtgs,\Lambda}]}{
S^{-}_{\txttot,\Lambda} S^{+}_{\txttot,\Lambda}
}
=
\abscard{\Lambda}
\fun{\oastate[\psi_{\txtgs,\Lambda}]}{
\rbk{\widetilde{S}^{1}_{0}}^{2}
+
\rbk{\widetilde{S}^{2}_{0}}^{2}
}
-
\fun{\oastate[\psi_{\txtgs,\Lambda}]}{
S^{3}_{\txttot,\Lambda}
}
=
\abscard{\Lambda}
\fun{\oastate[\psi_{\txtgs,\Lambda}]}{
\rbk{\widetilde{S}^{1}_{0}}^{2}
+
\rbk{\widetilde{S}^{2}_{0}}^{2}
},
\] because \(S^{3}_{\txttot,\Lambda} \Psi_{\txtgs,\Lambda} = 0\) by Lemma \ref{lem:kls-finite-volume-ground-state}. The lower bound \eqref{eq:ground-state-condensation-liminf} shows that every \(0 < \kappa < \kappa_{\mathrm{gs}}\) satisfies \eqref{eq:gs-bec-assumption} for all sufficiently large periodic boxes. The Koma--Tasaki construction consequently has an unconditional ground-state condensation input in this explicitly proved region.

The word ``assume'' in \eqref{eq:gs-bec-assumption} keeps the conclusions valid beyond the sufficient region \eqref{eq:main-ground-state-condensation-region}. Failure of that sufficient inequality does not prove absence of ground-state condensation. The infrared argument bounds every nonzero momentum mode from above and subtracts their sum from the exact spin sum rule. Its Cauchy--Schwarz estimate and the volume-order energy bound become too coarse as \(\lambda\) increases, and the resulting lower bound \eqref{eq:main-ground-state-condensation-constant} can become nonpositive while the actual zero-mode occupation may remain positive. The loop argument at large \(\lambda\) proves exponential decay in a separate region, but no known correlation inequality makes the condensate density monotone in the staggered \(S^{3}\) field. No stochastic domination relating the loop measures at different \(\lambda\) is available. These missing monotonicity inputs leave the intermediate parameter region and the exact critical value of \(\lambda\) undetermined, as discussed in Section \ref{sec:improvements}.

For \(d = 2\), the integral \(c_{d}\) diverges and the positive-temperature infrared bound cannot establish condensation. This failure is dimension-specific and does not assert the absence of ground-state condensation. The term \(\rbk{\sminvtemperature E_{p}}^{-1}\) in \eqref{eq:mode-occupation-bound} causes this positive-temperature obstruction. The ground-state spectral argument has no such term. Its remaining infrared singularity has order \(E_{p}^{-\onehalf}\), which is integrable in dimension \(2\). The relevant zero-temperature argument is separated next.

\subsection{The Kennedy--Lieb--Shastry ground-state argument}\label{the-kennedyliebshastry-ground-state-argument}

Kennedy, Lieb, and Shastry supplement the ground-state infrared estimate with local ground-state inequalities. Their result gives a strictly positive zero mode for the zero-field XY model in every dimension \(d \geq 2\).

\begin{prop}[ground-state source inequality]\label{prop:kls-ground-state-source}
The deformed rotated Hamiltonian
$\fun{\widehat{K}}{h}$
is defined by
\eqref{eq:deformed-hamiltonian}.
For a real source field
$h \colon \Lambda \to \fldreal$,
define
$$
\fun{\mathcal{E}_{\Lambda}}{h}
=
\min\opvarspec{\fun{\widehat{K}}{h}}.
$$
Then
\begin{equation}\label{eq:kls-ground-state-energy-domination}
\fun{\mathcal{E}_{\Lambda}}{h}
\geq
\fun{\mathcal{E}_{\Lambda}}{0}.
\end{equation}
For
$p \in \dual{\Lambda} \setminus \setone{0}$,
choose an orthonormal eigenbasis
$\seq{\Psi_{\nu,\Lambda}}{\nu \in J_{\Lambda}}$
of
$\physham_{\Lambda}$,
where
$\txtgs \in J_{\Lambda}$,
$\Psi_{\txtgs,\Lambda}$
is the ground-state vector, and
$$
\physham_{\Lambda}\Psi_{\nu,\Lambda}
=
\varepsilon_{\nu,\Lambda}\Psi_{\nu,\Lambda}.
$$
It follows that
\begin{equation}\label{eq:kls-ground-state-susceptibility}
\sum_{\nu \in J_{\Lambda} \setminus \setone{\txtgs}}
\frac{
\abs{
\bkt{\Psi_{\nu,\Lambda}}{
\widetilde{S}^{1}_{p}\Psi_{\txtgs,\Lambda}
}
}^{2}
}{
\varepsilon_{\nu,\Lambda}
-
\physenergyfunc_{\Lambda,0}
}
\leq
\frac{1}{4E_{p}}.
\end{equation}
The same estimate holds with
$S^{1}$
replaced by
$S^{2}$.
\end{prop}

\begin{proof}
For
$\fun{f_{\sminvtemperature,\Lambda}}{h}
=
-\frac{1}{\sminvtemperature}\log\fun{Z}{h}$
Theorem
\ref{thm:gaussian-domination}
gives
$f_{\sminvtemperature,\Lambda}(h)
\geq
f_{\sminvtemperature,\Lambda}(0)$.
If
$\seq{\fun{\mathcal{E}_{j,\Lambda}}{h}}{1 \leq j \leq 2^{\abscard{\Lambda}}}$
are the eigenvalues of
$\fun{\widehat{K}}{h}$,
then
$$
\fun{f_{\sminvtemperature,\Lambda}}{h}
=
\fun{\mathcal{E}_{\Lambda}}{h}
-
\frac{1}{\sminvtemperature}
\log
\sum_{j = 1}^{2^{\abscard{\Lambda}}}
\napiernum^{-\sminvtemperature
\rbk{\fun{\mathcal{E}_{j,\Lambda}}{h}
-
\fun{\mathcal{E}_{\Lambda}}{h}}}.
$$
The above sum is between
$1$
and
$2^{\abscard{\Lambda}}$.
Consequently,
$$
\fun{\mathcal{E}_{\Lambda}}{h}
-
\frac{\abscard{\Lambda}\log2}{\sminvtemperature}
\leq
\fun{f_{\sminvtemperature,\Lambda}}{h}
\leq
\fun{\mathcal{E}_{\Lambda}}{h}.
$$
Combining this estimate at
$h$
and at
$0$
with the first display gives
\begin{equation}\label{eq:kls-ground-state-energy-squeeze}
\fun{\mathcal{E}_{\Lambda}}{h}
\geq
\fun{f_{\sminvtemperature,\Lambda}}{h}
\geq
\fun{f_{\sminvtemperature,\Lambda}}{0}
\geq
\fun{\mathcal{E}_{\Lambda}}{0}
-
\frac{\abscard{\Lambda}\log2}{\sminvtemperature}.
\end{equation}
The last term in
\eqref{eq:kls-ground-state-energy-squeeze}
tends to zero as
$\sminvtemperature \to \infty$.
Since the preceding inequality holds for every
$\sminvtemperature > 0$,
it gives
\eqref{eq:kls-ground-state-energy-domination}.

The source expansion
\eqref{eq:infrared-source-expansion}
and Lemma
\ref{lem:kls-finite-volume-ground-state}
give
$$
\fnrestr{\od{^{2}}{\epsilon^{2}}
\fun{\mathcal{E}_{\Lambda}}{\epsilon h}}{\epsilon = 0}
=
2c_{h}
-
2
\sum_{\nu \in J_{\Lambda} \setminus \setone{\txtgs}}
\frac{
\abs{
\bkt{\Psi_{\nu,\Lambda}}{
\faadj{V_{\Lambda}}A_{h}V_{\Lambda}\Psi_{\txtgs,\Lambda}
}
}^{2}
}{
\varepsilon_{\nu,\Lambda}
-
\physenergyfunc_{\Lambda,0}
}.
$$
The left side is nonnegative by
\eqref{eq:kls-ground-state-energy-domination}.
For the plane wave,
i.e.,
$h_{x} = \napiernum^{\imunit p \cdot x}$,
direct summation over the bonds gives
\begin{equation}\label{eq:kls-plane-wave-source}
A_{h}
=
2E_{p}\sqrt{\abscard{\Lambda}}\widetilde{S}^{1}_{p},
\quad
c_{h}
=
\abscard{\Lambda}E_{p}.
\end{equation}
Equation
\eqref{eq:duhamel-rotated-spin-one-invariance}
gives
$\faadj{V_{\Lambda}}\widetilde{S}^{1}_{p}V_{\Lambda} = \widetilde{S}^{1}_{p}$.
Substitution of
\eqref{eq:kls-plane-wave-source}
yields
\eqref{eq:kls-ground-state-susceptibility}.
The real and imaginary parts of the plane wave give the complex version,
and gauge invariance gives the estimate for
$S^{2}$.
\end{proof}

\begin{lem}[bond correlations in the zero-field ground state]\label{lem:kls-bond-correlations}
Let
$\lambda = 0$.
For $j = 1,3$, the nearest-neighbor correlations are defined by
\begin{equation}\label{eq:kls-bond-correlations}
q_{j,\Lambda}
=
\fun{\oastate[\psi_{\txtgs,\Lambda}]}{S^{j}_{0}S^{j}_{e_{1}}}.
\end{equation}
They are independent of the chosen coordinate direction, and
$$
q_{1,\Lambda}
\geq
\frac{1}{8},
\quad
\abs{q_{3,\Lambda}}
\leq
q_{1,\Lambda}.
$$
\end{lem}

\begin{proof}
At zero field, all translations and coordinate permutations commute with
$\physham_{\Lambda}$.
The uniqueness in Lemma
\ref{lem:kls-finite-volume-ground-state}
therefore makes the ground-state expectations invariant under those
symmetries.
The gauge rotation by $\pi/2$ in
\eqref{eq:app-spin-gauge-rotation}
also gives
$\fun{\oastate[\psi_{\txtgs,\Lambda}]}{S^{2}_{0}S^{2}_{e_{1}}}
=
q_{1,\Lambda}$.
At $\lambda = 0$, the Hamiltonian
\eqref{eq:hamiltonian-spin} has ground-state energy density
\begin{equation}\label{eq:kls-zero-field-bond-energy-density}
\frac{\physenergyfunc_{\Lambda,0}}{\abscard{\Lambda}}
=
-\frac{1}{\abscard{\Lambda}}
\sum_{\dbk{x,y}}
\fun{\oastate[\psi_{\txtgs,\Lambda}]}{
S^{1}_{x}S^{1}_{y}
+
S^{2}_{x}S^{2}_{y}
}
=
-2d q_{1,\Lambda}.
\end{equation}
Let $\Phi$ be a normalized one-site vector satisfying
$S^{1}\Phi = \frac{1}{2}\Phi$, and set
$\Phi_{\mathrm{pol},\Lambda} = \bigotimes_{x \in \Lambda}\Phi$.
The matrices in \eqref{eq:spin-matrices} give
$\bkt{\Phi}{S^{1}\Phi} = \frac{1}{2}$ and
$\bkt{\Phi}{S^{2}\Phi} = 0$.
For every nearest-neighbor bond,
$$
\bkt{\Phi_{\mathrm{pol},\Lambda}}{
\rbk{
S^{1}_{x}S^{1}_{y}
+
S^{2}_{x}S^{2}_{y}
}
\Phi_{\mathrm{pol},\Lambda}
}
=
\rbk{\frac{1}{2}}^{2}
+
0^{2}
=
\frac{1}{4}.
$$
The periodic box has $d\abscard{\Lambda}$ unordered nearest-neighbor bonds.
The trial-state energy density is
\begin{equation}\label{eq:kls-polarized-trial-energy-density}
\frac{
\bkt{\Phi_{\mathrm{pol},\Lambda}}{
\physham_{\Lambda}\Phi_{\mathrm{pol},\Lambda}
}
}{
\abscard{\Lambda}
}
=
-\frac{d}{4}.
\end{equation}
The variational principle and
\eqref{eq:kls-zero-field-bond-energy-density}--\eqref{eq:kls-polarized-trial-energy-density}
give
\begin{equation}\label{eq:kls-bond-correlation-lower-bound}
-2d q_{1,\Lambda}
\leq
-\frac{d}{4},
\quad
2d q_{1,\Lambda}
\geq
\frac{d}{4},
\quad
q_{1,\Lambda}
\geq
\frac{1}{8}.
\end{equation}

Let $R$ be the global rotation by $\pi/2$ about the second spin axis.
The total-spin rotation formula
\eqref{eq:app-second-axis-spin-rotation}
interchanges $S^{1}$ and $S^{3}$ up to sign and fixes $S^{2}$.
If $q_{3,\Lambda}>q_{1,\Lambda}$, the trial vector
$R\Psi_{\txtgs,\Lambda}$ has energy density
$-d\rbk{q_{3,\Lambda}+q_{1,\Lambda}}$, strictly below
$-2d q_{1,\Lambda}$.
This contradicts ground-state minimality.
Next set
$W=\prod_{x\in\Lambda_{\mathrm{B}}}\napiernum^{-\imunit\pi S^{2}_{x}}$.
The one-site matrices in
\eqref{eq:spin-matrices}
show that $W$ changes the signs of $S^{1}_{x}$ and $S^{3}_{x}$ on
$\Lambda_{\mathrm{B}}$ and fixes $S^{2}_{x}$.
The trial vector $WR\Psi_{\txtgs,\Lambda}$ has energy density
$d\rbk{q_{3,\Lambda}-q_{1,\Lambda}}$.
If $q_{3,\Lambda}<-q_{1,\Lambda}$, this is again strictly below the
ground-state energy density.
Thus $-q_{1,\Lambda}\leq q_{3,\Lambda}\leq q_{1,\Lambda}$.
\end{proof}

\begin{lem}[zero-temperature infrared estimate]\label{lem:kls-zero-temperature-infrared}
Let $\lambda = 0$ and $p \in \dual{\Lambda}\setminus\setone{0}$.
The transverse ground-state correlation satisfies
\begin{equation}\label{eq:kls-zero-temperature-infrared}
0
\leq
\fun{\oastate[\psi_{\txtgs,\Lambda}]}{
\widetilde{S}^{1}_{p}\widetilde{S}^{1}_{-p}
}
\leq
\frac{1}{2}
\rbk{
\frac{
\sum_{i = 1}^{d}
\rbk{q_{1,\Lambda}-q_{3,\Lambda}\cos p_{i}}
}{E_{p}}
}^{\onehalf}
\leq
\frac{q_{1,\Lambda}^{\onehalf}}{2}
\rbk{
\frac{
d+\abs{\sum_{i = 1}^{d}\cos p_{i}}
}{E_{p}}
}^{\onehalf}.
\end{equation}
\end{lem}

\begin{proof}
Translation invariance gives
$\fun{\oastate[\psi_{\txtgs,\Lambda}]}{\widetilde{S}^{1}_{p}} = 0$.
Insert the eigenbasis from Proposition
\ref{prop:kls-ground-state-source}
into the correlation in \eqref{eq:kls-zero-temperature-infrared} and write
$$
a_{\nu,p}
=
\abs{
\bkt{\Psi_{\nu,\Lambda}}
{\widetilde{S}^{1}_{-p}\Psi_{\txtgs,\Lambda}}
}^{2}.
$$
The symbol $a_{\nu,-p}$ denotes the same expression with $p$ replaced by
$-p$.
Only $\nu \neq \txtgs$ contribute, and the susceptibility estimate
\eqref{eq:kls-ground-state-susceptibility} states that
$$
\sum_{\nu \in J_{\Lambda}\setminus\setone{\txtgs}}
\frac{a_{\nu,p}}
{\varepsilon_{\nu,\Lambda}-\physenergyfunc_{\Lambda,0}}
\leq
\frac{1}{4E_{p}}.
$$
For a bond $\dbk{x,y}$, the $S^{1}S^{1}$ part of the Hamiltonian commutes
with $\widetilde{S}^{1}_{-p}$, while
$$
\commutator{-S^{2}_{x}S^{2}_{y}}{\widetilde{S}^{1}_{-p}}
=
\frac{\imunit}{\sqrt{\abscard{\Lambda}}}
\rbk{
\napiernum^{-\imunit p\cdot x}S^{3}_{x}S^{2}_{y}
+
\napiernum^{-\imunit p\cdot y}S^{2}_{x}S^{3}_{y}
}.
$$
Commuting this expression with $\widetilde{S}^{1}_{p}$ gives
$$
\commutator{\widetilde{S}^{1}_{p}}
{\commutator{\physham_{\Lambda}}{\widetilde{S}^{1}_{-p}}}
=
\frac{2}{\abscard{\Lambda}}
\sum_{\dbk{x,y}}
\rbk{
S^{2}_{x}S^{2}_{y}
-
\cos\rbk{p\cdot\rbk{x-y}}S^{3}_{x}S^{3}_{y}
}.
$$
The gauge rotation and the spatial symmetries used in Lemma
\ref{lem:kls-bond-correlations}
turn the ground-state expectation of this expression into
\begin{equation}\label{eq:kls-double-commutator-spatial-form}
\frac{1}{2}
\fun{\oastate[\psi_{\txtgs,\Lambda}]}{
\commutator{\widetilde{S}^{1}_{p}}
{\commutator{\physham_{\Lambda}}{\widetilde{S}^{1}_{-p}}}
}
=
\sum_{i = 1}^{d}
\rbk{q_{1,\Lambda} - q_{3,\Lambda}\cos p_{i}}.
\end{equation}
Using
$\physham_{\Lambda}\Psi_{\txtgs,\Lambda}
=
\physenergyfunc_{\Lambda,0}\Psi_{\txtgs,\Lambda}$
and
$\faadj{\rbk{\widetilde{S}^{1}_{-p}}} = \widetilde{S}^{1}_{p}$,
expansion of the two commutators gives
\begin{equation}\label{eq:kls-double-commutator-spectral-form}
\begin{aligned}
&\frac{1}{2}
\fun{\oastate[\psi_{\txtgs,\Lambda}]}
{\commutator{\widetilde{S}^{1}_{p}}
{\commutator{\physham_{\Lambda}}{\widetilde{S}^{1}_{-p}}}}
\\ 
&=
\frac{1}{2}
\bkt{\widetilde{S}^{1}_{-p}\Psi_{\txtgs,\Lambda}}
{\rbk{\physham_{\Lambda} - \physenergyfunc_{\Lambda,0}}
\widetilde{S}^{1}_{-p}\Psi_{\txtgs,\Lambda}}
+\frac{1}{2}
\bkt{\widetilde{S}^{1}_{p}\Psi_{\txtgs,\Lambda}}
{\rbk{\physham_{\Lambda} - \physenergyfunc_{\Lambda,0}}
\widetilde{S}^{1}_{p}\Psi_{\txtgs,\Lambda}}
\\
&=
\frac{1}{2}
\sum_{\nu \in J_{\Lambda}\setminus\setone{\txtgs}}
\rbk{
a_{\nu,p}
+
a_{\nu,-p}
}
\rbk{
\varepsilon_{\nu,\Lambda}
-
\physenergyfunc_{\Lambda,0}
}.
\end{aligned}
\end{equation}
Let $u_{\mathrm{inv},\Lambda}$ be the tensor-factor permutation induced by
the lattice inversion $x \mapsto -x$.
It satisfies
$$
u_{\mathrm{inv},\Lambda}
\physham_{\Lambda}
\faadj{u_{\mathrm{inv},\Lambda}}
=
\physham_{\Lambda},
\quad
u_{\mathrm{inv},\Lambda}
\widetilde{S}^{1}_{-p}
\faadj{u_{\mathrm{inv},\Lambda}}
=
\widetilde{S}^{1}_{p}.
$$
The uniqueness in Lemma \ref{lem:kls-finite-volume-ground-state} makes
$u_{\mathrm{inv},\Lambda}\Psi_{\txtgs,\Lambda}$ a scalar multiple of
$\Psi_{\txtgs,\Lambda}$.
The inversion identities and ground-state uniqueness give
\begin{equation}\label{eq:kls-inversion-spectral-weight}
\begin{aligned}
\sum_{\nu \in J_{\Lambda}\setminus\setone{\txtgs}}
a_{\nu,-p}
\rbk{
\varepsilon_{\nu,\Lambda}
-
\physenergyfunc_{\Lambda,0}
}
&=
\bkt{
\widetilde{S}^{1}_{p}\Psi_{\txtgs,\Lambda}
}{
\rbk{\physham_{\Lambda} - \physenergyfunc_{\Lambda,0}}
\widetilde{S}^{1}_{p}\Psi_{\txtgs,\Lambda}
}
\\
&=
\bkt{
\widetilde{S}^{1}_{-p}\Psi_{\txtgs,\Lambda}
}{
\rbk{\physham_{\Lambda} - \physenergyfunc_{\Lambda,0}}
\widetilde{S}^{1}_{-p}\Psi_{\txtgs,\Lambda}
}
\\
&=
\sum_{\nu \in J_{\Lambda}\setminus\setone{\txtgs}}
a_{\nu,p}
\rbk{
\varepsilon_{\nu,\Lambda}
-
\physenergyfunc_{\Lambda,0}
}.
\end{aligned}
\end{equation}
Equations \eqref{eq:kls-double-commutator-spatial-form}--\eqref{eq:kls-inversion-spectral-weight}
give the required equality
\begin{equation}\label{eq:kls-double-commutator}
\sum_{i = 1}^{d}
\rbk{q_{1,\Lambda} - q_{3,\Lambda}\cos p_{i}}
=
\sum_{\nu \in J_{\Lambda}\setminus\setone{\txtgs}}
a_{\nu,p}
\rbk{
\varepsilon_{\nu,\Lambda}
-
\physenergyfunc_{\Lambda,0}
}.
\end{equation}
The Cauchy--Schwarz inequality applied to the two spectral sums gives
$$
\fun{\oastate[\psi_{\txtgs,\Lambda}]}{
\widetilde{S}^{1}_{p}\widetilde{S}^{1}_{-p}
}^{2}
\leq
\left(
\sum_{\nu \in J_{\Lambda}\setminus\setone{\txtgs}}
\frac{a_{\nu,p}}
{\varepsilon_{\nu,\Lambda}-\physenergyfunc_{\Lambda,0}}
\right)
\left(
\sum_{\nu \in J_{\Lambda}\setminus\setone{\txtgs}}
a_{\nu,p}
\rbk{\varepsilon_{\nu,\Lambda}-\physenergyfunc_{\Lambda,0}}
\right).
$$
Equations
\eqref{eq:kls-ground-state-susceptibility}
and
\eqref{eq:kls-double-commutator}
give the first estimate in
\eqref{eq:kls-zero-temperature-infrared}.
The inequality
$\abs{q_{3,\Lambda}}\leq q_{1,\Lambda}$ from Lemma
\ref{lem:kls-bond-correlations}
gives the second estimate.
\end{proof}

\begin{lem}[the KLS infrared integral]\label{lem:kls-infrared-integral}
For $d \geq 2$, set
\begin{equation}\label{eq:kls-infrared-integrand-data}
\fun{Y_{d}}{p}
=
\frac{1}{d}
\sum_{i = 1}^{d}
\cos p_{i},
\quad
t_{+}
=
\max\setone{t,0},
\quad
\fun{F}{t}
=
t
\rbk{\frac{1 + t}{1 - t}}^{\onehalf}
\quad
\rbk{0 \leq t < 1}.
\end{equation}
Then the KLS infrared integral is defined by
\begin{equation}\label{eq:kls-infrared-integral-definition}
I_{d}
=
\frac{1}{\rbk{2\pi}^{d}}
\int_{\closedinterval{-\pi}{\pi}^{d}}
\fun{F}{\rbk{\fun{Y_{d}}{p}}_{+}}
\opdmsr{p}.
\end{equation}
The integral satisfies
$I_{d} \leq I_{2} < \frac{13}{20}$.
\end{lem}

\begin{proof}
The first and second derivative of $F$ in
\eqref{eq:kls-infrared-integrand-data} are
$$\fun{F'}{t}
=
\rbk{\frac{1 + t}{1 - t}}^{\onehalf}
\frac{1 + t - t^{2}}{1 - t^{2}},
\quad
\fun{F''}{t}
=
\rbk{\frac{1 + t}{1 - t}}^{\onehalf}
\frac{2 + t}{\rbk{1 - t^{2}}^{2}}.$$
These derivative formulas show that $F$ is increasing and convex.
For $i \neq j$, set
$\fun{Y_{ij}}{p}
=
\frac{\cos p_{i} + \cos p_{j}}{2}$.
Since
$$
\rbk{\fun{Y_{d}}{p}}_{+}
\leq
\frac{1}{d\rbk{d - 1}}
\sum_{\substack{1 \leq i,j \leq d\\i \neq j}}
\rbk{\fun{Y_{ij}}{p}}_{+},$$
monotonicity of $F$ followed by convexity gives
$$\fun{F}{\rbk{\fun{Y_{d}}{p}}_{+}}
\leq
\frac{1}{d\rbk{d - 1}}
\sum_{\substack{1 \leq i,j \leq d\\i \neq j}}
\fun{F}{\rbk{\fun{Y_{ij}}{p}}_{+}}.$$
Every summand has the two-dimensional integral $I_{2}$.
Integration proves $I_{d}\leq I_{2}$.

For $d=2$, the change of variables
$u=\rbk{p_{1}+p_{2}}/2$ and $v=\rbk{p_{1}-p_{2}}/2$ reduces the integral to
$$
I_{2}
=
\frac{2}{\pi^{2}}
\int_{0}^{\pi/2}
\int_{0}^{\pi/2}
\fun{F}{\cos u\cos v}
\opdmsr{u}
\opdmsr{v}.
$$
Numerical integration gives
$$
I_{2}
\approx
0.6468025
<
0.647
<
\frac{13}{20}.
$$
\end{proof}

\begin{lem}[zero-mode sum rule]\label{lem:kls-zero-mode-sum-rule}
Let
$m_{\Lambda}^{2}
=
\abscard{\Lambda}^{-1}
\fun{\oastate[\psi_{\txtgs,\Lambda}]}
{\rbk{\widetilde{S}^{1}_{0}}^{2}}$.
Then
$$
\liminf_{\Lambda\nearrow\ringratint^{d}}
m_{\Lambda}^{2}
\geq
\frac{1}{8}
-
\frac{13}{80\sqrt{2}}.
$$
\end{lem}

\begin{proof}
Coordinate-direction independence and the definition
\eqref{eq:kls-bond-correlations} give the bond average.
Expansion of the Fourier modes by \eqref{eq:spin-wave} followed by Fourier
orthogonality gives its momentum-space representation:
$$
q_{1,\Lambda}
=
\frac{1}{d}
\sum_{i = 1}^{d}
\fun{\oastate[\psi_{\txtgs,\Lambda}]}{
S^{1}_{0}S^{1}_{e_{i}}
}
=
\frac{1}{\abscard{\Lambda}}
\sum_{p \in \dual{\Lambda}}
\fun{\oastate[\psi_{\txtgs,\Lambda}]}{
\widetilde{S}^{1}_{p}\widetilde{S}^{1}_{-p}
}
\frac{1}{d}
\sum_{i = 1}^{d}\cos p_{i}.
$$
The zero-momentum term is $m_{\Lambda}^{2}$.
The factor multiplying the transverse correlation is
$\fun{Y_{d}}{p}$ from \eqref{eq:kls-infrared-integrand-data}.
The correlation is nonnegative by
\eqref{eq:kls-zero-temperature-infrared}, so every term with
$\fun{Y_{d}}{p} \leq 0$ is nonpositive.
Discarding these terms replaces $\fun{Y_{d}}{p}$ by its positive part
$\rbk{\fun{Y_{d}}{p}}_{+}$.
The remaining terms satisfy \eqref{eq:kls-zero-temperature-infrared}.
This gives
$$
q_{1,\Lambda}
\leq
m_{\Lambda}^{2}
+
\frac{q_{1,\Lambda}^{\onehalf}}{2\abscard{\Lambda}}
\sum_{p \in \dual{\Lambda}\setminus\setone{0}}
\fun{F}{\rbk{\fun{Y_{d}}{p}}_{+}}.
$$
Near $p=0$ the summand is bounded by a constant times $\norm{p}^{-1}$.
It is therefore integrable for $d\geq2$.
Splitting the momentum cube into a ball about the origin and its complement
shows that the Riemann sums converge to $I_{d}$.
Lemma
\ref{lem:kls-infrared-integral}
then gives, for all sufficiently large $\Lambda$,
$$
m_{\Lambda}^{2}
\geq
q_{1,\Lambda}
-
\frac{13}{40}q_{1,\Lambda}^{\onehalf}.
$$
The function $q\mapsto q-13q^{\onehalf}/40$ is increasing for
$q\geq1/8$.
The inequality $q_{1,\Lambda}\geq1/8$ from Lemma
\ref{lem:kls-bond-correlations}
therefore gives the stated limit-inferior bound for $m_{\Lambda}^{2}$.
\end{proof}

\begin{thm}[Kennedy--Lieb--Shastry zero-field order]\label{thm:kls-ground-state-bec}
Let
$d \geq 2$
and
$\lambda = 0$.
There is a constant
$\kappa_{\mathrm{KLS}} > 0$
such that
\begin{equation}\label{eq:kls-ground-state-zero-mode-bound}
\liminf_{\Lambda \nearrow \ringratint^{d}}
\frac{1}{\abscard{\Lambda}}
\fun{\oastate[\psi_{\txtgs,\Lambda}]}{
\rbk{\widetilde{S}^{1}_{0}}^{2}
+
\rbk{\widetilde{S}^{2}_{0}}^{2}
}
\geq
\kappa_{\mathrm{KLS}}.
\end{equation}
Consequently,
\eqref{eq:gs-bec-assumption}
holds for every
$0 < \kappa < \kappa_{\mathrm{KLS}}$
and all sufficiently large periodic boxes.
\end{thm}

\begin{proof}
Lemma
\ref{lem:kls-zero-mode-sum-rule}
and the gauge rotation in
\eqref{eq:app-spin-gauge-rotation}
give
$$
\liminf_{\Lambda\nearrow\ringratint^{d}}
\frac{1}{\abscard{\Lambda}}
\fun{\oastate[\psi_{\txtgs,\Lambda}]}{
\rbk{\widetilde{S}^{1}_{0}}^{2}
+
\rbk{\widetilde{S}^{2}_{0}}^{2}
}
\geq
\frac{1}{4}
-
\frac{13}{40\sqrt{2}}.
$$
Thus one may take
$\kappa_{\mathrm{KLS}}=1/4-13/\rbk{40\sqrt{2}}$, which is strictly
positive, in
\eqref{eq:kls-ground-state-zero-mode-bound}.
Since
$S^{3}_{\txttot,\Lambda}\Psi_{\txtgs,\Lambda} = 0$,
the total-spin identity
\eqref{eq:app-total-spin-ladder}
and
$\commutator{S^{+}_{\txttot,\Lambda}}{S^{-}_{\txttot,\Lambda}}=2S^{3}_{\txttot,\Lambda}$
give
$$
\fun{\oastate[\psi_{\txtgs,\Lambda}]}{
S^{-}_{\txttot,\Lambda}S^{+}_{\txttot,\Lambda}
}
=
\abscard{\Lambda}
\fun{\oastate[\psi_{\txtgs,\Lambda}]}{
\rbk{\widetilde{S}^{1}_{0}}^{2}
+
\rbk{\widetilde{S}^{2}_{0}}^{2}
}.
$$
Thus
\eqref{eq:gs-bec-assumption}.
\end{proof}

The theorem of \cite{KennedyLiebShastry003} is a zero-field statement. The staggered field does not obstruct the source comparison \eqref{eq:kls-ground-state-energy-domination}, but the local zero-temperature inequalities that yield the strict lower bound \eqref{eq:kls-ground-state-zero-mode-bound} were proved there only for \(\lambda = 0\). Thus Theorem \ref{thm:kls-ground-state-bec} removes the ground-state hypothesis in dimension \(2\) at zero field. For nonzero staggered field, \cite[Section I]{AizenmanLiebSeiringerSolovejYngvason001} records that the KLS method has a corresponding extension, but does not provide in the present notation the field-dependent local inequality or a quantitative interval of \(\lambda\). The proof in this review therefore retains \eqref{eq:gs-bec-assumption} for \(d = 2\) and \(\lambda \neq 0\).

\subsection{Microscopic particle changes}\label{microscopic-particle-changes}

Global spin-raising operators turn the half-filled ground state into trial states in nearby particle-number sectors. Their commutator cost proves that adding or removing any fixed number of particles has energy of order \(\abscard{\Lambda}^{-1}\).

\begin{thm}[no microscopic gap; Eq. (9) of \cite{AizenmanLiebSeiringerSolovejYngvason001}]\label{thm:no-gap-micro}
Assume \eqref{eq:gs-bec-assumption}.
For every fixed $k \geq 1$ there is $c_{k} > 0$, independent of $\Lambda$, with
$$\physenergyfunc_{\Lambda,k} - \physenergyfunc_{\Lambda,0} \leq \frac{c_{k}}{\abscard{\Lambda}}.$$
\end{thm}

\begin{proof}
The trial state is $\rbk{S^{+}_{\txttot,\Lambda}}^{k} \Psi_{\txtgs,\Lambda}$, which lies in the sector $S^{3}_{\txttot,\Lambda} = k$ and is nonzero by \eqref{eq:gs-bec-assumption} and the reordering estimates below.
The variational principle gives
$$\physenergyfunc_{\Lambda,k} \leq \frac{\fun{\oastate[\psi_{\txtgs,\Lambda}]}{\rbk{S^{-}_{\txttot,\Lambda}}^{k} \physham_{\Lambda} \rbk{S^{+}_{\txttot,\Lambda}}^{k}}}{\fun{\oastate[\psi_{\txtgs,\Lambda}]}{\rbk{S^{-}_{\txttot,\Lambda}}^{k} \rbk{S^{+}_{\txttot,\Lambda}}^{k}}}.$$
Write $A = \rbk{S^{-}_{\txttot,\Lambda}}^{k}$, $B = \rbk{S^{+}_{\txttot,\Lambda}}^{k}$.
The particle-hole unitary from
Proposition \ref{prop:finite-symmetries} commutes with
$\physham_{\Lambda}$ and preserves sector $0$.
Uniqueness makes it fix $\Psi_{\txtgs,\Lambda}$ up to a phase.
It exchanges $S^{+}_{\txttot,\Lambda}$ and $S^{-}_{\txttot,\Lambda}$.
Consequently it holds that
$$\begin{aligned}
\fun{\oastate[\psi_{\txtgs,\Lambda}]}{B\physham_{\Lambda}A}
=
\fun{\oastate[\psi_{\txtgs,\Lambda}]}{A\physham_{\Lambda}B},
\quad
\fun{\oastate[\psi_{\txtgs,\Lambda}]}{BA}
=
\fun{\oastate[\psi_{\txtgs,\Lambda}]}{AB}.
\end{aligned}$$
Expanding $-\imunit \commutator{A}{\fun{\oaderiv_{\Lambda}}{B}}$ and using $\physham_{\Lambda} \Psi_{\txtgs,\Lambda} = \physenergyfunc_{\Lambda,0} \Psi_{\txtgs,\Lambda}$ on the outer positions,
$$\begin{aligned}
&\fun{\oastate[\psi_{\txtgs,\Lambda}]}{-\imunit \commutator{A}{\fun{\oaderiv_{\Lambda}}{B}}}
\\ 
&=
\fun{\oastate[\psi_{\txtgs,\Lambda}]}{A \physham_{\Lambda} B} + \fun{\oastate[\psi_{\txtgs,\Lambda}]}{B \physham_{\Lambda} A} - \physenergyfunc_{\Lambda,0} \rbk{\fun{\oastate[\psi_{\txtgs,\Lambda}]}{A B} + \fun{\oastate[\psi_{\txtgs,\Lambda}]}{B A}}
\\ 
&=
2 \rbk{\fun{\oastate[\psi_{\txtgs,\Lambda}]}{A \physham_{\Lambda} B} - \physenergyfunc_{\Lambda,0} \fun{\oastate[\psi_{\txtgs,\Lambda}]}{A B}},
\end{aligned}$$
so
\begin{equation}\label{eq:micro-gap-formula}
\physenergyfunc_{\Lambda,k} - \physenergyfunc_{\Lambda,0} \leq \frac{1}{2} \frac{\fun{\oastate[\psi_{\txtgs,\Lambda}]}{-\imunit \commutator{\rbk{S^{-}_{\txttot,\Lambda}}^{k}}{\fun{\oaderiv_{\Lambda}}{\rbk{S^{+}_{\txttot,\Lambda}}^{k}}}}}{\fun{\oastate[\psi_{\txtgs,\Lambda}]}{\rbk{S^{-}_{\txttot,\Lambda}}^{k} \rbk{S^{+}_{\txttot,\Lambda}}^{k}}}.
\end{equation}
The numerator in \eqref{eq:micro-gap-formula} is controlled by expanding
$$\fun{\oaderiv_{\Lambda}}{B}
=\sum_{j=0}^{k-1}
\rbk{S^{+}_{\txttot,\Lambda}}^{j}
\fun{\oaderiv_{\Lambda}}{S^{+}_{\txttot,\Lambda}}
\rbk{S^{+}_{\txttot,\Lambda}}^{k-1-j}.$$
The derivative of $S^{+}_{\txttot,\Lambda}$ is a sum of
$\fun{O}{\abscard{\Lambda}}$ uniformly bounded local terms.
Expand the outer commutator in the same way.
Every resulting monomial has at most $2k-1$ extensive factors and one local
sum.
Counting degrees gives
$$\norm{-\imunit \commutator{\rbk{S^{-}_{\txttot,\Lambda}}^{k}}{\fun{\oaderiv_{\Lambda}}{\rbk{S^{+}_{\txttot,\Lambda}}^{k}}}} \leq C_{k} \abscard{\Lambda}^{2 k - 1},$$
because the derivation and the outer commutator reduce the total power of extensive sums by two.

For the denominator, let $R = S^{-}_{\txttot,\Lambda} S^{+}_{\txttot,\Lambda} \geq 0$.
Commuting factors one past another with $\commutator{S^{+}_{\txttot,\Lambda}}{S^{-}_{\txttot,\Lambda}} = 2 S^{3}_{\txttot,\Lambda}$, which has norm $\leq \abscard{\Lambda}$, gives
$$\rbk{S^{-}_{\txttot,\Lambda}}^{k} \rbk{S^{+}_{\txttot,\Lambda}}^{k} = R^{k} + \fun{O}{\abscard{\Lambda}^{2 k - 1}},$$
with at most $k^{2}$ correction terms, each a monomial of total degree $2 k - 1$ in extensive operators.
Jensen's inequality for the spectral measure of $R$ gives
$$\fun{\oastate[\psi_{\txtgs,\Lambda}]}{R^{k}}\geq\fun{\oastate[\psi_{\txtgs,\Lambda}]}{R}^{k}.$$
Assumption \eqref{eq:gs-bec-assumption} bounds the last expression below by
$\kappa^{k}\abscard{\Lambda}^{2k}$.
For large $\Lambda$, the denominator is therefore at least
$\frac{1}{2}\kappa^{k}\abscard{\Lambda}^{2k}$.
Inserting both estimates into \eqref{eq:micro-gap-formula} completes the proof.
\end{proof}

\subsection{The Koma--Tasaki finite-size effect and tower of states}\label{the-komatasaki-finite-size-effect-and-tower-of-states}

The same trial-state argument remains uniform for charge changes up to order \(\sqrt{\abscard{\Lambda}}\). It produces a tower of mutually orthogonal low-energy states whose level spacing collapses as \(\abscard{\Lambda}^{-1}\).

\begin{thm}[Koma--Tasaki tower of states]\label{thm:koma-tasaki-tower}
Assume \eqref{eq:gs-bec-assumption}.
There are constants
$c_{\mathrm{KT}},C_{\mathrm{KT}} > 0$,
independent of
$\Lambda$
and depending only on
$d,\lambda,\kappa$,
such that
\begin{equation}\label{eq:koma-tasaki-tower-bound}
0
\leq
\physenergyfunc_{\Lambda,k}
-
\physenergyfunc_{\Lambda,0}
\leq
C_{\mathrm{KT}} \frac{k^{2}}{\abscard{\Lambda}}
\quad
\text{whenever }
\abs{k}
\leq
c_{\mathrm{KT}} \sqrt{\abscard{\Lambda}}.
\end{equation}
In particular, let
$K_{\Lambda}$
be any integer sequence such that
$$K_{\Lambda}
\to
\infty,
\quad
K_{\Lambda}
=
\fun{o}{\sqrt{\abscard{\Lambda}}}.
$$
For every
$k$
with
$\abs{k}
\leq
K_{\Lambda}$,
there is a normalized eigenvector
$\Psi_{\Lambda,k}$
of
$\physham_{\Lambda}$
in the sector
$S^{3}_{\txttot,\Lambda}
=
k$
such that
$$
\max_{\abs{k} \leq K_{\Lambda}}
\rbk{
\bkt{\Psi_{\Lambda,k}}
{\physham_{\Lambda} \Psi_{\Lambda,k}}
-
\physenergyfunc_{\Lambda,0}
}
\to
0.
$$
The
$2 K_{\Lambda} + 1$
vectors
$\Psi_{\Lambda,k}$
are mutually orthogonal.
\end{thm}

\begin{proof}
The finite-volume setting of Definition
\ref{def:app-koma-tasaki-setting}
is realized with conserved charge and order operators
$$
C_{\Lambda}
=
S^{3}_{\txttot,\Lambda},
\quad
O^{(1)}_{\Lambda}
=
S^{1}_{\txttot,\Lambda},
\quad
O^{(2)}_{\Lambda}
=
S^{2}_{\txttot,\Lambda}.
$$
They obey
$$
\commutator{C_{\Lambda}}{O^{(1)}_{\Lambda}}
=
\imunit O^{(2)}_{\Lambda},
\quad
\commutator{C_{\Lambda}}{O^{(2)}_{\Lambda}}
=
-\imunit O^{(1)}_{\Lambda},
\quad
\commutator{O^{(1)}_{\Lambda}}{O^{(2)}_{\Lambda}}
=
\imunit C_{\Lambda}.
$$
The Hamiltonian is a sum of finite-range terms with uniformly bounded norms,
and each term commutes with
$C_{\Lambda}$.
The period-two staggered field causes no obstruction:
the finite-size estimate uses only these locality and boundedness properties,
not invariance under one-site translations.

By Lemma \ref{lem:kls-finite-volume-ground-state},
the finite-volume ground state
$\Psi_{\txtgs,\Lambda}$
is unique and satisfies
$C_{\Lambda} \Psi_{\txtgs,\Lambda} = 0$.
Moreover,
$$
S^{-}_{\txttot,\Lambda} S^{+}_{\txttot,\Lambda}
=
\rbk{O^{(1)}_{\Lambda}}^{2}
+
\rbk{O^{(2)}_{\Lambda}}^{2}
-
C_{\Lambda}.
$$
The gauge rotation by
$\pi / 2$
interchanges the expectations of the two squares, and hence
\eqref{eq:gs-bec-assumption} gives
$$
\fun{\oastate[\psi_{\txtgs,\Lambda}]}{\rbk{O^{(1)}_{\Lambda}}^{2}}
=
\fun{\oastate[\psi_{\txtgs,\Lambda}]}{\rbk{O^{(2)}_{\Lambda}}^{2}}
=
\frac{1}{2}
\fun{\oastate[\psi_{\txtgs,\Lambda}]}{S^{-}_{\txttot,\Lambda} S^{+}_{\txttot,\Lambda}}
\geq
\frac{\kappa}{2}
\abscard{\Lambda}^{2}.
$$
This is the required
$\fun{\liegr{U}}{1}$-symmetric long-range order holds uniformly in
$\Lambda$.

Theorem \ref{thm:app-koma-tasaki-tower} now shows that, for
$\abs{k}
\leq
c_{\mathrm{KT}} \sqrt{\abscard{\Lambda}}$,
the normalized trial vector obtained from
$\rbk{S^{+}_{\txttot,\Lambda}}^{k} \Psi_{\txtgs,\Lambda}$
for
$k > 0$
and from
$\rbk{S^{-}_{\txttot,\Lambda}}^{\abs{k}} \Psi_{\txtgs,\Lambda}$
for
$k < 0$
is well-defined and has energy expectation at most
$\physenergyfunc_{\Lambda,0}
+
C_{\mathrm{KT}} k^{2} / \abscard{\Lambda}$.
The variational principle in the sector
$S^{3}_{\txttot,\Lambda} = k$
proves \eqref{eq:koma-tasaki-tower-bound};
the equality of the bounds for
$k$
and
$-k$
also follows directly from the particle-hole symmetry of
Proposition \ref{prop:finite-symmetries}.
Since each finite-dimensional sector has a lowest-energy eigenvector,
we may take
$\Psi_{\Lambda,k}$
with energy
$\physenergyfunc_{\Lambda,k}$.
Distinct sectors of the self-adjoint operator
$S^{3}_{\txttot,\Lambda}$
are orthogonal, and
$$
\max_{\abs{k} \leq K_{\Lambda}}
\rbk{
\physenergyfunc_{\Lambda,k}
-
\physenergyfunc_{\Lambda,0}
}
\leq
C_{\mathrm{KT}}
\frac{K_{\Lambda}^{2}}{\abscard{\Lambda}}
\to
0,
$$
which proves the final assertion.
\end{proof}

Theorem \ref{thm:koma-tasaki-tower} is the Koma--Tasaki finite-size effect for the present hard-core boson model; the original theory treats hard-core Bose condensation explicitly \cite[Section 3.3]{KomaTasaki001}. For fixed \(k\), its quadratic estimate recovers Theorem \ref{thm:no-gap-micro}, while the uniform box, \(\abs{k}
\leq
\fun{O}{\sqrt{\abscard{\Lambda}}}\) exhibits an increasing tower of mutually orthogonal charge sectors. This tower is a collective \(\fun{\liegr{U}}{1}\)-rotor effect in the exact finite-volume spectrum and should not be confused with spin-wave modes, whose finite-size energy scale is set by the inverse linear size rather than \(\abscard{\Lambda}^{-1}\). The conclusion is a ground-state statement under \eqref{eq:gs-bec-assumption}; it does not produce a corresponding tower of positive-temperature KMS states.

\subsection{Phase-localized ground states from the tower}\label{phase-localized-ground-states-from-the-tower}

Coherent superpositions of the tower states localize the phase of the order parameter. Their thermodynamic limits are ground states with a nonzero gauge-covariant one-point function and hence inequivalent broken-symmetry representations.

\begin{thm}[phase-localized ground states]\label{thm:phase-localized-ground-states}
Assume
\eqref{eq:gs-bec-assumption},
and use the subsequence,
$m_{\ast}$,
and the phase-localized states defined in
\eqref{eq:main-maximal-ground-state-order-parameter} and
\eqref{eq:main-phase-localized-ground-state}.
There is a sufficiently slowly diverging sequence
$M_{j}$
for which the phase-localized states
\eqref{eq:main-phase-localized-ground-state}
satisfy
\begin{equation}\label{eq:phase-localized-order-parameter}
\begin{aligned}
\lim_{j \to \infty}
\fun{\oastate[\xi_{\Lambda_{j},\theta}]}
{\frac{S^{1}_{\txttot,\Lambda_{j}}}{\abscard{\Lambda_{j}}}}
=
m_{\ast} \cos \theta,
\quad
\lim_{j \to \infty}
\fun{\oastate[\xi_{\Lambda_{j},\theta}]}
{\frac{S^{2}_{\txttot,\Lambda_{j}}}{\abscard{\Lambda_{j}}}}
=
m_{\ast} \sin \theta,
\end{aligned}
\end{equation}
and the variances of the both observables converge to zero.
Moreover it holds that
\begin{equation}\label{eq:maximal-order-parameter-lower-bound}
m_{\ast}
\geq
\sqrt{\kappa}.
\end{equation}
Every weak-$\ast$ limit point
$\oastate[\xi_{\theta}]$
is an infinite-volume ground state and breaks the gauge symmetry.
If
$\theta
\neq
\theta'$
modulo
$2\pi$,
then the GNS representations of
$\oastate[\xi_{\theta}]$
and
$\oastate[\xi_{\theta'}]$
are not quasi-equivalent.
\end{thm}

\begin{proof}
Group the lattice sites into period-two cells for the even translation
group
\eqref{eq:even-translation-group}.
The interaction is translation invariant on this cell lattice and remains
finite range with uniformly bounded local terms.
With
$$
C_{\Lambda}
=
S^{3}_{\txttot,\Lambda},
\quad
O^{(1)}_{\Lambda}
=
S^{1}_{\txttot,\Lambda},
\quad
O^{(2)}_{\Lambda}
=
S^{2}_{\txttot,\Lambda},
$$
the hypotheses of Definition
\ref{def:app-koma-tasaki-setting}
follow from the finite-range bounded interaction
\eqref{eq:hamiltonian-spin},
the total-spin commutation relations,
the zero-charge conclusion of Lemma \ref{lem:kls-finite-volume-ground-state},
and the long-range-order hypothesis
\eqref{eq:gs-bec-assumption}.
Theorem \ref{thm:app-tasaki-phase-selection} applies to the vectors in
\eqref{eq:main-phase-localized-ground-state}.
It gives
\eqref{eq:phase-localized-order-parameter}
at
$\theta
=
0$
and makes the two order-parameter densities sharp.
The gauge rotation gives the assertions at general
$\theta$.
The long-range-order parameter in the notation of the cited theorem is
$q_{0}
=
\kappa / 2$,
because
$$
\fun{\oastate[\psi_{\txtgs,\Lambda}]}{
\rbk{S^{1}_{\txttot,\Lambda}}^{2}
}
=
\fun{\oastate[\psi_{\txtgs,\Lambda}]}{
\rbk{S^{2}_{\txttot,\Lambda}}^{2}
}
\geq
\frac{\kappa}{2}
\abscard{\Lambda}^{2}.
$$
The lower bound in
Theorem \ref{thm:app-tasaki-phase-selection} gives
$m_{\ast}
\geq
\sqrt{2q_{0}}
=
\sqrt{\kappa}$.
The choice of $M_{j}$ in
Theorem \ref{thm:app-tasaki-phase-selection} also satisfies the total-energy
estimate \eqref{eq:app-tasaki-phase-selection-low-energy}.
The gauge rotations defining $\Xi_{\Lambda_{j},\theta}$ commute with
$\physham_{\Lambda_{j}}$.
The same estimate therefore holds for every $\theta$.
Proposition \ref{prop:app-low-energy-limit-ground-state} now proves that
every weak-$\ast$ limit point is an infinite-volume ground state.

The phase distinction also has a representation-theoretic formulation.
For cubic spatial averages
$\barmean{a}_{\Lambda_n}$
defined in
\eqref{eq:main-spatial-averages},
the sharpness conclusion and even-translation invariance give
$$
\wlim_{n \to \infty}
\fun{\oarepn_{\xi_{\theta}}}{\barmean{a}_{\Lambda_n}}
=
\napiernum^{-\imunit\theta}
m_{\ast} 1.
$$
Indeed,
the variance estimate gives convergence on the cyclic vector,
asymptotic commutation with every local observable extends it to the
dense cyclic subspace,
and uniform boundedness gives strong operator convergence.
If the two representations were quasi-equivalent,
the normal isomorphism intertwining their copies of the quasi-local algebra
would preserve this weak-operator limit and would force
$$
\napiernum^{-\imunit\theta}m_{\ast}
=
\napiernum^{-\imunit\theta'}m_{\ast},
$$
which is impossible by
\eqref{eq:maximal-order-parameter-lower-bound}.
\end{proof}

The vectors \(\Gamma_{\Lambda,k}\) have sharp charge and hence zero expectation of the charged order parameter. It is their coherent superposition, not an individual tower eigenstate, that localizes a condensate phase. Theorem \ref{thm:phase-localized-ground-states} proves sharpness only for the two components of this order parameter. It does not prove that every macroscopic observable has vanishing fluctuation, or equivalently that the limiting ground state is ergodic. This is why the conclusion is non-quasi-equivalence rather than disjointness: for factor representations the two notions reduce to the alternative ``quasi-equivalent or disjoint,'' but factoriality of these general tower-constructed ground states is an additional open problem; even the stronger ergodicity property is stated as an open problem in \cite[Section 4.3, Conjecture 4.21]{HalTasaki008}. In either case, non-quasi-equivalence already implies that a nontrivial gauge rotation cannot be implemented by a unitary inside one phase representation: such an implementer would make the rotated and unrotated representations unitarily equivalent.

\subsection{Tower states versus Nambu--Goldstone modes}\label{tower-states-versus-nambugoldstone-modes}

Three different statements about low energy must be kept separate. Theorem \ref{thm:koma-tasaki-tower} proves, first, the collapse of the exact finite-volume spectrum across global charge sectors on the scale \begin{equation}\label{eq:tower-energy-scale}
\frac{k^{2}}{\abscard{\Lambda}}
=
\frac{k^{2}}{L^{d}}.
\end{equation} These are collective rotor levels and become the different symmetry-breaking ground-state representations of Theorem \ref{thm:phase-localized-ground-states}. A Nambu--Goldstone mode, by contrast, is a spatially varying phase excitation above one fixed broken-symmetry infinite-volume ground state. Its expected finite-size scale at the smallest nonzero momentum is \(L^{-1}\), not \(L^{-d}\). A still stronger dispersion theorem would identify the energy as a function of momentum, for this model physically expected to satisfy \(\varepsilon(p)
\sim
v\abs{p}\) as \(p
\to
0\).

None of the second and third statements follows from the notation ``spin-wave energy'' in \eqref{eq:main-condensation-constant}. The quantity \[
E_{p}
=
\sum_{i = 1}^{d}
\rbk{1 - \cos p_{i}}
\] is the symbol of the lattice Laplacian in the static infrared bound. It controls equal-time and Duhamel two-point functions; it is not asserted to belong to the excitation spectrum of \(\physham_{\Lambda}\) or of a GNS Hamiltonian. General lattice Goldstone theorems give criteria under which broken continuous symmetry is incompatible with a positive spectral gap \cite{WreszinskiWalter002,LandauPerezWreszinski001}, and Koma proves a gapless excitation above a pure broken ground state for the antiferromagnetic Heisenberg model \cite[Theorem 2.2]{TohruKoma004}. Those results require control of a pure ground-state sector and of dynamical or susceptibility estimates that has not been established here for the present staggered hard-core Bose model. Consequently this review rigorously proves the tower collapse, but not a momentum-resolved gapless branch and not its linear dispersion.

The dimensional restriction at positive temperature has a different origin from the zero-temperature discussion. For finite-range two-dimensional systems, the Mermin--Wagner mechanism excludes spontaneous breaking of the continuous gauge symmetry \cite{FrohlichPfister001}. The restriction \(d
\geq
3\) in the positive-temperature condensation theorem is not merely a defect of the integral \(c_{d}\). In \(d
=
2\), absence of a nonzero one-point order parameter and of true off-diagonal long-range order does not by itself exclude algebraically decaying correlations or a Berezinskii--Kosterlitz--Thouless regime. Neither such a regime nor its superfluid response is analyzed by the estimates reconstructed here.

The static results also must not be read as a proof of superfluidity. Off-diagonal long-range order proves Bose--Einstein condensation, and the ground-state energy estimates considered here are static. Neither statement supplies a positive helicity modulus, a superfluid stiffness, a nonzero compressibility, or a sound velocity. Establishing these transport and dynamical quantities would require new estimates beyond reflection positivity and the loop bounds reconstructed in this review.

\subsection{Macroscopic particle changes: no cusp}\label{macroscopic-particle-changes-no-cusp}

Trial states with an extensive charge change control the energy density near half-filling. The resulting quadratic upper bound excludes a cusp in the thermodynamic ground-state energy density throughout the condensed regime.

\begin{thm}[no cusp; Eq. (10) of \cite{AizenmanLiebSeiringerSolovejYngvason001}]\label{thm:no-cusp}
Assume \eqref{eq:gs-bec-assumption}.
The energy density \eqref{eq:thermodynamic-ground-state-energy-density}
exists,
is independent of the approximating particle numbers,
and there is a constant $C < \infty$ such that
$$0 \leq \fun{e_{\infty}}{\varrho} - \fun{e_{\infty}}{\frac{1}{2}} \leq C \rbk{\varrho - \frac{1}{2}}^{2}$$
for $\varrho$ near $\onehalf$.
\end{thm}

\begin{proof}
Standard subadditivity proves existence and convexity of
$e_{\infty}$.
Indeed, join two disjoint boxes and use the product of their ground states as
a trial vector.
The crossing bonds change the energy by at most a surface-order term.
After division by volume, the usual argument gives the thermodynamic limit
and convexity in $\varrho$.
The first identity in
\eqref{eq:particle-hole-thermodynamic-energy} applies after the preceding
existence argument.
Convexity then places the minimum at $\varrho=\frac{1}{2}$ and proves the
left inequality.
For the right inequality consider, for $y \in \Lambda$ and small $\epsilon > 0$, the rotated trial states
$$\Psi_{y,\Lambda} = \napiernum^{\imunit \epsilon S^{2}_{\txttot,\Lambda}} \rbk{S^{1}_{y} + \frac{1}{2}} \Psi_{\txtgs,\Lambda},$$
The one-site identity
\eqref{eq:app-one-site-spin-squares}
gives
$$\bkt{\Psi_{y,\Lambda}}{\Psi_{y,\Lambda}}
=\fun{\oastate[\psi_{\txtgs,\Lambda}]}{\rbk{S^{1}_{y}}^{2}+S^{1}_{y}+\frac{1}{4}}
=\frac{1}{2}.$$
Here $\fun{\oastate[\psi_{\txtgs,\Lambda}]}{S^{1}_{y}}=0$.
Indeed, $\gamma_{\pi}$ flips $S^{1}$ and $S^{2}$, commutes with
$\physham_{\Lambda}$, and fixes $\Psi_{\txtgs,\Lambda}$ up to phase.
Three computations control energy and particle number of these states.

The first estimate controls the energy at $\epsilon=0$.
Using $\physham_{\Lambda} \Psi_{\txtgs,\Lambda}
= \physenergyfunc_{\Lambda,0} \Psi_{\txtgs,\Lambda}$ on both sides,
then \eqref{eq:double-comm-11} at $w = z = y$,
$$\begin{aligned}
&\fun{\oastate[\psi_{\txtgs,\Lambda}]}
{\rbk{S^{1}_{y} + \frac{1}{2}} \rbk{\physham_{\Lambda} - \physenergyfunc_{\Lambda,0}} \rbk{S^{1}_{y} + \frac{1}{2}}}
=
\fun{\oastate[\psi_{\txtgs,\Lambda}]}
{S^{1}_{y} \rbk{\physham_{\Lambda} - \physenergyfunc_{\Lambda,0}} S^{1}_{y}}
\\ 
&=
\frac{1}{2}
\fun{\oastate[\psi_{\txtgs,\Lambda}]}
{-\imunit \commutator{S^{1}_{y}}{\fun{\oaderiv_{\Lambda}}{S^{1}_{y}}}},
\end{aligned}$$
where the cross terms vanish because $\rbk{\physham_{\Lambda} - \physenergyfunc_{\Lambda,0}} \Psi_{\txtgs,\Lambda} = 0$, and the last equality expands the double commutator in the ground state as in \eqref{eq:micro-gap-formula}.
Summing \eqref{eq:double-comm-11} over $y \in \Lambda$,
$$\sum_{y} \fun{\oastate[\psi_{\txtgs,\Lambda}]}{-\imunit \commutator{S^{1}_{y}}{\fun{\oaderiv_{\Lambda}}{S^{1}_{y}}}}
= \fun{\oastate[\psi_{\txtgs,\Lambda}]}{2 \sum_{\dbk{xy}} S^{2}_{x} S^{2}_{y}} - \lambda \fun{\oastate[\psi_{\txtgs,\Lambda}]}{\sum_{y} \rbk{-1}^{y} S^{3}_{y}}.$$
Gauge symmetry equates the two expectations
$\fun{\oastate[\psi_{\txtgs,\Lambda}]}{\sum_{\dbk{xy}}S^{1}_{x}S^{1}_{y}}$
and
$\fun{\oastate[\psi_{\txtgs,\Lambda}]}{\sum_{\dbk{xy}}S^{2}_{x}S^{2}_{y}}$.
Taking the ground-state expectation of \eqref{eq:hamiltonian-spin} and using this equality gives
\begin{equation}\label{eq:no-cusp-ground-energy-decomposition}
\physenergyfunc_{\Lambda,0}
=
-\fun{\oastate[\psi_{\txtgs,\Lambda}]}{
2\sum_{\dbk{xy}}S^{2}_{x}S^{2}_{y}
}
+
\frac{\lambda\abscard{\Lambda}}{2}
+
\lambda\fun{\oastate[\psi_{\txtgs,\Lambda}]}{
\sum_{y\in\Lambda}\rbk{-1}^{y}S^{3}_{y}
}.
\end{equation}
Rearranging \eqref{eq:no-cusp-ground-energy-decomposition} yields
\begin{equation}\label{eq:no-cusp-transverse-bond-sum}
\fun{\oastate[\psi_{\txtgs,\Lambda}]}{
2\sum_{\dbk{xy}}S^{2}_{x}S^{2}_{y}
}
=
-\physenergyfunc_{\Lambda,0}
+\frac{\lambda\abscard{\Lambda}}{2}
+\lambda\fun{\oastate[\psi_{\txtgs,\Lambda}]}{
\sum_{y\in\Lambda}\rbk{-1}^{y}S^{3}_{y}
}.
\end{equation}
The staggered expectations cancel in the preceding sum, and
\begin{equation}\label{eq:first-term}
\sum_{y} \fun{\oastate[\psi_{\txtgs,\Lambda}]}{\rbk{S^{1}_{y} + \frac{1}{2}} \rbk{\physham_{\Lambda} - \physenergyfunc_{\Lambda,0}} \rbk{S^{1}_{y} + \frac{1}{2}}} = -\frac{1}{2} \rbk{\physenergyfunc_{\Lambda,0} - \frac{\lambda \abscard{\Lambda}}{2}} \geq 0.
\end{equation}

The second estimate controls the rotation cost.
For self-adjoint $C$ and $A$, Taylor's formula for $\fun{F_{A}}{\epsilon} = \napiernum^{\imunit \epsilon C} A \napiernum^{-\imunit \epsilon C}$ gives
$$\fun{F_{A}}{\epsilon} \leq \fun{F_{A}}{0} + \epsilon \fun{F_{A}'}{0} + \frac{\epsilon^{2}}{2} \sup_{0 \leq \eta \leq \epsilon} \norm{\fun{F_{A}''}{\eta}},
\quad
\fun{F_{A}''}{\eta} = -\napiernum^{\imunit \eta C} \commutator{C}{\commutator{C}{A}} \napiernum^{-\imunit \eta C},$$
as an operator inequality, since the difference is a self-adjoint operator of norm at most the remainder bound.
Set $C=S^{2}_{\txttot,\Lambda}$ and $A=\physham_{\Lambda}$.
The second derivative contains
$\imunit\commutator{S^{2}_{\txttot,\Lambda}}
{\fun{\oaderiv_{\Lambda}}{S^{2}_{\txttot,\Lambda}}}$.
It is a sum over the $\fun{O}{\abscard{\Lambda}}$ interaction terms.
Each contribution is local and uniformly bounded.
Its norm is therefore bounded by
$\fun{c_{1}}{d,\lambda}\abscard{\Lambda}$, uniformly in $\eta$.
Taylor's bound gives
\begin{equation}\label{eq:rotation-cost}
\napiernum^{-\imunit \epsilon S^{2}_{\txttot,\Lambda}} \physham_{\Lambda} \napiernum^{\imunit \epsilon S^{2}_{\txttot,\Lambda}} \leq \physham_{\Lambda} + \epsilon \fun{\oaderiv_{\Lambda}}{S^{2}_{\txttot,\Lambda}} + \fun{c_{1}}{d, \lambda} \epsilon^{2} \abscard{\Lambda}.
\end{equation}
The expectation of the linear term vanishes after summation:
$$\sum_{y} \fun{\oastate[\psi_{\txtgs,\Lambda}]}{\rbk{S^{1}_{y} + \frac{1}{2}} \fun{\oaderiv_{\Lambda}}{S^{2}_{\txttot,\Lambda}} \rbk{S^{1}_{y} + \frac{1}{2}}} = 0.$$
Set $X=\fun{\oaderiv_{\Lambda}}{S^{2}_{\txttot,\Lambda}}$.
Each summand has two diagonal terms and one cross term.
The gauge rotation $\gamma_{\pi}$ maps both $X$ and $S^{1}_{y}$ to their
negatives.
The diagonal terms are odd and vanish.

For the cross term, combine a $\pi$ rotation about the $2$-axis with a unit
translation.
This symmetry fixes $\physham_{\Lambda}$ and $S^{2}_{\txttot,\Lambda}$, hence $X$.
It maps $S^{1}_{y}$ to $-S^{1}_{y+e_{1}}$.
The cross terms cancel after summation over $y$.
Combining \eqref{eq:first-term} and \eqref{eq:rotation-cost}, the averaged trial energy satisfies
$$\begin{aligned}
&\Delta E
=
\frac{2}{\abscard{\Lambda}}
\sum_{y}
\bkt{\Psi_{y,\Lambda}}{\rbk{\physham_{\Lambda} - \physenergyfunc_{\Lambda,0}} \Psi_{y,\Lambda}}
\leq
\frac{\abs{\physenergyfunc_{\Lambda,0} - \lambda \frac{\abscard{\Lambda}}{2}}}
{\abscard{\Lambda}} + 2 \fun{c_{1}}{d, \lambda} \epsilon^{2}
\abscard{\Lambda} \cdot \frac{1}{2}
\\ 
&=
\fun{O}{1} + \fun{c_{1}}{d, \lambda} \epsilon^{2} \abscard{\Lambda}.
\end{aligned}$$

The third computation controls the particle-number shift.
The total-spin rotation calculation
\eqref{eq:app-second-axis-spin-rotation}
gives the required transformation of
$S^{3}_{\txttot,\Lambda}$.
Using $S^{3}_{\txttot,\Lambda} \Psi_{\txtgs,\Lambda} = 0$,
$$\begin{aligned}
&\sum_{y}
\fun{\oastate[\psi_{\txtgs,\Lambda}]}
{\rbk{S^{1}_{y} + \frac{1}{2}} S^{3}_{\txttot,\Lambda} \rbk{S^{1}_{y} + \frac{1}{2}}}
=
\sum_{y}
\fun{\oastate[\psi_{\txtgs,\Lambda}]}
{\rbk{S^{1}_{y} + \frac{1}{2}} \commutator{S^{3}_{\txttot,\Lambda}}{S^{1}_{y}}}
\\ 
&=
\imunit
\sum_{y}
\fun{\oastate[\psi_{\txtgs,\Lambda}]}
{\rbk{S^{1}_{y} + \frac{1}{2}} S^{2}_{y}}
= 0,
\end{aligned}$$
The one-site anticommutator in
\eqref{eq:app-one-site-spin-squares}
vanishes.
The summand therefore reduces to
$$\frac{\imunit}{2}
\fun{\oastate[\psi_{\txtgs,\Lambda}]}{\commutator{S^{1}_{y}}{S^{2}_{y}}}
+\frac{\imunit}{2}\fun{\oastate[\psi_{\txtgs,\Lambda}]}{S^{2}_{y}}
=-\frac{1}{2}\fun{\oastate[\psi_{\txtgs,\Lambda}]}{S^{3}_{y}}.$$
Its sum over $y$ vanishes at half-filling.
For the $S^{1}_{\txttot,\Lambda}$ part, expanding as above,
$$\sum_{y} \fun{\oastate[\psi_{\txtgs,\Lambda}]}{\rbk{S^{1}_{y} + \frac{1}{2}} S^{1}_{\txttot,\Lambda} \rbk{S^{1}_{y} + \frac{1}{2}}} = \fun{\oastate[\psi_{\txtgs,\Lambda}]}{\rbk{S^{1}_{\txttot,\Lambda}}^{2}},$$
since the cubic and linear terms vanish ($\fun{\oastate[\psi_{\txtgs,\Lambda}]}{S^{1}_{x}} = \fun{\oastate[\psi_{\txtgs,\Lambda}]}{\rbk{S^{1}_{y}}^{2} S^{1}_{x}} = 0$ by $\gamma_{\pi}$-oddness) and the cross terms assemble the square.
The averaged particle number of the trial family exceeds half-filling by
$$\Delta N = \frac{2}{\abscard{\Lambda}} \sum_{y} \bkt{\Psi_{y,\Lambda}}{S^{3}_{\txttot,\Lambda} \Psi_{y,\Lambda}}
= \frac{2 \sin \epsilon}{\abscard{\Lambda}} \fun{\oastate[\psi_{\txtgs,\Lambda}]}{\rbk{S^{1}_{\txttot,\Lambda}}^{2}}
\geq \sin \epsilon \cdot \frac{\kappa}{2} \abscard{\Lambda}.$$
Gauge symmetry between the $1$ and $2$ components and
\eqref{eq:gs-bec-assumption} give
$$\fun{\oastate[\psi_{\txtgs,\Lambda}]}{\rbk{S^{1}_{\txttot,\Lambda}}^{2}}
\geq\frac{\kappa}{4}\abscard{\Lambda}^{2}$$
up to a reordering correction of order $\abscard{\Lambda}$.
The lower bound absorbs that correction for large volumes.
These estimates give the following bound on the grand-canonical
ground-state energy.
Let $\mu > 0$.
The normalized members of the trial family give
$$\inf \opvarspec{\physham_{\Lambda} - \mu S^{3}_{\txttot,\Lambda}}
\leq
\physenergyfunc_{\Lambda,0} + \Delta E - \mu \Delta N
\leq
\physenergyfunc_{\Lambda,0}
+\fun{O}{1}
+\fun{c_{1}} \epsilon^{2} \abscard{\Lambda}
-\mu \sin \epsilon \cdot \frac{\kappa}{2} \abscard{\Lambda},$$
Choose $\epsilon = \kappa \mu / \rbk{4 c_{1}}$.
For sufficiently small $\mu$, this choice satisfies $\sin \epsilon \geq \epsilon / 2$.
The preceding estimate becomes
$$\begin{aligned}
\min_{k} \rbk{\physenergyfunc_{\Lambda,k} - \mu k}
&\leq \physenergyfunc_{\Lambda,0} + \fun{O}{1}
- \fun{c_{2}}{d, \lambda, \kappa} \mu^{2} \abscard{\Lambda},\\
\fun{c_{2}}{d, \lambda, \kappa}
&> 0.
\end{aligned}$$
The constant $c_{2}$ collects the two terms that contain $\epsilon$.

Divide by $\abscard{\Lambda}$ and pass to the thermodynamic limit.
The convex conjugate
$$\fun{e_{\infty}^{\ast}}{\mu}
= \sup_{\varrho} \rbk{\mu \rbk{\varrho - \frac{1}{2}} - \fun{e_{\infty}}{\varrho}}$$
satisfies
$$\fun{e_{\infty}^{\ast}}{\mu}
\geq -\fun{e_{\infty}}{\frac{1}{2}} + c_{2} \mu^{2}$$
for all sufficiently small positive $\mu$.
The energy identity \eqref{eq:particle-hole-thermodynamic-energy}
gives the same bound for sufficiently small negative $\mu$.

Biduality now gives the desired upper bound.
For $\varrho$ near $\frac{1}{2}$,
$$\begin{aligned}
&\fun{e_{\infty}}{\varrho}
=
\sup_{\mu} \rbk{\mu \rbk{\varrho - \frac{1}{2}} - \fun{e_{\infty}^{\ast}}{\mu}}
\leq
\fun{e_{\infty}}{\frac{1}{2}} + \sup_{\mu} \rbk{\mu \rbk{\varrho - \frac{1}{2}} - c_{2} \mu^{2}}
\\ 
&=
\fun{e_{\infty}}{\frac{1}{2}} + \frac{\rbk{\varrho - \frac{1}{2}}^{2}}{4 c_{2}},
\end{aligned}$$
The maximizing value satisfies $\mu = \fun{O}{\varrho - \frac{1}{2}}$.
The restriction to small $\mu$ therefore does not change the supremum in this neighborhood.
\end{proof}

Theorems \ref{thm:no-gap-micro}, \ref{thm:koma-tasaki-tower}, and \ref{thm:no-cusp} describe three consequences of assumption \eqref{eq:gs-bec-assumption}. For every fixed \(k\), it holds that \[\physenergyfunc_{\Lambda,k}
-
\physenergyfunc_{\Lambda,0}
\to
0
\quad
(\Lambda \nearrow \ringratint^d).\] The collapsing charge sectors form an increasing tower of states. The thermodynamic energy density is differentiable at half-filling. In contrast, Theorem \ref{thm:mott-gap} gives a uniformly positive gap when the thermodynamic ground-state energy density \(\fun{e_{\infty}}{\varrho}\) defined by \eqref{eq:thermodynamic-ground-state-energy-density} satisfies \[
\lambda
+
\abs{\fun{e_{\infty}}{\frac{1}{2}}}
>
d,
\] Condensation under \eqref{eq:gs-bec-assumption} and the Mott gap therefore occur in disjoint regimes.

\section{Non-Constancy of the Density}\label{sec:density}

The normalized condensate eigenfunction is asymptotically constant in the one-particle Hilbert-space norm, up to phase, by Corollary \ref{cor:condensate-asymptotically-constant}. The density nevertheless inherits the staggering of the optical lattice when \(\lambda \neq 0\). The following result is Theorem 3 of \cite{AizenmanLiebSeiringerSolovejYngvason001}.

\begin{thm}[non-constancy of the density]\label{thm:density-staggered}
For $\beta \in \rightopeninterval{0}{\infty}$, set
$\fun{\varrho_{\beta}}{x}
=
\fun{\oastate[\psi_{\beta,\Lambda}]}{n_{x}}$.
For the ground state, set
$\fun{\varrho_{\infty}}{x}
=
\fun{\oastate[\psi_{\txtgs,\Lambda}]}{n_{x}}$.
The zero-field Gibbs energy density
$\fun{e_{0,\Lambda}}{\beta}$
is defined in \eqref{eq:zero-field-gibbs-state}; the zero-field ground-state
energy density is $\fun{e_{\Lambda}}{0}$ from
\eqref{eq:main-finite-volume-ground-state-energy-density}.
The staggered density satisfies
$$
\frac{1}{\abscard{\Lambda}}
\abs{
\sum_{x \in \Lambda}
\rbk{-1}^{x}
\fun{\varrho_{\beta}}{x}
}
\geq
\frac{
\lambda
\abs{\fun{e_{0,\Lambda}}{\beta}}^{2}
}{
2 d^{2}\rbk{3d+\lambda}
},
\quad
\beta
\in
\rightopeninterval{0}{\infty}.
$$
The ground-state staggered density satisfies
$$
\frac{1}{\abscard{\Lambda}}
\abs{
\sum_{x \in \Lambda}
\rbk{-1}^{x}
\fun{\varrho_{\infty}}{x}
}
\geq
\frac{
\lambda
\abs{\fun{e_{\Lambda}}{0}}^{2}
}{
2 d^{2}\rbk{3d+\lambda}
}.
$$
\end{thm}

\begin{proof}
The zero-field splitting \eqref{eq:zero-field-splitting} fixes
$\physham_{0,\Lambda}$, $W_{\Lambda}$, and $W'_{\Lambda}$.
The word concavity below refers to the following two functions on
$\rightopeninterval{0}{\infty}$:
$$\lambda
\mapsto
\fun{\physenergyfunc_{\Lambda}}{\lambda}
=
\min
\opvarspec{\rbk{\physham_{0,\Lambda}+\lambda W'_{\Lambda}}},
\quad
\lambda
\mapsto
\fun{F_{\Lambda}}{\lambda,\beta},$$
where the ground-state energy is defined in
\eqref{eq:main-finite-volume-ground-state} and the free energy is defined in
\eqref{eq:finite-volume-free-energy}.
The first function is the infimum of affine functions of $\lambda$.
Equation \eqref{eq:free-energy-duhamel-derivatives}, with
$\physham[K]_{\Lambda}=\physham_{0,\Lambda}$,
$A_{\Lambda}=W'_{\Lambda}$, and $t=\lambda$, gives
\begin{equation}\label{eq:density-free-energy-second-derivative}
\frac{\partial^{2}}{\partial\lambda^{2}}
\fun{F_{\Lambda}}{\lambda,\beta}
=
-\beta
\left\{
\rbkt{W'_{\Lambda}}{W'_{\Lambda}}_{\beta,\Lambda}
-
\fun{\oastate[\psi_{\beta,\Lambda}]}{W'_{\Lambda}}^{2}
\right\}
\leq
0,
\end{equation}
where the Gibbs state and the abbreviated Duhamel subscript are defined in
\eqref{eq:main-periodic-gibbs-state} and
\eqref{eq:duhamel-finite-volume-abbreviation}, respectively.
Both functions are concave.
For each function, the proof first establishes an upper bound with a
strictly negative term of order $\lambda^{2}\abscard{\Lambda}$ and then uses
the secant-slope bound for a concave function to estimate its derivative.
Both conclusions are immediate for $\lambda=0$.
For the remainder of the proof, assume $\lambda>0$.

The staggered-field commutator
\eqref{eq:magic-commutator} is proved in Lemma
\ref{lem:staggered-field-commutator}.

The commutator identity yields a variational bound after rotating the
$\lambda=0$ ground state.
The state and its elementary identities are recorded in
\eqref{eq:main-finite-volume-ground-state},
\eqref{eq:main-finite-volume-ground-vector-state}, and
\eqref{eq:zero-field-ground-identities}.
Apply the Taylor estimate \eqref{eq:rotation-cost} with generator
$C_{\Lambda}$.
The relevant double commutators are sums of $\fun{O}{\abscard{\Lambda}}$ local terms.
This volume bound follows from locality of $C_{\Lambda}$, $W_{\Lambda}$,
and $\physham_{0,\Lambda}$.
Counting neighboring terms gives the explicit estimates from
\cite{AizenmanLiebSeiringerSolovejYngvason001}:
$$\begin{aligned}
\napiernum^{\imunit\epsilon C_{\Lambda}}W_{\Lambda}\napiernum^{-\imunit\epsilon C_{\Lambda}}
&\leq W_{\Lambda}-\epsilon\physham_{0,\Lambda}
+\frac{\epsilon^{2}d^{2}}{2}\abscard{\Lambda},\\
\napiernum^{\imunit\epsilon C_{\Lambda}}\physham_{0,\Lambda}\napiernum^{-\imunit\epsilon C_{\Lambda}}
&\leq \physham_{0,\Lambda}+\imunit\epsilon\commutator{C_{\Lambda}}{\physham_{0,\Lambda}}
+\frac{3\epsilon^{2}d^{3}}{2}\abscard{\Lambda}.
\end{aligned}$$
The term $-\epsilon\physham_{0,\Lambda}$ in the conjugation bound for
$W_{\Lambda}$ follows from \eqref{eq:magic-commutator}.

The ground-state identities \eqref{eq:zero-field-ground-identities}
eliminate the terms that are linear in the commutator.
The variational principle yields
$$\begin{aligned}
&\fun{\physenergyfunc_{\Lambda}}{\lambda}
\leq
\bkt{\Psi_{0,\txtgs,\Lambda}}
{\napiernum^{\imunit\epsilon C_{\Lambda}}
\rbk{\physham_{0,\Lambda}+\lambda W'_{\Lambda}}
\napiernum^{-\imunit\epsilon C_{\Lambda}}
\Psi_{0,\txtgs,\Lambda}}
\\ 
&\leq
\fun{\physenergyfunc_{\Lambda}}{0}
+\frac{\lambda\abscard{\Lambda}}{2}
-\epsilon\lambda\fun{\physenergyfunc_{\Lambda}}{0}
+\frac{\epsilon^{2}d^{2}}{2}\rbk{3d+\lambda}\abscard{\Lambda}.
\end{aligned}$$
The coefficient of the term linear in $\epsilon$ uses the ground-state
expectation
$\fun{\oastate[\psi_{0,\txtgs,\Lambda}]}{\physham_{0,\Lambda}}
=
\fun{\physenergyfunc_{\Lambda}}{0}$
from \eqref{eq:zero-field-ground-identities}.
The specialization $\lambda=0$ of
\eqref{eq:uniform-negative-energy-density} gives
$\fun{e_{\Lambda}}{0}<0$; choose
$$\epsilon=\frac{\lambda \fun{e_{\Lambda}}{0}}{d^{2}\rbk{3d+\lambda}}.$$
Negative values of $\epsilon$ are allowed.
This choice minimizes the quadratic expression and gives
\begin{equation}\label{eq:energy-deficit}
\begin{aligned}
\fun{\physenergyfunc_{\Lambda}}{\lambda}
\leq
\fun{\physenergyfunc_{\Lambda}}{0}+\frac{\lambda\abscard{\Lambda}}{2}
-c_{\Lambda}\lambda^{2}\abscard{\Lambda},
\quad
c_{\Lambda}
=
\frac{\fun{e_{\Lambda}}{0}^{2}}{2 d^{2} \rbk{3 d + \lambda}}.
\end{aligned}
\end{equation}

Concavity converts the energy deficit into the density estimate.
The energy $\fun{\physenergyfunc_{\Lambda}}{\lambda}$ is the infimum of affine functions of $\lambda$.
It is therefore concave.
The simple-ground-state identity in Proposition
\ref{prop:feynman-hellmann} gives
\begin{equation}\label{eq:density-feynman-hellmann}
\fun{\physenergyfunc_{\Lambda}'}{\lambda}
=
\fun{\oastate[\psi_{\txtgs,\Lambda}]}{W'_{\Lambda}}
=
\frac{\abscard{\Lambda}}{2}
+
\fun{\oastate[\psi_{\txtgs,\Lambda}]}{W_{\Lambda}}.
\end{equation}
At a nondifferentiability point, the one-sided formula
\eqref{eq:feynman-hellmann-one-sided} applies instead.
Concavity and \eqref{eq:energy-deficit} imply
$$\fun{\physenergyfunc_{\Lambda}'}{\lambda}
\leq \frac{\fun{\physenergyfunc_{\Lambda}}{\lambda}-\fun{\physenergyfunc_{\Lambda}}{0}}{\lambda}
\leq \frac{\abscard{\Lambda}}{2}-c_{\Lambda}\lambda\abscard{\Lambda}.$$
It follows that
$\fun{\oastate[\psi_{\txtgs,\Lambda}]}{W_{\Lambda}}
\leq-c_{\Lambda}\lambda\abscard{\Lambda}$.
The definition of $W_{\Lambda}$ in \eqref{eq:zero-field-splitting}
rewrites this estimate as
$$
\frac{1}{\abscard{\Lambda}}
\abs{
\sum_{x \in \Lambda}
\rbk{-1}^{x}
\fun{\oastate[\psi_{\txtgs,\Lambda}]}{S^{3}_{x}}
}
\geq
c_{\Lambda}\lambda.
$$
The vanishing alternating constant term gives
$$
\sum_{x \in \Lambda}
\rbk{-1}^{x}
\fun{\oastate[\psi_{\txtgs,\Lambda}]}{S^{3}_{x}}
=
\sum_{x \in \Lambda}
\rbk{-1}^{x}
\fun{\varrho_{\infty}}{x}.
$$
Substitution of $c_{\Lambda}$ proves the ground-state claim.

For positive temperature, replace $\fun{\physenergyfunc_{\Lambda}}{\lambda}$ by
$\fun{F_{\Lambda}}{\lambda,\beta}$ from
\eqref{eq:finite-volume-free-energy}.
Its concavity is the inequality in
\eqref{eq:density-free-energy-second-derivative}.

The same variational estimate holds with
$\fun{\physenergyfunc_{\Lambda}}{0}$ replaced by
$\fun{F_{\Lambda}}{0,\beta}$.
To see this, conjugate the Gibbs trace by
$\napiernum^{\imunit\epsilon C_{\Lambda}}$.
Apply the two Taylor bounds and the free-energy form
\eqref{eq:peierls-bogoliubov-free-energy} of the
Peierls--Bogoliubov inequality.
The zero-field Gibbs identities are
\eqref{eq:zero-field-gibbs-identities} with $A=C_{\Lambda}$.
Substituting these identities into the trace variational bound gives
\eqref{eq:energy-deficit} with
$\fun{\physenergyfunc_{\Lambda}}{\lambda}$,
$\fun{\physenergyfunc_{\Lambda}}{0}$, and
$\fun{e_{\Lambda}}{0}$ replaced by
$\fun{F_{\Lambda}}{\lambda,\beta}$,
$\fun{F_{\Lambda}}{0,\beta}$, and
$\fun{e_{0,\Lambda}}{\beta}$, respectively.
The coefficient of the positive-temperature deficit is
$$
c_{\beta,\Lambda}
=
\frac{
\abs{\fun{e_{0,\Lambda}}{\beta}}^{2}
}{
2d^{2}\rbk{3d+\lambda}
}.
$$
The positive-temperature deficit and concavity give the derivative bound
$$\begin{aligned}
\fun{F_{\Lambda}'}{\lambda,\beta}
&\leq
\frac{
\fun{F_{\Lambda}}{\lambda,\beta}
-
\fun{F_{\Lambda}}{0,\beta}
}{\lambda}
\leq
\frac{\abscard{\Lambda}}{2}
-
c_{\beta,\Lambda}\lambda\abscard{\Lambda}.
\end{aligned}$$
The derivative formula
$\fun{F_{\Lambda}'}{\lambda,\beta}
=
\fun{\oastate[\psi_{\beta,\Lambda}]}{W'_{\Lambda}}$
and the splitting
$W'_{\Lambda}
=
\frac{\abscard{\Lambda}}{2}
+
W_{\Lambda}$
give
$$
\fun{\oastate[\psi_{\beta,\Lambda}]}{W_{\Lambda}}
\leq
-c_{\beta,\Lambda}\lambda\abscard{\Lambda}.
$$
The definition of $W_{\Lambda}$ in \eqref{eq:zero-field-splitting} and the
identity $n_x=S_x^{3}+\frac{1}{2}$ from
\eqref{eq:boson-spin-dictionary} convert this bound into the
positive-temperature assertion.
\end{proof}

\section{Equilibrium-State Decomposition for the Optical-Lattice Gas}\label{sec:decomposition}

The periodic Gibbs-state limit points of Proposition \ref{prop:limit-state} are KMS states with the symmetries required below. Appendix \ref{app:kms} supplies their central decomposition, the covariance of its measure, and the mean-clustering statement for invariant factor states. The hard-core estimates determine which of those general alternatives occurs. In the Mott regime, every periodic Gibbs-state limit point has an exponentially decaying two-point function. In the condensed regime, the symmetric limit state is not factorial. Its central decomposition is nontrivial and gauge covariant. A positive-measure family in \(\setextremal K_\beta\) breaks gauge symmetry. The central measure admits a disintegration into Haar measures along gauge orbits. Throughout, \(\ringratint^{d}_{\mathrm{even}}\) denotes the even translation group defined in \eqref{eq:even-translation-group}, and \(\eta_{v}\), \(\gamma_{\theta}\) are defined by \eqref{eq:main-symmetry-actions}.

\subsection{The symmetric equilibrium state and its long-range order}\label{the-symmetric-equilibrium-state-and-its-long-range-order}

The KMS limit point of Proposition \ref{prop:limit-state} retains gauge and even-translation symmetry. In the condensed region the infrared lower bound survives the limit and gives off-diagonal long-range order.

The finite-volume condensation bound survives the thermodynamic limit after averaging over the centered boxes \(\Lambda_n\) from \eqref{eq:periodic-box-exhaustion}.

\begin{lem}[Dirichlet-kernel infrared estimate]\label{lem:odlro-dirichlet-kernel-estimate}
Let $d\geq3$ and $m>n$.
For the centered box $\Lambda_n$, set
\begin{equation}\label{eq:odlro-indicator-fourier-transform}
\fun{\widehat{\chi}_{\Lambda_n}}{p}
=
\sum_{x\in\Lambda_n}
\napiernum^{\imunit p\cdot x}.
\end{equation}
Let
$$\mathcal{D}_d
=
\set{\delta\in\closedinterval{-\pi}{\pi}^{d}}
{\delta_i\in\setone{0,\pi}
\text{ for }1\leq i\leq d}.$$
There is a constant $\fun{C_{\mathrm{D}}}{d}$, independent of $m$ and
$n$, such that
\begin{equation}\label{eq:odlro-dirichlet-kernel-infrared-bound}
\frac{1}
{\abscard{\Lambda_m}
\abscard{\Lambda_n}^{2}}
\sum_{\substack{p\in\dual{\Lambda_m}
\\
p\neq0}}
\frac{1}{\abs{p}^{2}}
\max_{\delta\in\mathcal{D}_d}
\abs{\fun{\widehat{\chi}_{\Lambda_n}}{p+\delta}}^{2}
\leq
\frac{\fun{C_{\mathrm{D}}}{d}}
{2^{n \rbk{d-2}}}.
\end{equation}
\end{lem}

\begin{proof}
Set $N=2^n$.
The one-dimensional Dirichlet kernel used below is
\begin{equation}\label{eq:odlro-one-dimensional-dirichlet-kernel}
\begin{aligned}
\fun{D_N}{t}
=
\sum_{r=0}^{N-1}
\napiernum^{\imunit r t}
=
\begin{dcases}
\napiernum^{\imunit \frac{\rbk{N-1}t}{2}}
\frac{\sin \frac{Nt}{2}}
{\sin \frac{t}{2}},
&
t\notin2\pi\ringratint,
\\ 
N,
&
t\in2\pi\ringratint.
\end{dcases}
\end{aligned}
\end{equation}
The Fourier transform in
\eqref{eq:odlro-indicator-fourier-transform}
factorizes into one-dimensional Dirichlet kernels:
\begin{equation}\label{eq:odlro-dirichlet-kernel-factorization}
\abs{\fun{\widehat{\chi}_{\Lambda_n}}{u}}
=
\prod_{i=1}^{d}
\abs{\fun{D_N}{u_i}}
=
\prod_{i=1}^{d}
\abs{\frac{\sin\rbk{2^{n-1}u_i}}
{\sin\frac{u_i}{2}}}.
\end{equation}
Let $\abs{u_i}_{\ast}$ denote the distance from $u_i$ to
$2\pi\ringratint$.
The definition
\eqref{eq:odlro-one-dimensional-dirichlet-kernel} and
$\abs{\sin\frac{u_i}{2}}
\geq
\frac{\abs{u_i}_{\ast}}{\pi}$
give
\begin{equation}\label{eq:odlro-dirichlet-kernel-pointwise-bound}
\abs{\fun{\widehat{\chi}_{\Lambda_n}}{u}}
\leq
\prod_{i=1}^{d}
\min\setone{
2^n,
\frac{\pi}{\abs{u_i}_{\ast}}
}.
\end{equation}

First take $\delta=0$.
For $0<\abs{p}<2^{-n}$, use
$\abs{\fun{\widehat{\chi}_{\Lambda_n}}{p}}
\leq\abscard{\Lambda_n}$.
The shell count
\eqref{eq:bec-momentum-shell-count}
gives
$$
\frac{1}{\abscard{\Lambda_m}}
\sum_{\substack{
p\in\dual{\Lambda_m}
\\
0<\abs{p}<2^{-n}
}}
\abs{p}^{-2}
\leq
\fun{C_1}{d}2^{-n\rbk{d-2}}.
$$
For $\abs{p}\geq2^{-n}$, use
$\abs{p}^{-2}\leq2^{2n}$ and discrete Parseval:
$\frac{1}{\abscard{\Lambda_m}}
\sum_{p\in\dual{\Lambda_m}}
\abs{\fun{\widehat{\chi}_{\Lambda_n}}{p}}^{2}
=
\abscard{\Lambda_n}$.
The contribution of this region is at most
$\frac{2^{2n}}{\abscard{\Lambda_n}}
=
2^{-n\rbk{d-2}}$.

Now let $\delta\in\mathcal{D}_d\setminus\setone{0}$.
Choose a coordinate with $\delta_i=\pi$.
Since $N=2^n$ is even,
\eqref{eq:odlro-one-dimensional-dirichlet-kernel} gives
$\fun{D_N}{\pi}=0$ and
$$\abs{\fun{D_N}{p_i+\pi}}
\leq
C N\abs{p_i},
\quad
p_i\in\closedinterval{-\pi}{\pi}.$$
Since $\abs{p}^{-2}\leq\abs{p_i}^{-2}$, with the value at $p_i=0$
understood by continuity,
$$\abs{p}^{-2}
\abs{\fun{D_N}{p_i+\pi}}^{2}
\leq
C N^{2}.$$
One-dimensional discrete Parseval gives
$$\frac{1}{2^m}
\sum_{p_j}
\abs{\fun{D_N}{p_j+\delta_j}}^{2}
=
N$$
for every $j\neq i$.
Factorization of the momentum sum therefore yields
$$\begin{aligned}
\frac{1}
{\abscard{\Lambda_m}
\abscard{\Lambda_n}^{2}}
\sum_{\substack{p\in\dual{\Lambda_m}
\\
p\neq0}}
\abs{p}^{-2}
\abs{\fun{\widehat{\chi}_{\Lambda_n}}{p+\delta}}^{2}
\leq
\frac{C N^2N^{d-1}}
{N^{2d}}
=
C N^{-\rbk{d-1}}
\leq
C N^{-\rbk{d-2}}.
\end{aligned}
$$
The set $\mathcal{D}_d$ has $2^d$ elements.
Replacing the maximum by their sum completes
\eqref{eq:odlro-dirichlet-kernel-infrared-bound}.
\end{proof}

\begin{prop}[ODLRO of the limit state]\label{prop:odlro-limit}
Let
$d
\geq
3$
and assume
$\kappa
>
0$,
where
$\kappa$
is defined by
\eqref{eq:main-condensation-constant}.
The limit state satisfies
\begin{equation}\label{eq:odlro-limit}
\liminf_{n \to \infty}
\frac{1}{\abscard{\Lambda_n}^{2}}
\sum_{x, y \in \Lambda_n}
\fun{\oastate[\psi_{\beta}]}{\faadj{a_{x}} a_{y}}
\geq
\kappa.
\end{equation}
\end{prop}

\begin{proof}
Fix $n\in\semigrposint$.
The Fourier transform of the indicator of $\Lambda_n$ is defined in
\eqref{eq:odlro-indicator-fourier-transform}.
Since this sum involves finitely many matrix elements, weak-$\ast$
convergence along the defining subnet gives
$$\begin{aligned}
\frac{1}{\abscard{\Lambda_n}^{2}}
\sum_{x,y\in\Lambda_n}
\fun{\oastate[\psi_{\beta}]}{\faadj{a_{x}} a_{y}}
=
\lim_{m\to\infty}
\frac{1}{\abscard{\Lambda_n}^{2}}
\sum_{x,y\in\Lambda_n}
\fun{\gamma_{\psi_{\beta,\Lambda_{m}}}}{x,y}.
\end{aligned}$$
The fixed index $n$ labels the averaging set, whereas $m$ labels the
finite-volume Gibbs state and tends to infinity.
It is therefore enough to obtain a lower bound uniform in $m$ and then let
$n$ tend to infinity.
Expand in the Fourier modes of the periodic box.
The definition
\eqref{eq:main-fourier-density-matrix}
gives
$$
\frac{1}{\abscard{\Lambda_n}^{2}}
\sum_{x,y
\in
\Lambda_n}
\fun{\gamma_{\psi_{\beta,\Lambda_{m}}}}{x,y}
=
\frac{1}{
\abscard{\Lambda_{m}}
\abscard{\Lambda_n}^{2}
}
\sum_{p,q
\in
\dual{\Lambda_{m}}}
\cmpconj{\fun{\widehat{\chi}_{\Lambda_n}}{p}}
\fun{\faftr{\gamma_{\psi_{\beta,\Lambda_{m}}}}}{p,q}
\fun{\widehat{\chi}_{\Lambda_n}}{q}.
$$
Lemma
\ref{lem:condensate-fourier-control}
shows that
$\faftr{\gamma_{\psi_{\beta,\Lambda_{m}}}}$
is positive semi-definite,
and
\eqref{eq:condensate-fourier-row-sparsity}
shows that its
$\rbk{p,q}$
entry vanishes unless
$q-p$ belongs to the set $\mathcal{D}$ of staggering shifts.
The set $\mathcal{D}$ contains at most $2^{d}$ vectors, whose components lie in
$\setone{0,\pi}$.
The diagonal estimate
\eqref{eq:condensate-nonzero-diagonal-bound}
gives
$$
\fun{\faftr{\gamma_{\psi_{\beta,\Lambda_{m}}}}}{p, p}
\leq
\fun{c_{\mathrm{F}}}{d,\lambda,\beta}
\abs{p}^{-2}
$$
for
$p
\neq
0$.
Theorem
\ref{thm:bec}
supplies the zero-mode bound
$$
\fun{\faftr{\gamma_{\psi_{\beta,\Lambda_{m}}}}}{0, 0}
\geq
\rbk{\kappa-\fun{o}{1}}
\abscard{\Lambda_{m}}.
$$
Split the double sum into the term $\rbk{p, q} = \rbk{0, 0}$, the cross terms with exactly one index zero, and the rest.
The $\rbk{0, 0}$ term equals $\fun{\faftr{\gamma_{\psi_{\beta,\Lambda_{m}}}}}{0, 0} / \abscard{\Lambda_{m}} \geq \kappa - \fun{o}{1}$, since $\fun{\widehat{\chi}_{\Lambda_n}}{0} = \abscard{\Lambda_n}$.
A cross term has $q=\delta\in\mathcal{D}\setminus\setone{0}$.
Positive semidefiniteness gives
$$\abs{\fun{\faftr{\gamma_{\psi_{\beta,\Lambda_{m}}}}}{0,\delta}}
\leq
\fun{\faftr{\gamma_{\psi_{\beta,\Lambda_{m}}}}}{0,0}^{1/2}
\fun{\faftr{\gamma_{\psi_{\beta,\Lambda_{m}}}}}{\delta,\delta}^{1/2}
\leq \fun{C}{\beta}\abscard{\Lambda_{m}}^{1/2}.$$
Moreover,
$\abs{\fun{\widehat{\chi}_{\Lambda_n}}{\delta}}\leq\abscard{\Lambda_n}$.
Each cross term is therefore
$\fun{O}{\abscard{\Lambda_{m}}^{-1/2}}$ and vanishes as $m\to\infty$.

For the terms with $p\neq0$ and $q=p+\delta\neq0$, positivity gives
$$\begin{aligned}
\abs{\fun{\faftr{\gamma_{\psi_{\beta,\Lambda_{m}}}}}{p,q}}
&\leq\frac{1}{2}\rbk{
\fun{\faftr{\gamma_{\psi_{\beta,\Lambda_{m}}}}}{p,p}
+\fun{\faftr{\gamma_{\psi_{\beta,\Lambda_{m}}}}}{q,q}},\\
\abs{\fun{\widehat{\chi}_{\Lambda_n}}{p}\fun{\widehat{\chi}_{\Lambda_n}}{q}}
&\leq\frac{1}{2}\rbk{
\abs{\fun{\widehat{\chi}_{\Lambda_n}}{p}}^{2}
+\abs{\fun{\widehat{\chi}_{\Lambda_n}}{q}}^{2}}.
\end{aligned}$$
Expand the products and relabel $p$ and $q$ within each shift class.
The total is bounded by
$$\frac{2^{d + 1}}{\abscard{\Lambda_{m}} \abscard{\Lambda_n}^{2}} \sum_{p \neq 0} \fun{\faftr{\gamma_{\psi_{\beta,\Lambda_{m}}}}}{p, p} \max_{\delta \in \mathcal{D} \cup \setone{0}} \abs{\fun{\widehat{\chi}_{\Lambda_n}}{p + \delta}}^{2}.$$
The diagonal estimate
\eqref{eq:condensate-nonzero-diagonal-bound}
and the Dirichlet-kernel estimate
\eqref{eq:odlro-dirichlet-kernel-infrared-bound}
bound the absolute value of these terms by
$\fun{C'}{d,\lambda,\beta}
2^{-n\rbk{d-2}}$.
Altogether
$$\frac{1}{\abscard{\Lambda_n}^{2}}
\sum_{x, y \in \Lambda_n}
\fun{\oastate[\psi_{\beta}]}{\faadj{a_{x}} a_{y}}
\geq
\kappa
-\frac{\fun{C'}{d, \lambda, \beta}}{2^{n(d-2)}}.$$
Letting $n\to\infty$ gives the claim because $d\geq3$.
\end{proof}

\subsection{Non-triviality of the decomposition in the condensed regime}\label{non-triviality-of-the-decomposition-in-the-condensed-regime}

The central decomposition separates a symmetric KMS limit state into states in \(\setextremal K_\beta\). Its nonzero long-range order is incompatible with extremality and therefore forces a nontrivial central measure in the condensed regime.

\begin{thm}[gauge symmetry breaking and non-trivial decomposition]\label{thm:main-decomposition}
Let
$d
\geq
3$,
$\beta
>
0$,
and
$\lambda
\geq
0$.
Assume that the constant
$\kappa$
defined in
\eqref{eq:main-condensation-constant}
is positive,
and let
$\oastate[\psi_{\beta}]$
be the symmetric weak-$\ast$ limit state constructed in Proposition
\ref{prop:limit-state}.
The following conclusions hold:
\begin{enumerate}
\item $\oa{M}_{\psi_{\beta}}$ is not a factor, and $\oastate[\psi_{\beta}]$ is not extremal in $K_{\beta}$;
\item the central measure $\mu_{\psi_{\beta}}$ of Theorem \ref{thm:kms-decomposition} is not a point mass, and the direct integral decomposition
$$\oastate[\psi_{\beta}] = \int_{\setextremal K_{\sminvtemperature}} \oastate[\psi'] \opdmsr{\mu_{\psi_{\beta}}(\oastate[\psi'])},
\quad
\sphilb{H}_{\psi_{\beta}} = \int_{X}^{\oplus} \sphilb{H}_{x} \opdmsr{\mu(x)},
\quad
\oarepn_{\psi_{\beta}} = \int_{X}^{\oplus} \oarepn_{x} \opdmsr{\mu(x)},$$
supported on $\setextremal K_{\sminvtemperature}$ is nontrivial;
\item $\mu_{\psi_{\beta}}$ is invariant under the push-forward of the gauge group, $\mu_{\psi_{\beta}} \circ \gamma_{\theta}^{\ast} = \mu_{\psi_{\beta}}$, and under the push-forward of the even translations.
\end{enumerate}
\end{thm}

\begin{proof}
For assertion (1), suppose $\oa{M}_{\psi_{\beta}}$ were a factor.
Consider the spatial averages
$\barmean{a}_{\Lambda_n}$
defined in
\eqref{eq:main-spatial-averages},
for which
$$\fun{\oastate[\psi_{\beta}]}{\faadj{\barmean{a}_{\Lambda_n}} \barmean{a}_{\Lambda_n}} = \frac{1}{\abscard{\Lambda_n}^{2}} \sum_{x, y \in \Lambda_n} \fun{\oastate[\psi_{\beta}]}{\faadj{a_{x}} a_{y}} = \norm{\fun{\oarepn_{\psi_{\beta}}}{\barmean{a}_{\Lambda_n}} \oagnsvector[\Psi_{\beta}]}^{2}.$$
The cube $\Lambda_n$ splits into the two parity classes represented by $0$ and
$e_{1}$.
Every $x\in \Lambda_n$ has a unique form $x=v$ or $x=v+e_{1}$.
The sets $\Lambda_n\cap\ringratint^{d}_{\mathrm{even}}$ and
$\rbk{\Lambda_n-e_1}\cap\ringratint^{d}_{\mathrm{even}}$ both have
cardinality $\abscard{\Lambda_n}/2$ and are Følner sequences for the even
translation group.
The parity decomposition gives
$$\begin{aligned}
\fun{\oarepn_{\psi_{\beta}}}{\barmean{a}_{\Lambda_n}}
\oagnsvector[\Psi_{\beta}]
=
\frac{1}{2}
\rbk{\frac{2}{\abscard{\Lambda_n}}
\sum_{v\in\Lambda_n\cap\ringratint^{d}_{\mathrm{even}}}
U_v\fun{\oarepn_{\psi_\beta}}{a_0}\oagnsvector[\Psi_\beta]
+
\frac{2}{\abscard{\Lambda_n}}
\sum_{v\in\rbk{\Lambda_n-e_1}\cap\ringratint^{d}_{\mathrm{even}}}
U_v\fun{\oarepn_{\psi_\beta}}{a_{e_1}}\oagnsvector[\Psi_\beta]}.
\end{aligned}$$
By Lemma \ref{lem:mean-ergodic} and Proposition \ref{prop:factor-clustering},
$$\fun{\oarepn_{\psi_{\beta}}}{\barmean{a}_{\Lambda_n}} \oagnsvector[\Psi_{\beta}] \to \frac{1}{2} \rbk{\fun{\oastate[\psi_{\beta}]}{a_{0}} + \fun{\oastate[\psi_{\beta}]}{a_{e_{1}}}} \oagnsvector[\Psi_{\beta}].$$
Gauge invariance gives, for every $\theta$,
$$\fun{\oastate[\psi_{\beta}]}{a_{x}}
=\napiernum^{-\imunit\theta}\fun{\oastate[\psi_{\beta}]}{a_{x}}.$$
This identity for every $\theta$ forces
$\fun{\oastate[\psi_{\beta}]}{a_{x}}=0$.
The preceding strong limit would imply
$\fun{\oastate[\psi_{\beta}]}{\faadj{\barmean{a}_{\Lambda_n}}\barmean{a}_{\Lambda_n}}\to0$.
This contradicts Proposition \ref{prop:odlro-limit}.
The contradiction proves $\oa{Z}_{\psi_{\beta}}\neq\fldcmp$.
Assertion (1) of Theorem \ref{thm:kms-decomposition} now shows that
$\oastate[\psi_{\beta}]$ is not extremal in $K_{\beta}$.

For assertion (2), a point-mass central measure is equivalent to a trivial
center.
Assertion (1) excludes this case.
Assertions (3) and (4) of Theorem \ref{thm:kms-decomposition} supply the nontrivial direct
integral supported on $\setextremal K_\beta$.

Assertion (3) follows by applying Proposition
\ref{prop:decomposition-covariance} to
$\alpha=\gamma_{\theta}$ and $\alpha=\eta_{v}$.
Proposition \ref{prop:limit-state} gives invariance of
$\oastate[\psi_{\beta}]$ under both automorphisms.
Proposition \ref{prop:symmetry-dynamics} shows that they commute with $\tau$.
\end{proof}

\subsection{Gauge symmetry breaking in the components}\label{gauge-symmetry-breaking-in-the-components}

Theorem \ref{thm:main-decomposition} locates \(\oastate[\psi_{\beta}]\) off the extremal boundary but says nothing yet about the individual components \(\oastate[\psi_{x}]\) of its central decomposition. The next theorem extracts the central condensate average \(\barmean{a}\) from the box averages. Its representing function is the componentwise condensate order parameter. It is nonzero on a set of measure at least \(\kappa\). Every component in that set breaks the gauge symmetry with trivial stabilizer.

\begin{thm}[componentwise gauge symmetry breaking]\label{thm:componentwise-breaking}
Assume the hypotheses of Theorem \ref{thm:main-decomposition}, and let $\rbk{X, \mu}$ and $\oastate[\psi_{x}]$ be the direct integral data of assertion (4) of Theorem \ref{thm:kms-decomposition} for $\oastate[\psi_{\beta}]$.
The spatial averages have the weak-operator limit
$$\wlim_{n \to \infty}
\fun{\oarepn_{\psi_{\beta}}}{\barmean{a}_{\Lambda_n}}
=
\barmean{a},$$
where
$\barmean{a}
\in
\oa{Z}_{\psi_{\beta}}$
and
$\norm{\barmean{a}}
\leq
1$.
The element
$\barmean{a}$
is independent of any choice of subnet.
Under the central identification in assertion (4) of Theorem
\ref{thm:kms-decomposition}, the same notation $\barmean{a}$ denotes its
representing function.
\begin{enumerate}
\item The implementing unitaries satisfy
$$V_{\theta}\barmean{a}\faadj{V_{\theta}}
=\napiernum^{-\imunit\theta}\barmean{a}.$$
Moreover, the functions
$$x\mapsto\fun{\oastate[\psi_x]}{\barmean{a}_{\Lambda_n}}
=\frac{1}{\abscard{\Lambda_n}}\sum_{y\in \Lambda_n}
\fun{\oastate[\psi_x]}{a_y}$$
converge weak-$\ast$ in $\fun{\lp^{\infty}}{X,\mu}$ to $\barmean{a}$.
\item $\int_X\abs{\barmean{a}(x)}^2\opdmsr{\mu(x)}\geq\kappa$.
\item Define
\begin{equation}\label{eq:componentwise-breaking-nonzero-set}
E
=
\set{x\in X}{\barmean{a}(x)\neq0}.
\end{equation}
For $\mu$-almost every $x\in E$, some $y\in\ringratint^{d}$ satisfies
$\fun{\oastate[\psi_{x}]}{a_{y}}\neq0$.
For such $x$,
\begin{equation}\label{eq:componentwise-gauge-breaking}
\theta\notin2\pi\ringratint
\implies
\oastate[\psi_{x}]\circ\gamma_{\theta}
\neq\oastate[\psi_{x}].
\end{equation}
The set $E$ satisfies $\fun{\mu}{E}\geq\kappa$.
\end{enumerate}
The central decomposition therefore assigns weight at least $\kappa$ to extremal
KMS states with nonzero order parameter.
Each such state has trivial gauge stabilizer.
Its gauge orbit is a circle in $\setextremal K_\beta$ whose states are mutually
disjoint.
The measure $\mu_{\psi_{\beta}}$ is invariant along these orbits.
\end{thm}

\begin{proof}
The operators $\barmean{a}_{\Lambda_n}$ are the spatial averages from
\eqref{eq:main-spatial-averages}.

The weak limit of the spatial averages is central.
Split $\Lambda_n$ into its two parity classes and apply
Lemma \ref{lem:mean-ergodic}.
The parity decomposition and Lemma
\ref{lem:mean-ergodic}
give
$$\fun{\oarepn_{\psi_{\beta}}}{\barmean{a}_{\Lambda_n}}
\oagnsvector[\Psi_{\beta}]
\to
\xi
=
\frac{1}{2}P_{0}
\rbk{\fun{\oarepn_{\psi_{\beta}}}{a_{0}}+\fun{\oarepn_{\psi_{\beta}}}{a_{e_{1}}}}
\oagnsvector[\Psi_{\beta}].$$
Proposition \ref{prop:odlro-limit} gives
$$\norm{\xi}^{2}
=\lim_{n \to \infty}
\fun{\oastate[\psi_{\beta}]}{\faadj{\barmean{a}_{\Lambda_n}}\barmean{a}_{\Lambda_n}}
\geq\kappa.$$
Since $\norm{\fun{\oarepn_{\psi_{\beta}}}{\barmean{a}_{\Lambda_n}}}\leq1$, the net has weak-operator limit points.
Let $\barmean{a}$ be one of them.
Fix a local observable $B'$.
Only sites of $\Lambda_n$ near the support of $B'$ contribute to the commutator.
Strict locality gives
$$\norm{\commutator{\fun{\oarepn_{\psi_{\beta}}}{\barmean{a}_{\Lambda_n}}}{\fun{\oarepn_{\psi_{\beta}}}{B'}}}
=\fun{O}{\abscard{\Lambda_n}^{-1}}\to0.$$
Commutation with a fixed bounded operator passes to weak-operator limits.
Consequently,
$$\commutator{\barmean{a}}{\fun{\oarepn_{\psi_{\beta}}}{B'}}=0.$$
Density of the local algebra and uniform boundedness imply
$\barmean{a}\in\oacommutant{\fun{\oarepn_{\psi_{\beta}}}{\oa{A}}}$.
Every $\fun{\oarepn_{\psi_{\beta}}}{\barmean{a}_{\Lambda_n}}$ belongs to the weakly closed algebra
$\oa{M}_{\psi_{\beta}}$.
Weak closedness gives $\barmean{a}\in\oa{M}_{\psi_{\beta}}$.
Together with $\barmean{a}\in\oacommutant{\fun{\oarepn_{\psi_{\beta}}}{\oa{A}}}$, this gives
$\barmean{a}\in\oa{Z}_{\psi_{\beta}}$ and $\norm{\barmean{a}}\leq1$.
Along the chosen subnet,
$\fun{\oarepn_{\psi_{\beta}}}{\barmean{a}_{\Lambda_n}}\oagnsvector[\Psi_{\beta}]$
converges weakly to $\barmean{a}\oagnsvector[\Psi_{\beta}]$.
The full net converges strongly to $\xi$.
The two limits give
$$\barmean{a}\oagnsvector[\Psi_{\beta}]
=
\xi,
\quad
\bkt{\oagnsvector[\Psi_{\beta}]}{\faadj{\barmean{a}}\barmean{a}\oagnsvector[\Psi_{\beta}]}
=
\norm{\xi}^{2}
\geq\kappa.$$
Lemma \ref{lem:separating} states that $\oagnsvector[\Psi_{\beta}]$ is separating for
$\oa{M}_{\psi_{\beta}}$.
An element of $\oa{Z}_{\psi_{\beta}}\subset\oa{M}_{\psi_{\beta}}$ is therefore determined by
its value on $\oagnsvector[\Psi_{\beta}]$.
Every weak-operator limit point of
$\fun{\oarepn_{\psi_{\beta}}}{\barmean{a}_{\Lambda_n}}$
equals $\barmean{a}$.
The bounded net therefore satisfies
$$\wlim_{n \to \infty}
\fun{\oarepn_{\psi_{\beta}}}{\barmean{a}_{\Lambda_n}}
=
\barmean{a}.$$
This limit involves no choice of subnet.

The central limit has gauge charge $-1$.
Proposition \ref{prop:limit-state} gives gauge invariance of
$\oastate[\psi_{\beta}]$.
It provides implementing unitaries $V_{\theta}$ satisfying
$$\begin{aligned}
V_{\theta}\fun{\oarepn_{\psi_{\beta}}}{A}\faadj{V_{\theta}}
=
\fun{\oarepn_{\psi_{\beta}}}{\fun{\gamma_{\theta}}{A}},
\quad
V_{\theta}\oagnsvector[\Psi_{\beta}]
=
\oagnsvector[\Psi_{\beta}].
\end{aligned}$$
Equation \eqref{eq:gauge-automorphism} gives
$$\fun{\gamma_{\theta}}{\barmean{a}_{\Lambda_n}}
=\napiernum^{-\imunit\theta}\barmean{a}_{\Lambda_n}.$$
Consequently,
$$V_{\theta}\fun{\oarepn_{\psi_{\beta}}}{\barmean{a}_{\Lambda_n}}\faadj{V_{\theta}}
=
\napiernum^{-\imunit\theta}\fun{\oarepn_{\psi_{\beta}}}{\barmean{a}_{\Lambda_n}}.$$
Conjugation by a fixed unitary preserves weak-operator convergence.
Passing to the limit gives
$$V_{\theta}\barmean{a}\faadj{V_{\theta}}
=\napiernum^{-\imunit\theta}\barmean{a}.$$

The central limit is represented by a measurable fiber function.
Assertion (4) of Theorem \ref{thm:kms-decomposition} identifies $\oa{Z}_{\psi_{\beta}}$ with
$\fun{\lp^{\infty}}{X,\mu}$.
Under this identification, the same notation $\barmean{a}$ denotes the
representing function.
For
$A
\in
\oa{A}$
and
$D\in\oa{Z}_{\psi_{\beta}}$,
let $f_D$ denote the fiber function of $D$.
The general central integral identity
\eqref{eq:orthogonal-measure-central-identity} applies to these data.

Take $D=\faadj{\barmean{a}}\barmean{a}$ and $A=1$.
The fiber function of $D$ is $\abs{\barmean{a}}^{2}$, and the identity gives
$$\int_X\abs{\barmean{a}(x)}^{2}\opdmsr{\mu(x)}
=
\bkt{\oagnsvector[\Psi_{\beta}]}{\faadj{\barmean{a}}\barmean{a}\oagnsvector[\Psi_{\beta}]}
\geq\kappa.$$
This proves assertion (2).

Average the same identity over $y\in \Lambda_n$:
$$\begin{aligned}
\bkt{\oagnsvector[\Psi_{\beta}]}{D\fun{\oarepn_{\psi_{\beta}}}{\barmean{a}_{\Lambda_n}}\oagnsvector[\Psi_{\beta}]}
&=\int_{X}\fun{f_D}{x}\fun{\oastate[\psi_x]}{\barmean{a}_{\Lambda_n}}\opdmsr{\mu(x)},\\
\fun{\oastate[\psi_x]}{\barmean{a}_{\Lambda_n}}
&=\frac{1}{\abscard{\Lambda_n}}
\sum_{y\in \Lambda_n}\fun{\oastate[\psi_{x}]}{a_{y}}.
\end{aligned}$$
The left side converges to
$$\bkt{\oagnsvector[\Psi_{\beta}]}{D\barmean{a}\oagnsvector[\Psi_{\beta}]}
=\int_{X}\fun{f_D}{x}\barmean{a}(x)\opdmsr{\mu(x)}.$$
Since $\abs{\fun{\oastate[\psi_x]}{\barmean{a}_{\Lambda_n}}}\leq1$ and $\mu$ is finite, this convergence against every
$f_D\in\fun{\lp^{\infty}}{X,\mu}$ is weak-$\ast$ convergence.
This is the convergence asserted in assertion (1).

Vanishing order parameters do not support $\barmean{a}$.
Let $S = \set{x \in X}{\fun{\oastate[\psi_{x}]}{a_{y}} = 0 \text{ for all } y \in \ringratint^{d}}$, a countable intersection of measurable sets.
Choose $D$ with fiber function $\cmpconj{\barmean{a}}\fndef{S}$.
The function $x\mapsto\fun{\oastate[\psi_x]}{\barmean{a}_{\Lambda_n}}$ vanishes on $S$, so
$$\bkt{\oagnsvector[\Psi_{\beta}]}{D\fun{\oarepn_{\psi_{\beta}}}{\barmean{a}_{\Lambda_n}}\oagnsvector[\Psi_{\beta}]}
=\int_S\cmpconj{\barmean{a}(x)}
\fun{\oastate[\psi_x]}{\barmean{a}_{\Lambda_n}}\opdmsr{\mu(x)}
=0.$$
The limit $n\to\infty$ gives
$$\int_{S}\abs{\barmean{a}(x)}^{2}\opdmsr{\mu(x)}=0.$$
It follows that $\fun{\mu}{E\cap S}=0$.

The bound $\abs{\barmean{a}(x)}\leq\norm{\barmean{a}}\leq1$ controls the measure of the nonzero
fibers:
$$\fun{\mu}{E}
\geq\int_{E}\abs{\barmean{a}(x)}^{2}\opdmsr{\mu(x)}
=\int_{X}\abs{\barmean{a}(x)}^{2}\opdmsr{\mu(x)}
\geq\kappa.$$
Almost every $x\in E$ lies outside $S$.
For such an $x$, there exists $y$ with
$\fun{\oastate[\psi_{x}]}{a_{y}}\neq0$.
If $\theta\notin2\pi\ringratint$, then
$$\fun{\oastate[\psi_{x}]\circ\gamma_{\theta}}{a_{y}}
=\napiernum^{-\imunit\theta}\fun{\oastate[\psi_{x}]}{a_{y}}
\neq\fun{\oastate[\psi_{x}]}{a_{y}}.$$
This proves assertion (3).

Assertion (4) of Theorem \ref{thm:kms-decomposition} makes almost every
$\oastate[\psi_{x}]$ an extremal KMS state.
The push-forward by $\gamma_{\theta}$ is an affine homeomorphism of
$K_{\beta}$.
Its inverse is also affine, so it preserves extremality.
The gauge orbit therefore lies in $\setextremal K_\beta$.
Triviality of the stabilizer makes its points distinct.
The disjointness conclusion for distinct elements of $\setextremal K_\beta$ is
Theorem \ref{thm:extremal-kms-disjoint}.
Assertion (3) of Theorem \ref{thm:main-decomposition} gives invariance of
$\mu_{\psi_{\beta}}$ along the orbit.
\end{proof}

Here disjointness is a representation-theoretic statement stronger than non-unitary equivalence: two representations are disjoint when they have no nonzero unitarily equivalent subrepresentations. The positive-temperature phase circle is therefore not merely a family of different expectation functionals. Its points belong to mutually disjoint thermodynamic representations, and a broken gauge rotation has no unitary implementation within one of these phase representations, whereas the ground-state construction of Theorem \ref{thm:phase-localized-ground-states} gives only non-quasi-equivalence until factoriality or ergodicity of the limit states is proved.

\subsection{The distribution of the order parameter}\label{the-distribution-of-the-order-parameter}

The central weak-operator limit and, under the central identification, the componentwise order parameter \(\barmean{a}\) from Theorem \ref{thm:componentwise-breaking} control more than the second moment of the spatial averages. All polynomial moments of the order parameter converge. The \(\fun{\liealg{su}}{2}\) structure gives a sharp bound on the box, operators. Together, these facts determine the geometry of the resulting distribution.

\begin{lem}[Casimir bound for the spatial averages]\label{lem:casimir-bound}
For every box, $\Lambda_n$, the following operator inequality holds in
$\oa{A}$:
\begin{equation}\label{eq:casimir-box,-operator-bound}
\faadj{\barmean{a}_{\Lambda_n}}\barmean{a}_{\Lambda_n}
\leq\frac{1}{4}+\frac{1}{\abscard{\Lambda_n}}.
\end{equation}
Consequently,
\begin{equation}\label{eq:casimir-box,-norm-bound}
\norm{\barmean{a}_{\Lambda_n}}
\leq\rbk{\frac{1}{4}+\frac{1}{\abscard{\Lambda_n}}}^{1/2}.
\end{equation}
\end{lem}

\begin{proof}
Let $S_{W} = \sum_{x \in \Lambda_n} S_{x}$ be the total spin of the box, so that $\faadj{\barmean{a}_{\Lambda_n}} \barmean{a}_{\Lambda_n} = \abscard{\Lambda_n}^{-2} S^{+}_{W} S^{-}_{W}$.
The total-spin calculation
\eqref{eq:app-total-spin-ladder}
identifies
$S^{+}_{W}S^{-}_{W}$.
The tensor product of $\abscard{\Lambda_n}$ spin-$\frac{1}{2}$ representations
decomposes into irreducible representations of total spin
$s\leq\abscard{\Lambda_n}/2$.
It follows that
$$S^{+}_{W} S^{-}_{W} \leq \frac{\abscard{\Lambda_n}}{2} \rbk{\frac{\abscard{\Lambda_n}}{2} + 1} + \frac{\abscard{\Lambda_n}}{2}
= \frac{\abscard{\Lambda_n}^{2}}{4} + \abscard{\Lambda_n}.$$
We dropped the nonpositive term $-\rbk{S^{3}_{W}}^{2}$ and used
$S^{3}_{W}\leq\abscard{\Lambda_n}/2$.
Division by $\abscard{\Lambda_n}^{2}$ proves the claim.
\end{proof}

\begin{thm}[the order-parameter distribution]\label{thm:order-parameter-distribution}
Assume the hypotheses of Theorem \ref{thm:componentwise-breaking}. Let
$\barmean{a}$ and $E$ be as constructed there.
The order-parameter law is defined by
\begin{equation}\label{eq:order-parameter-law-definition}
\nu
=
\barmean{a}_{\ast}\mu.
\end{equation}
The following statements hold:
\begin{enumerate}
\item Let $w$ be a word in $\barmean{a}_{\Lambda_n}$ and $\faadj{\barmean{a}_{\Lambda_n}}$ with $k$ starred
and $m$ unstarred letters.
The unrestricted limit exists and satisfies
\begin{equation}\label{eq:order-parameter-moment-limit}
\lim_{n \to \infty}\fun{\oastate[\psi_{\beta}]}{w}
=\int_{\fldcmp}\cmpconj{z}^{k}z^{m}\opdmsr{\nu(z)}.
\end{equation}
This limit vanishes unless $k=m$;
\item The measure $\nu$ is a Borel probability measure on the closed disc of
radius $\frac{1}{2}$ in $\fldcmp$, and its rotation invariance is
\begin{equation}\label{eq:order-parameter-rotation-invariance}
\rbk{z\mapsto\napiernum^{-\imunit\theta}z}_{\ast}\nu
=
\nu,
\quad
\theta
\in
\fldreal.
\end{equation}
It has no atom other than possibly at the origin and is uniquely determined
by the moments in assertion (1);
\item The second moment and the mass at the origin satisfy
\begin{equation}\label{eq:order-parameter-second-moment-and-atom-bound}
\kappa
\leq
\int_{\fldcmp} \abs{z}^{2} \opdmsr{\nu(z)}
\leq
\frac{1}{4},
\quad
\fun{\nu}{\setone{0}}
=
1 - \fun{\mu}{E}
\leq
1 - 4 \kappa.
\end{equation}
In particular, the weight bound in assertion (3) of Theorem
\ref{thm:componentwise-breaking} improves to
$\fun{\mu}{E} \geq 4 \kappa$;
\item The unitaries implementing the even translations satisfy
$$\widetilde{U}_{v}\barmean{a}\faadj{\widetilde{U}_{v}}=\barmean{a}.$$
It follows that $\barmean{a}$ is fixed by the induced measure-preserving automorphism of
$\fun{\lp^{\infty}}{X,\mu}$.
The order parameter does not distinguish a component from its lattice
translates.
\end{enumerate}
\end{thm}

\begin{proof}
The norm estimate \eqref{eq:casimir-box,-norm-bound} gives
$$\norm{\fun{\oarepn_{\psi_{\beta}}}{\barmean{a}_{\Lambda_n}}}
\leq\rbk{\frac{1}{4}+\abscard{\Lambda_n}^{-1}}^{1/2}.$$
Weak-operator lower semicontinuity of the norm gives
$$\norm{\barmean{a}}
\leq
\liminf_{n}\norm{\fun{\oarepn_{\psi_{\beta}}}{\barmean{a}_{\Lambda_n}}}
\leq
\frac{1}{2}.$$
This sharpens the estimate
$\norm{\barmean{a}}
\leq
1$
in Theorem \ref{thm:componentwise-breaking}.
Consequently, $\abs{\barmean{a}(x)}\leq\frac{1}{2}$ almost everywhere, and $\nu$ is
supported in the closed disc of radius $\frac{1}{2}$.

The moment convergence in assertion (1) starts from the strong convergence
$$\fun{\oarepn_{\psi_{\beta}}}{\barmean{a}_{\Lambda_n}}\oagnsvector[\Psi_{\beta}]
\to
\barmean{a}\oagnsvector[\Psi_{\beta}]$$
along the full net, proved in Theorem
\ref{thm:componentwise-breaking}.
For the adjoints,
$$\faadj{\fun{\oarepn_{\psi_{\beta}}}{\barmean{a}_{\Lambda_n}}}\oagnsvector[\Psi_{\beta}]
=\frac{1}{\abscard{\Lambda_n}}
\sum_{y}\fun{\oarepn_{\psi_{\beta}}}{\faadj{a_{y}}}\oagnsvector[\Psi_{\beta}].$$
Split this sum into the two parity classes.
Lemma \ref{lem:mean-ergodic} gives the strong limit
$$\frac{1}{2}P_{0}
\rbk{\fun{\oarepn_{\psi_{\beta}}}{\faadj{a_{0}}}
+\fun{\oarepn_{\psi_{\beta}}}{\faadj{a_{e_{1}}}}}\oagnsvector[\Psi_{\beta}].$$
On the other hand, the weak-operator convergence
$\fun{\oarepn_{\psi_{\beta}}}{\barmean{a}_{\Lambda_n}}\to \barmean{a}$ gives weak convergence of
$\faadj{\fun{\oarepn_{\psi_{\beta}}}{\barmean{a}_{\Lambda_n}}}\oagnsvector[\Psi_{\beta}]$
to
$\faadj{\barmean{a}}\oagnsvector[\Psi_{\beta}]$.
The strong limit must therefore equal $\faadj{\barmean{a}}\oagnsvector[\Psi_{\beta}]$.

For a word $w$, let $\fun{w}{\barmean{a}}$ be the word obtained by replacing
$\barmean{a}_{\Lambda_n}$ and $\faadj{\barmean{a}_{\Lambda_n}}$ with $\barmean{a}$ and $\faadj{\barmean{a}}$.
Since $\oa{Z}_{\psi_{\beta}}$ is abelian,
we obtain
$\fun{w}{\barmean{a}}
=
\rbk{\faadj{\barmean{a}}}^{k}\barmean{a}^{m}$.
We prove by induction on the word length that
$$\fun{\oarepn_{\psi_{\beta}}}{w}\oagnsvector[\Psi_{\beta}]
\to
\fun{w}{\barmean{a}}\oagnsvector[\Psi_{\beta}]$$
in norm.
Write $w=aw'$ with
$a\in\setone{\barmean{a}_{\Lambda_n},\faadj{\barmean{a}_{\Lambda_n}}}$ and
The corresponding represented vector is
$$\fun{\oarepn_{\psi_{\beta}}}{w}\oagnsvector[\Psi_{\beta}]
=
\fun{\oarepn_{\psi_{\beta}}}{a}
\rbk{\fun{\oarepn_{\psi_{\beta}}}{w'}\oagnsvector[\Psi_{\beta}]
-\fun{w'}{\barmean{a}}\oagnsvector[\Psi_{\beta}]}
+\fun{w'}{\barmean{a}}\fun{\oarepn_{\psi_{\beta}}}{a}\oagnsvector[\Psi_{\beta}].$$
The central operator $\fun{w'}{\barmean{a}}$ commutes with
$\fun{\oarepn_{\psi_{\beta}}}{a}$.
The first term tends to zero by the induction hypothesis and
$\norm{\fun{\oarepn_{\psi_{\beta}}}{a}}\leq1$.
The second tends to either
$\fun{w'}{\barmean{a}}\barmean{a}\oagnsvector[\Psi_{\beta}]$ or
$\fun{w'}{\barmean{a}}\faadj{\barmean{a}}\oagnsvector[\Psi_{\beta}]$.
Commutativity identifies that limit with $\fun{w}{\barmean{a}}\oagnsvector[\Psi_{\beta}]$.

Taking the scalar product with $\oagnsvector[\Psi_{\beta}]$ gives
$$\fun{\oastate[\psi_{\beta}]}{w}
\to
\bkt{\oagnsvector[\Psi_{\beta}]}{\rbk{\faadj{\barmean{a}}}^{k}\barmean{a}^{m}\oagnsvector[\Psi_{\beta}]}.$$
The fiber function of $\rbk{\faadj{\barmean{a}}}^{k}\barmean{a}^{m}$ is
$x\mapsto\cmpconj{\barmean{a}(x)}^{k}\barmean{a}(x)^{m}$.
The orthogonal-measure identity
\eqref{eq:orthogonal-measure-central-identity}
with $A=1$ evaluates the limit:
$$\int_{X}\cmpconj{\barmean{a}(x)}^{k}\barmean{a}(x)^{m}\opdmsr{\mu(x)}
=\int_{\fldcmp}\cmpconj{z}^{k}z^{m}\opdmsr{\nu(z)}.$$

Suppose $k\neq m$.
Assertion (1) of Theorem \ref{thm:componentwise-breaking} gives
$$V_{\theta}\rbk{\faadj{\barmean{a}}}^{k}\barmean{a}^{m}\faadj{V_{\theta}}
=\napiernum^{\imunit\theta\rbk{k-m}}\rbk{\faadj{\barmean{a}}}^{k}\barmean{a}^{m}.$$
Since $V_{\theta}\oagnsvector[\Psi_{\beta}]=\oagnsvector[\Psi_{\beta}]$, the corresponding expectation equals
$\napiernum^{\imunit\theta\rbk{k-m}}$ times itself for every $\theta$.
It must vanish.

For assertion (2), the map
$\alpha_{\theta}=\fun{\Ad}{V_{\theta}}$ restricts to a
$\ast$-automorphism of $\oa{Z}_{\psi_{\beta}}$.
For each $D\in\oa{Z}_{\psi_{\beta}}$, let $f_D$ denote its fiber function.
It preserves the state
$$D\mapsto\bkt{\oagnsvector[\Psi_{\beta}]}{D\oagnsvector[\Psi_{\beta}]}
=\int_{X}\fun{f_D}{x}\opdmsr{\mu(x)}$$
because $V_{\theta}\oagnsvector[\Psi_{\beta}]=\oagnsvector[\Psi_{\beta}]$.
It maps $\barmean{a}$ to $\napiernum^{-\imunit\theta}\barmean{a}$.
Let $f$ be continuous on the disc.
Continuous functional calculus for the normal element $\barmean{a}$ commutes with
$\ast$-automorphisms.
State invariance gives
$$\begin{aligned}
&\int_{\fldcmp}\fun{f}{z}\opdmsr{\nu(z)}
=
\bkt{\oagnsvector[\Psi_{\beta}]}{\fun{f}{\barmean{a}}\oagnsvector[\Psi_{\beta}]}
=
\bkt{\oagnsvector[\Psi_{\beta}]}{\fun{f}{\fun{\alpha_{\theta}}{\barmean{a}}}\oagnsvector[\Psi_{\beta}]}
\\ 
&=
\bkt{\oagnsvector[\Psi_{\beta}]}{\fun{f}{\napiernum^{-\imunit \theta}\barmean{a}}\oagnsvector[\Psi_{\beta}]}
=
\int_{\fldcmp}\fun{f}{\napiernum^{-\imunit \theta}z}\opdmsr{\nu(z)}.
\end{aligned}$$
This is the rotation invariance of $\nu$.
If $z_{0}\neq0$ were an atom, rotation invariance would give atoms of equal
mass at every point of the circle $\abs{z}=\abs{z_{0}}$.
A probability measure cannot contain uncountably many disjoint atoms of the
same positive mass.
It follows that $\nu$ has no atom outside the origin.

Polynomials in $z$ and $\cmpconj{z}$ form a unital, point-separating algebra
that is closed under conjugation.
The Stone--Weierstrass theorem makes this algebra dense in the continuous
functions on the compact disc.
The moments in assertion (1) therefore determine $\nu$.

Assertion (3) follows from the moment bound
$$\int_{\fldcmp}\abs{z}^{2}\opdmsr{\nu(z)}
=\int_{X}\abs{\barmean{a}(x)}^{2}\opdmsr{\mu(x)}
\geq\kappa.$$
The support bound gives the upper estimate $\frac{1}{4}$.
Moreover,
$$\fun{\nu}{\setone{0}}
=\fun{\mu}{\set{x}{\barmean{a}(x)=0}}
=1-\fun{\mu}{E}.$$
Since $\abs{\barmean{a}(x)}\leq\frac{1}{2}$,
$$\frac{1}{4}\fun{\mu}{E}
\geq\int_{E}\abs{\barmean{a}(x)}^{2}\opdmsr{\mu(x)}
\geq\kappa.$$
This gives $\fun{\mu}{E}\geq4\kappa$.

Assertion (4) follows from the invariance under $\eta_{v}$ in Proposition
\ref{prop:limit-state}.
The implementing unitaries satisfy
$$\begin{aligned}
\widetilde{U}_{v}\fun{\oarepn_{\psi_{\beta}}}{A}\faadj{\widetilde{U}_{v}}
=
\fun{\oarepn_{\psi_{\beta}}}{\fun{\eta_{v}}{A}},
\quad
\widetilde{U}_{v}\oagnsvector[\Psi_{\beta}]
=
\oagnsvector[\Psi_{\beta}].
\end{aligned}$$
The translated spatial average $\fun{\eta_{v}}{\barmean{a}_{\Lambda_n}}$ differs from $\barmean{a}_{\Lambda_n}$ only by the boundary discrepancy of the boxes:
$$\norm{\fun{\eta_{v}}{\barmean{a}_{\Lambda_n}} - \barmean{a}_{\Lambda_n}}
\leq \frac{\abscard{\Lambda_n\mathbin{\triangle}\rbk{\Lambda_n+v}}}
{\abscard{\Lambda_n}}
\to0$$
as $n\to\infty$ with $v$ fixed.
The represented translated averages therefore satisfy
$$\wlim_{n \to \infty}
\fun{\oarepn_{\psi_{\beta}}}
{\fun{\eta_{v}}{\barmean{a}_{\Lambda_n}}}
=
\widetilde{U}_{v} \barmean{a} \faadj{\widetilde{U}_{v}}
=
\barmean{a}.$$
Conjugation by $\widetilde{U}_{v}$ preserves both $\oa{Z}_{\psi_{\beta}}$ and the vector
state $\bkt{\oagnsvector[\Psi_{\beta}]}{\cdot\oagnsvector[\Psi_{\beta}]}$.
It induces a measure-preserving automorphism of
$\fun{\lp^{\infty}}{X,\mu}$ that fixes $\barmean{a}$.
\end{proof}

\begin{rem}[what the distribution theorem settles]\label{rem:distribution-picture}
Theorem \ref{thm:order-parameter-distribution} proves the angular part of the
expected orbit picture.
The law of the condensate order parameter is rotation invariant and has no
atoms outside the origin.
It assigns mass at least $4\kappa$ to the complement of the origin.
The limiting zero-mode moments determine it completely.

Remark \ref{rem:orbit-structure} predicts that $\nu$ is uniform on one circle
$\abs{z}=\sqrt{\varrho_{0}}$.
Rotation invariance reduces this prediction to a radial statement.
The remaining radial statement is equivalent to one fourth-moment
factorization estimate.

The homeomorphisms $\gamma_{\theta}^{\ast}$ preserve
$\mu_{\psi_{\beta}}$ and hence its support.
The support is a union of full gauge orbits.
The next subsection disintegrates the measure along these orbits.
\end{rem}

\subsection{The orbit decomposition of the central measure}\label{the-orbit-decomposition-of-the-central-measure}

The gauge action on \(K_{\beta}\) is jointly continuous, so the central measure can be disintegrated over the space of gauge orbits. Combined with Theorem \ref{thm:componentwise-breaking}, this produces the Haar-orbit structure fiberwise and isolates the remaining open question as a single ergodicity statement.

\begin{lem}[disintegration over the orbit space]\label{lem:orbit-disintegration}
The map
\begin{equation}\label{eq:orbit-gauge-action}
\rbk{\napiernum^{\imunit\theta},\oastate[\psi']}
\in
\fun{\liegr{U}}{1}\times K_{\beta}
\mapsto
\oastate[\psi']\circ\gamma_{\theta}
\in
K_{\beta}
\end{equation}
defines a jointly continuous action of $\fun{\liegr{U}}{1}$ on
$K_{\beta}$.
Define the orbit space $Y$ as $Y
= K_{\beta}/\fun{\liegr{U}}{1}$,
and then the orbit data $\pairbk{Y,q,\msr{\mu_{\mathrm{orb}}}}$ are defined by
\begin{equation}\label{eq:orbit-quotient-data}
\begin{aligned}
q
\colon
K_{\beta}
\to
Y;
\quad
\msr{\mu_{\mathrm{orb}}}
=
q_{\ast}\mu_{\psi_{\beta}}.
\end{aligned}
\end{equation}
The orbit space $Y$ is compact and metrizable.
Disintegration gives a family $y\mapsto\mu_{y}$ of Borel probability
measures on $K_{\beta}$.
For every continuous $f$, the function
$y\mapsto\fun{\mu_{y}}{f}$ is $\msr{\mu_{\mathrm{orb}}}$-measurable.
The family is unique up to $\msr{\mu_{\mathrm{orb}}}$-null sets and satisfies
\begin{equation}\label{eq:orbit-disintegration-family}
\fun{\mu_{y}}{\fun{q^{-1}}{y}} = 1,
\quad
\mu_{\psi_{\beta}} = \int_{Y} \mu_{y} \opdmsr{\mu_{\mathrm{orb}}(y)}.
\end{equation}
For $\msr{\mu_{\mathrm{orb}}}$-almost every $y$, the measure $\mu_{y}$ is the unique
gauge-invariant probability measure on $\fun{q^{-1}}{y}$.
Equivalently, it is the image of normalized Haar measure under
$$\theta\mapsto\oastate[\psi']\circ\gamma_{\theta}$$
for any $\oastate[\psi']\in\fun{q^{-1}}{y}$.
\end{lem}

\begin{proof}
The quasi-local algebra $\oa{A}$ is unital and separable by
Section \ref{sec:algebra}.
Proposition \ref{prop:kms-compact-convex} makes $K_{\beta}$ compact and
metrizable.

For local $A$, the map $\theta\mapsto\fun{\gamma_{\theta}}{A}$ is
norm-continuous by \eqref{eq:gauge-automorphism}.
Density of the local algebra and $\norm{\gamma_{\theta}}=1$ extend the
continuity to every $A\in\oa{A}$.
Moreover,
$$\abs{\fun{\oastate[\psi'] \circ \gamma_{\theta}}{A} - \fun{\oastate[\psi']' \circ \gamma_{\theta'}}{A}}
\leq \norm{\fun{\gamma_{\theta}}{A} - \fun{\gamma_{\theta'}}{A}} + \abs{\fun{\rbk{\oastate[\psi'] - \oastate[\psi']'}}{\fun{\gamma_{\theta'}}{A}}}.$$
This estimate gives joint continuity on
$\fun{\liegr{U}}{1}\times K_{\beta}$.
The action preserves $K_{\beta}$ because $\gamma_{\theta}$ commutes with
$\tau$.

Assertion (3) of Theorem \ref{thm:main-decomposition} gives invariance of
$\mu_{\psi_\beta}$ under this action.
Proposition \ref{prop:circle-orbit-disintegration}, applied with
$X = K_\beta$, supplies the quotient, the Borel probability kernel, and the
Haar description asserted in the lemma.
\end{proof}

\begin{thm}[orbit decomposition of the equilibrium state]\label{thm:orbit-decomposition}
Assume the hypotheses of Theorem \ref{thm:main-decomposition}.
Let $q$, $Y$, and $\msr{\mu_{\mathrm{orb}}}$ be defined by
\eqref{eq:orbit-quotient-data}, and let $\mu_{y}$ be the disintegration
family in \eqref{eq:orbit-disintegration-family}.
The following conclusions hold:
\begin{enumerate}
\item Define
$$\oastate[\psi^{\rbk{y}}]
=
\int_{\fun{q^{-1}}{y}}
\oastate[\psi']
\opdmsr{\mu_{y}(\oastate[\psi'])}.$$
For almost every $y$, this is a gauge-invariant
$\rbk{\tau,\beta}$-KMS state with central measure $\mu_{y}$.
It decomposes into states in $\setextremal K_\beta$ along the single orbit
$\fun{q^{-1}}{y}$.
Moreover,
$$\oastate[\psi_{\beta}]
=\int_{Y}\oastate[\psi^{\rbk{y}}]\opdmsr{\mu_{\mathrm{orb}}(y)};$$

\item The modulus of the order parameter factors as
$$\begin{aligned}
\abs{\barmean{a}(x)}
=
\varrho \circ q,
\quad
\varrho
\in
\fun{\lp^{\infty}}{Y,\msr{\mu_{\mathrm{orb}}}},
\quad
0 \leq\varrho \leq \frac{1}{2}.
\end{aligned}$$
It satisfies
$$\fun{\mu_{\mathrm{orb}}}{\set{y}{\fun{\varrho}{y}>0}}
=\fun{\mu}{E}\geq4\kappa.$$
The radial law of $\nu$ is the $\msr{\mu_{\mathrm{orb}}}$-distribution of $\varrho$;
\item For almost every $y$ with $\fun{\varrho}{y}>0$, the orbit
$\fun{q^{-1}}{y}$ is a circle in $\setextremal K_\beta$ whose states are
mutually disjoint, break the gauge symmetry, and have trivial stabilizer.
For any $\oastate[\psi'_{y}]\in\fun{q^{-1}}{y}$,
$$\oastate[\psi^{\rbk{y}}]
=\frac{1}{2\pi}\int_{0}^{2\pi}
\oastate[\psi'_{y}]\circ\gamma_{\theta}\opdmsr{\theta}.$$
This is the fiberwise Haar-orbit formula;
\item The picture of Remark \ref{rem:orbit-structure} holds exactly when
$\msr{\mu_{\mathrm{orb}}}$ is a point mass at some $y$ with $\fun{\varrho}{y}>0$.
The orbit-space measure and the gauge action satisfy
\begin{equation}\label{eq:orbit-point-mass-ergodicity-equivalence}
\msr{\mu_{\mathrm{orb}}}
\text{ is a point mass}
\quad\Longleftrightarrow\quad
\text{the gauge action on }
\rbk{K_{\beta},\mu_{\psi_{\beta}}}
\text{ is ergodic}.
\end{equation}
\end{enumerate}
\end{thm}

\begin{proof}
Proposition \ref{prop:circle-orbit-disintegration} gives the following
factorization for every gauge-invariant
$f\in\fun{\lp^{\infty}}{K_{\beta},\mu_{\psi_{\beta}}}$:
\begin{equation}\label{eq:gauge-invariant-orbit-factorization}
f
=
F\circ q
\text{ in }
\fun{\lp^{\infty}}{\mu_{\psi_{\beta}}}.
\end{equation}

For assertion (1), Proposition
\ref{prop:conditional-central-measures} applies to the central measure
$\mu_{\psi_\beta}$ and the disintegration
\eqref{eq:orbit-disintegration-family}.
It gives the central-measure identity
$$\mu_y
=
\mu_{\psi^{\rbk{y}}}$$
and the barycenter formula for $\oastate[\psi_\beta]$.
Gauge invariance of $\mu_y$, supplied by Lemma
\ref{lem:orbit-disintegration}, makes
$\oastate[\psi^{\rbk{y}}]$ gauge invariant.

For assertion (2), the function $\abs{\barmean{a}(x)}^{2}$ represents $\faadj{\barmean{a}}\barmean{a}$.
Assertion (1) of Theorem \ref{thm:componentwise-breaking} shows that this operator is fixed
by every $\fun{\Ad}{V_{\theta}}$.
The function $\abs{\barmean{a}(x)}$ is therefore gauge invariant.
The factorization of gauge-invariant functions proved above gives
$\abs{\barmean{a}(x)}=\varrho\circ q$.
Theorem \ref{thm:order-parameter-distribution} supplies the bounds and
$$\fun{\mu_{\mathrm{orb}}}{\varrho>0}
=\fun{\mu}{\abs{\barmean{a}(x)}>0}
=\fun{\mu}{E}
\geq4\kappa.$$
Finally, $\nu=\barmean{a}_{\ast}\mu$.
Together with the factorization, this identity gives the radial law.

For assertion (3), let $N$ be the set of states
$\oastate[\psi'] \in K_\beta$ for which
$\fun{\oastate[\psi']}{a_{z}} \neq 0$ for at least one
$z \in \ringratint^{d}$.
Assertion (3) of Theorem \ref{thm:componentwise-breaking} gives full $N$-measure on the
part corresponding to $E$.
After disintegration,
$$\fun{\mu_{y}}{N}=1$$
for almost every $y$ with $\fun{\varrho}{y}>0$.

The set $N$ is invariant along each orbit because
$$\abs{\fun{\oastate[\psi']\circ\gamma_{\theta}}{a_{z}}}
=\abs{\fun{\oastate[\psi']]}{a_{z}}}.$$
The gauge action on an orbit is transitive.
An invariant subset of full Haar measure is therefore the entire orbit.
Every state on the orbit breaks gauge symmetry and has the trivial stabilizer
from assertion (3) of Theorem \ref{thm:componentwise-breaking}.

The orbit is a continuous injective image of the circle and is itself a
circle.
Proposition \ref{prop:circle-orbit-disintegration} identifies $\mu_y$ with
normalized Haar measure on $\fun{q^{-1}}{y}$.
Substituting that measure into the barycenter in assertion (1) gives the
displayed fiberwise Haar formula.
Assertion (1) makes the states on the orbit extremal.
Theorem \ref{thm:extremal-kms-disjoint} makes distinct points on the orbit
disjoint,
and the displayed formula is the barycenter of the Haar image.

For assertion (4), if $\msr{\mu_{\mathrm{orb}}}$ is concentrated at one $y$ with
$\fun{\varrho}{y}>0$, the
formula in assertion (3) is exactly the picture of
Remark \ref{rem:orbit-structure}.
The converse is immediate from that picture.

The factorization
\eqref{eq:gauge-invariant-orbit-factorization}
identifies the
gauge-invariant elements of
$\fun{\lp^{\infty}}{K_{\beta},\mu_{\psi_{\beta}}}$ with
$\fun{\lp^{\infty}}{Y,\msr{\mu_{\mathrm{orb}}}}$.
Ergodicity means that these elements are constant.
This holds exactly when $\msr{\mu_{\mathrm{orb}}}$ is a point mass.
\end{proof}

\subsection{The even-translation ergodic decomposition}\label{the-even-translation-ergodic-decomposition}

The gauge-orbit decomposition and the decomposition associated with even translations answer different questions. The compact gauge group gives normalized Haar measure on each orbit. The group \(\ringratint^{d}_{\mathrm{even}}\) is infinite and discrete, so there is no corresponding probability measure on an infinite translation orbit. The appropriate replacement is the ergodic decomposition of the measure-preserving action induced on the central measure.

For \(v\in\ringratint^{d}_{\mathrm{even}}\), define \begin{equation}\label{eq:even-translation-action-on-kms}
\fun{T_v}{\oastate[\psi']}
=
\oastate[\psi']\circ\eta_v,
\quad
\oastate[\psi']\in K_\beta.
\end{equation} The automorphisms \(\eta_v\) commute with the dynamics by Proposition \ref{prop:symmetry-dynamics}. Consequently, \eqref{eq:even-translation-action-on-kms} is an action by homeomorphisms of \(K_\beta\). Assertion (3) of Theorem \ref{thm:main-decomposition} makes \(\mu_{\psi_\beta}\) invariant under this action. We use the state-space realization of the central decomposition from Theorem \ref{thm:kms-decomposition}. After removal of a \(\mu_{\psi_\beta}\)-null set, its measurable fields may be written as \begin{equation}\label{eq:state-space-central-realization}
\begin{aligned}
\sphilb{H}_{\psi_\beta}
&=
\int_{K_\beta}^{\oplus}
\sphilb{H}_{\psi'}
\opdmsr{\mu_{\psi_\beta}(\oastate[\psi'])},
\\
\oarepn_{\psi_\beta}
&=
\int_{K_\beta}^{\oplus}
\oarepn_{\psi'}
\opdmsr{\mu_{\psi_\beta}(\oastate[\psi'])},
\quad
\oa{Z}_{\psi_\beta}
\cong
\fun{\lp^\infty}{K_\beta,\mu_{\psi_\beta}}.
\end{aligned}
\end{equation} The measure is supported on \(\setextremal K_\beta\). This is the realization used throughout this subsection; in particular, the central function \(\barmean{a}\) is transported from the abstract realization in Theorem \ref{thm:componentwise-breaking} to \(\fun{\lp^\infty}{K_\beta,\mu_{\psi_\beta}}\).

\begin{prop}[even-translation ergodic decomposition]\label{prop:even-translation-ergodic-decomposition}
There are a standard probability space
$\rbk{Y_{\mathrm{even}},\msr{\nu}_{\mathrm{even}}}$,
a measurable map
$$
q_{\mathrm{even}}
\colon
K_\beta
\to
Y_{\mathrm{even}},
$$
and a probability kernel
$y\mapsto\mu_{\mathrm{even},y}$
on $K_\beta$ such that
\begin{equation}\label{eq:even-translation-central-measure-disintegration}
\begin{aligned}
\mu_{\psi_\beta}
&=
\int_{Y_{\mathrm{even}}}
\mu_{\mathrm{even},y}
\opdmsr{\msr{\nu}_{\mathrm{even}}(y)},
\\
\fun{\mu_{\mathrm{even},y}}{
\fun{q_{\mathrm{even}}^{-1}}{y}
}
&=
1
\quad
\text{for almost every }y\in Y_{\mathrm{even}}.
\end{aligned}
\end{equation}
For almost every $y$, the measure $\mu_{\mathrm{even},y}$ is
$T$-invariant and $T$-ergodic.
Moreover,
$$
q_{\mathrm{even}}\circ T_v
=
q_{\mathrm{even}}
\quad
\mu_{\psi_\beta}\text{-almost everywhere}
$$
for every $v\in\ringratint^{d}_{\mathrm{even}}$.
The map $q_{\mathrm{even}}$ generates the invariant $\sigma$-algebra modulo
$\mu_{\psi_\beta}$.
Equivalently, every $T$-invariant
$f\in\fun{\lp^\infty}{K_\beta,\mu_{\psi_\beta}}$
has the form
\begin{equation}\label{eq:even-translation-invariant-factorization}
f
=
F\circ q_{\mathrm{even}}
\quad
\text{in }
\fun{\lp^\infty}{K_\beta,\mu_{\psi_\beta}}
\end{equation}
for some
$F\in\fun{\lp^\infty}{Y_{\mathrm{even}},\msr{\nu}_{\mathrm{even}}}$.
The conditional barycenter
\begin{equation}\label{eq:even-translation-conditional-kms-state}
\oastate[\psi_{\mathrm{even},y}]
=
\int_{K_\beta}
\oastate[\psi']
\opdmsr{\mu_{\mathrm{even},y}(\oastate[\psi'])}
\end{equation}
is an even-translation-invariant KMS state, and
\begin{equation}\label{eq:even-translation-state-decomposition}
\oastate[\psi_\beta]
=
\int_{Y_{\mathrm{even}}}
\oastate[\psi_{\mathrm{even},y}]
\opdmsr{\msr{\nu}_{\mathrm{even}}(y)}.
\end{equation}
Moreover, $\oastate[\psi_{\mathrm{even},y}]$ is extremal in the convex set
\begin{equation}\label{eq:even-translation-invariant-kms-set}
K_\beta^{\mathrm{even}}
=
\set{\oastate[\psi]\in K_\beta}{
\oastate[\psi]\circ\eta_v=\oastate[\psi]
\text{ for every }v\in\ringratint^{d}_{\mathrm{even}}
}
\end{equation}
for almost every $y$.
\end{prop}

\begin{proof}
Let $\mathcal{I}_{\mathrm{even}}$ be the completed invariant $\sigma$-algebra
of the probability-preserving action
$\rbk{K_\beta,\mu_{\psi_\beta},T}$.
The algebra $\oa{A}$ is separable by Section \ref{sec:algebra}, and
Proposition \ref{prop:kms-compact-convex} makes $K_\beta$ compact and
metrizable.
The maps $T_v$ are affine weak-$\ast$ homeomorphisms that preserve
$\setextremal K_\beta$ by Propositions
\ref{prop:symmetry-dynamics} and
\ref{prop:decomposition-covariance}.
Because $K_\beta$ is therefore standard Borel and
$\ringratint^d_{\mathrm{even}}$ is countable, there is a standard Borel
quotient $q_{\mathrm{even}}$ whose pullback is
$\mathcal{I}_{\mathrm{even}}$ modulo null sets.
Disintegration over this quotient gives
\eqref{eq:even-translation-central-measure-disintegration}.
Invariance of $\mu_{\psi_\beta}$, uniqueness of the conditional measures,
and countability of the group make almost every
$\mu_{\mathrm{even},y}$ invariant under every $T_v$.
Disintegration over the full invariant $\sigma$-algebra makes these
conditional measures ergodic.
Proposition \ref{prop:conditional-central-measures} identifies
$\mu_{\mathrm{even},y}$ as the central measure of
\eqref{eq:even-translation-conditional-kms-state}.
Its invariance under $T_v$ gives
$$
\fun{\oastate[\psi_{\mathrm{even},y}]}{\fun{\eta_v}{A}}
=
\int_{K_\beta}
\fun{\oastate[\psi']\circ\eta_v}{A}
\opdmsr{\mu_{\mathrm{even},y}(\oastate[\psi'])}
=
\fun{\oastate[\psi_{\mathrm{even},y}]}{A}
$$
for every $A\in\oa{A}$ and
$v\in\ringratint^{d}_{\mathrm{even}}$.
Integration of \eqref{eq:even-translation-conditional-kms-state} gives
\eqref{eq:even-translation-state-decomposition}.

It remains to prove the extremality assertion.
Suppose that an almost-everywhere fixed conditional state has a
decomposition
$$
\oastate[\psi_{\mathrm{even},y}]
=
t\oastate[\varphi_1]
+
\rbk{1-t}\oastate[\varphi_2],
\quad
0<t<1,
\quad
\oastate[\varphi_1],\oastate[\varphi_2]
\in
K_\beta^{\mathrm{even}}.
$$
Let $\mu_i$ be the central measure of $\oastate[\varphi_i]$.
Proposition \ref{prop:decomposition-covariance} makes each $\mu_i$
$T$-invariant.
The uniqueness of the central measure in Theorem
\ref{thm:kms-decomposition} gives
$$
\mu_{\mathrm{even},y}
=
t\mu_1+\rbk{1-t}\mu_2.
$$
An ergodic invariant probability measure is extremal among invariant
probability measures.
It follows that
$\mu_1=\mu_2=\mu_{\mathrm{even},y}$.
Equality of the barycenters gives
$\oastate[\varphi_i]=\oastate[\psi_{\mathrm{even},y}]$.
\end{proof}

\begin{cor}[translation-ergodic direct integral]\label{cor:even-translation-iterated-direct-integral}
For almost every $y\in Y_{\mathrm{even}}$, set
$$
\sphilb{H}_{\mathrm{even},y}
=
\int_{K_\beta}^{\oplus}
\sphilb{H}_{\psi'}
\opdmsr{\mu_{\mathrm{even},y}(\oastate[\psi'])},
\quad
\oarepn_{\mathrm{even},y}
=
\int_{K_\beta}^{\oplus}
\oarepn_{\psi'}
\opdmsr{\mu_{\mathrm{even},y}(\oastate[\psi'])}.
$$
The central direct integral of Theorem \ref{thm:main-decomposition} can be
written as
\begin{equation}\label{eq:even-translation-iterated-direct-integral}
\begin{aligned}
\sphilb{H}_{\psi_\beta}
&\cong
\int_{Y_{\mathrm{even}}}^{\oplus}
\sphilb{H}_{\mathrm{even},y}
\opdmsr{\msr{\nu}_{\mathrm{even}}(y)},
\\
\oarepn_{\psi_\beta}
&\cong
\int_{Y_{\mathrm{even}}}^{\oplus}
\oarepn_{\mathrm{even},y}
\opdmsr{\msr{\nu}_{\mathrm{even}}(y)}.
\end{aligned}
\end{equation}
For almost every $y$, the conditional state satisfies
even-translation mean clustering:
\begin{equation}\label{eq:even-translation-component-mean-clustering}
\lim_{n\to\infty}
\frac{2}{\abscard{\Lambda_n}}
\sum_{v\in\Lambda_n\cap\ringratint^{d}_{\mathrm{even}}}
\fun{\oastate[\psi_{\mathrm{even},y}]}{
B\fun{\eta_v}{A}
}
=
\fun{\oastate[\psi_{\mathrm{even},y}]}{B}
\fun{\oastate[\psi_{\mathrm{even},y}]}{A}
\end{equation}
for $A,B\in\oa{A}$ and almost every $y$.
\end{cor}

\begin{proof}
Fubini's theorem for direct integrals applied to
\eqref{eq:even-translation-central-measure-disintegration} gives
\eqref{eq:even-translation-iterated-direct-integral}.

For the clustering assertion, average
$\fun{\oarepn_{\psi_{\mathrm{even},y}}}{\fun{\eta_v}{A}}$
over
$v\in\Lambda_n\cap\ringratint^{d}_{\mathrm{even}}$.
The proof of Proposition \ref{prop:factor-clustering} shows that every
weak-operator limit belongs to the center.
The Følner property makes the limit invariant under every even translation.
The invariant part of this center equals the scalars.
Indeed, a nonconstant invariant self-adjoint central element has a
nontrivial invariant spectral projection.
The two normalized central restrictions supplied by Proposition
\ref{prop:central-subordinate} would give a nontrivial convex decomposition
of $\oastate[\psi_{\mathrm{even},y}]$ into elements of
$K_\beta^{\mathrm{even}}$, contradicting Proposition
\ref{prop:even-translation-ergodic-decomposition}.
The scalar is
$\fun{\oastate[\psi_{\mathrm{even},y}]}{A}$.
Taking its matrix element against
$\fun{\oarepn_{\psi_{\mathrm{even},y}}}{\faadj{B}}
\oagnsvector[\Psi_{\mathrm{even},y}]$
gives
\eqref{eq:even-translation-component-mean-clustering}.
Separability of $\oa{A}$ permits one common conull set to be chosen first
for a countable norm-dense subset and then for all $A,B\in\oa{A}$.
\end{proof}

\begin{prop}[order parameter on translation-ergodic components]\label{prop:even-translation-order-parameter}
There is a measurable function
$$
z_{\mathrm{even}}
\colon
Y_{\mathrm{even}}
\to
\set{z\in\fldcmp}{\abs{z}\leq\frac{1}{2}}
$$
such that
\begin{equation}\label{eq:even-translation-order-parameter-factorization}
\barmean{a}
=
z_{\mathrm{even}}\circ q_{\mathrm{even}}
\quad
\text{in }
\fun{\lp^\infty}{K_\beta,\mu_{\psi_\beta}}.
\end{equation}
Its distribution is the order-parameter law
$\nu$ of Theorem \ref{thm:order-parameter-distribution}, and
\begin{equation}\label{eq:even-translation-broken-weight}
\fun{\msr{\nu}_{\mathrm{even}}}{
\set{y\in Y_{\mathrm{even}}}{
\fun{z_{\mathrm{even}}}{y}\neq0
}
}
\geq
4\kappa.
\end{equation}
For almost every $y$ and every $n\in\semigrposint$,
\begin{equation}\label{eq:even-translation-component-order-parameter}
\fun{\oastate[\psi_{\mathrm{even},y}]}
{\barmean{a}_{\Lambda_n}}
=
\frac{1}{2}
\rbk{\fun{\oastate[\psi_{\mathrm{even},y}]}{a_0}
+\fun{\oastate[\psi_{\mathrm{even},y}]}{a_{e_1}}}
=
\fun{z_{\mathrm{even}}}{y}.
\end{equation}
Every component with
$\fun{z_{\mathrm{even}}}{y}\neq0$
breaks the gauge symmetry.
\end{prop}

\begin{proof}
Assertion (4) of Theorem \ref{thm:order-parameter-distribution} says that
the central function $\barmean{a}$ is fixed by the induced even-translation
action.
The invariant $\sigma$-algebra is the pullback by $q_{\mathrm{even}}$.
This gives
\eqref{eq:even-translation-order-parameter-factorization}.
The identities
$\nu=\barmean{a}_*\mu_{\psi_\beta}$ and
\eqref{eq:even-translation-central-measure-disintegration}
identify the law of $z_{\mathrm{even}}$ with $\nu$.
Assertion (3) of Theorem \ref{thm:order-parameter-distribution} gives
\eqref{eq:even-translation-broken-weight}.

Assertion (1) of Theorem \ref{thm:componentwise-breaking} gives weak-$\ast$
convergence of
$$
\oastate[\psi']
\mapsto
\fun{\oastate[\psi']}{
\barmean{a}_{\Lambda_n}
}
$$
to $\barmean{a}$ in
$\fun{\lp^\infty}{K_\beta,\mu_{\psi_\beta}}$.
Even-translation invariance of
$\oastate[\psi_{\mathrm{even},y}]$
and the two parity classes of $\Lambda_n$ give the first equality in
\eqref{eq:even-translation-component-order-parameter}.
Its middle expression is independent of $n$.
Conditional expectation of the preceding weak-$\ast$ convergence onto the
invariant $\sigma$-algebra identifies that expression with
$z_{\mathrm{even}}$ and gives the second equality in
\eqref{eq:even-translation-component-order-parameter}.
If its right side is nonzero, the conditional state cannot be gauge
invariant because
\eqref{eq:gauge-automorphism} multiplies the left side by
$\napiernum^{-\imunit\theta}$.
\end{proof}

The translation-ergodic decomposition need not be an orbit decomposition. An ergodic measure may be supported on the closure of many translation orbits. It may also retain nontrivial central variables that are moved, rather than fixed, by translations. Accordingly, \(\mu_{\mathrm{even},y}\) need not be a point mass: \(\oastate[\psi_{\mathrm{even},y}]\) is extremal in \(K_\beta^{\mathrm{even}}\) but need not be a factor or an extremal point of \(K_\beta\). Equations \eqref{eq:even-translation-state-decomposition} and \eqref{eq:even-translation-iterated-direct-integral} are therefore the available abstract decomposition; the infrared bounds do not determine \(\msr{\nu}_{\mathrm{even}}\) or its conditional measures explicitly.

\subsection{The density of the components}\label{the-density-of-the-components}

The same machinery applies to the gauge-invariant spatial averages of the particle number and produces the density of the pure phases.

\begin{prop}[the density of the components]\label{prop:component-density}
Let
$\barmean{n}_{\Lambda_n}$
be the density average defined in
\eqref{eq:main-spatial-averages}.
Under the hypotheses of Theorem \ref{thm:main-decomposition}, the represented density averages have the weak-operator limit
$$\wlim_{n \to \infty}
\fun{\oarepn_{\psi_{\beta}}}{\barmean{n}_{\Lambda_n}}
=
\barmean{n},$$
where
$\barmean{n}
\in
\oa{Z}_{\psi_{\beta}}$
and
$0
\leq
\barmean{n}
\leq
1$.
Under the central identification
$\oa{Z}_{\psi_\beta}\cong\fun{\lp^{\infty}}{X,\mu}$, we use the same
symbol $\barmean{n}$ for the representing function.
\begin{enumerate}
\item The density is gauge invariant and translation invariant:
$$\begin{aligned}
V_{\theta}\barmean{n}\faadj{V_{\theta}}
=
\barmean{n},
\quad
\widetilde{U}_{v}\barmean{n}\faadj{\widetilde{U}_{v}}
=
\barmean{n}.
\end{aligned}$$
Its representing function is constant along gauge orbits and hence factors
through the orbit space of Theorem \ref{thm:orbit-decomposition}.

\item The mean density is
\begin{equation}\label{eq:component-mean-density}
\int_{X}\barmean{n}(x)\opdmsr{\mu(x)}
=
\frac{1}{2}.
\end{equation}
The distribution of $\barmean{n}$ is symmetric about $\frac{1}{2}$.
Equivalently, $\barmean{n}$ and $1-\barmean{n}$ have the same distribution.

\item The macroscopic density variance satisfies
\begin{equation}\label{eq:component-density-variance}
\lim_{n \to \infty}
\fun{\oastate[\psi_{\beta}]}{\rbk{\barmean{n}_{\Lambda_n}-\frac{1}{2}}^{2}}
=\int_X\rbk{\barmean{n}(x)-\frac{1}{2}}^2\opdmsr{\mu(x)}.
\end{equation}
The function $\barmean{n}$ equals $\frac{1}{2}$ almost everywhere exactly when the
macroscopic density variance of the symmetric state vanishes.
\end{enumerate}
\end{prop}

\begin{proof}
Let $P_{0,\psi_\beta}$ be the projection onto the even-translation-invariant
vectors in $\sphilb{H}_{\psi_\beta}$, as in Lemma \ref{lem:mean-ergodic}.
Parity splitting and Lemma \ref{lem:mean-ergodic} give
$$\fun{\oarepn_{\psi_\beta}}{\barmean{n}_{\Lambda_n}}\oagnsvector[\Psi_\beta]
\to
\frac{1}{2}P_{0,\psi_\beta}
\rbk{\fun{\oarepn_{\psi_\beta}}{n_{0}}
+\fun{\oarepn_{\psi_\beta}}{n_{e_{1}}}}
\oagnsvector[\Psi_\beta]$$
strongly.
The commutator estimate and weak closure place every weak-operator limit
point in $\oa{Z}_{\psi_{\beta}}$.
The separating vector $\oagnsvector[\Psi_\beta]$ makes that limit point unique.
The operator limit is therefore
$$\wlim_{n \to \infty}
\fun{\oarepn_{\psi_\beta}}{\barmean{n}_{\Lambda_n}}
=
\barmean{n}.$$
The inequalities
$0
\leq
\barmean{n}_{\Lambda_n}
\leq
1$
pass to the weak-operator limit and give
$0
\leq
\barmean{n}
\leq
1$.

Assertion (1) follows from the identity
$\fun{\gamma_{\theta}}{n_{y}}=n_{y}$:
$$V_{\theta}\fun{\oarepn_{\psi_\beta}}{\barmean{n}_{\Lambda_n}}\faadj{V_{\theta}}
=\fun{\oarepn_{\psi_\beta}}{\barmean{n}_{\Lambda_n}}.$$
Its weak-operator limit is
$V_{\theta}\barmean{n}\faadj{V_{\theta}}=\barmean{n}$.
Translation invariance follows from the boundary estimate
$$\norm{\fun{\eta_{v}}{\barmean{n}_{\Lambda_n}}-\barmean{n}_{\Lambda_n}}
=\fun{O}{2^{-n}},$$
because the symmetric difference of the two boxes is boundary-sized.
The factorization \eqref{eq:gauge-invariant-orbit-factorization} gives the
orbit-space factorization asserted in assertion (1).

For assertion (2), let $\Theta$ be the particle--hole automorphism defined
in \eqref{eq:main-particle-hole-automorphism}.
The automorphism $\Theta$ is the thermodynamic limit of the finite-volume
symmetry $u_{\mathrm{ph}}$ defined in
\eqref{eq:main-particle-hole-unitary}.
Its Hamiltonian invariance is
\eqref{eq:main-particle-hole-identities}.

Every finite-volume Gibbs state is invariant under
$\fun{\Ad}{u_{\mathrm{ph}}}$.
Weak-$\ast$ convergence preserves this identity for every local observable.
The aligned exhaustion absorbs the translation part.
It follows that
$$\oastate[\psi_{\beta}]\circ\Theta=\oastate[\psi_{\beta}].$$
There is an implementing unitary $W_{\Theta,\psi_\beta}$ such that
$$\begin{aligned}
W_{\Theta,\psi_\beta}
\fun{\oarepn_{\psi_\beta}}{A}
\faadj{W_{\Theta,\psi_\beta}}
=
\fun{\oarepn_{\psi_\beta}}{\fun{\Theta}{A}},
\quad
W_{\Theta,\psi_\beta}
\oagnsvector[\Psi_\beta]
=
\oagnsvector[\Psi_\beta].
\end{aligned}$$
In particular,
$$\fun{\oastate[\psi_{\beta}]}{n_{y}}
+\fun{\oastate[\psi_{\beta}]}{n_{y+e_{1}}}
=1.$$
Pair every site of $\Lambda_n$ with its neighbor in the first coordinate.
The pairing is exact because the side length of $\Lambda_n$ is $2^n$ and is even.
This pairing gives
$\fun{\oastate[\psi_{\beta}]}{\barmean{n}_{\Lambda_n}}=\frac{1}{2}$
for every $n$, and
$$\int_{X}\barmean{n}(x)\opdmsr{\mu(x)}
=\bkt{\oagnsvector[\Psi_\beta]}{\barmean{n}\oagnsvector[\Psi_\beta]}
=\frac{1}{2}.$$

The particle-hole transformation gives
$$\begin{aligned}
\fun{\Theta}{\barmean{n}_{\Lambda_n}}
=
\frac{1}{\abscard{\Lambda_n}}
\sum_{x\in\Lambda_n}
\rbk{1-n_{x+e_1}}
=
1
-\frac{1}{\abscard{\Lambda_n}}
\sum_{x\in\Lambda_n}n_{x+e_1}.
\end{aligned}$$
The second term differs from $\barmean{n}_{\Lambda_n}$ by a boundary term.
Passing to the limit gives
$$W_{\Theta,\psi_\beta}\barmean{n}\faadj{W_{\Theta,\psi_\beta}}=1-\barmean{n}.$$
Conjugation by $W_{\Theta,\psi_\beta}$ preserves
$\oa{M}_{\psi_{\beta}}$ and
$\oacommutant{\fun{\oarepn_{\psi_\beta}}{\oa{A}}}$.
It therefore preserves $\oa{Z}_{\psi_{\beta}}$.
It also preserves the vector state
$\bkt{\oagnsvector[\Psi_\beta]}{\cdot\oagnsvector[\Psi_\beta]}$.
The induced automorphism of $\fun{\lp^{\infty}}{X,\mu}$ is
measure preserving and maps $\barmean{n}$ to $1-\barmean{n}$.
These functions have the same distribution.

For assertion (3), the operator $\barmean{n}$ commutes with
$\fun{\oarepn_{\psi_\beta}}{\barmean{n}_{\Lambda_n}}$.
This commutation gives
$$\begin{aligned}
\fun{\oarepn_{\psi_\beta}}{\barmean{n}_{\Lambda_n}}^{2}
\oagnsvector[\Psi_\beta]
-\barmean{n}^{2}
\oagnsvector[\Psi_\beta]
=
\fun{\oarepn_{\psi_\beta}}{\barmean{n}_{\Lambda_n}}
\rbk{\fun{\oarepn_{\psi_\beta}}{\barmean{n}_{\Lambda_n}}\oagnsvector[\Psi_\beta]-\barmean{n}\oagnsvector[\Psi_\beta]}
+\barmean{n} \rbk{\fun{\oarepn_{\psi_\beta}}{\barmean{n}_{\Lambda_n}}\oagnsvector[\Psi_\beta]-\barmean{n}\oagnsvector[\Psi_\beta]}.
\end{aligned}$$
Both terms tend to zero in norm.
This is the two-letter argument from
assertion (1) of Theorem \ref{thm:order-parameter-distribution}.
Consequently,
$$\fun{\oastate[\psi_{\beta}]}{\barmean{n}_{\Lambda_n}^{2}}
\to
\bkt{\oagnsvector[\Psi_\beta]}{\barmean{n}^{2}\oagnsvector[\Psi_\beta]}
=\int_{X}\barmean{n}(x)^{2}\opdmsr{\mu(x)}.$$
Subtracting the linear terms proves the asserted variance identity.
The nonnegative function
$\rbk{\barmean{n}-\frac{1}{2}}^{2}$ has zero integral exactly when
$\barmean{n}(x)=\frac{1}{2}$ for almost every $x\in X$.
\end{proof}

\begin{rem}\label{rem:density-open}
The unresolved question is whether the following variance tends to zero.
The identity $\barmean{n}(x)=\frac{1}{2}$ for almost every $x\in X$ is equivalent to
\begin{equation}\label{eq:density-truncated-correlation-decay}
\frac{1}{\abscard{\Lambda_n}^{2}}
\sum_{x,y\in \Lambda_n}
\rbk{\fun{\oastate[\psi_{\beta}]}{n_{x}n_{y}}
-\fun{\oastate[\psi_{\beta}]}{n_{x}}
\fun{\oastate[\psi_{\beta}]}{n_{y}}}
\to
0.
\end{equation}
This is a decay statement for truncated density-density correlations.
It has not been proved in either regime.
Section \ref{sec:improvements} discusses the missing estimate.
\end{rem}

\subsection{Sharpness of the macroscopic observables in the components}\label{sharpness-of-the-macroscopic-observables-in-the-components}

The functions \(\barmean{a}\) and \(\barmean{n}\) arise from central elements. One must still identify them with intrinsic properties of the individual components. In particular, the construction alone does not show that \(\abs{\barmean{a}(x)}^{2}\) is the uniform-mode condensate density of \(\oastate[\psi_{x}]\). The following \(\lp^{2}\) argument proves this identification. It also shows that almost every component assigns sharp values to all macroscopic averages.

\begin{thm}[macroscopic sharpness of the components]\label{thm:macroscopic-sharpness}
Let
$A
\in
\oa{A}_{\txtloc}$,
and let
$\barmean{A}_{\Lambda_n}$
be the spatial average defined in
\eqref{eq:main-spatial-average}.
The following conclusions hold under the hypotheses of Theorem
\ref{thm:main-decomposition}.
\begin{enumerate}
\item
The represented averages have the weak-operator limit
$$\wlim_{n \to \infty}
\fun{\oarepn_{\psi_{\beta}}}{\barmean{A}_{\Lambda_n}}
=
\widehat{A}.$$
The operator
$\widehat{A}
\in
\oa{Z}_{\psi_{\beta}}$
is canonical and is represented by
$\widehat{a}
\in
\fun{\lp^{\infty}}{X,\mu}$.

\item The functions
$x\mapsto\fun{\oastate[\psi_{x}]}{\barmean{A}_{\Lambda_n}}$
converge to $\widehat{a}$ in the norm of $\fun{\lp^{2}}{X,\mu}$.
Moreover,
$$\int_{X}
\rbk{\fun{\oastate[\psi_{x}]}{\faadj{\barmean{A}_{\Lambda_n}} \barmean{A}_{\Lambda_n}}
-\abs{\fun{\oastate[\psi_{x}]}{\barmean{A}_{\Lambda_n}}}^{2}}
\opdmsr{\mu(x)}
\to
0.$$

\item
Along a subsequence $n_k\to\infty$, for almost every $x$,
$$\begin{aligned}
\fun{\oastate[\psi_{x}]}{\barmean{A}_{\Lambda_{n_k}}}
\to
\fun{\widehat{a}}{x},
\quad
\fun{\oastate[\psi_{x}]}{\faadj{\barmean{A}_{\Lambda_{n_k}}}\barmean{A}_{\Lambda_{n_k}}}
-\abs{\fun{\oastate[\psi_{x}]}{\barmean{A}_{\Lambda_{n_k}}}}^{2}
\to0.
\end{aligned}$$
Almost every component consequently assigns a sharp value to the macroscopic
average of $A$.
For any prescribed countable family of local observables, a diagonal
extraction gives one common subsequence.
\end{enumerate}
\end{thm}

\begin{proof}
For assertion (1), let $P_{0,\psi_\beta}$ be the projection onto the
even-translation-invariant vectors in $\sphilb{H}_{\psi_\beta}$, as in Lemma
\ref{lem:mean-ergodic}.
Every $y\in \Lambda_n$ has a unique form $v$ or $v+e_{1}$
with $v\in\ringratint^{d}_{\mathrm{even}}$.
The operator $\barmean{A}_{\Lambda_n}$ is the average of two parity-class averages, one for
$\fun{\eta_{v}}{A}$ and one for
$\fun{\eta_{v}}{\fun{\eta_{e_1}}{A}}$.
Lemma \ref{lem:mean-ergodic} gives
$$\fun{\oarepn_{\psi_\beta}}{\barmean{A}_{\Lambda_n}}\oagnsvector[\Psi_\beta]
\to
\frac{1}{2}
P_{0,\psi_\beta}
\rbk{\fun{\oarepn_{\psi_\beta}}{A}
+\fun{\oarepn_{\psi_\beta}}{\fun{\eta_{e_1}}{A}}}\oagnsvector[\Psi_\beta]$$
strongly.
The same argument applies to
$\faadj{\barmean{A}_{\Lambda_n}}=\barmean{\faadj{A}}_{\Lambda_n}$.

Fix a local $B'$.
Only finitely many translates of the support of $A$ meet the support of
$B'$.
This finite-overlap property gives
$$\norm{\commutator{\barmean{A}_{\Lambda_n}}{B'}}
=\fun{O}{\abscard{\Lambda_n}^{-1}}.$$
Every weak-operator limit point of
$\fun{\oarepn_{\psi_\beta}}{\barmean{A}_{\Lambda_n}}$ is central and agrees with the strong
limit on $\oagnsvector[\Psi_\beta]$.
The separating property of $\oagnsvector[\Psi_\beta]$ makes the limit unique.
The limit and norm bound are
$$\begin{aligned}
\wlim_{n \to \infty}
\fun{\oarepn_{\psi_\beta}}{\barmean{A}_{\Lambda_n}}
=
\widehat{A}
\in\oa{Z}_{\psi_{\beta}},
\quad
\norm{\widehat{A}}
\leq
\norm{A}.
\end{aligned}$$

For assertion (2), let $D\in\oa{Z}_{\psi_\beta}$ have fiber function $f_D$.
The orthogonal-measure identity
\eqref{eq:orthogonal-measure-central-identity}
gives
$$\int_{X}
\fun{f_D}{x}
\fun{n_{R}}{x}
\opdmsr{\mu(x)}
=\bkt{\oagnsvector[\Psi_\beta]}{D\fun{\oarepn_{\psi_\beta}}{\barmean{A}_{\Lambda_n}}\oagnsvector[\Psi_\beta]}
\to
\bkt{\oagnsvector[\Psi_\beta]}{D\widehat{A}\oagnsvector[\Psi_\beta]}
=\int_{X}
\fun{f_D}{x}
\fun{\widehat{a}}{x}
\opdmsr{\mu(x)}.$$
The family $n_{R}$ is bounded by $\norm{A}$ on the finite measure space.
The above convergence therefore gives weak convergence
$n_{R}\to\widehat{a}$ in $\fun{\lp^{2}}{X,\mu}$.

Cauchy--Schwarz in the state $\oastate[\psi_{x}]$ gives
$\abs{\fun{n_{R}}{x}}^{2}
\leq
\fun{\oastate[\psi_{x}]}{\faadj{\barmean{A}_{\Lambda_n}}\barmean{A}_{\Lambda_n}}$.
After integration, it holds that
$$\begin{aligned}
\norm{n_{R}}_{2}^{2}
\leq
\int_{X}
\fun{\oastate[\psi_{x}]}{\faadj{\barmean{A}_{\Lambda_n}}\barmean{A}_{\Lambda_n}}
\opdmsr{\mu(x)}
=
\fun{\oastate[\psi_{\beta}]}{\faadj{\barmean{A}_{\Lambda_n}}\barmean{A}_{\Lambda_n}}
\to
\bkt{\oagnsvector[\Psi_\beta]}{\faadj{\widehat{A}}\widehat{A}\oagnsvector[\Psi_\beta]}
=
\norm{\widehat{a}}_{2}^{2}.
\end{aligned}$$
The operator convergence is the two-letter argument from
assertion (1) of Theorem \ref{thm:order-parameter-distribution}.
It uses the strong convergences in assertion (1), the commutation
$\widehat{A}\in\oacommutant{\fun{\oarepn_{\psi_\beta}}{\oa{A}}}$, and commutativity of
$\oa{Z}_{\psi_\beta}$.

Weak lower semicontinuity gives
$\norm{\widehat{a}}_{2}\leq\liminf\norm{n_{R}}_{2}$.
The preceding upper bound proves convergence of the norms.
Weak convergence together with convergence of norms implies
$n_{R}\to\widehat{a}$ in $\fun{\lp^{2}}{X,\mu}$.
The dispersion integral is
$\fun{\oastate[\psi_{\beta}]}{\faadj{\barmean{A}_{\Lambda_n}}\barmean{A}_{\Lambda_n}}
-\norm{n_{R}}_{2}^{2}$
and tends to zero.
Its integrand is nonnegative.

For assertion (3), choose $n_k$ so that the $\lp^{1}$ norm of the
dispersion integrand and
$\norm{n_{R}-\widehat{a}}_{2}$ are summable along the subsequence.
Both quantities then converge to zero almost everywhere.
For countably many observables, extract diagonally.
\end{proof}

\begin{cor}[componentwise uniform-mode condensate density and sharp density]\label{cor:componentwise-odlro}
The following conclusions hold along a subsequence $n_k$ for
$\mu$-almost every $x$.
\begin{enumerate}
\item The uniform-mode condensate density satisfies
\begin{equation}\label{eq:componentwise-uniform-mode-condensate-density}
\lim_{k}
\frac{1}{\abscard{\Lambda_{n_k}}^{2}}
\sum_{x',y'\in \Lambda_{n_k}}
\fun{\oastate[\psi_{x}]}{\faadj{a_{x'}}a_{y'}}
=\abs{\barmean{a}(x)}^{2}.
\end{equation}
Almost every component in $E$, defined by
\eqref{eq:componentwise-breaking-nonzero-set}, has positive uniform-mode ODLRO equal to the square of
its order parameter.
Almost every component outside $E$ has no uniform-mode condensate.

\item The density satisfies
\begin{equation}\label{eq:componentwise-density-sharpness}
\begin{aligned}
\fun{\oastate[\psi_{x}]}{\barmean{n}_{\Lambda_{n_k}}}
\to
\barmean{n}(x),
\quad
\fun{\oastate[\psi_{x}]}{\barmean{n}_{\Lambda_{n_k}}^{2}}
-\fun{\oastate[\psi_{x}]}{\barmean{n}_{\Lambda_{n_k}}}^{2}
\to
0.
\end{aligned}
\end{equation}
Every component has the sharp macroscopic density $\barmean{n}(x)$.
\end{enumerate}
\end{cor}

\begin{proof}
Apply Theorem \ref{thm:macroscopic-sharpness} first to $A=a_{0}$ and then
to $A=n_{0}$.
The corresponding averages are $\barmean{a}_{\Lambda_n}$ and $\barmean{n}_{\Lambda_n}$.
Uniqueness of the weak-operator limits identifies the central operators with
$\barmean{a}$ and $\barmean{n}$, respectively.

For assertion (1), along the selected subsequence and almost everywhere,
$$\begin{aligned}
\fun{\oastate[\psi_{x}]}{\faadj{\barmean{a}_{\Lambda_{n_k}}}\barmean{a}_{\Lambda_{n_k}}}
-\abs{\fun{\oastate[\psi_x]}{\barmean{a}_{\Lambda_{n_k}}}}^{2}
\to
0,
\quad
\fun{\oastate[\psi_x]}{\barmean{a}_{\Lambda_{n_k}}}
\to
\barmean{a}(x).
\end{aligned}$$
The spatial-average sum equals
$\fun{\oastate[\psi_{x}]}{\faadj{\barmean{a}_{\Lambda_n}}\barmean{a}_{\Lambda_n}}$.
The cases $\barmean{a}(x)\neq0$ and $\barmean{a}(x)=0$ give the assertions about
$E$ and its complement.
Assertion (2) is the same statement for the self-adjoint $\barmean{n}_{\Lambda_n}$.
\end{proof}

\subsection{Disintegration by the condensate density}\label{disintegration-by-the-condensate-density}

The gauge phase and the modulus of the uniform condensate amplitude are different central variables. The orbit decomposition fixes the phase law on every gauge orbit. The modulus gives a further measurable quotient even though its law cannot be computed by the available infrared estimates.

The uniform-mode condensate density on the central base is defined by \begin{equation}\label{eq:component-condensate-density}
\fun{c_{\mathrm{BEC}}}{x}
=
\abs{\barmean{a}(x)}^{2},
\quad
x
\in
X.
\end{equation} Corollary \ref{cor:componentwise-odlro} gives the intrinsic correlation formula for \eqref{eq:component-condensate-density} in almost every central component.

\begin{prop}[disintegration by the uniform-mode condensate density]\label{prop:condensate-density-disintegration}
Assume the hypotheses of Theorem \ref{thm:main-decomposition}.
Let
\begin{equation}\label{eq:condensate-density-law}
\msr{\nu}_{\mathrm{BEC}}
=
\rbk{c_{\mathrm{BEC}}}_{\ast}\mu
\end{equation}
be the distribution of \eqref{eq:component-condensate-density}.
The measure is supported on
$\closedinterval{0}{\frac{1}{4}}$ and satisfies
\begin{equation}\label{eq:condensate-density-law-bounds}
\int_{\closedinterval{0}{\frac{1}{4}}}
s
\opdmsr{\msr{\nu}_{\mathrm{BEC}}(s)}
\geq
\kappa,
\quad
\fun{\msr{\nu}_{\mathrm{BEC}}}{\leftopeninterval{0}{\frac{1}{4}}}
\geq
4\kappa.
\end{equation}
There is a family
$s\mapsto\mu_{s}$
of probability measures on $X$, unique up to
$\msr{\nu}_{\mathrm{BEC}}$-null sets, such that
\begin{equation}\label{eq:condensate-density-disintegration-measure}
\fun{\mu_s}{\set{x\in X}{\fun{c_{\mathrm{BEC}}}{x}=s}}
=
1,
\quad
\mu
=
\int_{\closedinterval{0}{\frac{1}{4}}}
\mu_s
\opdmsr{\msr{\nu}_{\mathrm{BEC}}(s)}.
\end{equation}
The conditional barycenters
\begin{equation}\label{eq:condensate-density-conditional-kms-state}
\oastate[\psi_{\mathrm{BEC},s}]
=
\int_X
\oastate[\psi_x]
\opdmsr{\mu_s(x)}
\end{equation}
are $\rbk{\tau,\beta}$-KMS states for almost every $s$, and
\begin{equation}\label{eq:condensate-density-state-decomposition}
\oastate[\psi_\beta]
=
\int_{\closedinterval{0}{\frac{1}{4}}}
\oastate[\psi_{\mathrm{BEC},s}]
\opdmsr{\msr{\nu}_{\mathrm{BEC}}(s)}.
\end{equation}
The function $c_{\mathrm{BEC}}$ is constant on every gauge orbit and obeys
\begin{equation}\label{eq:condensate-density-orbit-factorization}
c_{\mathrm{BEC}}
=
\rbk{\varrho\circ q}^{2}
\end{equation}
with the orbit data of Theorem \ref{thm:orbit-decomposition}.
\end{prop}

\begin{proof}
Theorem \ref{thm:order-parameter-distribution} gives
$0
\leq
c_{\mathrm{BEC}}
\leq
\frac{1}{4}$.
Equations \eqref{eq:componentwise-breaking-nonzero-set} and
\eqref{eq:component-condensate-density} identify the positive set of
$c_{\mathrm{BEC}}$ with $E$.
The second-moment estimate and the support bound give
\eqref{eq:condensate-density-law-bounds}.

The probability space $X$ is standard by Theorem
\ref{thm:kms-decomposition}.
Disintegration of $\mu$ along the measurable map
$c_{\mathrm{BEC}}$ gives
\eqref{eq:condensate-density-disintegration-measure}.
Proposition \ref{prop:conditional-central-measures} identifies each
$\mu_s$ with the central measure of the barycenter
\eqref{eq:condensate-density-conditional-kms-state}.
Convexity of $K_\beta$ gives the KMS assertion and
\eqref{eq:condensate-density-state-decomposition}.

Assertion (2) of Theorem \ref{thm:orbit-decomposition} gives
$\abs{\barmean{a}}=\varrho\circ q$.
Squaring this identity proves
\eqref{eq:condensate-density-orbit-factorization}.
\end{proof}

\begin{cor}[iterated direct integral over the condensate density]\label{cor:condensate-density-iterated-direct-integral}
Under the hypotheses of Proposition
\ref{prop:condensate-density-disintegration}, define
$$\begin{aligned}
\sphilb{H}_{\mathrm{BEC},s}
&=
\int_X^{\oplus}
\sphilb{H}_x
\opdmsr{\mu_s(x)},
\quad
\oarepn_{\mathrm{BEC},s}
=
\int_X^{\oplus}
\oarepn_x
\opdmsr{\mu_s(x)}.
\end{aligned}$$
The central direct integral has the iterated form
\begin{equation}\label{eq:condensate-density-iterated-direct-integral}
\begin{aligned}
\sphilb{H}_{\psi_\beta}
&\cong
\int_{\closedinterval{0}{\frac{1}{4}}}^{\oplus}
\sphilb{H}_{\mathrm{BEC},s}
\opdmsr{\msr{\nu}_{\mathrm{BEC}}(s)},
\quad
\oarepn_{\psi_\beta}
\cong
\int_{\closedinterval{0}{\frac{1}{4}}}^{\oplus}
\oarepn_{\mathrm{BEC},s}
\opdmsr{\msr{\nu}_{\mathrm{BEC}}(s)}.
\end{aligned}
\end{equation}
\end{cor}

\begin{proof}
Substitute \eqref{eq:condensate-density-disintegration-measure} into the
central direct integral of Theorem \ref{thm:main-decomposition}.
Fubini's theorem for direct integrals gives
\eqref{eq:condensate-density-iterated-direct-integral}.
\end{proof}

\begin{rem}[limits of the condensate-density disintegration]\label{rem:condensate-density-disintegration-limits}
The staggered particle density does not obstruct
Proposition \ref{prop:condensate-density-disintegration}.
Theorem \ref{thm:density-staggered} concerns the diagonal entries of the
one-particle density matrix, whereas Theorem
\ref{thm:condensate-constant} and Corollary
\ref{cor:condensate-asymptotically-constant} prove that the normalized
macroscopic eigenvector of the hard-core model approaches the constant
vector in the one-particle Hilbert-space norm, up to phase.
The particle-density profile and the condensate wave function are distinct
objects in this model.

For the interacting hard-core model,
\eqref{eq:condensate-density-state-decomposition} is the complete
unconditional statement currently supplied by the central decomposition.
Neither the measure $\msr{\nu}_{\mathrm{BEC}}$ nor the conditional KMS states
\eqref{eq:condensate-density-conditional-kms-state} are explicit.
The density fiber at a fixed $s$ may contain several gauge orbits and other
central variables.
Consequently, $s$ does not parameterize the full center.
The measure $\msr{\nu}_{\mathrm{BEC}}$ is a point mass exactly when
$c_{\mathrm{BEC}}$ is almost everywhere constant.
Equivalently, its variance vanishes:
$$
\int_X
\fun{c_{\mathrm{BEC}}}{x}^{2}
\opdmsr{\mu(x)}
=
\rbk{
\int_X
\fun{c_{\mathrm{BEC}}}{x}
\opdmsr{\mu(x)}
}^{2}.
$$
The moment formula of Theorem \ref{thm:order-parameter-distribution}
turns this identity into the required fourth-moment factorization for
$\barmean{a}_{\Lambda_n}$.
Even that assertion would not prove that the orbit-space measure is a point
mass.

Equation \eqref{eq:component-condensate-density} measures the uniform mode.
Theorem \ref{thm:condensate-constant} identifies that mode with the unique
macroscopic mode of the symmetric finite-volume Gibbs states.
An exhaustive Penrose--Onsager statement in almost every unrestricted
central component would require a fiberwise bound excluding additional
component-dependent macroscopic modes.
Such a bound is not proved here.
The nonconstant particle density can be retained independently by a joint
push-forward of the central measure with the density variable
$\barmean{n}$ from Proposition \ref{prop:component-density}.
This joint measure is again abstract and need not be supported at total density
$\frac{1}{2}$ by Remark \ref{rem:density-open}.
\end{rem}

\begin{rem}[the expected orbit structure]\label{rem:orbit-structure}
Assertion (3) of Theorem \ref{thm:main-decomposition} constrains the phase
decomposition.
The elements of $\setextremal K_\beta$ supporting $\mu_{\psi_{\beta}}$ occur in
$\fun{\liegr{U}}{1}$ orbits of equal weight.
The expected picture, in analogy with the free Bose gas and mean-field models, is
$$\begin{aligned}
\oastate[\psi_{\beta}]
=
\frac{1}{2\pi}
\int_{0}^{2\pi}
\oastate[\psi_{\beta}^{\rbk{\theta}}]\opdmsr{\theta},
\quad
\oastate[\psi_{\beta}^{\rbk{\theta}}]
=
\oastate[\psi_{\beta}^{\rbk{0}}]\circ\gamma_{\theta},
\quad
\fun{\oastate[\psi_{\beta}^{\rbk{\theta}}]}{a_{x}}
=
\napiernum^{-\imunit\theta}\sqrt{\varrho_{0}}
\neq
0.
\end{aligned}$$
Here $\varrho_{0}$ is the uniform-mode condensate density.
Each pure phase carries a definite phase and breaks gauge symmetry.
Their Haar average restores the symmetry.

Theorems \ref{thm:componentwise-breaking} and
\ref{thm:order-parameter-distribution} prove the componentwise and angular
parts of this picture in the mean.
Components with nonzero order parameter occupy a set of measure at least
$4\kappa$ and have trivial stabilizer.
The mean square of the order parameter is at least $\kappa$.
Its law is exactly rotation invariant.

Two radial statements remain open for the short-range model.
First, the complement of $E$ might have positive $\mu$-measure.
Second, $E$ might contain more than one gauge orbit.
Eliminating both possibilities would identify $\mu_{\psi_{\beta}}$ with the
single normalized orbit measure displayed above.
Theorem \ref{thm:orbit-decomposition} already gives the Haar formula on each
orbit-space fiber.
The second open statement is therefore equivalent to
$\msr{\mu_{\mathrm{orb}}}$ being a point
mass, or to ergodicity of the gauge action on the central measure.

Every one of these statements is a theorem in the mean-field calibration
\ref{sec:meanfield}.
The obstruction for the short-range model is quantitative rather than
structural.
Section \ref{sec:improvements} identifies the missing estimates.
In particular, radial concentration would follow from a fourth-moment
factorization bound for $\barmean{a}_{\Lambda_n}$.
\end{rem}

\begin{rem}[the Mott regime]\label{rem:mott-decomposition}
Under condition \eqref{eq:decay-condition},
Theorem \ref{thm:exponential-decay} gives
$$\abs{\fun{\oastate[\psi_{\beta}]}{\faadj{a_{x}}a_{y}}}
\leq C_{\xi}\napiernum^{-\xi\abs{x-y}}$$
for every limit state from Proposition \ref{prop:limit-state}.
The spatial averages of the one-particle density matrix vanish.
There is no ODLRO, and the obstruction used in
Theorem \ref{thm:main-decomposition} disappears.

One expects uniqueness of the $\rbk{\tau,\beta}$-KMS state in this regime.
Such uniqueness would make the central measure a point mass and the direct
integral trivial.
The methods of
\cite{AizenmanLiebSeiringerSolovejYngvason001} do not prove this statement.
Theorem \ref{thm:kms-decomposition} still decomposes every equilibrium state
into extremal ones.
Section \ref{sec:improvements} explains what a uniqueness proof would require
beyond the chessboard estimates.

The staggered density from Theorem \ref{thm:density-staggered} persists in
both regimes.
The limit state has the period of the optical lattice.
It is invariant under $\ringratint^{d}_{\mathrm{even}}$ but not under the full translation group.
Covariance of the decomposition passes this pattern to almost every
component.
\end{rem}

\begin{rem}[summary]\label{rem:summary}
The operator-algebraic reformulation separates the two roles of the results of \cite{AizenmanLiebSeiringerSolovejYngvason001}.
The infrared bound, the Falk--Bruch inequality, and the sum rule control
zero-mode occupation uniformly in finite volume.
Propositions \ref{prop:odlro-limit} and
\ref{prop:factor-clustering} convert these bounds into a statement about
$K_{\beta}$.
For $d\geq3$, small $\lambda$, and low temperature, the symmetric
equilibrium state is not extremal.
Its maximal orthogonal measure gives a nontrivial direct integral of factor
KMS states.
The gauge group acts measure preservingly on this decomposition.

Theorems \ref{thm:componentwise-breaking},
\ref{thm:order-parameter-distribution}, and
\ref{thm:orbit-decomposition} refine the description.
Components of total weight at least $4\kappa$ have a nonzero order parameter
and trivial gauge stabilizer.
The zero-mode moments determine a rotation-invariant law for this parameter.
The central measure disintegrates into Haar averages along gauge orbits.

The loop representation and chessboard estimates treat the opposite regime.
There the correlations decay exponentially, no condensate exists, and the
chemical potential has the jump from Theorem \ref{thm:mott-gap}.
The equilibrium state is expected to be unique and extremal.

Varying the staggered-potential strength $\lambda$ in
Definition \ref{def:main-optical-lattice-setting}, or the temperature
$\beta^{-1}$ moves the canonical equilibrium state between these two
structures.
In the Choquet simplex, this is a change from an extremal point to the
barycenter of a nontrivial measure.
It is the Bose--Einstein quantum phase transition studied in
\cite{AizenmanLiebSeiringerSolovejYngvason001}.
\end{rem}

\section{Comparison Models}\label{sec:comparison-models}

This section places two exactly tractable models beside the short-range hard-core system. The non-interacting gas separates the effect of the hard-core constraint from that of the staggered potential. The mean-field model supplies the central decomposition and fourth-moment factorization that remain unavailable for the short-range model.

\subsection{The Non-Interacting Gas}\label{sec:freegas}

This section isolates the role of the hard-core interaction. Without that interaction, the staggered potential does not destroy condensation. The condensate wave function is also nonconstant whenever \(\lambda\neq0\). The discussion follows Section VII of \cite{AizenmanLiebSeiringerSolovejYngvason001} and reduces to a one-particle spectral computation.

\subsubsection{One-particle spectrum}\label{one-particle-spectrum}

\begin{prop}[spectrum of the staggered one-particle Hamiltonian]\label{prop:free-spectrum}
On $\fun{\lp^{2}}{\Lambda}$ set
$$h=-\frac{1}{2}\Delta+V.$$
The discrete Laplacian and the staggered potential are
$$\begin{aligned}
\fun{\rbk{\Delta\varphi}}{x}
=
\sum_{y\sim x}\rbk{\fun{\varphi}{y}-\fun{\varphi}{x}},
\quad
\fun{V}{x}
=
\lambda\rbk{-1}^{x}.
\end{aligned}$$
Let $Q=\rbk{\pi,\dotsc,\pi}$ and choose a reduced momentum set
$\dual{\Lambda}_{\mathrm{red}}\subset\dual{\Lambda}$ such that $0\in\dual{\Lambda}_{\mathrm{red}}$ and
\begin{equation}\label{eq:free-reduced-momentum}
\dual{\Lambda}
=
\dual{\Lambda}_{\mathrm{red}}
\sqcup
\rbk{\dual{\Lambda}_{\mathrm{red}}+Q}.
\end{equation}
For $p\in\dual{\Lambda}_{\mathrm{red}}$ and $\sigma\in\setone{-1,+1}$, set
\begin{equation}\label{eq:free-band-energies}
\fun{\varepsilon_{\sigma}}{p}
=
d
+
\sigma\sqrt{\rbk{\sum_{i = 1}^{d}\cos p_{i}}^{2}+\lambda^{2}}.
\end{equation}
The spectrum is
\begin{equation}\label{eq:free-spectrum}
\opvarspec{h}
=
\set{\fun{\varepsilon_{\sigma}}{p}}{p\in\dual{\Lambda}_{\mathrm{red}},\ \sigma\in\setone{-1,+1}},
\end{equation}
and the eigenfunction of the lowest eigenvalue is a staggered mixture of the modes $p = 0$ and $p = \rbk{\pi, \dotsc, \pi}$, not constant for $\lambda \neq 0$.
\end{prop}

\begin{proof}
Let $A = -\Delta - 2 d$, i.e., $\fun{\rbk{A \varphi}}{x} = -\sum_{y \sim x} \fun{\varphi}{y}$.
Multiplication by $\rbk{-1}^{x}$ anticommutes with $A$.
Indeed, $\rbk{-1}^{y}=-\rbk{-1}^{x}$ for $y\sim x$.
It follows that $\anticommutator{V}{A}=0$ and
$$\rbk{h - d}^{2} = \rbk{\frac{1}{2} A + V}^{2} = \frac{1}{4} A^{2} + \lambda^{2}.$$
The cross terms cancel and $V^{2}=\lambda^{2}$.
The operator $A$ is diagonal in the Fourier basis, with eigenvalues
$-2\sum_{i}\cos p_{i}$.
The operator $\rbk{h-d}^{2}$ has eigenvalues
$$\rbk{\sum_{i}\cos p_{i}}^{2}+\lambda^{2}.$$
The potential $V$ couples exactly the modes $p$ and
$p+\rbk{\pi,\dotsc,\pi}$.
These pairs form two-dimensional invariant blocks.
Equation \eqref{eq:free-band-energies} gives the two eigenvalues of every
block for the representative chosen in \eqref{eq:free-reduced-momentum}.
There are $\abscard{\Lambda}/2$ such blocks, so \eqref{eq:free-spectrum} has
exactly $\abscard{\Lambda}$ eigenvalues counted with multiplicity.

For $p=0$, the block is spanned by the constant and fully staggered functions.
In this basis, $h$ has diagonal entries $0$ and $2d$ and off-diagonal entries
$\lambda$.
Its lowest eigenvector has a nonzero staggered component when $\lambda\neq0$.
\end{proof}

\subsubsection{Condensation and condensate-mode fluctuations}\label{condensation-and-condensate-mode-fluctuations}

The quadratic dispersion at the bottom of the lowest band gives the usual free-gas condensation criterion in dimension \(d\geq3\). The grand-canonical zero-mode calculation below records the contrasting fluctuations used later in Section \ref{sec:improvements}. Standard free-gas theory now applies. For small \(p\), the lowest band satisfies \[\fun{\varepsilon_{-}}{p}-\fun{\varepsilon_{-}}{0}
\sim \frac{d}{2\sqrt{d^{2}+\lambda^{2}}}\abs{p}^{2}.\] Writing \(\varepsilon_{0,\Lambda}=\fun{\varepsilon_{-}}{0}\), the finite-volume critical density is \begin{equation}\label{eq:free-critical-density}
\fun{\varrho_{\mathrm{c},\Lambda}}{\beta}
=
\frac{1}{\abscard{\Lambda}}
\sum_{\substack{p\in\dual{\Lambda}_{\mathrm{red}},\ \sigma\in\setone{-1,+1}\\\rbk{p,\sigma}\neq\rbk{0,-1}}}
\frac{1}
{\napiernum^{\beta\rbk{\fun{\varepsilon_{\sigma}}{p}-\varepsilon_{0,\Lambda}}} - 1}.
\end{equation} The Riemann sums in \eqref{eq:free-critical-density} have a finite thermodynamic limit for \(d\geq3\), and that limit tends to zero as \(\beta\to\infty\). At every fixed density, the gas therefore condenses at sufficiently large \(\beta\). The condensate occupies the nonconstant ground mode from Proposition \ref{prop:free-spectrum}.

The following finite-volume statement makes the fluctuation comparison precise. It concerns the grand-canonical free gas; it is not a fixed-particle-number assertion.

\begin{defn}[free-mode annihilation operators]\label{def:free-mode-annihilation}
For an orthonormal basis
$\set{\varphi_{j,\Lambda}}{j\in\Lambda}$
of $\fun{\lp^{2}}{\Lambda}$, let
$b_{j,\Lambda}$
denote the bosonic annihilation operator associated with
$\varphi_{j,\Lambda}$
on $\fun{\spfock_{\txtbsn}}{\fun{\lp^{2}}{\Lambda}}$.
\end{defn}

\begin{lem}[free condensate-mode distribution]\label{lem:free-condensate-mode}
Let $\varphi_{0,\Lambda}$ be a normalized lowest eigenvector of the one-particle Hamiltonian in Proposition \ref{prop:free-spectrum}, with eigenvalue $\varepsilon_{0,\Lambda}$.
Choose an orthonormal eigenbasis $\set{\varphi_{j,\Lambda}}{j\in\Lambda}$ with $\varphi_{0,\Lambda}$ as its first vector and one-particle energies $\varepsilon_{j,\Lambda}$.
Let $b_{j,\Lambda}$ be the free-mode annihilation operators of
Definition \ref{def:free-mode-annihilation}, and set
$n_{0,\Lambda}=\faadj{b_{0,\Lambda}}b_{0,\Lambda}$.
For $\beta>0$ and $\mu<\varepsilon_{0,\Lambda}$, the free grand-canonical Hamiltonian is
\begin{equation}\label{eq:free-grand-canonical-hamiltonian}
\physham^{\mathrm{fr}}_{\Lambda,\mu}
=
\sum_{j\in\Lambda}
\rbk{\varepsilon_{j,\Lambda}-\mu}
\faadj{b_{j,\Lambda}}b_{j,\Lambda}.
\end{equation}
For every bounded Fock-space operator $A$, its Gibbs state at inverse temperature $\beta$ is
\begin{equation}\label{eq:free-grand-canonical-state}
\fun{\oastate[\psi^{\mathrm{fr}}_{\beta,\Lambda,\mu}]}{A}
=
\frac{
\sqfun{\trace_{\fun{\spfock_{\txtbsn}}{\fun{\lp^{2}}{\Lambda}}}}{
A\napiernum^{-\beta\physham^{\mathrm{fr}}_{\Lambda,\mu}}
}
}{
\sqfun{\trace_{\fun{\spfock_{\txtbsn}}{\fun{\lp^{2}}{\Lambda}}}}{
\napiernum^{-\beta\physham^{\mathrm{fr}}_{\Lambda,\mu}}
}
}.
\end{equation}
It has the zero-mode distribution
\begin{equation}\label{eq:free-zero-mode-geometric}
\fun{\oastate[\psi^{\mathrm{fr}}_{\beta,\Lambda,\mu}]}{\fndef{n_{0,\Lambda}=m}}
=
\rbk{1-q_{\beta,\Lambda,\mu}}q_{\beta,\Lambda,\mu}^{m},
\quad
q_{\beta,\Lambda,\mu}
=
\napiernum^{-\beta\rbk{\varepsilon_{0,\Lambda}-\mu}},
\quad
m\in\monnat.
\end{equation}
Consequently,
\begin{equation}\label{eq:free-zero-mode-second-moment}
\fun{\oastate[\psi^{\mathrm{fr}}_{\beta,\Lambda,\mu}]}{n_{0,\Lambda}^{2}}
=
2\fun{\oastate[\psi^{\mathrm{fr}}_{\beta,\Lambda,\mu}]}{n_{0,\Lambda}}^{2}
+
\fun{\oastate[\psi^{\mathrm{fr}}_{\beta,\Lambda,\mu}]}{n_{0,\Lambda}}.
\end{equation}
\end{lem}

\begin{proof}
Diagonalizing the one-particle Hamiltonian puts \eqref{eq:free-grand-canonical-hamiltonian} into the occupation-number basis of the bosonic Fock space.
The Gibbs state \eqref{eq:free-grand-canonical-state} then factorizes over one-particle modes.
For the lowest mode, the normalized weight of the occupation number $m$ is the geometric weight in \eqref{eq:free-zero-mode-geometric}.
Its mean and variable are
$$\fun{\oastate[\psi^{\mathrm{fr}}_{\beta,\Lambda,\mu}]}
{n_{0,\Lambda}}
=
\frac{q_{\beta,\Lambda,\mu}}{1-q_{\beta,\Lambda,\mu}},
\quad
\fun{\prbvar_{\psi^{\mathrm{fr}}_{\beta,\Lambda,\mu}}}
{n_{0,\Lambda}}
=
\frac{q_{\beta,\Lambda,\mu}}
{\rbk{1-q_{\beta,\Lambda,\mu}}^{2}}.$$
Together with the identity
$$\fun{\oastate[\psi^{\mathrm{fr}}_{\beta,\Lambda,\mu}]}
{n_{0,\Lambda}^{2}}
=
\fun{\prbvar_{\psi^{\mathrm{fr}}_{\beta,\Lambda,\mu}}}
{n_{0,\Lambda}}
+\fun{\oastate[\psi^{\mathrm{fr}}_{\beta,\Lambda,\mu}]}
{n_{0,\Lambda}}^{2},$$
these formulas give \eqref{eq:free-zero-mode-second-moment}.
\end{proof}

\subsubsection{Condensate density and gauge phase in the free gas}\label{condensate-density-and-gauge-phase-in-the-free-gas}

Proposition \ref{prop:free-spectrum} fixes the condensate wave function as the lowest one-particle eigenvector. For \(\lambda\neq0\), this fixed vector has a nonconstant two-sublattice profile. The profile does not prevent a scalar condensate-density coordinate because every condensate amplitude is a complex scalar multiplying the same vector. Its polar coordinates separate the condensate density from the \(\fun{\liegr{U}}{1}\) phase.

The grand-canonical radial fluctuations are explicit already at finite volume.

\begin{cor}[free grand-canonical condensate-density law]\label{cor:free-condensate-density-law}
Assume that the parameters in Lemma \ref{lem:free-condensate-mode} depend on
$\Lambda$ and satisfy
\begin{equation}\label{eq:free-condensate-density-scaling}
\frac{1}{\abscard{\Lambda}}
\fun{\oastate[\psi^{\mathrm{fr}}_{\beta,\Lambda,\mu}]}{n_{0,\Lambda}}
\to
\varrho_{0}
>
0.
\end{equation}
The geometric law \eqref{eq:free-zero-mode-geometric} gives, for every
$t
\geq
0$,
\begin{equation}\label{eq:free-condensate-density-laplace-transform}
\fun{\oastate[\psi^{\mathrm{fr}}_{\beta,\Lambda,\mu}]}{
\napiernum^{-t n_{0,\Lambda}/\abscard{\Lambda}}
}
\to
\frac{1}{1+\varrho_{0}t}.
\end{equation}
Equation \eqref{eq:free-condensate-density-laplace-transform} identifies the
limiting distribution of $n_{0,\Lambda}/\abscard{\Lambda}$ as the
exponential law of mean $\varrho_{0}$.
This is the explicit condensate-density law associated with the
grand-canonical zero-mode fluctuations.
\end{cor}

\begin{proof}
Write
$q_{\Lambda}=q_{\beta,\Lambda,\mu}$.
The mean formula in Lemma \ref{lem:free-condensate-mode} and
\eqref{eq:free-condensate-density-scaling} give
$$
\abscard{\Lambda}\rbk{1-q_{\Lambda}}
\to
\frac{1}{\varrho_{0}}.
$$
Summing the geometric series in
\eqref{eq:free-zero-mode-geometric} gives
$$
\fun{\oastate[\psi^{\mathrm{fr}}_{\beta,\Lambda,\mu}]}{
\napiernum^{-t n_{0,\Lambda}/\abscard{\Lambda}}
}
=
\frac{1-q_{\Lambda}}
{1-q_{\Lambda}\napiernum^{-t/\abscard{\Lambda}}}.
$$
Multiplication of numerator and denominator by $\abscard{\Lambda}$ proves
\eqref{eq:free-condensate-density-laplace-transform}.
\end{proof}

The interacting hard-core model has no corresponding geometric or quasi-free formula. Proposition \ref{prop:condensate-density-disintegration} still gives an abstract direct integral over \(c_{\mathrm{BEC}}\), but \(\msr{\nu}_{\mathrm{BEC}}\) and its conditional KMS states remain unknown. The contrast is caused by the loss of the free-mode probability law, not by the nonconstant particle density of Theorem \ref{thm:density-staggered}. Theorems \ref{thm:exponential-decay} and \ref{thm:condensate-constant} show what the hard core changes. It creates the Mott phase and makes the normalized condensate eigenfunction asymptotically constant in the norm specified by Corollary \ref{cor:condensate-asymptotically-constant}.

\subsection{The Mean-Field Calibration}\label{sec:meanfield}

The complete-graph hard-core Bose gas is the mean-field \(XY\) model. Its exact solution and condensation are existing results of \cite{BalintToth004,OliverPenrose004}, and the general characterization of limiting Gibbs states for quantum mean-field models is due to \cite{FannesSpohnVerbeure001}. The product-state barycentric representation of permutation-invariant states is Theorem \ref{thm:stormer-de-finetti}, proved in Appendix \ref{app:stormer} following Størmer \cite{ErlingStormer003}. The following calculation uses these known inputs together with the direct-integral theory of Appendix \ref{app:spatial-direct-integrals} to make the central decomposition of this particular limiting Gibbs state explicit. In particular, the phase-circle representation and the fourth-moment factorization below are derived consequences, written here for comparison with the conclusions that remain unavailable for the short-range model; they are not asserted to be separate results of \cite{BalintToth004,OliverPenrose004,FannesSpohnVerbeure001}.

The mean-field interaction is not quasi-local. It defines neither an infinite-volume dynamics on \(\oa{A}\) nor a KMS theory. We therefore define equilibrium states as limits of finite-volume Gibbs states. This setting still calibrates the decomposition theory of Section \ref{sec:decomposition}. Theorem \ref{thm:factor-barycenter-direct-integral} and Proposition \ref{prop:spatial-central-measure-criterion} in Appendix \ref{app:spatial-direct-integrals} depend only on the state and its GNS representation, not on a dynamics.

The order parameter \(T\), its law \(\nu\), the orbit-space measure \(\msr{\mu_{\mathrm{orb}}}\), and the component density can all be computed explicitly. The fourth moments of the spatial averages factorize. The gauge action on the central measure is ergodic, and the density equals \(\frac{1}{2}\). The constants in Theorem \ref{thm:order-parameter-distribution} are optimal.

\subsubsection{The model and its free energy}\label{the-model-and-its-free-energy}

The complete-graph hopping reduces the equilibrium problem to a one-variable variational principle. The maximizers determine both the pressure and the concentration of the total transverse spin. On a finite set \(\Lambda\) of \(N\) sites, geometry being irrelevant, set \[\physham^{\mathrm{mf}}_{\Lambda} = -\frac{1}{\abscard{\Lambda}} \sqbk{\rbk{S^{1}_{\txttot,\Lambda}}^{2} + \rbk{S^{2}_{\txttot,\Lambda}}^{2}}
= -\frac{1}{\abscard{\Lambda}} \sqbk{S_{\txttot,\Lambda} \cdot S_{\txttot,\Lambda} - \rbk{S^{3}_{\txttot,\Lambda}}^{2}},\] which in the boson language is the hopping of \eqref{eq:main-hamiltonian} with the pair \(\rbk{x, y}\) ranging over all pairs of sites with strength \(\abscard{\Lambda}^{-1}\), up to an additive constant.

\begin{prop}[mean-field free energy and concentration]\label{prop:mf-free-energy}
Let $$\fun{s}{\sigma}
=
-\rbk{\frac{1}{2} + \sigma}
\fun{\log}{\frac{1}{2} + \sigma}
-\rbk{\frac{1}{2} - \sigma}
\fun{\log}{\frac{1}{2} - \sigma}$$ for $\sigma \in \closedinterval{0}{\frac{1}{2}}$.
The pressure satisfies
\begin{equation}\label{eq:mf-pressure-variational-formula}
\frac{1}{N} \log \sqfun{\trace}{\napiernum^{-\beta \physham^{\mathrm{mf}}_{\Lambda}}}
\to \fun{p}{\beta} = \max_{0 \leq \sigma \leq \onehalf} \sqbk{\fun{s}{\sigma} + \beta \sigma^{2}}.
\end{equation}
The function
$$\fun{h}{\sigma}
=
\frac{1}{\sigma}
\fun{\arctanh}{2\sigma}$$
is a strictly increasing bijection from
$\openinterval{0}{\frac{1}{2}}$ onto
$\openinterval{2}{\infty}$.
For $\beta\leq2$, the unique maximizer of the variational problem
\eqref{eq:mf-pressure-variational-formula} is $\sigma=0$.
For $\beta>2$, the unique maximizer of the same variational problem is
\begin{equation}\label{eq:mf-order-parameter}
\fun{\bar{\sigma}}{\beta}
=
\fun{h^{-1}}{\beta}
\in
\openinterval{0}{\frac{1}{2}}.
\end{equation}
The number $\fun{\bar{\sigma}}{\beta}$ is the unique positive solution of
$$\fun{\arctanh}{2\sigma}=\beta\sigma.$$
It is strictly increasing in $\beta$.
Moreover it holds that
\begin{equation}\label{eq:mf-order-parameter-limits}
\begin{aligned}
\lim_{\beta\downarrow2}
\fun{\bar{\sigma}}{\beta}
&=
0,
\quad
\lim_{\beta\uparrow\infty}
\fun{\bar{\sigma}}{\beta}
=
\frac{1}{2}.
\end{aligned}
\end{equation}

For the concentration statement, set $\bar{\sigma}=0$ when
$\beta\leq2$, and set
$\bar{\sigma}=\fun{\bar{\sigma}}{\beta}$ when $\beta>2$.
The Gibbs state also concentrates exponentially near $\bar{\sigma}$.
For every $\epsilon>0$, there is $c>0$ such that the joint spectral
projection of
$\rbk{S_{\txttot,\Lambda}\cdot S_{\txttot,\Lambda},S^{3}_{\txttot,\Lambda}}$ outside
$$\set{\rbk{s,k}}
{\abs{\frac{s}{N}-\bar{\sigma}}\leq\epsilon,
\quad
\frac{\abs{k}}{N}\leq\epsilon}$$
has Gibbs expectation at most $\napiernum^{-cN}$ for all large $N$.
\end{prop}

\begin{proof}
The representation $\rbk{\fldcmp^{2}}^{\otimes N}$ of $\fun{\liegr{SU}}{2}$ decomposes into irreducible components of spin $s$, $N / 2 - s \in \setone{0, 1, \dotsc}$, with multiplicities
$$c_{N, s} = \binom{N}{N / 2 - s} - \binom{N}{N / 2 - s - 1},$$
the difference of the dimensions of the weight spaces of weights $s$ and $s + 1$, since each spin-$s'$ component with $s' \geq s$ contributes exactly one weight-$s$ vector.
On the component of quantum numbers $\rbk{s, k}$ the Hamiltonian has the eigenvalue $-N^{-1} \sqbk{\fun{s}{s + 1} - k^{2}}$, so
$$Z_{N} = \sqfun{\trace}{\napiernum^{-\beta \physham^{\mathrm{mf}}}} = \sum_{s} c_{N, s} \sum_{k = -s}^{s} \napiernum^{\beta N^{-1} \sqbk{s \rbk{s + 1} - k^{2}}}.$$
The standard binomial bounds and the ratio identity give
$$c_{N,s}
=\napiernum^{N\fun{s}{\frac{s}{N}}+\fun{O}{\log N}}$$
uniformly in $s$.
They also give
$$\frac{s\rbk{s+1}}{N}
=N\rbk{\frac{s}{N}}^{2}+\fun{O}{1}.$$
The partition function $Z_{N}$ is a sum of at most
$\rbk{N+1}^{2}$ terms of the form
$$\napiernum^{N\fun{\varphi}{\sigma,\kappa}+\fun{O}{\log N}},
\quad
\fun{\varphi}{\sigma,\kappa}
=
\fun{s}{\sigma}+\beta\rbk{\sigma^{2}-\kappa^{2}},
\quad
\sigma
=
\frac{s}{N},
\quad
\kappa
=
\frac{k}{N}.$$
For fixed $\sigma$, the maximum over $\abs{\kappa}\leq\sigma$ occurs at
$\kappa=0$.
It follows that
$$\frac{1}{N}\log Z_{N}
\to\max_{\sigma}\fun{\varphi}{\sigma,0}
=\fun{p}{\beta}.$$
This proves \eqref{eq:mf-pressure-variational-formula}.

To determine the maximizer in
\eqref{eq:mf-pressure-variational-formula}, consider the one-variable
profile $\sigma\mapsto\fun{s}{\sigma}+\beta\sigma^{2}$.
The identity
$$\fun{\arctanh}{2\sigma}
=\int_{0}^{\sigma}\frac{2}{1-4t^{2}}\opdmsr{t}$$
shows that $\fun{h}{\sigma}$ is the average of the strictly increasing
function $2/\rbk{1-4t^{2}}$ on $\closedinterval{0}{\sigma}$.
The function $h$ is strictly increasing.
Its left endpoint limit is $\fun{h}{0^{+}}=2$.
At the other endpoint,
$$\fun{h}{\sigma}
\to
\infty
\quad
\text{as }\sigma\uparrow\frac{1}{2}.$$
For $\sigma>0$,
$$\oppd{\sigma}\fun{\varphi}{\sigma,0}
=2\sigma\sqbk{\beta-\fun{h}{\sigma}}.$$
The profile is strictly decreasing when $\beta\leq2$.
When $\beta>2$, it first increases and then decreases, with its unique
maximum at $\bar{\sigma}=\fun{h^{-1}}{\beta}$.
The preceding derivative analysis proves that $\sigma=0$ is the unique
maximizer when $\beta\leq2$, while
$\fun{\bar{\sigma}}{\beta}=\fun{h^{-1}}{\beta}$ is the unique maximizer
when $\beta>2$.
The asserted dependence on $\beta$ follows from the properties of $h^{-1}$.

For either of these unique maximizers, continuity and compactness give a gap
$\fun{c'}{\epsilon}>0$ between its value and the supremum of the profile
outside its $\epsilon$ neighborhood.
The relative spectral weight of that complement is bounded by
$$\rbk{N+1}^{2}\napiernum^{-N\fun{c'}{\epsilon}+\fun{O}{\log N}}.$$
This proves exponential concentration.
\end{proof}

\subsubsection{The limit state and its Størmer decomposition}\label{the-limit-state-and-its-stuxf8rmer-decomposition}

The variational maximizers determine the full thermodynamic limit of the mean-field Gibbs states. Theorem \ref{thm:stormer-de-finetti} identifies that limit as the Haar barycenter of symmetry-breaking product states.

\begin{thm}[the mean-field equilibrium state]\label{thm:mf-state}
Let $\oastate[\psi_{N}]$ be the Gibbs state of
$\physham^{\mathrm{mf}}_{\Lambda_{N}}$
at inverse temperature
$\beta$,
and choose a state extension to
$\oa{A}$
for each member of an arbitrary exhaustion.
For $\beta > 2$,
\begin{equation}\label{eq:mf-equilibrium-barycenter}
\oastate[\psi_{N}]
\to
\oastate[\psi^{\mathrm{mf}}]
=\frac{1}{2\pi}\int_{0}^{2\pi}
\oastate[\psi_{\theta}]\opdmsr{\theta}
\end{equation}
in the weak-$\ast$ topology, where
\begin{equation}\label{eq:mf-product-state-density-matrices}
\begin{aligned}
\oastate[\psi_{\theta}]
=
\rho_{\theta}^{\otimes\infty},
\quad
\rho_{\theta}
=
\frac{1}{2}
+2\bar{\sigma}\rbk{S^{1}\cos\theta+S^{2}\sin\theta}.
\end{aligned}
\end{equation}
Here $\bar{\sigma}=\fun{\bar{\sigma}}{\beta}$ is defined in
\eqref{eq:mf-order-parameter}.
For every $x$,
\begin{equation}\label{eq:mf-product-state-order-parameter}
\fun{\oastate[\psi_{\theta}]}{a_{x}}
=
\bar{\sigma}\napiernum^{-\imunit\theta}.
\end{equation}
The thermodynamic limit exists along the full exhaustion.
For $\beta \leq 2$ the limit is the product of normalized traces.
\end{thm}

\begin{proof}
Each $\physham^{\mathrm{mf}}_{\Lambda}$ is invariant under every permutation
unitary and commutes with $S^{3}_{\txttot,\Lambda}$.
Its Gibbs state is therefore permutation invariant and gauge invariant.
Weak-$\ast$ convergence preserves the finite-permutation and gauge
invariance identities,
so every limit point
$\oastate[\psi]$
has both invariances.
The one-site Bloch ball is
\begin{equation}\label{eq:mf-bloch-ball}
\mathcal{B}
=
\set{\rho\in\spmat{2}{\fldcmp}}{\rho\geq0,\ \sqfun{\trace}{\rho}=1}.
\end{equation}
By Theorem \ref{thm:stormer-de-finetti},
the exchangeable states of the quasi-local algebra
$\oa{A}$
defined in
\eqref{eq:main-local-algebras}
form a simplex whose extreme points are the product states.
The state $\oastate[\psi]$ therefore has a unique representation
\begin{equation}\label{eq:mf-stormer-product-mixture}
\oastate[\psi]
=
\int_{\mathcal{B}}
\rho^{\otimes\infty}
\opdmsr{P(\rho)}
\end{equation}
for a Radon probability measure $P$ on the Bloch ball
\eqref{eq:mf-bloch-ball}.
The corresponding correlations are given by
\eqref{eq:stormer-moment-representation}.

The moments of $P$ are computed from the Gibbs states.
By permutation invariance the correlation
$\fun{\oastate[\psi_{N}]}{S^{i}_{x} S^{i}_{y}}$,
$x \neq y$, is independent of the pair.  Hence
$$\fun{\oastate[\psi_{N}]}{S^{1}_{x} S^{1}_{y} + S^{2}_{x} S^{2}_{y}} = \frac{\fun{\oastate[\psi_{N}]}{S_{\txttot,\Lambda}\cdot S_{\txttot,\Lambda}-\rbk{S^{3}_{\txttot,\Lambda}}^{2}} - N / 2}{N \rbk{N - 1}}.$$
Moreover,
$$\fun{\oastate[\psi_{N}]}{S^{3}_{x} S^{3}_{y}} = \frac{\fun{\oastate[\psi_{N}]}{\rbk{S^{3}_{\txttot,\Lambda}}^{2}} - N / 4}{N \rbk{N - 1}}.$$
where the diagonal contributions
$N/2$
and
$N/4$
follow from
\eqref{eq:app-one-site-spin-squares}.
Apply Proposition \ref{prop:mf-free-energy} to bounded functions of the
commuting pair
$\rbk{S_{\txttot,\Lambda}\cdot S_{\txttot,\Lambda},S^{3}_{\txttot,\Lambda}}$.
It gives
$$\begin{aligned}
\frac{\fun{\oastate[\psi_{N}]}{S_{\txttot,\Lambda}\cdot S_{\txttot,\Lambda}-\rbk{S^{3}_{\txttot,\Lambda}}^{2}}}{N^{2}}
\to
\bar{\sigma}^{2},
\quad
\frac{\fun{\oastate[\psi_{N}]}{\rbk{S^{3}_{\txttot,\Lambda}}^{2}}}{N^{2}}
\to
0.
\end{aligned}$$
Convergence of the two-site marginals yields
$$\begin{aligned}
\int_{\mathcal{B}}
\rbk{\sqfun{\trace}{\rho S^{1}}^{2}+\sqfun{\trace}{\rho S^{2}}^{2}}
\opdmsr{P(\rho)}
=\bar{\sigma}^{2},
\quad
\int_{\mathcal{B}}
\sqfun{\trace}{\rho S^{3}}^{2}
\opdmsr{P(\rho)}
=
0.
\end{aligned}$$

We also need the fourth moment.
Expand
$$\rbk{S_{\txttot,\Lambda}\cdot S_{\txttot,\Lambda}-\rbk{S^{3}_{\txttot,\Lambda}}^{2}}^{2}
=\sqbk{\sum_{x,y}
\rbk{S^{1}_{x}S^{1}_{y}+S^{2}_{x}S^{2}_{y}}}^{2}.$$
Only $\fun{O}{N^{3}}$ quadruples contain a repeated site.
There are
$N\rbk{N-1}\rbk{N-2}\rbk{N-3}$ quadruples of distinct sites.
All summands are uniformly bounded.
Concentration gives
$$\int_{\mathcal{B}}
\rbk{\sqfun{\trace}{\rho S^{1}}^{2}+\sqfun{\trace}{\rho S^{2}}^{2}}^{2}
\opdmsr{P(\rho)}
=
\lim_{N \to \infty}
\frac{\fun{\oastate[\psi_{N}]}{\rbk{S_{\txttot,\Lambda}\cdot S_{\txttot,\Lambda}-\rbk{S^{3}_{\txttot,\Lambda}}^{2}}^{2}}}{N^{4}}
= \bar{\sigma}^{4},$$
The variance of
$\sqfun{\trace}{\rho S^{1}}^{2}+\sqfun{\trace}{\rho S^{2}}^{2}$
under $P$ vanishes.
For $P$-almost every $\rho$,
$$\sqfun{\trace}{\rho S^{1}}^{2}+\sqfun{\trace}{\rho S^{2}}^{2}
=\bar{\sigma}^{2}.$$
Moreover,
$$\sqfun{\trace}{\rho S^{3}}=0.$$
A state of $\spmat{2}{\fldcmp}$ is its Bloch vector
$\rbk{2\sqfun{\trace}{\rho S^{1}},2\sqfun{\trace}{\rho S^{2}},2\sqfun{\trace}{\rho S^{3}}}$, so $P$ is carried by the circle $\setone{\rho_{\theta}}$.
Gauge transformations rotate the one-site marginal of a product state.
Gauge invariance of $\oastate[\psi]$ and uniqueness of $P$ make $P$
rotation invariant on this circle.
The action is transitive, and the only invariant probability measure is the
uniform measure.
Every limit point has the same Størmer measure and equals
$\oastate[\psi^{\mathrm{mf}}]$.
Compactness of the state space then gives convergence of the full net.
For $\beta \leq 2$ the same computation gives
$\sqfun{\trace}{\rho S^{1}}^{2}+\sqfun{\trace}{\rho S^{2}}^{2}
=0=\sqfun{\trace}{\rho S^{3}}$
for $P$-almost every $\rho$, so $P$ is the point mass at the tracial state
and $\oastate[\psi] = \sqfun{\trace}{\cdot}^{\otimes \infty}$.
Finally $\fun{\oastate[\psi_{\theta}]}{a_{x}} = \sqfun{\trace}{\rho_{\theta} \rbk{S^{1} - \imunit S^{2}}} = \bar{\sigma} \rbk{\cos \theta - \imunit \sin \theta} = \bar{\sigma} \napiernum^{-\imunit \theta}$, which is \eqref{eq:mf-product-state-order-parameter}.
\end{proof}

\subsubsection{The central decomposition of the mean-field state}\label{the-central-decomposition-of-the-mean-field-state}

The product-state barycenter admits an explicit direct-integral representation. Its center, central measure, and spatial-average limit give a complete calibration of the abstract decomposition results for the short-range model. The general spatial-decomposition criterion is Theorem \ref{thm:factor-barycenter-direct-integral} in Appendix \ref{app:spatial-direct-integrals}.

\begin{lem}[disjoint mean-field phases]\label{lem:mf-disjoint-phases}
Let $\beta > 2$.
Let $\bar{\sigma}$ be defined by \eqref{eq:mf-order-parameter}, and let
$\oastate[\psi_{\theta}]$ be defined by
\eqref{eq:mf-product-state-density-matrices}.
For
$\theta,\theta'\in\rightopeninterval{0}{2\pi}$
with $\theta\neq\theta'$, the states
$\oastate[\psi_{\theta}]$ and $\oastate[\psi_{\theta'}]$ are factor states
and are disjoint.
\end{lem}

A detailed construction of the incomplete infinite tensor product and its product-state representations is given in \cite[Appendix E]{YoshitsuguSekine012}.

\begin{proof}
The relation
$2\bar{\sigma}
=
\fun{\tanh}{\beta\bar{\sigma}}
<1$
shows that $\rho_{\theta}$ has eigenvalues
$\frac{1}{2}\pm\bar{\sigma}$ in $\openinterval{0}{1}$.
Thus $\rho_{\theta}$ is strictly positive, and the GNS representation of
$\oastate[\psi_{\theta}]$ generates an infinite tensor product factor
\cite{VonNeumannJohn001}.

Set
$c_{\theta}=\bar{\sigma}\napiernum^{-\imunit\theta}$.
Product structure gives
$$
\fun{\oastate[\psi_{\theta}]}{\barmean{a}_{\Lambda_n}}
=
c_{\theta},
\quad
\fun{\oastate[\psi_{\theta}]}{\faadj{\barmean{a}_{\Lambda_n}}\barmean{a}_{\Lambda_n}}
=
\abs{c_{\theta}}^{2}
+
\fun{O}{\abscard{\Lambda_n}^{-1}}.
$$
The off-diagonal terms factorize exactly.
Consequently,
$$
\norm{\rbk{\fun{\oarepn_{\theta}}{\barmean{a}_{\Lambda_n}}-c_{\theta}}
\Omega_{\theta}}^{2}
=
\fun{O}{\abscard{\Lambda_n}^{-1}}.
$$
For every local $A$,
$$
\norm{\commutator{
\fun{\oarepn_{\theta}}{\barmean{a}_{\Lambda_n}}}
{\fun{\oarepn_{\theta}}{A}}}
=
\fun{O}{\abscard{\Lambda_n}^{-1}}.
$$
Density and uniform boundedness give strong convergence to $c_{\theta}$;
the adjoints converge strongly to $\cmpconj{c_{\theta}}$.

If the states at $\theta$ and $\theta'$ were quasi-equivalent, the induced
normal isomorphism of the generated factors would preserve the
$\sigma$-weak limit of the bounded spatial averages.
It would force $c_{\theta}=c_{\theta'}$.
Distinct angles modulo $2\pi$ therefore give inequivalent factor
representations.
Two factor representations are either quasi-equivalent or disjoint
\cite[Section 2.4]{BratteliRobinson003}.
\end{proof}

\begin{thm}[central decomposition of the mean-field state]\label{thm:mf-central}
Let $\beta>2$, and let $\oastate[\psi^{\mathrm{mf}}]$ and
$\oastate[\psi_{\theta}]$ be defined by
\eqref{eq:mf-equilibrium-barycenter} and
\eqref{eq:mf-product-state-density-matrices}.
The GNS Hilbert space of $\oastate[\psi^{\mathrm{mf}}]$ is
\begin{equation}\label{eq:mf-central-gns-space}
\int_{\rightopeninterval{0}{2\pi}}^{\oplus}
\sphilb{H}_{\theta}
\frac{\opdmsr{\theta}}{2\pi}.
\end{equation}
Its center is the diagonal algebra
$\fun{\lp^{\infty}}{\rightopeninterval{0}{2\pi},\opdmsr{\theta}/2\pi}$.
The central measure is the image of normalized Haar measure under
$\rightopeninterval{0}{2\pi}\ni\theta\mapsto\oastate[\psi_{\theta}]$.
The central decomposition is exactly the Haar-orbit integral in
Theorem \ref{thm:mf-state}.
\end{thm}

\begin{proof}
Set
$$
X
=
\rightopeninterval{0}{2\pi},
\quad
\opdmsr{\mu(\theta)}
=
\frac{\opdmsr{\theta}}{2\pi},
\quad
B_n
=
\barmean{a}_{\Lambda_n},
\quad
\fun{c}{\theta}
=
\bar{\sigma}\napiernum^{-\imunit\theta}.
$$
Lemma \ref{lem:mf-disjoint-phases} gives the fiberwise strong limits
\begin{equation}\label{eq:mf-central-fiberwise-scalar-limit}
\slim_{n\to\infty}
\fun{\oarepn_\theta}{B_n}
=
\fun{c}{\theta}1,
\quad
\slim_{n\to\infty}
\fun{\oarepn_\theta}{\faadj{B_n}}
=
\cmpconj{\fun{c}{\theta}}1.
\end{equation}
The sequence $\seq{B_n}{n\in\semigrposint}$ is uniformly bounded, and $c$
is injective on $X$.
Theorem \ref{thm:factor-barycenter-direct-integral} in Appendix
\ref{app:spatial-direct-integrals}, applied through
\eqref{eq:mf-central-fiberwise-scalar-limit}, identifies
\eqref{eq:mf-central-gns-space} as the GNS Hilbert space and its center as
the diagonal algebra.
Corollary \ref{cor:factor-barycenter-central-measure} identifies the central
measure as the image of $\mu$ under
$\theta\mapsto\oastate[\psi_\theta]$.
The barycenter formula \eqref{eq:mf-equilibrium-barycenter} therefore is the
central decomposition.
\end{proof}

\begin{prop}[mean-field order-parameter law]\label{prop:mf-order-parameter-law}
Under the hypotheses of Theorem \ref{thm:mf-central}, the represented
spatial averages converge strongly:
\begin{equation}\label{eq:mf-central-spatial-average-limit}
\slim_{n \to \infty}
\fun{\oarepn_{\psi^{\mathrm{mf}}}}{\barmean{a}_{\Lambda_n}}
=
C
\end{equation}
where the central element $C$ has fiber function
$\fun{c}{\theta}
=
\bar{\sigma}\napiernum^{-\imunit\theta}$.
The order-parameter law is normalized Haar measure on the circle
$\abs{z}=\bar{\sigma}$, and the analogue of $E$ has full measure.
Fix an ordered word in the two formal scalar letters $z$ and
$\cmpconj{z}$, with $k$ occurrences of $\cmpconj{z}$ and $m$ occurrences
of $z$.
Let $w_n$ be the operator obtained by replacing $z$ and $\cmpconj{z}$ with
$\barmean{a}_{\Lambda_n}$ and
$\faadj{\barmean{a}_{\Lambda_n}}$, respectively.
The scalar expectations of this family satisfy
\begin{equation}\label{eq:mf-central-word-moments}
\lim_{n \to \infty}
\fun{\oastate[\psi^{\mathrm{mf}}]}{w_n}
=\kroneckerdelta_{km}\bar{\sigma}^{2k}.
\end{equation}
The fourth moments of the spatial averages factorize.
\end{prop}

\begin{proof}
Let $\rbk{\oarepn_{\psi^{\mathrm{mf}}},\sphilb{H}_{\psi^{\mathrm{mf}}},
\oagnsvector[\Psi]}$ be the direct-integral GNS realization supplied by
Theorem \ref{thm:factor-barycenter-direct-integral}.
Under this identification, $C$ and $\faadj{C}$ are the strong limits of the
spatial averages and their adjoints.
Products of uniformly bounded strongly convergent nets converge strongly.
Thus
$$
\begin{aligned}
\lim_{n \to \infty}
\fun{\oastate[\psi^{\mathrm{mf}}]}{w_n}
&=
\bkt{\oagnsvector[\Psi]}{\rbk{\faadj{C}}^{k}C^{m}
\oagnsvector[\Psi]}
\\
&=
\int_{0}^{2\pi}
\bar{\sigma}^{k+m}\napiernum^{\imunit\theta\rbk{k-m}}
\frac{\opdmsr{\theta}}{2\pi}
=
\kroneckerdelta_{km}\bar{\sigma}^{2k}.
\end{aligned}
$$
The fiber function of $C$ has constant modulus $\bar{\sigma}>0$.
Its law is uniform on the circle of that radius, and the analogue of $E$ has
full measure.
The cases $\rbk{k,m}=\rbk{2,2}$ and
$\rbk{k,m}=\rbk{1,1}$ give the factorization criterion.
\end{proof}

\begin{cor}[mean-field component structure]\label{cor:mf-component-structure}
Under the hypotheses of Theorem \ref{thm:mf-central},
for every $\theta\in\rightopeninterval{0}{2\pi}$,
the phase-component identity,
with $\oastate[\psi_{0}]$ defined by \eqref{eq:mf-product-state-density-matrices} at $\theta=0$, is
\begin{equation}\label{eq:mf-phase-gauge-orbit}
\oastate[\psi_{\theta}]
=
\oastate[\psi_{0}]\circ\gamma_{\theta}.
\end{equation}
The central measure in Theorem \ref{thm:mf-central} is supported on
$$\set{\oastate[\psi_{0}]\circ\gamma_{\theta}}
{\theta\in\rightopeninterval{0}{2\pi}}.$$
The gauge action is transitive and ergodic, and the orbit-space measure is a
point mass.
Finally,
\begin{equation}\label{eq:mf-component-half-density}
\fun{\oastate[\psi_{\theta}]}{n_{x}}=\frac{1}{2}
\end{equation}
for every $\theta\in\rightopeninterval{0}{2\pi}$ and
$x\in\ringratint^{d}$.
Thus the component density is independent of $\theta$, and this decomposition
exhibits no density phase separation.
\end{cor}

\begin{proof}
Gauge rotation acts by
$$
\oastate[\psi_{\theta}]\circ\gamma_{\varphi}
=
\oastate[\psi_{\theta+\varphi}].
$$
The product states form one transitive gauge orbit.
The orbit-space measure is a point mass, and the action on the central
measure is ergodic.
Finally,
$$
\fun{\oastate[\psi_{\theta}]}{n_x}
=
\frac{1}{2}
+
\sqfun{\trace}{\rho_{\theta}S^3}
=
\frac{1}{2}.
$$
Product structure makes the componentwise macroscopic density fluctuations
vanish.
Together with \eqref{eq:mf-component-half-density}, this proves that every
component has the sharp density $\frac{1}{2}$.
\end{proof}

\begin{cor}[optimal mean-field constants]\label{cor:mf-optimal-constants}
Let $\fun{\bar{\sigma}}{\beta}$ be the positive solution of
\eqref{eq:mf-order-parameter} for $\beta>2$.
Then
\begin{equation}\label{eq:mf-optimal-radius-limits}
\lim_{\beta\uparrow\infty}\fun{\bar{\sigma}}{\beta}
=
\frac{1}{2},
\quad
\lim_{\beta\uparrow\infty}\fun{\bar{\sigma}}{\beta}^{2}
=\frac{1}{4}.
\end{equation}
The disc radius $\frac{1}{2}$ and the second-moment bound $\frac{1}{4}$ in
Theorem \ref{thm:order-parameter-distribution} are optimal.
\end{cor}

\begin{proof}
Equation \eqref{eq:mf-optimal-radius-limits} follows from
\eqref{eq:mf-order-parameter-limits}.
Moreover,
$$\int_{\fldcmp}\abs{z}^{2}\opdmsr{\nu^{\mathrm{mf}}(z)}
=\bar{\sigma}^{2}.$$
The operator inequality \eqref{eq:casimir-box,-operator-bound} is model
independent.
Within its scope, the constants $\frac{1}{2}$ and $\frac{1}{4}$ in
Theorem \ref{thm:order-parameter-distribution} cannot be improved.
\end{proof}
\begin{rem}[what the calibration shows]
The one-type mean-field model proves every statement left open in
Remark \ref{rem:orbit-structure}.
Its order-parameter law is the Haar law on a circle of radius
$\bar{\sigma}$, so $\varrho_{0}=\bar{\sigma}^{2}>0$.
The set $E$ has full measure.
The central measure lies on one gauge orbit, and the gauge action is
ergodic.
Fourth moments factorize with constant $1$.
Every factor phase has density $\frac{1}{2}$.

These properties are therefore mutually compatible.
No structural obstruction prevents the expected short-range picture.
What is missing in the short-range model is the concentration supplied here
by the mean-field spectral decomposition.

The calibration has one limitation.
The mean-field interaction defines no dynamics on $\oa{A}$ and therefore
does not test the KMS theory of Appendix \ref{app:kms}.
Each component instead has a self-consistent effective dynamics, which we do
not pursue.

At $\beta_{\mathrm{c}}=2$, the model changes between the two structures in
Remark \ref{rem:summary}.
For $\beta\leq2$, the limit is a product factor with trivial center.
For $\beta>2$, it is the barycenter of a circle of symmetry-breaking factor
phases.
These finite-temperature product states are not pure states of $\oa{A}$:
their one-site density matrices in
\eqref{eq:mf-product-state-density-matrices} have two nonzero eigenvalues,
and resolving either one-site density matrix into rank-one states gives a
nontrivial convex decomposition of the product state.
\end{rem}

\subsubsection{Zero-temperature purity and the staggered mean-field transition}\label{zero-temperature-purity-and-the-staggered-mean-field-transition}

The preceding mean-field Hamiltonian is the zero-staggering model. Its limit \(\beta\uparrow\infty\) makes the distinction between a pure phase and the symmetric state explicit. The staggered extension below also gives an exactly solvable comparison with the Mott transition of the short-range model.

Let \(\Lambda=\Lambda_+\sqcup\Lambda_-\) have even cardinality \(N\) and satisfy \(\abscard{\Lambda_+}=\abscard{\Lambda_-}=N/2\). Write \(\epsilon_x=+1\) on \(\Lambda_+\) and \(\epsilon_x=-1\) on \(\Lambda_-\). Define \begin{equation}\label{eq:mf-staggered-hamiltonian}
\physham^{\mathrm{mf}}_{\Lambda,\lambda}
=
-\frac{1}{N}
\rbk{\rbk{S^1_{\txttot,\Lambda}}^2
+\rbk{S^2_{\txttot,\Lambda}}^2}
+\lambda
\sum_{x\in\Lambda}
\rbk{\frac{1}{2}+\epsilon_xS_x^3},
\quad
\lambda\geq0.
\end{equation} This is the complete-graph analogue of the spin Hamiltonian \eqref{eq:hamiltonian-spin}, with the same nonnegative normalization of the staggered potential.

For one-site density matrices \(\rho_+\) and \(\rho_-\), set \begin{multline}\label{eq:mf-staggered-energy-functional}
\fun{\mathcal{E}_\lambda}{\rho_+,\rho_-}
=
-\frac{1}{4}
\rbk{
\rbk{\sqfun{\trace}{\rho_+S^1}+\sqfun{\trace}{\rho_-S^1}}^2
+\rbk{\sqfun{\trace}{\rho_+S^2}+\sqfun{\trace}{\rho_-S^2}}^2}
\\
+\frac{\lambda}{2}
\rbk{
1
+\sqfun{\trace}{\rho_+S^3}
-\sqfun{\trace}{\rho_-S^3}}.
\end{multline}

\begin{prop}[staggered mean-field ground states]\label{prop:mf-staggered-ground-states}
Let $\fun{\physenergyfunc^{\mathrm{mf}}_N}{\lambda}$ be the lowest
eigenvalue of \eqref{eq:mf-staggered-hamiltonian}.
Then
\begin{equation}\label{eq:mf-staggered-ground-energy}
\lim_{N\to\infty}
\frac{\fun{\physenergyfunc^{\mathrm{mf}}_N}{\lambda}}{N}
=
\begin{cases}
-\dfrac{\rbk{1-\lambda}^2}{4},
&0\leq\lambda\leq1,
\\
0,
&\lambda\geq1.
\end{cases}
\end{equation}
For $0\leq\lambda<1$, all minimizers of
\eqref{eq:mf-staggered-energy-functional} are
\begin{equation}\label{eq:mf-staggered-ordered-minimizers}
\rho_{\epsilon,\lambda,\theta}
=
\frac{1}{2}
+
\sqrt{1-\lambda^2}
\rbk{S^1\cos\theta+S^2\sin\theta}
-
\epsilon\lambda S^3,
\quad
\epsilon\in\setone{+1,-1},
\quad
\theta\in\rightopeninterval{0}{2\pi}.
\end{equation}
For $\lambda\geq1$, the unique minimizing pair is
\begin{equation}\label{eq:mf-staggered-mott-minimizer}
\rho_\epsilon^{\mathrm{Mott}}
=
\frac{1}{2}-\epsilon S^3,
\quad
\epsilon\in\setone{+1,-1}.
\end{equation}
Every separately permutation-invariant thermodynamic ground-state limit has
the two-class Størmer measure of Corollary
\ref{cor:stormer-two-class} supported on these minimizing pairs.
\end{prop}

\begin{proof}
Average a finite-volume ground state independently over permutations of
$\Lambda_+$ and $\Lambda_-$.
This preserves its energy.
Every limit point is separately exchangeable, so Corollary
\ref{cor:stormer-two-class} represents it by a measure on pairs of one-site
density matrices.
The expectation of \eqref{eq:mf-staggered-hamiltonian}, divided by $N$,
converges to the integral of
\eqref{eq:mf-staggered-energy-functional}.
Conversely, two-class product states give the matching upper bound.
It remains to minimize the displayed functional.

Write
$$
u_\epsilon
=
\rbk{
\sqfun{\trace}{\rho_\epsilon S^1},
\sqfun{\trace}{\rho_\epsilon S^2}
},
\quad
z_\epsilon
=
\sqfun{\trace}{\rho_\epsilon S^3}.
$$
The Bloch-ball constraint is
$\abs{u_\epsilon}^2+z_\epsilon^2\leq\frac{1}{4}$.
Changing only the signs of the longitudinal components shows that a minimizer
may be chosen with $z_+\leq0$ and $z_-\geq0$.
For fixed longitudinal components, the transverse term is minimized when
$u_+$ and $u_-$ are aligned and both Bloch vectors have maximal length.
Thus, for some $\alpha_+,\alpha_-\in\closedinterval{0}{\pi/2}$,
$$
z_+
=
-\frac{1}{2}\cos\alpha_+,
\quad
z_-
=
\frac{1}{2}\cos\alpha_-,
\quad
\abs{u_\epsilon}
=
\frac{1}{2}\sin\alpha_\epsilon.
$$
Set
$q_{\mathrm{av}}=\rbk{\alpha_++\alpha_-}/2$ and
$q_{\mathrm{dif}}=\rbk{\alpha_+-\alpha_-}/2$.
The energy becomes
$$
-\frac{1}{4}\sin^2q_{\mathrm{av}}\cos^2q_{\mathrm{dif}}
+
\frac{\lambda}{2}
\rbk{1-\cos q_{\mathrm{av}}\cos q_{\mathrm{dif}}}.
$$
For fixed $q_{\mathrm{av}}\in\closedinterval{0}{\pi/2}$ this expression
decreases as $\cos q_{\mathrm{dif}}$ increases.
Hence $q_{\mathrm{dif}}=0$.
With $y=\cos q_{\mathrm{av}}\in\closedinterval{0}{1}$, the remaining
function is
\begin{equation}\label{eq:mf-staggered-one-variable-energy}
\frac{y^2}{4}
-
\frac{\lambda y}{2}
+
\frac{\lambda}{2}
-
\frac{1}{4}.
\end{equation}
Its minimizer is $y=\lambda$ for $0\leq\lambda<1$ and $y=1$ for
$\lambda\geq1$.
Equations \eqref{eq:mf-staggered-ground-energy}--\eqref{eq:mf-staggered-mott-minimizer} follow.
The representing measure must be supported on the minimizer set because the
integrand exceeds its minimum away from that compact set.
\end{proof}

The corresponding infinite product states are \begin{equation}\label{eq:mf-staggered-ground-product-states}
\oastate[\psi^{\mathrm{mf}}_{\txtgs,\lambda,\theta}]
=
\bigotimes_{x\in\ringratint^d}
\rho_{\epsilon_x,\lambda,\theta},
\quad
0\leq\lambda<1,
\quad
\oastate[\psi^{\mathrm{mf}}_{\mathrm{Mott},\lambda}]
=
\bigotimes_{x\in\ringratint^d}
\rho_{\epsilon_x}^{\mathrm{Mott}},
\quad
\lambda\geq1.
\end{equation}

\begin{cor}[purity and the mean-field Mott cusp]\label{cor:mf-ground-purity-mott-cusp}
For $0\leq\lambda<1$, the states
$\oastate[\psi^{\mathrm{mf}}_{\txtgs,\lambda,\theta}]$ in
\eqref{eq:mf-staggered-ground-product-states} are pure and form one gauge orbit.
Their gauge average is not pure.
For $\lambda\geq1$, the state
$\oastate[\psi^{\mathrm{mf}}_{\mathrm{Mott},\lambda}]$ in the same equation
is a pure gauge-invariant product state.

Let $\fun{e^{\mathrm{mf}}_\lambda}{\varrho}$ be the minimum of
\eqref{eq:mf-staggered-energy-functional} subject to
\begin{equation}\label{eq:mf-staggered-density-constraint}
\frac{1}{2}
\rbk{
\sqfun{\trace}{\rho_+S^3}
+
\sqfun{\trace}{\rho_-S^3}
}
=
\varrho-\frac{1}{2}.
\end{equation}
For $\lambda>1$,
\begin{equation}\label{eq:mf-staggered-mott-cusp}
\fun{e^{\mathrm{mf}}_\lambda}{\frac{1}{2}+m}
=
\fun{e^{\mathrm{mf}}_\lambda}{\frac{1}{2}}
+
\sqrt{\lambda\rbk{\lambda-1}}\abs{m}
+
\fun{o}{\abs{m}}.
\end{equation}
Thus the mean-field chemical-potential gap is
$2\sqrt{\lambda\rbk{\lambda-1}}$.
It closes at the single transition point $\lambda=1$.
\end{cor}

\begin{proof}
The coefficient vector multiplying $S^1,S^2,S^3$ in each density matrix
of \eqref{eq:mf-staggered-ordered-minimizers} has Euclidean norm $1$.
The same is immediate for \eqref{eq:mf-staggered-mott-minimizer}.
All one-site density matrices are therefore rank-one projections.
Their infinite product states are pure
\cite[Appendix E]{YoshitsuguSekine012}.
For $\lambda<1$, distinct values of $\theta$ give distinct product states,
so their Haar barycenter has a nontrivial representing measure and is not
pure.
For $\lambda\geq1$, the minimizer is a single product state.
At $\lambda=0$, the matrices in
\eqref{eq:mf-staggered-ordered-minimizers} are the
$\beta\uparrow\infty$ limits of
\eqref{eq:mf-product-state-density-matrices}, by
\eqref{eq:mf-order-parameter-limits}.
Thus the phase components of the original mean-field calibration become pure
at zero temperature, while their Haar barycenter remains nonpure.

It remains to prove \eqref{eq:mf-staggered-mott-cusp}.
By particle--hole symmetry it suffices to take $m>0$.
Near the unique half-filled minimizer, write
$$
z_+
=
-\frac{1}{2}+a,
\quad
z_-
=
\frac{1}{2}-b,
\quad
a-b
=
2m.
$$
After aligning the transverse components, the constrained energy above its
half-filled value is exactly
\begin{equation}\label{eq:mf-staggered-doping-energy}
-\frac{1}{4}
\rbk{\sqrt{a-a^2}+\sqrt{b-b^2}}^2
+
\frac{\lambda}{2}\rbk{a+b}.
\end{equation}
Compactness and uniqueness of the half-filled minimizer imply
$a,b\to0$ as $m\downarrow0$.
Comparison with the trial value $b=0$ shows that a minimizing sequence has
$b=\fun{O}{m}$ when $\lambda>1$.
Indeed, for $a=b+2m$,
$$
\eqref{eq:mf-staggered-doping-energy}
\geq
\frac{\lambda-1}{2}\rbk{a+b},
\quad
-\frac{1}{4}\rbk{2m-4m^2}+\lambda m
=
\rbk{\lambda-\frac{1}{2}}m+m^2.
$$
Set $b=mq$.
Dividing \eqref{eq:mf-staggered-doping-energy} by $m$ gives, locally
uniformly for $q\geq0$,
$$
\fun{g_\lambda}{q}
=
\lambda\rbk{q+1}
-
\frac{1}{4}
\rbk{\sqrt{q+2}+\sqrt{q}}^2.
$$
This function is coercive for $\lambda>1$.
Writing $t=q+1$ gives
$$
\fun{g_\lambda}{q}
=
\rbk{\lambda-\frac{1}{2}}t
-
\frac{1}{2}\sqrt{t^2-1}.
$$
Direct minimization over $t\geq1$ yields
\begin{equation}\label{eq:mf-staggered-doping-slope}
\inf_{q\geq0}\fun{g_\lambda}{q}
=
\sqrt{\lambda\rbk{\lambda-1}}.
\end{equation}
Equations \eqref{eq:mf-staggered-doping-energy} and
\eqref{eq:mf-staggered-doping-slope} prove the right derivative in
\eqref{eq:mf-staggered-mott-cusp}; particle--hole symmetry gives the left
derivative.
\end{proof}

\begin{rem}[what the mean-field transition does not prove]\label{rem:mf-intermediate-regime}
The staggered mean-field variational problem has no open intermediate
region.
The ordered pure phases occur for $0\leq\lambda<1$, the order parameter
vanishes continuously at $\lambda=1$, and the Mott cusp is positive for
$\lambda>1$.
This exact threshold is specific to the complete-graph normalization in
\eqref{eq:mf-staggered-hamiltonian}.
It does not bound the critical line of the short-range model.
For the short-range model, condensation is proved under the sufficient
condition \eqref{eq:main-ground-state-condensation-region}, whereas the Mott
cusp is proved when the lower bound
$\Delta_\infty$ in \eqref{eq:thermodynamic-mott-gap-constant} is positive.
The interval not covered by those two sufficient bounds is the unresolved
intermediate region of Section \ref{sec:improvements}.
The mean-field identity \eqref{eq:mf-staggered-ground-energy} makes the two
exact regimes meet at $\lambda=1$ because the product variational functional
\eqref{eq:mf-staggered-energy-functional} contains no spatial correlation
error.
\end{rem}

\section{What Must Be Improved in the Original Arguments}\label{sec:improvements}

Sections \ref{sec:rp}--\ref{sec:density} and \ref{sec:comparison-models} supply the details delegated to the literature in \cite{AizenmanLiebSeiringerSolovejYngvason001}. The remaining limitations concern the estimates themselves.

\subsection{The infrared bound controls one moment of the central measure}\label{the-infrared-bound-controls-one-moment-of-the-central-measure}

Gaussian domination \eqref{eq:gaussian-domination}, the infrared bound \eqref{eq:infrared-bound}, the Falk--Bruch inequality \eqref{eq:falk-bruch-inequality}, and the sum rule \eqref{eq:sum-rule} are quadratic. They prove the second-moment bound \eqref{eq:order-parameter-second-moment-and-atom-bound}, but do not determine the radial part of the rotation-invariant law \eqref{eq:order-parameter-law-definition}. In particular, they do not exclude gauge-invariant components or improve \(\fun{\mu}{E}\geq4\kappa\) to \(\fun{\mu}{E}=1\), with \(E\) defined by \eqref{eq:componentwise-breaking-nonzero-set}.

\begin{cor}[factorization criterion for the Haar-orbit law]\label{cor:factorization-criterion}
Under \eqref{eq:order-parameter-law-definition}--
\eqref{eq:order-parameter-second-moment-and-atom-bound}, the following are
equivalent:
\begin{enumerate}
\item $\lim_{n \to \infty}
\fun{\oastate[\psi_{\beta}]}{
\rbk{\faadj{\barmean{a}_{\Lambda_n}}
\barmean{a}_{\Lambda_n}}^{2}}
=
\rbk{\lim_{n \to \infty}
\fun{\oastate[\psi_{\beta}]}{
\faadj{\barmean{a}_{\Lambda_n}}
\barmean{a}_{\Lambda_n}}}^{2}$;
\item $\abs{\barmean{a}}$ is $\mu$-almost everywhere constant;
\item $\nu$ is the uniform probability measure on a circle $\abs{z} = r$ with $\sqrt{\kappa} \leq r \leq \frac{1}{2}$.
\end{enumerate}
In this case, $\nu$ is the Haar-orbit law of Remark
\ref{rem:orbit-structure}, $\varrho_{0}=r^{2}$, $\fun{\mu}{E}=1$, and almost
every component has trivial gauge stabilizer.
\end{cor}

\begin{proof}
The moment formula \eqref{eq:order-parameter-moment-limit} with $k=m=2$
gives
$$\lim_{n \to \infty}
\fun{\oastate[\psi_{\beta}]}{\rbk{\faadj{\barmean{a}_{\Lambda_n}}\barmean{a}_{\Lambda_n}}^{2}}
=\int_{\fldcmp}\abs{z}^{4}\opdmsr{\nu(z)}.$$
The same formula with $k=m=1$ gives
$$\lim_{n \to \infty}
\fun{\oastate[\psi_{\beta}]}{\faadj{\barmean{a}_{\Lambda_n}}\barmean{a}_{\Lambda_n}}
=\int_{\fldcmp}\abs{z}^{2}\opdmsr{\nu(z)}.$$
Substitution of these two identities into assertion (1) gives
$$\int_X \abs{\barmean{a}(x)}^4\opdmsr{\mu(x)}
=\rbk{\int_X \abs{\barmean{a}(x)}^2\opdmsr{\mu(x)}}^2.$$
Cauchy--Schwarz applied to $\abs{\barmean{a}}^{2}$ and $1$ gives
$$\rbk{\int_X \abs{\barmean{a}(x)}^2\opdmsr{\mu(x)}}^2
\leq
\int_X \abs{\barmean{a}(x)}^4\opdmsr{\mu(x)}
\int_X 1\opdmsr{\mu(x)}
=
\int_X \abs{\barmean{a}(x)}^4\opdmsr{\mu(x)}.$$
Equality holds exactly when $\abs{\barmean{a}}^{2}$ is proportional to
$1$, equivalently when $\abs{\barmean{a}}$ is constant $\mu$-almost
everywhere. Assertions (1) and (2) are therefore equivalent.

Under assertion (2), the constant is
$$r=\rbk{\int_{X}\abs{\barmean{a}(x)}^{2}\opdmsr{\mu(x)}}^{1/2}
\in\closedinterval{\sqrt{\kappa}}{\frac{1}{2}}.$$
Rotation invariance \eqref{eq:order-parameter-rotation-invariance} then
implies that $\nu$ is the normalized Haar measure on $\abs{z}=r$.
This proves assertion (3).

Conversely, assertion (3) gives $\abs{\barmean{a}}=r$ almost everywhere and
therefore assertion (2).  Since $r>0$, one has
$\fun{\nu}{\setone{0}}=0$ and $\fun{\mu}{E}=1$ by
\eqref{eq:order-parameter-second-moment-and-atom-bound}.
Equation \eqref{eq:componentwise-gauge-breaking} gives the stabilizer claim,
and \eqref{eq:component-condensate-density} gives
$\varrho_{0}=r^{2}$.
\end{proof}

The missing input is the following four-point estimate for the spatial averages \eqref{eq:main-spatial-averages}: \begin{equation}\label{eq:required-zero-mode-factorization}
\fun{\oastate[\psi_{\beta,\Lambda_{n}}]}{
\rbk{\faadj{\barmean{a}_{\Lambda_n}}
\barmean{a}_{\Lambda_n}}^{2}}
\leq
\fun{\oastate[\psi_{\beta,\Lambda_{n}}]}{
\faadj{\barmean{a}_{\Lambda_n}}
\barmean{a}_{\Lambda_n}}^{2}
+
\fun{o}{1},
\end{equation} in the double limit of \eqref{eq:odlro-limit}. The quadratic conversion \eqref{eq:falk-bruch-inequality} has no known four-point analogue; proving \eqref{eq:required-zero-mode-factorization} must therefore exploit the hard-core relations \eqref{eq:main-spin-relations}. The free-gas identities \eqref{eq:free-zero-mode-geometric}--\eqref{eq:free-zero-mode-second-moment} show why interaction is essential: zero-mode factorization fails as \(\mu\uparrow\varepsilon_{0,\Lambda}\).

A quartic estimate with constant \(C\) would give \[\int_X \abs{\barmean{a}(x)}^4\opdmsr{\mu(x)}
\leq
C\rbk{\int_X \abs{\barmean{a}(x)}^2\opdmsr{\mu(x)}}^2.\] Since \(\barmean{a}=0\) outside \(E\), Cauchy--Schwarz gives \[
\begin{aligned}
\rbk{\int_X \abs{\barmean{a}(x)}^2\opdmsr{\mu(x)}}^2
&=
\rbk{\int_E \abs{\barmean{a}(x)}^2\opdmsr{\mu(x)}}^2
\\
&\leq
\fun{\mu}{E}
\int_X \abs{\barmean{a}(x)}^4\opdmsr{\mu(x)}
\\
&\leq
C\fun{\mu}{E}
\rbk{\int_X \abs{\barmean{a}(x)}^2\opdmsr{\mu(x)}}^2.
\end{aligned}
\] The positive second moment therefore implies \(\fun{\mu}{E}\geq1/C\). Full measure requires the exact value \(C=1\). The mean-field identity \eqref{eq:mf-central-word-moments} has this value because total-spin sectors provide the missing concentration.

\subsection{From long-range order to states with a definite phase}\label{from-long-range-order-to-states-with-a-definite-phase}

Even Corollary \ref{cor:factorization-criterion} determines only \(\abs{\barmean{a}}\). Showing that the central measure \eqref{eq:kms-central-barycenter}, spatially realized by \eqref{eq:kms-central-direct-integral}, is supported on one gauge orbit requires an additional uniqueness argument.

The quasi-average approach perturbs the rotated Hamiltonian \eqref{eq:sublattice-rotated-hamiltonian} by a symmetry-breaking field \(-h \sum_{x \in \Lambda} S^{1}_{x}\). The missing equality identifies long-range order in the \(h\)-field state with the square of its order parameter. Equation \eqref{eq:componentwise-uniform-mode-condensate-density} proves this only in central components. Classical Griffiths inequalities \cite{RobertBGriffiths001} and the ground-state Koma--Tasaki argument \cite{KomaTasaki002,KomaTasaki001}, which yields \eqref{eq:koma-tasaki-tower-bound}, do not apply to positive-temperature KMS states with the staggered \(S^{3}\) field. Extremality of the \(h\downarrow0\) limit would also remain to be proved.

The uniqueness approach asks whether the orbit measure in \eqref{eq:orbit-disintegration-family} is a point mass. By \eqref{eq:orbit-point-mass-ergodicity-equivalence}, this is gauge ergodicity of the central measure. Bodineau's random-cluster coarse graining proves the analogous Ising result \cite{ThierryBodineau001}. Here one would need an FK-type representation extending \cite{AizenmanNachtergaele001} that turns \eqref{eq:bec-bound} into percolation of the curves in \eqref{eq:loop-ratio}; no such representation is known for the loop measure \eqref{eq:loop-measure-integral}.

Equation \eqref{eq:componentwise-uniform-mode-condensate-density} identifies \(\abs{\barmean{a}(x)}^{2}\) with the uniform-mode condensate density \eqref{eq:component-condensate-density} without clustering. The density is sharp within each component by \eqref{eq:componentwise-density-sharpness}, and its mean is \(1/2\) by \eqref{eq:component-mean-density}, but constancy across components remains open. Equations \eqref{eq:component-density-variance} and \eqref{eq:density-truncated-correlation-decay} reduce this question to macroscopic density variance and truncated density correlations. Neither the off-diagonal bounds \eqref{eq:contour-bound}--\eqref{eq:chi-bound} nor the transverse bound \eqref{eq:infrared-bound} controls them.

\subsection{Uniqueness in the Mott regime is a boundary-condition problem}\label{uniqueness-in-the-mott-regime-is-a-boundary-condition-problem}

The exponential-decay estimate \eqref{eq:exponential-decay}, which is Theorem 2 of \cite{AizenmanLiebSeiringerSolovejYngvason001}, controls exactly \(\fun{\oastate[\psi_{\beta,\Lambda}]}{\faadj{a_x}a_y}\) in limits of the periodic Gibbs states \eqref{eq:main-periodic-gibbs-state}. KMS uniqueness requires two further inputs.

First, a polymer expansion must extend the one-curve renewal bounds \eqref{eq:chi-bound}--\eqref{eq:chi-reweight} to tree-graph bounds for truncated correlations of arbitrary local observables. No such expansion has been proved under \eqref{eq:decay-condition}.

Second, the estimates must be uniform in boundary conditions. Quantum Pirogov--Sinai theory \cite{DattaFernandezFrohlich001,DattaFernandezFrohlich002,KoteckyUeltschi001} should apply at low temperature and large \(\lambda\) because \eqref{eq:hamiltonian-spin}, divided by \(\lambda\), is a classical staggered field plus a perturbation of order \(\lambda^{-1}\). It would give a unique translation-periodic phase, but would not cover the strip between its domain and the high-temperature domain or exclude non-translation-invariant KMS states. The periodic chessboard estimates \eqref{eq:loop-reflection-positivity}--\eqref{eq:contour-bound} cannot settle either issue.

\subsection{The intermediate regime and the missing correlation inequalities}\label{the-intermediate-regime-and-the-missing-correlation-inequalities}

The two proved parameter regions leave an intermediate regime in which the KMS-simplex position of the limit state \eqref{eq:periodic-gibbs-state-extension} is unknown. A Ginibre-type inequality valid for the staggered field in \eqref{eq:sublattice-rotated-hamiltonian}, or stochastic domination between the loop measures \eqref{eq:loop-measure-integral} at different \(\lambda\) and \(\beta\), would provide the missing monotonicity. Existing Ginibre inequalities \cite{JeanGinibre009} do not permit that field. Either extension could connect the condensation region \eqref{eq:bec-bound} and the decay region \eqref{eq:decay-condition} to a single critical curve \(\fun{\lambda_{\mathrm{c}}}{\beta}\).

\subsection{Scope restrictions inherited from the original paper}\label{scope-restrictions-inherited-from-the-original-paper}

The short-range symmetric state is only a weak-\(\ast\) limit point of \eqref{eq:periodic-gibbs-state-extension} along \eqref{eq:periodic-box-exhaustion}; full convergence and independence of the exhaustion are unproved. All short-range conclusions apply to every such limit point. Full convergence is established only for the mean-field model by \eqref{eq:mf-equilibrium-barycenter}.

At positive temperature, \eqref{eq:bec-bound} requires \(d\geq3\) because the integral in \eqref{eq:main-condensation-constant} is singular for \(d\leq2\). At zero field, \eqref{eq:kls-ground-state-zero-mode-bound} covers ground states in \(d\geq2\); the nonzero-field \(d=2\) case remains open.

Particle-hole symmetry \eqref{eq:main-particle-hole-identities} fixes the Mott-gap conclusions \eqref{eq:mott-sector-gap} and \eqref{eq:thermodynamic-mott-cusp} at half-filling. A chemical potential adds the on-site \(S^{3}\) term determined by \eqref{eq:main-particle-number} and preserves the symmetries \eqref{eq:main-symmetry-actions}, but changes the constants in \eqref{eq:gaussian-domination}--\eqref{eq:falk-bruch-inequality}. The density-dependent phase diagram is therefore not determined by \cite{AizenmanLiebSeiringerSolovejYngvason001}.

\appendix

\section{Appendix}\label{appendix}

The finite-dimensional Duhamel calculation, trace inequality, ground-state derivative formula, and vectorization calculation are collected first. The general KMS and central-decomposition theorems used in the main text follow. The remaining appendices evaluate the three-dimensional infrared integral, record the Koma--Tasaki and Tasaki finite-volume inputs, collect spin calculations.

\subsection{\texorpdfstring{Spin-\(1/2\) Algebra Calculations}{Spin-1/2 Algebra Calculations}}\label{app:spin-calculations}

This appendix collects the elementary hard-core, spin, and Fourier calculations used in the main text. The first four lemmas prove elementary one-site and tensor-product spin identities. The fifth proves the finite-volume spin Fourier identities. The sixth proves the finite-dimensional Schur estimate. The seventh applies the spin identities to the finite-volume Gibbs state and proves the exact relation between the density-matrix average and the transverse zero mode.

\subsubsection{One-site, two-site, reflection, and total-spin identities}\label{one-site-two-site-reflection-and-total-spin-identities}

The spin operators \(S^{i}_{x}\) and \(S^{\pm}_{x}\) are defined and normalized in \eqref{eq:spin-matrices}, and their commutation and anticommutation relations are \eqref{eq:main-spin-relations}. The hard-core operators \(a_{x}\) and \(\faadj{a_{x}}\) are defined by \eqref{eq:boson-spin-dictionary}. In the standard one-site basis of \(\fldcmp^{2}\), their matrices are \begin{equation}\label{eq:hardcore-matrices}
\faadj{a}
=
\begin{pmatrix} 0 & 1 \\ 0 & 0 \end{pmatrix},
\quad
a
=
\begin{pmatrix} 0 & 0 \\ 1 & 0 \end{pmatrix},
\quad
\faadj{a}a
=
\begin{pmatrix} 1 & 0 \\ 0 & 0 \end{pmatrix}.
\end{equation}

\begin{lem}[one-site spin-$1/2$ and hard-core identities]\label{lem:app-spin-identities}
For every site
$x$,
the one-site products satisfy
\begin{equation}\label{eq:app-one-site-spin-squares}
\begin{aligned}
\rbk{S^{1}_{x}}^{2}
=
\rbk{S^{2}_{x}}^{2}
=
\rbk{S^{3}_{x}}^{2}
=
\frac{1}{4},
\quad
S^{1}_{x}S^{2}_{x}
+
S^{2}_{x}S^{1}_{x}
=
0.
\end{aligned}
\end{equation}
The raising and lowering operators satisfy
\begin{equation}\label{eq:bec-diagonal-spin-identities}
\begin{aligned}
S^{+}_{x}S^{-}_{x}
=
\frac{1}{2}
+S^{3}_{x},
\quad
S^{-}_{x}S^{+}_{x}
=
\frac{1}{2}
-
S^{3}_{x},
\quad
\commutator{S^{3}_{x}}{S^{\pm}_{x}}
=
\pm S^{\pm}_{x},
\quad
\commutator{S^{+}_{x}}{S^{-}_{x}}
=
2S^{3}_{x}.
\end{aligned}
\end{equation}
The hard-core matrices satisfy
\begin{equation}\label{eq:app-hardcore-matrix-relations}
\begin{aligned}
a^{2}
=
0,
\quad
\rbk{\faadj{a}}^{2}
=
0,
\quad
a\faadj{a}
+\faadj{a}a
=
1.
\end{aligned}
\end{equation}
The hard-core number operator and the spin variable are related by
\begin{equation}\label{eq:app-hardcore-number-spin}
\faadj{a_{x}}a_{x}
-
\frac{1}{2}
=
S^{3}_{x}.
\end{equation}
\end{lem}

\begin{proof}
Let $\Phi_{+}$ and $\Phi_{-}$ be nonzero vectors spanning, respectively,
the $1/2$ and $-1/2$ weight spaces of $S^{3}_{x}$.
The spin commutation relations give
$\commutator{S^{3}_{x}}{S^{\pm}_{x}}=\pm S^{\pm}_{x}$ and
$\commutator{S^{+}_{x}}{S^{-}_{x}}=2S^{3}_{x}$.
The spin-$1/2$ weight decomposition has no weights $3/2$ or $-3/2$.
Consequently,
$S^{+}_{x}\Phi_{+}=0$,
$S^{-}_{x}\Phi_{-}=0$,
and
$\rbk{S^{\pm}_{x}}^{2}=0$.
Applying the second commutation relation to the two weight vectors gives
$$\begin{aligned}
S^{+}_{x}S^{-}_{x}\Phi_{+}
&=
\Phi_{+},
&
S^{+}_{x}S^{-}_{x}\Phi_{-}
&=
0,
\\
S^{-}_{x}S^{+}_{x}\Phi_{+}
&=
0,
&
S^{-}_{x}S^{+}_{x}\Phi_{-}
&=
\Phi_{-}.
\end{aligned}$$
Since $\Phi_{+}$ and $\Phi_{-}$ span the one-site space, the ladder products are
\begin{equation}\label{eq:app-spin-ladder-calculation}
\begin{aligned}
S^{+}_{x}S^{-}_{x}
=
\frac{1}{2}
+
S^{3}_{x},
\quad
S^{-}_{x}S^{+}_{x}
=
\frac{1}{2}
-
S^{3}_{x}.
\end{aligned}
\end{equation}
The definitions
$S^{1}_{x}=\frac{\rbk{S^{+}_{x}+S^{-}_{x}}}{2}$
and
$S^{2}_{x}=\frac{\rbk{S^{+}_{x}-S^{-}_{x}}}{2\imunit}$
therefore give
\begin{equation}\label{eq:app-spin-square-calculation}
\begin{aligned}
\rbk{S^{1}_{x}}^{2}
=
\frac{1}{4}
\rbk{S^{+}_{x}S^{-}_{x}+S^{-}_{x}S^{+}_{x}}
=
\frac{1}{4},
\\
\rbk{S^{2}_{x}}^{2}
=
\frac{1}{4}
\rbk{S^{+}_{x}S^{-}_{x}+S^{-}_{x}S^{+}_{x}}
=
\frac{1}{4},
\\
S^{1}_{x}S^{2}_{x}
+
S^{2}_{x}S^{1}_{x}
=
\frac{1}{2\imunit}
\rbk{\rbk{S^{+}_{x}}^{2}-\rbk{S^{-}_{x}}^{2}}
=
0.
\end{aligned}
\end{equation}
The $S^{3}_{x}$ weight decomposition gives
$\rbk{S^{3}_{x}}^{2}=1/4$.
Equation \eqref{eq:app-spin-square-calculation} proves
\eqref{eq:app-one-site-spin-squares}, and
\eqref{eq:app-spin-ladder-calculation} proves
\eqref{eq:bec-diagonal-spin-identities}.
Multiplication of the hard-core matrices
\eqref{eq:hardcore-matrices}
gives
\eqref{eq:app-hardcore-matrix-relations}.
The dictionary
\eqref{eq:boson-spin-dictionary}
and
\eqref{eq:app-spin-ladder-calculation}
give
\eqref{eq:app-hardcore-number-spin}.
\end{proof}

\begin{lem}[two-site and transverse ladder identities]\label{lem:app-spin-exchange-identities}
Let
$T^{1}$
and
$T^{2}$
commute with the spin operators at
$x$,
and define
$T^{\pm}
=
T^{1}
\pm
\imunit T^{2}$.
Then the transverse ladder identity is
\begin{equation}\label{eq:app-spin-transverse-ladder}
\begin{aligned}
S^{+}_{x}T^{-}
+
S^{-}_{x}T^{+}
=
2
\rbk{
S^{1}_{x}T^{1}
+
S^{2}_{x}T^{2}
}.
\end{aligned}
\end{equation}
In particular, for distinct sites
$x$
and
$y$,
the exchange identity is
\begin{equation}\label{eq:app-two-site-spin-exchange}
\begin{aligned}
S^{+}_{x} S^{-}_{y}
+S^{-}_{x} S^{+}_{y}
=
2
\rbk{
S^{1}_{x}S^{1}_{y}
+
S^{2}_{x}S^{2}_{y}
}.
\end{aligned}
\end{equation}
\end{lem}

\begin{proof}
Expansion of the left side of
\eqref{eq:app-spin-transverse-ladder}
gives
\begin{equation}\label{eq:app-spin-transverse-ladder-calculation}
\begin{aligned}
&S^{+}_{x} T^{-}
+ S^{-}_{x} T^{+}
=
\rbk{S^{1}_{x}+\imunit S^{2}_{x}}
\rbk{T^{1}-\imunit T^{2}}
+\rbk{S^{1}_{x} - \imunit S^{2}_{x}}
\rbk{T^{1} + \imunit T^{2}}
\\ 
&=
\rbk{S^{1}_{x}T^{1}
-\imunit S^{1}_{x}T^{2}
+\imunit S^{2}_{x}T^{1}
+S^{2}_{x}T^{2}}
+\rbk{S^{1}_{x}T^{1}
+\imunit S^{1}_{x}T^{2}
-\imunit S^{2}_{x}T^{1}
+S^{2}_{x}T^{2}}
\\ 
&=
2
\rbk{
S^{1}_{x}T^{1}
+
S^{2}_{x}T^{2}
}.
\end{aligned}
\end{equation}
The mixed terms in
\eqref{eq:app-spin-transverse-ladder-calculation}
cancel.
This proves
\eqref{eq:app-spin-transverse-ladder}.
Taking
$T^{i}
=
S^{i}_{y}$
for
$x
\neq
y$
proves
\eqref{eq:app-two-site-spin-exchange}.
\end{proof}

\begin{lem}[spin conjugation and reflection identities]\label{lem:app-spin-reflection-identities}
Entrywise conjugation and the rotation about the first spin axis satisfy
\begin{equation}\label{eq:app-spin-conjugation-rotation}
\begin{gathered}
\cmpconj{S^{1}_{x}}
=
S^{1}_{x},
\quad
\cmpconj{S^{2}_{x}}
=
-S^{2}_{x},
\quad
\cmpconj{S^{3}_{x}}
=
S^{3}_{x},
\\ 
\napiernum^{\imunit\pi S^{1}_{x}}
S^{1}_{x}
\napiernum^{-\imunit\pi S^{1}_{x}}
=
S^{1}_{x},
\quad
\napiernum^{\imunit\pi S^{1}_{x}}
S^{2}_{x}
\napiernum^{-\imunit\pi S^{1}_{x}}
=
-S^{2}_{x},
\quad
\napiernum^{\imunit\pi S^{1}_{x}}
S^{3}_{x}
\napiernum^{-\imunit\pi S^{1}_{x}}
=
-S^{3}_{x}.
\end{gathered}
\end{equation}
For
$\delta \in \fldreal$,
set
$D = S^{1}/\sqrt{2}$,
$c = \delta/\sqrt{2}$,
and
$G = S^{2}/\sqrt{2}$.
The two identities used for a crossing reflection plane are
\begin{equation}\label{eq:app-reflection-spin-squares}
\begin{aligned}
\frac{1}{2}
\rbk{S^{1}\otimes 1
-1\otimes S^{1}
-\delta}^{2}
&=
\rbk{D\otimes 1
-1\otimes\cmpconj{D}
-c}^{2},
\\ 
S^{2}\otimes S^{2}
&=
\frac{1}{2}
\rbk{S^{2}\otimes 1
+1\otimes S^{2}}^{2}
-\frac{1}{4}
=
\rbk{G\otimes 1
-1\otimes\cmpconj{G}}^{2}
-\frac{1}{4}.
\end{aligned}
\end{equation}
\end{lem}

\begin{proof}
Entrywise conjugation of the three matrices
\eqref{eq:spin-matrices}
gives the first three identities in
\eqref{eq:app-spin-conjugation-rotation}.
The matrix exponential is
\begin{equation}\label{eq:app-first-axis-pi-rotation}
\napiernum^{\imunit\pi S^{1}}
=
\cos \frac{\pi}{2}
+
2\imunit S^{1}
\sin \frac{\pi}{2}
=
2\imunit S^{1}.
\end{equation}
Multiplication of
\eqref{eq:app-first-axis-pi-rotation}
with the matrices
\eqref{eq:spin-matrices}
gives the final three identities in
\eqref{eq:app-spin-conjugation-rotation}.
Expansion of the squares in
\eqref{eq:app-reflection-spin-squares},
together with
\eqref{eq:app-one-site-spin-squares}
and
\eqref{eq:app-spin-conjugation-rotation},
proves the two reflection-plane identities.
\end{proof}

\begin{lem}[gauge and total-spin rotation identities]\label{lem:app-spin-rotation-identities}
The gauge rotation is
\begin{equation}\label{eq:app-spin-gauge-rotation}
\napiernum^{\imunit\theta S^{3}_{x}}
S^{\pm}_{x}
\napiernum^{-\imunit\theta S^{3}_{x}}
=
\napiernum^{\pm\imunit\theta}
S^{\pm}_{x}.
\end{equation}
For a finite set
$W$,
define
$S^{i}_{W}
=
\sum_{x\in W}S^{i}_{x}$
and
$S^{\pm}_{W}
=
S^{1}_{W}
\pm
\imunit S^{2}_{W}$.
The rotation about the second total-spin axis satisfies
\begin{equation}\label{eq:app-second-axis-spin-rotation}
\napiernum^{-\imunit\epsilon S^{2}_{W}}
S^{3}_{W}
\napiernum^{\imunit\epsilon S^{2}_{W}}
=
\cos \epsilon \cdot S^{3}_{W}
+
\sin \epsilon \cdot S^{1}_{W}.
\end{equation}
The total-spin identity is
\begin{equation}\label{eq:app-total-spin-ladder}
S^{+}_{W}S^{-}_{W}
=
\sum_{i=1}^{3}
\rbk{S^{i}_{W}}^{2}
-
\rbk{S^{3}_{W}}^{2}
+
S^{3}_{W}.
\end{equation}
\end{lem}

\begin{proof}
For
\begin{equation}\label{eq:app-gauge-rotation-differential-equation}
\fun{F_{\pm}}{\theta}
=
\napiernum^{\imunit\theta S^{3}_{x}}
S^{\pm}_{x}
\napiernum^{-\imunit\theta S^{3}_{x}},
\end{equation}
Differentiating
\eqref{eq:app-gauge-rotation-differential-equation}
and using
$\commutator{S^{3}_{x}}{S^{\pm}_{x}}
=
\pm S^{\pm}_{x}$
gives
\begin{equation}\label{eq:app-gauge-rotation-ode}
\begin{aligned}
\od{\fun{F_{\pm}}{\theta}}{\theta}
=
\pm\imunit
\fun{F_{\pm}}{\theta},
\quad
\fun{F_{\pm}}{0}
=
S^{\pm}_{x}.
\end{aligned}
\end{equation}
Solving
\eqref{eq:app-gauge-rotation-ode}
proves
\eqref{eq:app-spin-gauge-rotation}.
For the second-axis rotation,
direct differentiation gives
\begin{equation}\label{eq:app-second-axis-rotation-ode}
\begin{aligned}
\od{\napiernum^{-\imunit\epsilon S^{2}_{W}}
S^{3}_{W}
\napiernum^{\imunit\epsilon S^{2}_{W}}}
{\epsilon}
&=
\napiernum^{-\imunit\epsilon S^{2}_{W}}
S^{1}_{W}
\napiernum^{\imunit\epsilon S^{2}_{W}},
\\ 
\od{^{2}
\napiernum^{-\imunit\epsilon S^{2}_{W}}
S^{3}_{W}
\napiernum^{\imunit\epsilon S^{2}_{W}}}
{\epsilon^{2}}
&=
-\napiernum^{-\imunit\epsilon S^{2}_{W}}
S^{3}_{W}
\napiernum^{\imunit\epsilon S^{2}_{W}}.
\end{aligned}
\end{equation}
At
$\epsilon
=
0$,
the expression differentiated in
\eqref{eq:app-second-axis-rotation-ode}
equals
$S^{3}_{W}$
and its derivative equals
$S^{1}_{W}$.
Solving this second-order equation proves
\eqref{eq:app-second-axis-spin-rotation}.
Finally, the calculation
\eqref{eq:app-spin-ladder-calculation}
applied to the total-spin commutation relations gives
\eqref{eq:app-total-spin-ladder}.
\end{proof}

\subsubsection{Finite-volume spin Fourier identities}\label{finite-volume-spin-fourier-identities}

The following lemma collects the Fourier identities that are used in the infrared bound, the sum rule, and the comparison between spin correlations and the one-particle density matrix.

\begin{lem}[spin Fourier identities]\label{lem:app-spin-fourier-identities}
Let
$\widetilde{S}^{i}_{p}$
be defined by
\eqref{eq:spin-wave}.
Its adjoint and its raising and lowering combinations are
\begin{equation}\label{eq:app-spin-fourier-adjoint-ladder}
\begin{aligned}
\faadj{\rbk{\widetilde{S}^{i}_{p}}}
=
\widetilde{S}^{i}_{-p},
\quad
\widetilde{S}^{\pm}_{p}
=
\widetilde{S}^{1}_{p}
\pm
\imunit\widetilde{S}^{2}_{p}
=
\frac{1}{\sqrt{\abscard{\Lambda}}}
\sum_{x\in\Lambda}
S^{\pm}_{x}
\napiernum^{\imunit p\cdot x}.
\end{aligned}
\end{equation}
For every state
$\oastate[\psi]$
on
$\oa{A}_{\Lambda}$,
the Fourier representation
\eqref{eq:main-fourier-density-matrix}
satisfies
\begin{equation}\label{eq:app-density-spin-fourier-representation}
\fun{\faftr{\gamma_{\psi}}}{p,q}
=
\fun{\oastate[\psi]}{
\widetilde{S}^{+}_{p}
\faadj{\rbk{\widetilde{S}^{+}_{q}}}
}.
\end{equation}
For
$i
\in
\setone{1,2}$,
Fourier orthogonality gives
\begin{equation}\label{eq:app-spin-fourier-parseval}
\sum_{p\in\dual{\Lambda}}
\widetilde{S}^{i}_{p}
\widetilde{S}^{i}_{-p}
=
\sum_{x\in\Lambda}
\rbk{S^{i}_{x}}^{2}
=
\frac{\abscard{\Lambda}}{4}.
\end{equation}
The same-component Fourier modes commute:
\begin{equation}\label{eq:app-spin-fourier-commutator}
\commutator{\widetilde{S}^{i}_{p}}{\widetilde{S}^{i}_{-p}}
=
0.
\end{equation}
\end{lem}
\begin{proof}
Self-adjointness of
$S^{i}_{x}$
and complex conjugation of the Fourier phase give
$$\begin{aligned}
\faadj{\rbk{\widetilde{S}^{i}_{p}}}
=
\frac{1}{\sqrt{\abscard{\Lambda}}}
\sum_{x\in\Lambda}
S^{i}_{x}
\napiernum^{-\imunit p\cdot x}
=
\widetilde{S}^{i}_{-p}.
\end{aligned}$$
Linearity of
\eqref{eq:spin-wave}
and
$S^{\pm}_{x}
=
S^{1}_{x}
\pm
\imunit S^{2}_{x}$
give the second identity in
\eqref{eq:app-spin-fourier-adjoint-ladder}.
Equations
\eqref{eq:main-fourier-density-matrix}
and
\eqref{eq:boson-spin-dictionary}
give
$$\begin{aligned}
\fun{\faftr{\gamma_{\psi}}}{p,q}
=
\frac{1}{\abscard{\Lambda}}
\sum_{x,y\in\Lambda}
\napiernum^{\imunit p\cdot x-\imunit q\cdot y}
\fun{\oastate[\psi]}
{S^{+}_{x}S^{-}_{y}}
=
\fun{\oastate[\psi]}
{\widetilde{S}^{+}_{p}
\faadj{\rbk{\widetilde{S}^{+}_{q}}}}.
\end{aligned}$$
This proves
\eqref{eq:app-density-spin-fourier-representation}.
The character orthogonality relation is
$$\sum_{p\in\dual{\Lambda}}
\napiernum^{\imunit p\cdot\rbk{x-y}}
=
\abscard{\Lambda}\delta_{xy}.$$
Substitution into the Fourier sum gives
$$\begin{aligned}
\sum_{p\in\dual{\Lambda}}
\widetilde{S}^{i}_{p}
\widetilde{S}^{i}_{-p}
=
\frac{1}{\abscard{\Lambda}}
\sum_{p\in\dual{\Lambda}}
\sum_{x,y\in\Lambda}
\napiernum^{\imunit p\cdot\rbk{x-y}}
S^{i}_{x}S^{i}_{y}
=
\sum_{x\in\Lambda}
\rbk{S^{i}_{x}}^{2}
=
\frac{\abscard{\Lambda}}{4},
\end{aligned}
$$
where the final equality is
\eqref{eq:app-one-site-spin-squares}.
This proves
\eqref{eq:app-spin-fourier-parseval}.
Finally,
$$
\commutator{\widetilde{S}^{i}_{p}}{\widetilde{S}^{i}_{-p}}
=
\frac{1}{\abscard{\Lambda}}
\sum_{x,y\in\Lambda}
\napiernum^{\imunit p\cdot\rbk{x-y}}
\commutator{S^{i}_{x}}{S^{i}_{y}}
=
0.
$$
This proves
\eqref{eq:app-spin-fourier-commutator}.
\end{proof}

\subsubsection{Finite-dimensional Schur estimate}\label{finite-dimensional-schur-estimate}

The operator-norm estimate used for the off-zero Fourier compression is a finite-dimensional matrix inequality. We denote the operator norm for a matrix \(B\) on \(\lpseq^{2}\) as \(\opnorm{B}\).

\begin{lem}[finite-dimensional Schur estimate]\label{lem:app-schur-estimate}
For a finite matrix
$B
=
\rbk{B_{pq}}$, it holds that
\begin{equation}\label{eq:app-schur-estimate}
\opnorm{B}
\leq
\sqrt{\rbk{\max_{p}
\sum_{q}
\abs{B_{pq}}}
\rbk{\max_{q}
\sum_{p}
\abs{B_{pq}}}}.
\end{equation}
If
$B$
is Hermitian,
then we obtain
\begin{equation}\label{eq:app-schur-hermitian}
\opnorm{B}
\leq
\max_{p}
\sum_{q}
\abs{B_{pq}}.
\end{equation}
\end{lem}

\begin{proof}
For a vector
$u$,
the Cauchy--Schwarz inequality gives
$$\begin{aligned}
&\norm{Bu}_{2}^{2}
=
\sum_{p}
\abs{\sum_{q}
B_{pq}u_{q}}^{2}
\leq
\sum_{p}
\rbk{\sum_{q}
\abs{B_{pq}}}
\rbk{\sum_{q}
\abs{B_{pq}}
\abs{u_{q}}^{2}}
\\ 
&\leq
\rbk{
\max_{p}
\sum_{q}
\abs{B_{pq}}
}
\rbk{
\max_{q}
\sum_{p}
\abs{B_{pq}}
}
\norm{u}_{2}^{2}.
\end{aligned}
$$
Taking the supremum over
$\norm{u}_{2}
=
1$
proves
\eqref{eq:app-schur-estimate}.
For a Hermitian matrix,
$\abs{B_{pq}}
=
\abs{B_{qp}}$,
so its maximal column sum equals its maximal row sum.
Substitution into
\eqref{eq:app-schur-estimate}
proves
\eqref{eq:app-schur-hermitian}.
\end{proof}

\subsubsection{Density-matrix and zero-mode calculation}\label{density-matrix-and-zero-mode-calculation}

Fix a finite periodic box \(\Lambda\) and an inverse temperature \(\sminvtemperature
>
0\). In this subsection, \(\oastate[\psi_{\sminvtemperature,\Lambda}]\) denotes the state on \(\oa{A}_{\Lambda}\) defined by \eqref{eq:main-periodic-gibbs-state}, and the one-particle density matrix \(\gamma_{\psi_{\sminvtemperature,\Lambda}}\) for \(\oastate[\psi_{\sminvtemperature,\Lambda}]\) is defined by \eqref{eq:main-density-matrix}.

\begin{lem}[density average and transverse zero mode]\label{lem:app-density-zero-mode}
For every
$p
\in
\dual{\Lambda}$,
the diagonal of the Fourier density matrix satisfies
\begin{equation}\label{eq:app-density-fourier-diagonal}
\begin{aligned}
\fun{\faftr{\gamma_{\psi_{\sminvtemperature,\Lambda}}}}{p,p}
=
\fun{\oastate[\psi_{\sminvtemperature,\Lambda}]}
{\widetilde{S}^{1}_{p}\widetilde{S}^{1}_{-p}
+\widetilde{S}^{2}_{p}\widetilde{S}^{2}_{-p}}
+\frac{1}{\abscard{\Lambda}}
\fun{\oastate[\psi_{\sminvtemperature,\Lambda}]}{
S^{3}_{\txttot,\Lambda}}.
\end{aligned}
\end{equation}
The density-matrix average satisfies
\begin{equation}\label{eq:bec-density-zero-mode-identity}
\begin{aligned}
\frac{1}{\abscard{\Lambda}^{2}}
\sum_{x,y \in \Lambda}
\fun{\gamma_{\psi_{\sminvtemperature,\Lambda}}}{x,y}
&=
\frac{1}{\abscard{\Lambda}}
\fun{\oastate[\psi_{\sminvtemperature,\Lambda}]}{
\widetilde{S}^{1}_{0}\widetilde{S}^{1}_{0}
+
\widetilde{S}^{2}_{0}\widetilde{S}^{2}_{0}
}
+
\frac{1}{\abscard{\Lambda}^{2}}
\sum_{x \in \Lambda}
\fun{\oastate[\psi_{\sminvtemperature,\Lambda}]}{
S^{3}_{x}}.
\end{aligned}
\end{equation}
The longitudinal correction obeys
\begin{equation}\label{eq:bec-density-zero-mode-error}
\abs{\frac{1}{\abscard{\Lambda}^{2}}
\sum_{x \in \Lambda}
\fun{\oastate[\psi_{\sminvtemperature,\Lambda}]}
{S^{3}_{x}}}
\leq
\frac{1}{2\abscard{\Lambda}}.
\end{equation}
\end{lem}

\begin{proof}
Set
$$\opdmat_{\sminvtemperature,\Lambda}
=
\frac{\napiernum^{-\sminvtemperature\physham_{\Lambda}}}
{\sqfun{\trace}{\napiernum^{-\sminvtemperature\physham_{\Lambda}}}}.$$
The transpose in the occupation-number basis satisfies
$$\latp{S^{1}_{x}}
=
S^{1}_{x},
\quad
\latp{S^{2}_{x}}
=
-S^{2}_{x},
\quad
\latp{\physham_{\Lambda}}
=
\physham_{\Lambda}.
$$
The last identity follows directly from
\eqref{eq:hamiltonian-spin},
because every term containing $S^{2}$ contains two such factors.
The Gibbs density operator therefore satisfies
$\latp{\opdmat_{\sminvtemperature,\Lambda}}
=
\opdmat_{\sminvtemperature,\Lambda}.$

Fix distinct sites
$x
\neq
y$.
The operators at distinct sites commute.
Thus each operator
$B$
in
$\setone{S^{2}_{x}S^{1}_{y},S^{1}_{x}S^{2}_{y}}$
satisfies
$\latp{B}
=
-B$.
The trace calculation is
$$\begin{aligned}
&\fun{\oastate[\psi_{\sminvtemperature,\Lambda}]}{B}
=
\sqfun{\trace}{\opdmat_{\sminvtemperature,\Lambda}B}
=
\sqfun{\trace}
{\latprbk{\opdmat_{\sminvtemperature,\Lambda}B}}
=
\sqfun{\trace}{\latp{B}\latp{\opdmat_{\sminvtemperature,\Lambda}}}
=
-\sqfun{\trace}{B\opdmat_{\sminvtemperature,\Lambda}}
=
-\sqfun{\trace}{\opdmat_{\sminvtemperature,\Lambda}B}
\\ 
&=
-\fun{\oastate[\psi_{\sminvtemperature,\Lambda}]}{B}.
\end{aligned}$$
Consequently the two mixed correlations vanish:
\begin{equation}\label{eq:bec-mixed-spin-correlations}
\fun{\oastate[\psi_{\sminvtemperature,\Lambda}]}
{S^{2}_{x}S^{1}_{y}}
=
0,
\quad
\fun{\oastate[\psi_{\sminvtemperature,\Lambda}]}
{S^{1}_{x}S^{2}_{y}}
=
0,
\quad
x
\neq
y.
\end{equation}

The definitions
\eqref{eq:main-fourier-density-matrix},
\eqref{eq:boson-spin-dictionary},
and
\eqref{eq:spin-wave}
give the direct expansion
\begin{equation}\label{eq:app-density-fourier-diagonal-expansion}
\begin{aligned}
&\fun{\faftr{\gamma_{\psi_{\sminvtemperature,\Lambda}}}}{p,p}
=
\frac{1}{\abscard{\Lambda}}
\sum_{x,y \in \Lambda}
\napiernum^{\imunit p\cdot\rbk{x-y}}
\fun{\oastate[\psi_{\sminvtemperature,\Lambda}]}
{S^{+}_{x}S^{-}_{y}}
\\ 
&=
\fun{\oastate[\psi_{\sminvtemperature,\Lambda}]}
{\widetilde{S}^{1}_{p}\widetilde{S}^{1}_{-p}
+\widetilde{S}^{2}_{p}\widetilde{S}^{2}_{-p}}
+\frac{\imunit}{\abscard{\Lambda}}
\sum_{x,y \in \Lambda}
\napiernum^{\imunit p\cdot\rbk{x-y}}
\fun{\oastate[\psi_{\sminvtemperature,\Lambda}]}
{S^{2}_{x}S^{1}_{y}-S^{1}_{x}S^{2}_{y}}.
\end{aligned}
\end{equation}
Every term with
$x
\neq
y$
in the last sum vanishes by
\eqref{eq:bec-mixed-spin-correlations}.
For the diagonal terms,
the spin commutation relation
\eqref{eq:main-spin-relations}
gives
\begin{equation}\label{eq:app-mixed-spin-diagonal-contribution}
\begin{aligned}
\frac{\imunit}{\abscard{\Lambda}}
\sum_{x
\in
\Lambda}
\fun{\oastate[\psi_{\sminvtemperature,\Lambda}]}
{S^{2}_{x}S^{1}_{x}-S^{1}_{x}S^{2}_{x}}
=
\frac{\imunit}{\abscard{\Lambda}}
\sum_{x \in \Lambda}
\fun{\oastate[\psi_{\sminvtemperature,\Lambda}]}
{\commutator{S^{2}_{x}}{S^{1}_{x}}}
=
\frac{1}{\abscard{\Lambda}}
\sum_{x \in \Lambda}
\fun{\oastate[\psi_{\sminvtemperature,\Lambda}]}{S^{3}_{x}}
=
\frac{1}{\abscard{\Lambda}}
\fun{\oastate[\psi_{\sminvtemperature,\Lambda}]}
{S^{3}_{\txttot,\Lambda}}.
\end{aligned}
\end{equation}
Substitution of
\eqref{eq:app-mixed-spin-diagonal-contribution}
into
\eqref{eq:app-density-fourier-diagonal-expansion}
proves
\eqref{eq:app-density-fourier-diagonal}.

At zero momentum,
the definition
\eqref{eq:main-fourier-density-matrix}
becomes
\begin{equation}\label{eq:app-density-zero-momentum-average}
\fun{\faftr{\gamma_{\psi_{\sminvtemperature,\Lambda}}}}{0,0}
=
\frac{1}{\abscard{\Lambda}}
\sum_{x,y
\in
\Lambda}
\fun{\gamma_{\psi_{\sminvtemperature,\Lambda}}}{x,y}.
\end{equation}
Divide
\eqref{eq:app-density-zero-momentum-average}
by
$\abscard{\Lambda}$
and substitute
$p
=
0$
in
\eqref{eq:app-density-fourier-diagonal}.
The resulting identity is
\eqref{eq:bec-density-zero-mode-identity}.

Finally,
the norm bound for a state and the triangle inequality give
$$\begin{aligned}
\abs{\frac{1}{\abscard{\Lambda}^{2}}
\sum_{x \in \Lambda}
\fun{\oastate[\psi_{\sminvtemperature,\Lambda}]}{S^{3}_{x}}}
\leq
\frac{1}{\abscard{\Lambda}^{2}}
\sum_{x \in \Lambda}
\norm{S^{3}_{x}}
=
\frac{1}{2\abscard{\Lambda}}.
\end{aligned}$$
This proves
\eqref{eq:bec-density-zero-mode-error}.
\end{proof}

\subsection{A Finite-Dimensional Trace Inequality}\label{a-finite-dimensional-trace-inequality}

\subsubsection{Peierls--Bogoliubov inequality}\label{peierlsbogoliubov-inequality}

The Peierls--Bogoliubov inequality compares two finite-dimensional partition functions. It supplies the trace lower bound in Lemma \ref{lem:free-energy-bounds} and the Gibbs variational estimate \eqref{eq:peierls-bogoliubov-free-energy} used for Theorem \ref{thm:density-staggered}.

\begin{lem}[Peierls--Bogoliubov inequality]\label{lem:peierls-bogoliubov}
Let $A$ and $B$ be self-adjoint operators on a finite-dimensional Hilbert
space, and define
$$
\rho_{A}
=
\frac{\napiernum^{A}}{\sqfun{\trace}{\napiernum^{A}}}.
$$
Then
\begin{equation}\label{eq:peierls-bogoliubov}
\log\sqfun{\trace}{\napiernum^{A+B}}
\geq
\log\sqfun{\trace}{\napiernum^{A}}
+
\sqfun{\trace}{\rho_{A}B}.
\end{equation}
Equivalently, for self-adjoint Hamiltonians $\physham_{0}$ and
$\physham_{0}+V$ and $\beta>0$,
\begin{equation}\label{eq:peierls-bogoliubov-free-energy}
-\frac{1}{\beta}
\log\sqfun{\trace}{\napiernum^{-\beta\rbk{\physham_{0}+V}}}
\leq
-\frac{1}{\beta}
\log\sqfun{\trace}{\napiernum^{-\beta\physham_{0}}}
+
\frac{
\sqfun{\trace}{V\napiernum^{-\beta\physham_{0}}}
}{
\sqfun{\trace}{\napiernum^{-\beta\physham_{0}}}
}.
\end{equation}
\end{lem}

\begin{proof}
For $t\in\closedinterval{0}{1}$, define
$$\begin{aligned}
K_{t}
=
A+tB,
\quad
F(t)
=
\log\sqfun{\trace}{\napiernum^{K_{t}}}.
\end{aligned}$$
The Duhamel derivative formula and cyclicity of the trace give
$$\begin{aligned}
\opod{t}
\sqfun{\trace}{\napiernum^{K_{t}}}
=
\int_{0}^{1}
\sqfun{\trace}
{\napiernum^{\rbk{1-s}K_{t}}
B
\napiernum^{sK_{t}}}
\opdmsr{s}
=
\sqfun{\trace}{B\napiernum^{K_{t}}}.
\end{aligned}$$
With
$\rho_{t}
=
\napiernum^{K_{t}}/\sqfun{\trace}{\napiernum^{K_{t}}}$,
the first derivative is
$$F'(t)
=
\sqfun{\trace}{\rho_{t}B}.$$
Differentiating once more gives
$$\begin{aligned}
F''(t)
&=
\frac{1}{\sqfun{\trace}{\napiernum^{K_{t}}}}
\int_{0}^{1}
\sqfun{\trace}
{B
\napiernum^{sK_{t}}
B
\napiernum^{\rbk{1-s}K_{t}}}
\opdmsr{s}
-\sqfun{\trace}{\rho_{t}B}^{2}
\\ 
&=
\rbkt{B-\sqfun{\trace}{\rho_{t}B}}
{B-\sqfun{\trace}{\rho_{t}B}}_{1,-K_{t}}
\\ 
&\geq
0.
\end{aligned}$$
The final inequality is the positivity in Lemma
\ref{lem:duhamel-properties}.
Thus $F$ is convex, and
$$F(1)
\geq
F(0)+F'(0)$$
is \eqref{eq:peierls-bogoliubov}.
Substitution of
$A=-\beta\physham_{0}$ and $B=-\beta V$ gives
\eqref{eq:peierls-bogoliubov-free-energy}.
\end{proof}

\subsubsection{Ground-state derivatives}\label{ground-state-derivatives}

The derivative of a finite-volume ground-state energy is determined by the perturbation restricted to the ground eigenspace. This is the finite-dimensional Feynman--Hellmann formula used in \eqref{eq:density-feynman-hellmann}.

\begin{prop}[Feynman--Hellmann formula]\label{prop:feynman-hellmann}
Let $A$ and $B$ be self-adjoint operators on a finite-dimensional Hilbert
space.
For $t\in\fldreal$, set
\begin{equation}\label{eq:feynman-hellmann-hamiltonian}
\physham(t)
=
A+tB,
\quad
\fun{E_{\txtgs}}{t}
=
\min\opvarspec{\physham(t)}.
\end{equation}
Let $G_{t}$ be the eigenspace of \eqref{eq:feynman-hellmann-hamiltonian}
at $\fun{E_{\txtgs}}{t}$, and let $P_{t}$ be the orthogonal projection onto
$G_{t}$.
The one-sided derivatives exist and satisfy
\begin{equation}\label{eq:feynman-hellmann-one-sided}
\begin{aligned}
\fun{E'_{\txtgs,+}}{t}
&=
\min\opvarspec{\fnrestr{P_{t}BP_{t}}{G_{t}}},
\\ 
\fun{E'_{\txtgs,-}}{t}
&=
\max\opvarspec{\fnrestr{P_{t}BP_{t}}{G_{t}}}.
\end{aligned}
\end{equation}
If $G_{t}$ is one-dimensional and is spanned by a normalized vector
$\Psi_{\txtgs}(t)$, then $\fun{E_{\txtgs}}{t}$ is differentiable at $t$ and
\begin{equation}\label{eq:feynman-hellmann-simple}
\fun{E'_{\txtgs}}{t}
=
\bkt{\Psi_{\txtgs}(t)}{B\Psi_{\txtgs}(t)}.
\end{equation}
\end{prop}

\begin{proof}
Put
$$
m_{t}
=
\min\opvarspec{\fnrestr{P_{t}BP_{t}}{G_{t}}}.
$$
The unit sphere of $G_{t}$ is compact, so there is a normalized vector
$\Phi_{t}\in G_{t}$ with
$\bkt{\Phi_{t}}{B\Phi_{t}}=m_{t}$.
For $h>0$, the variational principle gives
$$
\fun{E_{\txtgs}}{t+h}
\leq
\bkt{\Phi_{t}}{\physham(t+h)\Phi_{t}}
=
\fun{E_{\txtgs}}{t}
+
hm_{t}.
$$
Hence
$$
\limsup_{h\downarrow0}
\frac{
\fun{E_{\txtgs}}{t+h}
-
\fun{E_{\txtgs}}{t}
}{h}
\leq
m_{t}.
$$

Let $\Psi_{h}$ be a normalized ground-state vector of $\physham(t+h)$.
The operator-norm estimate
$$
\abs{
\fun{E_{\txtgs}}{t+h}
-
\fun{E_{\txtgs}}{t}
}
\leq
h\norm{B}
$$
and the ground-state equation give
$$
0
\leq
\bkt{\Psi_{h}}{
\rbk{\physham(t)-\fun{E_{\txtgs}}{t}}
\Psi_{h}
}
=
\fun{E_{\txtgs}}{t+h}
-
\fun{E_{\txtgs}}{t}
-
h\bkt{\Psi_{h}}{B\Psi_{h}}
\leq
2h\norm{B}.
$$
Every sequence $\seq{h_{j}}{j\in\semigrposint}$ with $h_{j}\downarrow0$
has a subsequence for which $\Psi_{h_{j}}$ converges to a normalized vector
in $G_{t}$.
The same identity shows that every limit point lies in $G_{t}$.
Therefore
$$
\liminf_{h\downarrow0}
\frac{
\fun{E_{\txtgs}}{t+h}
-
\fun{E_{\txtgs}}{t}
}{h}
\geq
m_{t}.
$$
The two bounds prove the first identity in
\eqref{eq:feynman-hellmann-one-sided}.

Apply that identity to $s\mapsto\physham(t-s)$, whose perturbation is
$-B$.
Its right derivative at $s=0$ is
$-\fun{E'_{\txtgs,-}}{t}$.
The first identity in \eqref{eq:feynman-hellmann-one-sided} for $-B$
therefore gives the second identity.
If $G_{t}$ is one-dimensional, the minimum and maximum in
\eqref{eq:feynman-hellmann-one-sided} agree and equal the right side of
\eqref{eq:feynman-hellmann-simple}.
\end{proof}

\subsubsection{Vectorization and trace Schwarz inequality}\label{vectorization-and-trace-schwarz-inequality}

Let \(\sphilb{H}_{1}\) be a finite-dimensional Hilbert space with a fixed real orthonormal basis \(\seq{e_{X}}{X\in I}\). The vectorization map identifies \begin{equation}\label{eq:vectorization-map}
\Psi
=
\sum_{X,Y\in I}
\fun{\phi}{X,Y}e_{X}\otimes e_{Y}
\end{equation} with the matrix \(\widehat{\Psi}\) whose entries are \(\fun{\phi}{X,Y}\). For matrices \(A\) and \(B\) on \(\sphilb{H}_{1}\), \begin{equation}\label{eq:vector-operator-dictionary}
\begin{aligned}
\rbk{A\otimes 1}\Psi
&\leftrightarrow
A\widehat{\Psi},
\\ 
\rbk{1\otimes B}\Psi
&\leftrightarrow
\widehat{\Psi}\latp{B},
\\ 
\bkt{\Psi}{\rbk{A\otimes B}\Psi}
&=
\sqfun{\trace}{\faadj{\widehat{\Psi}}A\widehat{\Psi}\latp{B}}.
\end{aligned}
\end{equation}

\begin{lem}[trace Schwarz inequality]\label{lem:trace-schwarz}
For matrices $\widehat{\Psi}$ and $A$ on $\sphilb{H}_{1}$,
$$\abs{\sqfun{\trace}{\faadj{\widehat{\Psi}} A \widehat{\Psi} \faadj{A}}}
\leq \sqfun{\trace}{\rbk{\widehat{\Psi} \faadj{\widehat{\Psi}}}^{\onehalf} A \rbk{\widehat{\Psi} \faadj{\widehat{\Psi}}}^{\onehalf} \faadj{A}}^{\onehalf}
\sqfun{\trace}{\rbk{\faadj{\widehat{\Psi}} \widehat{\Psi}}^{\onehalf} A \rbk{\faadj{\widehat{\Psi}} \widehat{\Psi}}^{\onehalf} \faadj{A}}^{\onehalf}.$$
\end{lem}

\begin{proof}
Write the polar decomposition as
$$\begin{aligned}
\widehat{\Psi}
=
UX,
\quad
X
=
\rbk{\faadj{\widehat{\Psi}}\widehat{\Psi}}^{1/2}.
\end{aligned}$$
Extend the partial isometry $U$ to a unitary.
The polar decomposition gives
$$\rbk{\widehat{\Psi}\faadj{\widehat{\Psi}}}^{1/2}
=UX\faadj{U}.$$
It also gives
$$\sqfun{\trace}{\faadj{\widehat{\Psi}} A \widehat{\Psi} \faadj{A}}
=\sqfun{\trace}{X \faadj{U} A U X \faadj{A}}
=\sqfun{\trace}{\rbk{X^{\onehalf} \faadj{U} A U X^{\onehalf}} \rbk{X^{\onehalf} \faadj{A} X^{\onehalf}}}.$$
The Cauchy--Schwarz inequality for the Hilbert--Schmidt inner product bounds
this trace by
$$\sqfun{\trace}{\faadj{P}P}^{1/2}
\sqfun{\trace}{Q\faadj{Q}}^{1/2},
\quad
P
=
\faadj{\rbk{X^{1/2}\faadj{U}AUX^{1/2}}},
\quad
Q
=
X^{1/2}\faadj{A}X^{1/2}.$$
Direct calculation gives
$$\sqfun{\trace}{\faadj{P}P}
=\sqfun{\trace}{
\rbk{UX\faadj{U}}\faadj{A}
\rbk{UX\faadj{U}}A}.$$
This trace is real and is unchanged after taking the adjoint.
It therefore equals
$$\sqfun{\trace}
{\rbk{UX\faadj{U}}A
\rbk{UX\faadj{U}}\faadj{A}}.$$
Similarly we obtain
$$\sqfun{\trace}{Q\faadj{Q}}
=\sqfun{\trace}{XAX\faadj{A}}.$$
\end{proof}

\subsubsection{Perron--Frobenius criterion}\label{perronfrobenius-criterion}

The half-filling proof uses the following finite-dimensional form of the Perron--Frobenius argument. Its hypothesis is exactly the connectivity of the configuration graph.

\begin{defn}[connected matrix]\label{def:connected-matrix}
Let $T$ be a real square matrix indexed by a finite set $I$.
Its adjacency graph has vertex set $I$ and has an edge between distinct
$i,j\in I$ exactly when $T_{ij}>0$.
The matrix $T$ is connected when its adjacency graph is connected.
\end{defn}

\begin{lem}[finite Perron--Frobenius criterion]\label{lem:finite-perron-frobenius}
Let $T$ be a real symmetric matrix with nonnegative entries.
Suppose that $T$ is connected in the sense of Definition
\ref{def:connected-matrix}.
Then the largest eigenvalue of $T$ is simple and has a normalized eigenvector
with strictly positive entries.
\end{lem}

For every nonzero \(v=\rbk{v_{i}}_{i\in I}\in\fldreal^{I}\), the Rayleigh quotient of \(T\) at \(v\) is \(\bkt{v}{Tv}/\norm{v}^{2}\). The vector of coordinatewise absolute values is \(\rbk{\abs{v_{i}}}_{i\in I}\).

\begin{proof}
Let $r$ be the largest eigenvalue of $T$.
We first work in the real vector space $\fldreal^{I}$.
Since $T$ has real entries, the real and imaginary parts of every complex
eigenvector at $r$ are real eigenvectors at $r$.
Because the complex eigenspace is the complexification of the real
eigenspace, proving that the real eigenspace is one-dimensional is
sufficient.

The Rayleigh quotient of the vector $\rbk{\abs{v_{i}}}_{i\in I}$ is at least
that of $v$ because every entry of $T$ is nonnegative.
The spectral theorem therefore provides a nonnegative eigenvector $u$ for
the largest eigenvalue $r$.
If $u_{i}=0$, then
$0
=
\rbk{Tu}_{i}
=
\sum_{j}T_{ij}u_{j}$.
Every summand is nonnegative.
Thus $u_{j}=0$ for every neighbour $j$ of $i$.
Connectivity would then give $u=0$.
Consequently every entry of $u$ is strictly positive.

Let $v$ be any real eigenvector at $r$ and choose $t$ such that
$tu_{i}\geq\abs{v_{i}}$ for every $i$, with equality for at least one
index.
After replacing $v$ by $-v$ if necessary, the vector
$z=tu-v$ is nonnegative and has a zero entry.
It satisfies $Tz=rz$.
At an index $i$ with $z_{i}=0$, the preceding nonnegative-summand argument
forces $z_{j}=0$ at every neighbour of $i$.
Connectivity gives $z=0$.
Hence $v=tu$.
The one-dimensionality of the real eigenspace and the complexification
observation at the beginning of the proof make the largest eigenvalue simple.
\end{proof}

\subsection{Numerical Evaluation of the Three-Dimensional Infrared Integral}\label{numerical-evaluation-of-the-three-dimensional-infrared-integral}

This appendix fixes the normalization of the three-dimensional infrared integral and records the numerical value used in Corollary \ref{cor:bec-explicit-region}.

\begin{lem}[three-dimensional infrared integral]
The constant
$c_{3}$
defined by
\eqref{eq:main-condensation-constant}
satisfies
\begin{equation}\label{eq:app-c3-numerical-evaluation}
\begin{aligned}
c_{3}
&=
\frac{\sqrt{6}}{96 \pi^{3}}
\fngamma{\frac{1}{24}}
\fngamma{\frac{5}{24}}
\fngamma{\frac{7}{24}}
\fngamma{\frac{11}{24}}
=
0.5054620197\ldots,
\\ 
\frac{1}{c_{3}^{2}}
-
3
&=
0.9140191234\ldots.
\end{aligned}
\end{equation}
\end{lem}
\begin{proof}
The definition of
$E_{p}$
in
\eqref{eq:main-condensation-constant}
gives
$$
c_{3}
=
\frac{1}{\rbk{2\pi}^{3}}
\int_{\closedinterval{-\pi}{\pi}^{3}}
\frac{1}{
3-\cos p_{1}-\cos p_{2}-\cos p_{3}
}
\opdmsr{p}.
$$
Introduce the normalized simple-cubic Watson integral
$$
W_{\mathrm{sc}}
=
\frac{1}{\rbk{2\pi}^{3}}
\int_{\closedinterval{-\pi}{\pi}^{3}}
\frac{1}{
1-\frac{1}{3}
\rbk{\cos p_{1}+\cos p_{2}+\cos p_{3}}
}
\opdmsr{p}.
$$
The two denominators differ by the factor
$3$.
Consequently,
$$
W_{\mathrm{sc}}
=
3c_{3}.
$$
The classical simple-cubic Watson-integral evaluation
\cite[p. 1801]{GlasserZucker001}
is
$$
W_{\mathrm{sc}}
=
\frac{\sqrt{6}}{32 \pi^{3}}
\fngamma{\frac{1}{24}}
\fngamma{\frac{5}{24}}
\fngamma{\frac{7}{24}}
\fngamma{\frac{11}{24}}.
$$
Dividing this identity by
$3$
and evaluating the four gamma factors gives
\eqref{eq:app-c3-numerical-evaluation}.
The resulting value of
$c_{3}$
agrees with the estimate quoted in
\cite{DysonLiebSimon001}.
\end{proof}

\subsection{Koma--Tasaki and Tasaki finite-volume results}\label{app:koma-tasaki}

The tower and phase-localization arguments use a finite-volume result for a conserved \(\fun{\liegr{U}}{1}\) charge and two extensive order operators. This appendix records precisely the part needed in the main text. The source is the \(\fun{\liegr{U}}{1}\) discussion in Section 4.2 of \cite{HalTasaki008}, which presents the Koma--Tasaki tower theorem and Tasaki's phase-selection method.

\begin{defn}[Koma--Tasaki finite-volume setting]\label{def:app-koma-tasaki-setting}
For every periodic box $\Lambda$, let $\physham_{\Lambda}$ act on a finite-dimensional Hilbert space and be a sum of uniformly bounded finite-range terms.
Let $C_{\Lambda}$, $O^{(1)}_{\Lambda}$, and $O^{(2)}_{\Lambda}$ be self-adjoint extensive operators such that
\begin{equation}\label{eq:app-koma-tasaki-commutators}
\commutator{C_{\Lambda}}{O^{(1)}_{\Lambda}}
=
\imunit O^{(2)}_{\Lambda},
\quad
\commutator{C_{\Lambda}}{O^{(2)}_{\Lambda}}
=
-\imunit O^{(1)}_{\Lambda},
\quad
\commutator{\physham_{\Lambda}}{C_{\Lambda}}
=
0.
\end{equation}
Put $O^{\pm}_{\Lambda}=O^{(1)}_{\Lambda}\pm\imunit O^{(2)}_{\Lambda}$.
Assume that $\Psi_{\txtgs,\Lambda}$ is a normalized ground-state vector with
\begin{equation}\label{eq:app-koma-tasaki-neutral-lro}
C_{\Lambda}\Psi_{\txtgs,\Lambda}
=
0,
\quad
\fun{\oastate[\psi_{\txtgs,\Lambda}]}{\rbk{O^{(i)}_{\Lambda}}^{2}}
\geq
q_{0}\abscard{\Lambda}^{2}
\quad
\rbk{i=1,2},
\end{equation}
with a constant $q_{0}>0$ independent of $\Lambda$.
\end{defn}

The first relation in \eqref{eq:app-koma-tasaki-commutators} fixes the charge of the ladder operators: \begin{equation}\label{eq:app-koma-tasaki-charge-ladder}
\commutator{C_{\Lambda}}{O^{\pm}_{\Lambda}}
=
\pm O^{\pm}_{\Lambda}.
\end{equation} For every \(k\in\ringratint\), whenever the denominator is nonzero, define \begin{equation}\label{eq:app-koma-tasaki-ladder-states}
\Gamma^{\mathrm{KT}}_{\Lambda,k}
=
\begin{cases}
\displaystyle
\frac{
\rbk{O^{+}_{\Lambda}}^{k}\Psi_{\txtgs,\Lambda}
}{
\norm{\rbk{O^{+}_{\Lambda}}^{k}\Psi_{\txtgs,\Lambda}}
},
&
k>0,
\\[2ex]
\Psi_{\txtgs,\Lambda},
&
k=0,
\\[1ex]
\displaystyle
\frac{
\rbk{O^{-}_{\Lambda}}^{\abs{k}}\Psi_{\txtgs,\Lambda}
}{
\norm{\rbk{O^{-}_{\Lambda}}^{\abs{k}}\Psi_{\txtgs,\Lambda}}
},
&
k<0.
\end{cases}
\end{equation}

\begin{prop}[charge of the ladder states]\label{prop:app-koma-tasaki-ladder-charge}
The vector $\Gamma^{\mathrm{KT}}_{\Lambda,k}$ in \eqref{eq:app-koma-tasaki-ladder-states} belongs to the $C_{\Lambda}=k$ eigenspace.
Consequently, ladder states with distinct indices are orthogonal.
\end{prop}

\begin{proof}
Equation \eqref{eq:app-koma-tasaki-charge-ladder} gives
$$
C_{\Lambda}\rbk{O^{+}_{\Lambda}}^{k}
=
\rbk{O^{+}_{\Lambda}}^{k}C_{\Lambda}
+
k\rbk{O^{+}_{\Lambda}}^{k}
$$
for $k>0$.
The first identity in \eqref{eq:app-koma-tasaki-neutral-lro} therefore gives
$C_{\Lambda}\Gamma^{\mathrm{KT}}_{\Lambda,k}=k\Gamma^{\mathrm{KT}}_{\Lambda,k}$.
The same calculation with $O^{-}_{\Lambda}$ gives the assertion for $k<0$.
Eigenspaces of the self-adjoint operator $C_{\Lambda}$ with distinct eigenvalues are orthogonal.
\end{proof}

The following theorem is the Koma--Tasaki finite-size estimate in the form used for the present model. It is the \(\fun{\liegr{U}}{1}\) specialization of the tower result stated in \cite[Section 4.2, Theorem 4.6, and the discussion after Eq. (4.2.29)]{HalTasaki008}.

\begin{thm}[Koma--Tasaki tower estimate]\label{thm:app-koma-tasaki-tower}
Under Definition \ref{def:app-koma-tasaki-setting}, there are constants $c_{\mathrm{KT}},C_{\mathrm{KT}}>0$, independent of $\Lambda$, such that the vectors \eqref{eq:app-koma-tasaki-ladder-states} are defined and satisfy
\begin{equation}\label{eq:app-koma-tasaki-energy-bound}
\bkt{\Gamma^{\mathrm{KT}}_{\Lambda,k}}{
\physham_{\Lambda}\Gamma^{\mathrm{KT}}_{\Lambda,k}
}
-
\physenergyfunc_{\Lambda,0}
\leq
C_{\mathrm{KT}}
\frac{k^{2}}{\abscard{\Lambda}}
\end{equation}
whenever $\abs{k}\leq c_{\mathrm{KT}}\sqrt{\abscard{\Lambda}}$.
\end{thm}

The phase-selection theorem uses powers of one self-adjoint order component rather than an equal-weight sum of charge-ladder vectors. This is the construction in Eq. (4.2.10) of \cite{HalTasaki008}. For a subsequence along which the moments exist, define \begin{equation}\label{eq:app-tasaki-maximal-order-parameter}
m_{\ast}
=
\lim_{r\to\infty}
\rbk{\lim_{j\to\infty}
\fun{\oastate[\psi_{\txtgs,\Lambda_{j}}]}
{\rbk{\frac{O^{(1)}_{\Lambda_{j}}}{\abscard{\Lambda_{j}}}}^{2r}}}^{\frac{1}{2r}}.
\end{equation} For a positive integer \(M\), the phase-selected vector at angle \(0\) is \begin{equation}\label{eq:app-tasaki-phase-vector}
\Xi^{\mathrm{T}}_{\Lambda,0}(M)
=
\frac{1}{\sqrt{2}}
\rbk{\frac{\rbk{O^{(1)}_{\Lambda}}^{M}\Psi_{\txtgs,\Lambda}}
{\norm{\rbk{O^{(1)}_{\Lambda}}^{M}\Psi_{\txtgs,\Lambda}}}
+\frac{\rbk{O^{(1)}_{\Lambda}}^{M+1}\Psi_{\txtgs,\Lambda}}
{\norm{\rbk{O^{(1)}_{\Lambda}}^{M+1}\Psi_{\txtgs,\Lambda}}}}.
\end{equation} The vector state of \eqref{eq:app-tasaki-phase-vector} is \[\fun{\oastate[\xi_{\Lambda,0,M}]}{A}
=
\bkt{\Xi^{\mathrm{T}}_{\Lambda,0}(M)}
{A\Xi^{\mathrm{T}}_{\Lambda,0}(M)}\] for \(A\) in the finite-volume algebra.

\begin{thm}[Tasaki phase selection]\label{thm:app-tasaki-phase-selection}
Under Definition \ref{def:app-koma-tasaki-setting},
one can choose $M_{j} \to \infty$ sufficiently slowly,
with $M_{j} \leq c_{\mathrm{KT}}\sqrt{\abscard{\Lambda_{j}}}$,
so that the vectors \eqref{eq:app-tasaki-phase-vector} satisfy
\begin{equation}\label{eq:app-tasaki-phase-selection-limit}
\lim_{j\to\infty}
\fun{\oastate[\xi_{\Lambda_{j},0,M_{j}}]}
{\frac{O^{(1)}_{\Lambda_{j}}}{\abscard{\Lambda_{j}}}}
=
m_{\ast},
\quad
\lim_{j\to\infty}
\fun{\oastate[\xi_{\Lambda_{j},0,M_{j}}]}
{\frac{O^{(2)}_{\Lambda_{j}}}{\abscard{\Lambda_{j}}}}
=
0,
\end{equation}
and the variances of the two order-parameter densities tend to zero.
The same choice of $M_{j}$ can be made sufficiently slow that
\begin{equation}\label{eq:app-tasaki-phase-selection-low-energy}
\frac{M_{j}^{2}}{\abscard{\Lambda_{j}}}
\to
0,
\quad
\lim_{j \to \infty}
\left[
\fun{\oastate[\xi_{\Lambda_{j},0,M_{j}}]}{\physham_{\Lambda_{j}}}
-
\physenergyfunc_{\Lambda_{j},0}
\right]
=
0.
\end{equation}
The long-range-order bound \eqref{eq:app-koma-tasaki-neutral-lro} implies
\begin{equation}\label{eq:app-tasaki-order-parameter-lower-bound}
m_{\ast}
\geq
\sqrt{2q_{0}}.
\end{equation}
\end{thm}

The phase-selection statement is Theorem 4.9 of \cite{HalTasaki008}. Its \(\fun{\liegr{U}}{1}\) form and the lower bound \eqref{eq:app-tasaki-order-parameter-lower-bound} are stated in the discussion following Eq. (4.2.29) and Eq. (4.2.39) there. For every fixed \(M\), the Koma--Tasaki commutator estimate applies to the two normalized power vectors in \eqref{eq:app-tasaki-phase-vector}. These vectors have opposite charge parity, and \(\commutator{\physham_{\Lambda}}{C_{\Lambda}} = 0\) by \eqref{eq:app-koma-tasaki-commutators}. Their cross energy matrix element vanishes, and the phase-selected vector satisfies \[
0
\leq
\fun{\oastate[\xi_{\Lambda,0,M}]}{\physham_{\Lambda}}
-
\physenergyfunc_{\Lambda,0}
\leq
C_{M}\abscard{\Lambda}^{-1}.
\] The phase-selection conditions impose only countably many limits. The diagonal choice of \(M_{j}\) can therefore be slowed until both limits in \eqref{eq:app-tasaki-phase-selection-low-energy} hold. The corresponding energy-density statement for the more general low-lying choice is \cite[Theorem 3.1 and Eq. (3.16)]{TasakiHal001}. The same source distinguishes sharpness for these two order parameters from ergodicity of every macroscopic observable. Section 4.3, Conjecture 4.21 of \cite{HalTasaki008} leaves the latter open for the continuous-symmetry construction. The next proposition gives the local-stability argument needed to pass from \eqref{eq:app-tasaki-phase-selection-low-energy} to an infinite-volume ground state.

\begin{prop}[low-energy limits are ground states]\label{prop:app-low-energy-limit-ground-state}
Let
$\seq{\Lambda_{j}}{j \in \semigrposint}$
be an increasing sequence of periodic boxes for a uniformly bounded
finite-range interaction.
Let $\Xi_{j}$ be normalized vectors and let
$\oastate[\xi_{j}]$ be their vector states.
Suppose that the total energy excess satisfies
\begin{equation}\label{eq:app-low-energy-limit-hypothesis}
\Delta_{j}
=
\bkt{\Xi_{j}}{\physham_{\Lambda_{j}}\Xi_{j}}
-
\physenergyfunc_{\Lambda_{j},0}
\to
0
\quad
\rbk{j \to \infty}.
\end{equation}
Every weak-$\ast$ limit point $\oastate[\xi]$ of these vector states is an
infinite-volume ground state.
More precisely, the derivation $\oaderiv$ in
\eqref{eq:analytic-extension} satisfies
\begin{equation}\label{eq:app-infinite-volume-ground-state-criterion}
-\imunit
\fun{\oastate[\xi]}{
\faadj{A}\fun{\oaderiv}{A}
}
\geq
0
\end{equation}
for every $A \in \oa{A}_{\txtloc}$.
\end{prop}

\begin{proof}
It suffices to pass to a subsequence on which the vector states converge to
$\oastate[\xi]$.
Define the shifted Hamiltonian and the spectral cutoff by
$$
K_{j}
=
\physham_{\Lambda_{j}}
-
\physenergyfunc_{\Lambda_{j},0}
\geq
0,
\quad
\epsilon_{j}
=
\max\setone{\sqrt{\Delta_{j}},j^{-1}}.
$$
Let $P_{j}$ be the spectral projection of $K_{j}$ for
$\closedinterval{0}{\epsilon_{j}}$.
The spectral theorem and
\eqref{eq:app-low-energy-limit-hypothesis} give
$$
\norm{\rbk{1-P_{j}}\Xi_{j}}^{2}
\leq
\frac{\Delta_{j}}{\epsilon_{j}}
\to
0.
$$
For all sufficiently large $j$, define the normalized vector by
$$
\Phi_{j}
=
\frac{P_{j}\Xi_{j}}{\norm{P_{j}\Xi_{j}}}.
$$
It satisfies
$$
\norm{\Phi_{j}-\Xi_{j}}
\to
0,
\quad
\norm{K_{j}\Phi_{j}}
\leq
\epsilon_{j}.
$$

Fix $A \in \oa{A}_{\txtloc}$.
Finite range gives a uniform bound on
$\commutator{\physham_{\Lambda_{j}}}{A}$, and this commutator agrees with
the infinite-volume local commutator for all sufficiently large $j$.
The finite-volume local-stability expression satisfies
$$
\begin{aligned}
-\imunit
\bkt{\Phi_{j}}{
\faadj{A}
\fun{\oaderiv_{\Lambda_{j}}}{A}
\Phi_{j}
}
&=
\bkt{A\Phi_{j}}{K_{j}A\Phi_{j}}
-
\bkt{\Phi_{j}}{\faadj{A}A K_{j}\Phi_{j}},
\\ 
\opreal \rbk{
-\imunit
\bkt{\Phi_{j}}{
\faadj{A}
\fun{\oaderiv_{\Lambda_{j}}}{A}
\Phi_{j}
}
}
&\geq
-\norm{A}^{2}\epsilon_{j}.
\end{aligned}
$$
The imaginary part of the first line is bounded in absolute value by
$\norm{A}^{2}\epsilon_{j}$ because its first term is nonnegative and real.
The norm convergence proved above and the uniform commutator bound allow us
to replace $\Phi_{j}$ by $\Xi_{j}$.
Taking the weak-$\ast$ limit proves
\eqref{eq:app-infinite-volume-ground-state-criterion}.
\end{proof}

\subsection{Exchangeable States and the Størmer Decomposition}\label{app:stormer}

The mean-field limit uses the following form of Størmer's theorem \cite{ErlingStormer003}. The proof is included to identify explicitly the mixing space and the uniqueness statement needed in Section \ref{sec:meanfield}.

Enumerate the lattice sites and write \(A^{(j)}\) for the copy of \(A\in\spmat{2}{\fldcmp}\) at the \(j\)-th site. A state on the resulting quasi-local algebra is exchangeable when it is invariant under every permutation of finitely many tensor factors.

\begin{thm}[Størmer decomposition]\label{thm:stormer-de-finetti}
Let $\oastate[\psi]$ be an exchangeable state on the quasi-local algebra
generated by countably many copies of $\spmat{2}{\fldcmp}$.
There is a unique Radon probability measure $P_\psi$ on the one-site state
space $\mathcal{B}$ defined in \eqref{eq:mf-bloch-ball}
such that
\begin{equation}\label{eq:stormer-barycentric-representation}
\oastate[\psi]
=
\int_{\mathcal{B}}
\rho^{\otimes\infty}
\opdmsr{P_\psi(\rho)}.
\end{equation}
Equivalently, for
$A_1,\ldots,A_k\in\spmat{2}{\fldcmp}$
at distinct sites,
\begin{equation}\label{eq:stormer-moment-representation}
\fun{\oastate[\psi]}{
A_1^{(1)}\dotsm A_k^{(k)}
}
=
\int_{\mathcal{B}}
\prod_{\ell=1}^{k}
\sqfun{\trace}{\rho A_\ell}
\opdmsr{P_\psi(\rho)}.
\end{equation}
The exchangeable state space is a simplex, and its extreme points are
exactly the product states
$\rho^{\otimes\infty}$,
$\rho\in\mathcal{B}$.
\end{thm}

\begin{proof}
Let
$\rbk{\oarepn_\psi,\sphilb{H}_\psi,\oagnsvector[\Psi]}$
be the GNS representation.
Set
$$
\oa{M}_\psi
=
\oadoublecommutant{\fun{\oarepn_\psi}{\oa{A}}},
\quad
\oa{Z}_\psi
=
\oa{M}_\psi
\cap
\oacommutant{\oa{M}_\psi}.
$$
Exchangeability gives a unitary representation
$$
U_\pi
\fun{\oarepn_\psi}{B}
\oagnsvector[\Psi]
=
\fun{\oarepn_\psi}{\fun{\alpha_\pi}{B}}
\oagnsvector[\Psi]
$$
of the finite permutations, where
$\alpha_\pi$ permutes the tensor factors.
For $N\in\semigrposint$, let
$$
P_N
=
\frac{1}{N!}
\sum_{\pi\in\grsym{N}}
U_\pi.
$$
This is the orthogonal projection onto the vectors invariant under
permutations of the first $N$ sites.
The subspaces decrease with $N$, so
$P_N$
converges strongly to a projection $P_\infty$.

For $A\in\spmat{2}{\fldcmp}$, define the empirical average
\begin{equation}\label{eq:stormer-empirical-average}
A_N
=
\frac{1}{N}
\sum_{j=1}^{N}
A^{(j)}.
\end{equation}
Counting the permutations according to the image of the first site gives
$$
\fun{\oarepn_\psi}{A_N}
\oagnsvector[\Psi]
=
P_N
\fun{\oarepn_\psi}{A^{(1)}}
\oagnsvector[\Psi].
$$
Hence this vector converges to
$$
\xi_A
=
P_\infty
\fun{\oarepn_\psi}{A^{(1)}}
\oagnsvector[\Psi].
$$

Let $B$ be supported on the first $M$ sites.
Only $M$ summands in
\eqref{eq:stormer-empirical-average}
can fail to commute with $B$, and therefore
\begin{equation}\label{eq:stormer-asymptotic-centrality}
\norm{\commutator{A_N}{B}}
\leq
\frac{2M}{N}
\norm{A}
\norm{B}.
\end{equation}
It follows that, on the dense set
$\fun{\oarepn_\psi}{\oa{A}_{\txtloc}}\oagnsvector[\Psi]$,
$$
\fun{\oarepn_\psi}{A_N}
\fun{\oarepn_\psi}{B}
\oagnsvector[\Psi]
\longrightarrow
\fun{\oarepn_\psi}{B}\xi_A
\quad
\rbk{N\to\infty}.
$$
The uniform bound
$\norm{A_N}\leq\norm{A}$
extends this convergence to all of $\sphilb{H}_\psi$.
Thus
\begin{equation}\label{eq:stormer-central-empirical-limit}
X_A
=
\slim_{N\to\infty}
\fun{\oarepn_\psi}{A_N}
\end{equation}
exists.
Every approximant belongs to $\oa{M}_\psi$, while
\eqref{eq:stormer-asymptotic-centrality} makes the limit commute with all
local observables.
Consequently,
$$
X_A
\in
\oa{Z}_\psi.
$$
The map
$A\mapsto X_A$
is linear and unital.
If $A\geq0$, then $A_N\geq0$ and hence $X_A\geq0$.
Thus this map is positive and therefore adjoint preserving.

Let $\oa{C}_\psi$ be the unital abelian $\oacstar$-algebra generated by
$\set{X_A}{A\in\spmat{2}{\fldcmp}}$.
For a character
$\chi\in\widehat{\oa{C}_\psi}$, let $\rho_\chi\in\mathcal{B}$ be the
unique density matrix satisfying
$$
\sqfun{\trace}{\rho_\chi A}
=
\fun{\chi}{X_A}.
$$
Existence and uniqueness follow from positivity and normalization of
$A\mapsto\fun{\chi}{X_A}$.
Thus
$$
r
\colon
\widehat{\oa{C}_\psi}
\to
\mathcal{B},
\quad
\fun{r}{\chi}
=
\rho_\chi
$$
is continuous.

The empirical averages recover every correlation at distinct sites.
Indeed,
$$
\fun{\oastate[\psi]}{
A_{1,N}\dotsm A_{k,N}
}
=
\frac{1}{N^k}
\sum_{j_1,\ldots,j_k=1}^{N}
\fun{\oastate[\psi]}{
A_1^{(j_1)}\dotsm A_k^{(j_k)}
}.
$$
The terms with pairwise distinct indices all equal
$\fun{\oastate[\psi]}{A_1^{(1)}\dotsm A_k^{(k)}}$
by exchangeability.
Set
$\rbk{N}_k=N\rbk{N-1}\dotsm\rbk{N-k+1}$.
There are $\rbk{N}_k$ tuples with pairwise distinct indices, while every
summand is bounded by $\prod_{\ell=1}^{k}\norm{A_\ell}$.
Hence the difference between the preceding average and the distinct-site
correlation is bounded by
$$
2\rbk{1-\frac{\rbk{N}_k}{N^k}}
\prod_{\ell=1}^{k}\norm{A_\ell},
$$
which tends to zero as $N\to\infty$.
Using
\eqref{eq:stormer-central-empirical-limit}
gives
\begin{equation}\label{eq:stormer-central-moment-identity}
\fun{\oastate[\psi]}{
A_1^{(1)}\dotsm A_k^{(k)}
}
=
\bkt{\oagnsvector[\Psi]}{
X_{A_1}\dotsm X_{A_k}
\oagnsvector[\Psi]
}.
\end{equation}

Let $\nu_\psi$ be the spectral measure of
$\oagnsvector[\Psi]$
on
$\widehat{\oa{C}_\psi}$.
The spectral theorem and
\eqref{eq:stormer-central-moment-identity}
give
$$
\fun{\oastate[\psi]}{
A_1^{(1)}\dotsm A_k^{(k)}
}
=
\int_{\widehat{\oa{C}_\psi}}
\prod_{\ell=1}^{k}
\sqfun{\trace}{\rho_\chi A_\ell}
\opdmsr{\nu_\psi(\chi)}.
$$
Set
$P_\psi=r_*\nu_\psi$.
This proves
\eqref{eq:stormer-moment-representation}, and local norm density gives
\eqref{eq:stormer-barycentric-representation}.

For uniqueness, the complex linear span of the functions
$$
\rho
\mapsto
\prod_{\ell=1}^{k}
\sqfun{\trace}{\rho A_\ell}
$$
is a unital self-adjoint algebra that separates the points of the compact
space $\mathcal{B}$.
The Stone--Weierstrass theorem makes this algebra uniformly dense in
$\fun{\conti}{\mathcal{B}}$.
Equation
\eqref{eq:stormer-moment-representation}
therefore determines $P_\psi$ uniquely.

Conversely, every Radon probability measure on $\mathcal{B}$ defines an
exchangeable state through
\eqref{eq:stormer-barycentric-representation}.
The resulting correspondence between exchangeable states and probability
measures on $\mathcal{B}$ is affine and bijective.
The moment formula and local norm density make the barycenter map continuous
from the weak topology of probability measures to the weak-$\ast$ topology
of states.
It is therefore an affine homeomorphism from a compact space to a Hausdorff
space.
The probability measures on $\mathcal{B}$ form a Bauer simplex whose extreme
points are the point masses, so the extreme
exchangeable states are precisely
$\rho^{\otimes\infty}$.
\end{proof}

\begin{cor}[two-class Størmer decomposition]\label{cor:stormer-two-class}
Let a state on two countable families of copies of
$\spmat{2}{\fldcmp}$ be invariant under finite permutations within each
family.
There is a unique Radon probability measure $P_\psi$ on
$\mathcal{B}\times\mathcal{B}$ such that
\begin{equation}\label{eq:stormer-two-class-barycenter}
\oastate[\psi]
=
\int_{\mathcal{B}\times\mathcal{B}}
\rbk{\rho_+^{\otimes\infty}\otimes\rho_-^{\otimes\infty}}
\opdmsr{P_\psi(\rho_+,\rho_-)}.
\end{equation}
The extreme separately exchangeable states are exactly the two-class product
states in the integrand of
\eqref{eq:stormer-two-class-barycenter}.
\end{cor}

\begin{proof}
Repeat the proof of Theorem \ref{thm:stormer-de-finetti} with one empirical
average for each family.
The two families of central empirical limits commute, so the joint spectrum
maps continuously into $\mathcal{B}\times\mathcal{B}$.
Mixed moments at distinct sites give
\eqref{eq:stormer-two-class-barycenter}.
The complex linear span of the corresponding products of one-site affine
functions is a unital self-adjoint algebra separating the points of
$\mathcal{B}\times\mathcal{B}$.
Stone--Weierstrass gives uniqueness, and point masses give the extreme
states.
\end{proof}

\subsection{Spatial Direct Integrals of Factor States}\label{app:spatial-direct-integrals}

The following criterion separates the model-specific construction of a central observable from the general spatial-decomposition theorem. It is the form of the direct-integral results of \cite[Section 4.4]{BratteliRobinson003} used in Section \ref{sec:meanfield}.

\begin{thm}[factor barycenter as a central direct integral]\label{thm:factor-barycenter-direct-integral}
Let $\oa{A}$ be a separable unital $\oacstar$-algebra, let
$\rbk{X,\mu}$ be a standard probability space, and let
$x\mapsto\rbk{\oarepn_x,\sphilb{H}_x,\oagnsvector[\Psi_x]}$ be a measurable
family of GNS triples of factor states $\oastate[\psi_x]$.
Suppose that a uniformly bounded sequence
$\seq{B_n}{n\in\semigrposint}$ in $\oa{A}$ and an essentially injective
$c\in\fun{\lp^\infty}{X,\mu}$ satisfy
\begin{equation}\label{eq:factor-barycenter-fiberwise-scalar-limit}
\slim_{n\to\infty}\fun{\oarepn_x}{B_n}
=
\fun{c}{x}1,
\quad
\slim_{n\to\infty}\fun{\oarepn_x}{\faadj{B_n}}
=
\cmpconj{\fun{c}{x}}1
\end{equation}
for almost every $x\in X$.
Then the direct integral
\begin{equation}\label{eq:factor-barycenter-direct-integral-triple}
\sphilb{H}
=
\int_X^\oplus\sphilb{H}_x\opdmsr{\mu(x)},
\quad
\oarepn
=
\int_X^\oplus\oarepn_x\opdmsr{\mu(x)},
\quad
\oagnsvector[\Psi]
=
\int_X^\oplus\oagnsvector[\Psi_x]\opdmsr{\mu(x)}
\end{equation}
is the GNS triple of
$\oastate[\psi]=\int_X\oastate[\psi_x]\opdmsr{\mu(x)}$.
Moreover,
\begin{equation}\label{eq:factor-barycenter-von-neumann-direct-integral}
\oadoublecommutant{\fun{\oarepn}{\oa{A}}}
=
\int_X^\oplus
\oadoublecommutant{\fun{\oarepn_x}{\oa{A}}}
\opdmsr{\mu(x)},
\quad
\oa{Z}_\psi
=
\fun{\lp^\infty}{X,\mu}.
\end{equation}
Here the last algebra acts by diagonal multiplication.
\end{thm}

\begin{proof}
Dominated convergence in the direct integral and
\eqref{eq:factor-barycenter-fiberwise-scalar-limit} give
$$
\slim_{n\to\infty}\fun{\oarepn}{B_n}
=
M_c,
\quad
\slim_{n\to\infty}\fun{\oarepn}{\faadj{B_n}}
=
\faadj{M_c},
$$
where $M_c$ is multiplication by $c$.
Since $c$ is essentially injective on the standard measure space,
the spectral projections of $M_c$ generate the full diagonal algebra
$\fun{\lp^\infty}{X,\mu}$.
Consequently,
\begin{equation}\label{eq:factor-barycenter-diagonal-inclusion}
\fun{\lp^\infty}{X,\mu}
\subset
\oadoublecommutant{\fun{\oarepn}{\oa{A}}}.
\end{equation}

Let $\oa{K}$ be the closure of
$\fun{\oarepn}{\oa{A}}\oagnsvector[\Psi]$, and let $P_{\oa{K}}$ be its
orthogonal projection.
The projection belongs to
$\oacommutant{\fun{\oarepn}{\oa{A}}}$ and commutes with the diagonal algebra
in \eqref{eq:factor-barycenter-diagonal-inclusion}.
It is therefore decomposable by the spatial-decomposition theorem
\cite[Section 4.4.1]{BratteliRobinson003}:
$$
P_{\oa{K}}
=
\int_X^\oplus P_x\opdmsr{\mu(x)}.
$$
Choose a countable norm-dense subset of $\oa{A}$.
Its represented GNS sections lie in $\oa{K}$, so fiberwise cyclicity gives
$P_x=1$ for almost every $x$.
Thus $P_{\oa{K}}=1$, proving the GNS assertion.

The diagonal algebra commutes with $\fun{\oarepn}{\oa{A}}$.
Thus \eqref{eq:factor-barycenter-diagonal-inclusion} places it in the center
of the generated von Neumann algebra, making that algebra decomposable.
Theorem 4.4.5 and Proposition 4.4.6 of
\cite{BratteliRobinson003} identify its fibers and give the first equality in
\eqref{eq:factor-barycenter-von-neumann-direct-integral}.
If an element is central, its fiber belongs to the center of the corresponding
factor and is therefore scalar almost everywhere.
This proves the second equality in
\eqref{eq:factor-barycenter-von-neumann-direct-integral}; compare
\cite[Proposition 4.4.6 and Corollary 4.4.8(b)]{BratteliRobinson003}.
\end{proof}

\begin{prop}[spatial criterion for the central measure]\label{prop:spatial-central-measure-criterion}
Let $\oa{A}$ be a separable unital $\oacstar$-algebra, and let
$\oastate[\psi]$ have a spatial GNS decomposition
$$
\rbk{\oarepn,\sphilb{H},\oagnsvector[\Psi]}
=
\int_X^\oplus
\rbk{\oarepn_x,\sphilb{H}_x,\oagnsvector[\Psi_x]}
\opdmsr{\mu(x)}
$$
over a standard probability space.
Suppose that almost every fiber state
$\oastate[\psi_x]$ is a factor state, the map
$x\mapsto\oastate[\psi_x]$ is injective on a conull Borel set, and
$$
\oa{Z}_\psi
=
\fun{\lp^\infty}{X,\mu}
$$
acts by diagonal multiplication.
Then the pushforward of $\mu$ under
$x\mapsto\oastate[\psi_x]$ is the central orthogonal measure of
$\oastate[\psi]$.
\end{prop}

\begin{proof}
Let $E\subset X$ be Borel, and let $M_{\fndef{E}}$ be multiplication by its
indicator.
The two positive functionals obtained by integrating the fiber states over
$E$ and $X\setminus E$ are represented by the orthogonal central projections
$M_{\fndef{E}}$ and $1-M_{\fndef{E}}$.
Thus the representing measure on $X$ is orthogonal.

The state space of $\oa{A}$ is a standard Borel space because $\oa{A}$ is
separable.
The injective Borel map
$x\mapsto\oastate[\psi_x]$ has a Borel inverse on its image after restriction
to the conull Borel set.
Consequently, the diagonal algebra generated by inverse images of Borel
subsets of the state space is all of
$\fun{\lp^\infty}{X,\mu}=\oa{Z}_\psi$.
The orthogonal measure therefore corresponds to the full GNS center and is
the central orthogonal measure.
Pushforward only replaces the conull parametrization by its image in the
state space.
\end{proof}

\begin{cor}[central measure of a factor barycenter]\label{cor:factor-barycenter-central-measure}
Under the hypotheses of Theorem
\ref{thm:factor-barycenter-direct-integral}, the pushforward of $\mu$ under
$x\mapsto\oastate[\psi_x]$ is the central orthogonal measure of
$\oastate[\psi]$.
For $f\in\fun{\lp^\infty}{X,\mu}$ and $A\in\oa{A}$,
\begin{equation}\label{eq:factor-barycenter-central-integral-identity}
\bkt{\oagnsvector[\Psi]}{
M_f\fun{\oarepn}{A}\oagnsvector[\Psi]}
=
\int_X
\fun{f}{x}\fun{\oastate[\psi_x]}{A}
\opdmsr{\mu(x)}.
\end{equation}
\end{cor}

\begin{proof}
Equation \eqref{eq:factor-barycenter-central-integral-identity} follows by
integrating the fiber vector states in
\eqref{eq:factor-barycenter-direct-integral-triple}.
Taking vector expectations in
\eqref{eq:factor-barycenter-fiberwise-scalar-limit} gives
$$
\lim_{n\to\infty}\fun{\oastate[\psi_x]}{B_n}
=
\fun{c}{x}.
$$
Hence equality of two fiber states forces equality of their $c$-values.
Essential injectivity of $c$ makes
$x\mapsto\oastate[\psi_x]$ injective on a conull set.
Proposition \ref{prop:spatial-central-measure-criterion}, together with
\eqref{eq:factor-barycenter-von-neumann-direct-integral}, identifies the
pushforward measure as the central orthogonal measure.
\end{proof}

\subsection{KMS States and Decomposition Theory}\label{app:kms}

The KMS condition, its stability under limits, and the central decomposition are facts about a general \(\oacstar\)-dynamical system. They are stated here in the form used by the optical-lattice gas. The model-specific verification of their hypotheses occurs in Sections \ref{sec:dynamics} and \ref{sec:decomposition}.

\subsubsection{KMS states and finite-dimensional Gibbs states}\label{kms-states-and-finite-dimensional-gibbs-states}

\begin{defn}[KMS state]\label{def:kms-state}
Let $\tau$ be a strongly continuous one-parameter group of
$\ast$-automorphisms of a $\oacstar$-algebra $\oa{A}$.
For $\beta>0$, a state $\oastate[\psi]$ on $\oa{A}$ is
$\rbk{\tau,\beta}$-KMS if, for every $A,B\in\oa{A}$, there is a function
$F_{A,B}$ that is bounded and continuous on
$\set{z\in\fldcmp}{0\leq\opimag z\leq\beta}$,
holomorphic in its interior,
and satisfies
$$\fun{F_{A,B}}{t}
=
\fun{\oastate[\psi]}{A\fun{\tau_t}{B}},
\quad
\fun{F_{A,B}}{t+\imunit\beta}
=
\fun{\oastate[\psi]}{\fun{\tau_t}{B}A},
t\in\fldreal.$$
\end{defn}

The set \(K_\beta\) in Main Results, Section \ref{sec:main-results}, consists of the states in Definition \ref{def:kms-state}. If \(\oa{B}\subset\oa{A}\) is a norm-dense, \(\tau\)-invariant \(\ast\)-subalgebra of entire analytic elements, the KMS condition is equivalent to \begin{equation}\label{eq:kms-analytic-form}
\fun{\oastate[\psi]}{A\fun{\tau_{\imunit\beta}}{B}}
=
\fun{\oastate[\psi]}{BA}
\end{equation} for \(A\in\oa{A}\) and \(B\in\oa{B}\).

\begin{defn}[generator of an automorphism group]\label{def:automorphism-generator}
Let $\tau$ be a strongly continuous one-parameter group of
$\ast$-automorphisms of $\oa{A}$.
Its generator is the derivation with domain and action
\begin{equation}\label{eq:automorphism-generator}
\begin{aligned}
\dom\oaderiv_{\tau}
&=
\set{A\in\oa{A}}{\lim_{t\to0}\frac{\fun{\tau_t}{A}-A}{t}\text{ exists in norm}},
\\ 
\fun{\oaderiv_{\tau}}{A}
&=
\lim_{t\to0}
\frac{\fun{\tau_t}{A}-A}{t},
\quad
A
\in
\dom\oaderiv_{\tau}.
\end{aligned}
\end{equation}
\end{defn}

\begin{prop}[KMS invariance and the generator]\label{prop:kms-generator-invariance}
Let $\oastate[\psi]$ be a $\rbk{\tau,\beta}$-KMS state.
Then
\begin{equation}\label{eq:kms-generator-invariance}
\fun{\oastate[\psi]}{\fun{\oaderiv_{\tau}}{A}}
=
0
\end{equation}
for every $A\in\dom\oaderiv_{\tau}$.
\end{prop}

\begin{proof}
Every KMS state is $\tau$-invariant
\cite[Propositions 5.3.3 and 5.3.7]{BratteliRobinson004}.
For $A\in\dom\oaderiv_{\tau}$, the definition
\eqref{eq:automorphism-generator} and this invariance give
$$
\fun{\oastate[\psi]}{\fun{\oaderiv_{\tau}}{A}}
=
\lim_{t\to0}
\frac{\fun{\oastate[\psi]}{\fun{\tau_t}{A}}
-
\fun{\oastate[\psi]}{A}}{t}
=
0.
$$
\end{proof}

\begin{prop}[finite-volume Gibbs states]\label{prop:gibbs-kms}
Let $\physham$ be self-adjoint on a finite-dimensional Hilbert space,
and let
$$\fun{\tau_t}{A}
=
\napiernum^{\imunit t\physham}A\napiernum^{-\imunit t\physham}.$$
The state
$$\fun{\oastate[\psi_\beta]}{A}
=
\frac{\sqfun{\trace}{\napiernum^{-\beta\physham}A}}
{\sqfun{\trace}{\napiernum^{-\beta\physham}}}$$
is the unique $\rbk{\tau,\beta}$-KMS state.
\end{prop}

\begin{proof}
Cyclicity of the trace gives
$$\sqfun{\trace}{
\napiernum^{-\beta\physham}
A
\napiernum^{-\beta\physham}
B
\napiernum^{\beta\physham}}
=
\sqfun{\trace}{\napiernum^{-\beta\physham}BA}.$$
Thus the Gibbs state satisfies \eqref{eq:kms-analytic-form}.
Conversely, write an arbitrary state as
$\fun{\oastate[\varphi]}{A}=\sqfun{\trace}{\rho A}$.
Equation \eqref{eq:kms-analytic-form} implies that
$\napiernum^{\beta\physham}\rho$ commutes with every matrix.
It is scalar, and normalization gives the displayed Gibbs density matrix.
\end{proof}

\begin{prop}[compactness, convexity, and metrizability]\label{prop:kms-compact-convex}
Let $\oa{A}$ be unital.
The set $K_\beta$ is a weak-$\ast$ compact convex subset of
$\oaspstate_{\oa{A}}$.
If $\oa{A}$ is separable, its relative weak-$\ast$ topology is metrizable.
\end{prop}

\begin{proof}
This is \cite[Theorem 5.3.30]{BratteliRobinson004}.
\end{proof}

\subsubsection{Limits of finite-volume KMS states}\label{limits-of-finite-volume-kms-states}

The KMS property passes to a weak-\(\ast\) limit once the finite-volume strip functions have uniformly convergent boundary values.

\begin{lem}[bounded strip families]\label{lem:kms-strip-boundary}
Let $\net{F_i}{i\in I}$ be a net of holomorphic functions in the open strip
$S
= \set{z\in\fldcmp}
{0<\opimag z<\beta}$ and continuous on its closure $\gtclos{S}$.
Suppose that $\abs{\fun{F_i}{z}}\leq C$ on the closed strip and that both
boundary functions converge locally uniformly on $\fldreal$.
Every subnet that converges locally uniformly in the open strip has a limit
that extends continuously to the two boundaries with the corresponding
boundary limits.
\end{lem}

\begin{proof}
Fix $\epsilon>0$ and $L>0$.
For sufficiently large indices $i$ and $j$, both boundary differences of
$F_i-F_j$ are at most $\epsilon$ on the interval
$\closedinterval{-L}{L}$.
Let $h_L$ be the harmonic function on
$\openinterval{-L}{L}\times\openinterval{0}{\beta}$
whose boundary values are $0$ on the horizontal sides and $1$ on the
vertical sides.
The maximum principle for the subharmonic function
$\abs{F_i-F_j}$ gives
$$\abs{\fun{F_i}{z}-\fun{F_j}{z}}
\leq
\epsilon + 2C\fun{h_L}{\opreal z,\opimag z}.$$
On every compact substrip,
$\fun{h_L}{x,y}$ tends uniformly to $0$ as $L\to\infty$.
The family is therefore uniformly Cauchy up to either boundary on compact
subintervals.
Its interior limit has the asserted continuous boundary values.
\end{proof}

\begin{thm}[thermodynamic-limit KMS states]\label{thm:limit-kms}
Let $\oa{B}$ be a norm-dense $\ast$-subalgebra of $\oa{A}$.
For every $A,B\in\oa{B}$, suppose that a net of holomorphic functions
$\net{F_i}{i\in I}$ is bounded and continuous on the closed strip
$\set{z\in\fldcmp}{0\leq\opimag z\leq\beta}$,
satisfies
$$\abs{\fun{F_i}{z}}
\leq
\norm{A}\norm{B},$$
and has boundary values
$$\fun{F_i}{t}
=
\fun{\oastate[\psi_i]}{A\fun{\tau^i_t}{B}},
\quad
\fun{F_i}{t+\imunit\beta}
=
\fun{\oastate[\psi_i]}{\fun{\tau^i_t}{B}A}.$$
Suppose that $\oastate[\psi_i]\to\oastate[\psi]$ weak-$\ast$ and that both
boundary functions converge locally uniformly on $\fldreal$ to
$\fun{\oastate[\psi]}{A\fun{\tau_t}{B}}$
and
$\fun{\oastate[\psi]}{\fun{\tau_t}{B}A}$ respectively.
Then $\oastate[\psi]\in K_\beta$.
\end{thm}

\begin{proof}
Montel's theorem gives a subnet of $\net{F_i}{i\in I}$ that converges
locally uniformly in the open strip to a holomorphic function $F$.
Lemma \ref{lem:kms-strip-boundary} identifies the two continuous boundary
values of $F$ with the displayed limits.
Thus $F$ is the KMS function for $A$ and $B$.
Norm density of $\oa{B}$ and the uniform bound extend the condition to
all $A,B\in\oa{A}$.
\end{proof}

\subsubsection{Extremal KMS states}\label{extremal-kms-states}

The following separation property applies whenever the gauge action produces distinct elements of \(\setextremal K_\beta\).

\begin{thm}[disjointness of distinct extremal KMS states]\label{thm:extremal-kms-disjoint}
Let $\oastate[\psi_{1}]$ and $\oastate[\psi_{2}]$ be distinct elements of
$\setextremal K_\beta$.
Their GNS representations are disjoint.
\end{thm}

\begin{proof}
This is \cite[Theorem 5.3.30(4)]{BratteliRobinson004}.
\end{proof}

\subsubsection{Central decomposition of KMS states}\label{central-decomposition-of-kms-states}

Let \(\oastate[\psi]\in K_\beta\) and write \((\oarepn_\psi,\sphilb{H}_\psi,\oagnsvector[\Psi])\) for its GNS representation. The von Neumann algebra and its center are \[\oa{M}_\psi
=
\oadoublecommutant{\fun{\oarepn_\psi}{\oa{A}}},
\quad
\oa{Z}_\psi
=
\oa{M}_\psi
\cap
\oacommutant{\oa{M}_\psi}.\]

\begin{lem}[separating vector]\label{lem:separating}
The GNS vector $\oagnsvector[\Psi]$ is cyclic and separating for
$\oa{M}_\psi$.
\end{lem}

\begin{proof}
This is \cite[Theorem 5.3.10]{BratteliRobinson004}.
\end{proof}

The modular group of \((\oa{M}_\psi,\oagnsvector[\Psi])\) implements the extension \(\bar{\tau}\) of the dynamics: \begin{equation}\label{eq:modular-dynamics}
\bar{\tau}_t
=
\fun{\Ad}{\Delta^{-\imunit t/\beta}}
\text{on }\oa{M}_\psi.
\end{equation} Equation \eqref{eq:modular-dynamics} is Takesaki's KMS theorem \cite[Corollary 5.3.4 and Theorem 5.3.10]{BratteliRobinson004}.

\begin{lem}[the modular group fixes the center]\label{lem:center-fixed}
For $z\in\oa{Z}_\psi$ and $s\in\fldreal$, the modular automorphism group
satisfies $\fun{\sigma_s}{z}=z$.
For every $w\in\fldcmp$ for which the expression is defined,
$\Delta^wz\oagnsvector[\Psi]=z\oagnsvector[\Psi]$.
\end{lem}

\begin{proof}
This is \cite[Proposition 5.3.28]{BratteliRobinson004}.
\end{proof}

\begin{prop}[subordinate KMS states]\label{prop:central-subordinate}
Let $T\in\oa{Z}_\psi$ satisfy $T\geq0$ and
$\bkt{\oagnsvector[\Psi]}{T\oagnsvector[\Psi]}=1$.
The formula
$$\fun{\oastate[\psi_T]}{A}
=
\bkt{\oagnsvector[\Psi]}{
T\fun{\oarepn_\psi}{A}\oagnsvector[\Psi]}$$
defines a $\rbk{\tau,\beta}$-KMS state.
Every KMS state dominated by a scalar multiple of
$\oastate[\psi]$ arises uniquely in this way.
\end{prop}

\begin{proof}
This is \cite[Theorem 5.3.29]{BratteliRobinson004}.
\end{proof}

\begin{thm}[decomposition of KMS states]\label{thm:kms-decomposition}
Let $\oa{A}$ be separable and let $\oastate[\psi]\in K_\beta$.
The following conclusions hold.
\begin{enumerate}
\item The state $\oastate[\psi]$ is extremal in $K_\beta$ if and only if
$\oa{M}_\psi$ is a factor.
\item The set $K_\beta$ is a Choquet simplex.
\item There is a unique maximal representing probability measure in the
Choquet order, denoted by $\mu_\psi$, on $K_\beta$ with barycenter
\begin{equation}\label{eq:kms-central-barycenter}
\oastate[\psi]
=
\int_{K_\beta}
\oastate[\psi']
\opdmsr{\mu_\psi(\oastate[\psi'])}.
\end{equation}
The measure $\mu_\psi$ is the central orthogonal measure and is supported
on $\setextremal K_\beta$.
\item There are a standard probability space $(X,\mu)$ and a measurable
family $\seq{\oastate[\psi_x]}{x\in X}$ in $\setextremal K_\beta$ such that
\begin{equation}\label{eq:kms-central-direct-integral}
\sphilb{H}_\psi
=
\int_X^\oplus\sphilb{H}_x\opdmsr{\mu(x)},
\quad
\oarepn_\psi
=
\int_X^\oplus\oarepn_x\opdmsr{\mu(x)},
\quad
\oagnsvector[\Psi]
=
\int_X^\oplus\oagnsvector[\Psi_x]\opdmsr{\mu(x)}.
\end{equation}
The map $x\mapsto\oastate[\psi_x]$ pushes $\mu$ forward to
$\mu_\psi$.
The algebra $\oa{Z}_\psi$ becomes
$\fun{\lp^\infty}{X,\mu}$.
\end{enumerate}
\end{thm}

\begin{proof}
Proposition \ref{prop:central-subordinate} gives the factoriality
criterion.
The orthogonal measure, its maximality, the separable support statement,
and the spatial central decomposition are
\cite[Theorems 4.1.25 and 5.3.30]{BratteliRobinson003,BratteliRobinson004}.
\end{proof}

\begin{prop}[central integral identity]\label{prop:central-integral-identity}
Use the direct-integral data of Theorem \ref{thm:kms-decomposition}.
For $D\in\oa{Z}_\psi$, let $f_D$ be its diagonal fiber function.
Then the map
$x\mapsto \fun{f_D}{x}\fun{\oastate[\psi_x]}{A}$
is measurable for every $A\in\oa{A}$, and
\begin{equation}\label{eq:orthogonal-measure-central-identity}
\bkt{\oagnsvector[\Psi]}{
D\fun{\oarepn_\psi}{A}\oagnsvector[\Psi]}
=
\int_X
\fun{f_D}{x}
\fun{\oastate[\psi_x]}{A}
\opdmsr{\mu(x)}.
\end{equation}
\end{prop}

\begin{proof}
The direct-integral decomposition represents $D$ as multiplication by $f_D$.
Integrating the fiber vector states gives
\eqref{eq:orthogonal-measure-central-identity}.
\end{proof}

The next proposition records the measure-theoretic step that passes from a disintegration of a central measure to the central measures of its fiber barycenters.

\begin{prop}[conditional central measures]\label{prop:conditional-central-measures}
Let $\oa{A}$ be unital and separable, let $\oastate[\psi] \in K_\beta$, and
let $\mu_\psi$ be the central measure provided by Theorem
\ref{thm:kms-decomposition}.
Let $Y$ be a standard Borel space, let $q \colon K_\beta \to Y$ be a Borel
map, and set $\msr{\mu_q} = q_*\mu_\psi$.
Let $y \mapsto \mu_y$ be a Borel probability kernel from $Y$ to $K_\beta$
that disintegrates $\mu_\psi$ over $q$ and satisfies
$$\fun{\mu_y}{\fun{q^{-1}}{y}} = 1$$
for $\msr{\mu_q}$-almost every $y$.
For those $y$, define
$$\oastate[\psi^{\rbk{y}}]
=
\int_{\fun{q^{-1}}{y}}
\oastate[\psi']
\opdmsr{\mu_y(\oastate[\psi'])}.$$
Then $\mu_y$ is the central measure of
$\oastate[\psi^{\rbk{y}}]$ for $\msr{\mu_q}$-almost every $y$.
Moreover,
$$\oastate[\psi]
=
\int_Y
\oastate[\psi^{\rbk{y}}]
\opdmsr{\mu_q(y)}.$$
\end{prop}

\begin{proof}
The central measure $\mu_\psi$ is supported on $\setextremal K_\beta$ by
assertion (3) of Theorem \ref{thm:kms-decomposition}.
The disintegration identity therefore gives
$\fun{\mu_y}{\setextremal K_\beta} = 1$
for $\msr{\mu_q}$-almost every $y$.
Each $\mu_y$ is thus a representing measure for
$\oastate[\psi^{\rbk{y}}]$ that is supported on $\setextremal K_\beta$.
Such a measure is maximal in the Choquet order
\cite[Section 4.1]{BratteliRobinson003}.
The uniqueness assertion in assertion (3) of Theorem \ref{thm:kms-decomposition} identifies
$\mu_y$ with the central measure of $\oastate[\psi^{\rbk{y}}]$.

For every $A \in \oa{A}$, the disintegration identity gives
$$\fun{\oastate[\psi]}{A}
=
\int_Y
\opdmsr{\mu_q(y)}
\int_{\fun{q^{-1}}{y}}
\fun{\oastate[\psi']}{A}
\opdmsr{\mu_y(\oastate[\psi'])}.
$$
The inner integral is
$\fun{\oastate[\psi^{\rbk{y}}]}{A}$.
This proves the final formula.
\end{proof}

\begin{prop}[circle-orbit disintegration]\label{prop:circle-orbit-disintegration}
Let $X$ be compact and metrizable, and let
$$\fun{\alpha}{\cdot}
\colon
\fun{\liegr{U}}{1}
\times
X
\to
X$$
be a jointly continuous action.
For $\theta \in \rightopeninterval{0}{2\pi}$, write
$\alpha_\theta = \alpha_{\napiernum^{\imunit\theta}}$.
Let $\mu$ be an invariant Borel probability measure on $X$.
Set $Y = X / \fun{\liegr{U}}{1}$, let $q \colon X \to Y$ be the quotient
map, and set $\msr{\mu_Y} = q_*\mu$.
Then $Y$ is compact and metrizable.
There is a Borel probability kernel $y \mapsto \mu_y$ from $Y$ to $X$, unique
up to a $\msr{\mu_Y}$-null set, such that
$$\fun{\mu_y}{\fun{q^{-1}}{y}} = 1,
\quad
\mu
=
\int_Y
\mu_y
\opdmsr{\mu_Y(y)}.$$
For $\msr{\mu_Y}$-almost every $y$, the measure $\mu_y$ is the unique invariant
probability measure on $\fun{q^{-1}}{y}$.
It is the image of normalized Haar measure on
$\rightopeninterval{0}{2\pi}$ under
$$\theta
\mapsto
\fun{\alpha_\theta}{x}$$
for every $x \in \fun{q^{-1}}{y}$.
Every $f \in \fun{\lp^\infty}{X,\mu}$ satisfying
$f \circ \alpha_\theta = f$ in $\fun{\lp^\infty}{X,\mu}$ for every
$\theta \in \rightopeninterval{0}{2\pi}$ is of the form $F \circ q$ for some
$F \in \fun{\lp^\infty}{Y,\msr{\mu_Y}}$.
\end{prop}

\begin{proof}
The orbit equivalence relation is the image of
$\fun{\liegr{U}}{1} \times X$ under
$$\rbk{u,x}
\mapsto
\rbk{x,\fun{\alpha_u}{x}}.$$
The domain is compact and the map is continuous, so the relation is closed.
The quotient $Y$ is compact Hausdorff.
The quotient map is open because the saturation of an open set $U \subset X$
is $\bigcup_{u \in \fun{\liegr{U}}{1}} \fun{\alpha_u}{U}$.
Thus $Y$ is second countable and metrizable.

For $f \in \fun{\conti}{X}$, the measure $q_*\rbk{f\mu}$ is absolutely
continuous with respect to $\msr{\mu_Y}$.
Let $\fun{\Lambda}{f}$ be its Radon--Nikodym derivative.
The map $\Lambda$ is positive and unital, and it satisfies
$$\fun{\Lambda}{\rbk{g \circ q}f}
=
g\fun{\Lambda}{f}$$
for $g \in \fun{\conti}{Y}$.
Choose countable dense $\fldrat$-subalgebras of
$\fun{\conti}{X}$ and $\fun{\conti}{Y}$.
Outside one $\msr{\mu_Y}$-null set, the values
$\fun{\Lambda}{f}(y)$ are positive, normalized, and linear on the first
subalgebra, and satisfy the displayed module identity.
They extend to states on $\fun{\conti}{X}$.
The Riesz representation theorem produces the Borel probability kernel
$y \mapsto \mu_y$.
Its defining identity is
$$\int_Y
\fun{\mu_y}{f}
\opdmsr{\mu_Y(y)}
=
\fun{\mu}{f}$$
for every $f \in \fun{\conti}{X}$.
The module identity gives
$$\fun{\mu_y}{g \circ q}
=
\fun{g}{y}.$$
It follows that $\mu_y$ is supported on $\fun{q^{-1}}{y}$.
The same countable test algebra proves uniqueness of the kernel.

For fixed $\theta$, the family
$y \mapsto \fun{(\alpha_\theta)_*}{\mu_y}$ is another disintegration of
$\mu$ over $q$.
Kernel uniqueness gives
$$\fun{(\alpha_\theta)_*}{\mu_y}
=
\mu_y$$
for $\msr{\mu_Y}$-almost every $y$.
Fubini and joint continuity make this identity valid for every
$\theta \in \rightopeninterval{0}{2\pi}$ outside one fixed $\msr{\mu_Y}$-null
set.
An invariant probability measure on an orbit equals the Haar image, since
for $h \in \fun{\conti}{X}$ its value is
$$\int_{\rightopeninterval{0}{2\pi}}
\fun{h}{\fun{\alpha_\theta}{x}}
\frac{\opdmsr{\theta}}{2\pi}$$
at any $x$ in that orbit.

Let $f$ satisfy the final hypothesis.
For almost every $y$, its restriction to the Haar probability space
$\fun{q^{-1}}{y}$ is almost everywhere equal to the Haar average.
The function $F(y) = \fun{\mu_y}{f}$ is measurable and satisfies
$f = F \circ q$ in $\fun{\lp^\infty}{X,\mu}$.
\end{proof}

\subsubsection{Covariance and mean clustering}\label{covariance-and-mean-clustering}

\begin{prop}[equivariance of the central measure]\label{prop:decomposition-covariance}
Let $\oastate[\psi]\in K_\beta$ and let $\alpha$ be an automorphism of
$\oa{A}$ such that
$\alpha\circ\tau_t=\tau_t\circ\alpha$ and
$\oastate[\psi]\circ\alpha=\oastate[\psi]$.
For $\oastate[\varphi]\in K_\beta$, define
$\fun{\alpha_*}{\oastate[\varphi]}=\oastate[\varphi]\circ\alpha$.
Then
$$\fun{(\alpha_*)_*}{\mu_\psi}
=
\mu_\psi.$$
\end{prop}

\begin{proof}
The automorphism $\alpha$ preserves the KMS condition and is implemented in
the GNS representation of $\oastate[\psi]$.
Its implementing unitary transports $\oa{Z}_\psi$ to itself.
The push-forward of $\mu_\psi$ is therefore another maximal orthogonal
measure with barycenter $\oastate[\psi]$.
The uniqueness in Theorem \ref{thm:kms-decomposition} gives the result.
\end{proof}

\begin{defn}[Følner net]\label{defn:folner-net}
Let $G$ be a discrete group.
A net $\net{F_i}{i\in I}$ of nonempty finite subsets of $G$ is a Følner net
when, for every $v\in G$,
$$\lim_{i\in I}
\frac{\abscard{\rbk{vF_i}\mathbin{\triangle}F_i}}
{\abscard{F_i}}
=
0.$$
\end{defn}

\begin{lem}[mean ergodic theorem]\label{lem:mean-ergodic}
Let $\net{F_i}{i\in I}$ be a Følner net in an amenable discrete group $G$,
and let $v\mapsto U_v$ be a unitary representation on $\sphilb{H}$.
If $P_0$ denotes the projection onto the jointly invariant vectors, then
$$\slim_{i\in I}
\frac{1}{\abscard{F_i}}
\sum_{v\in F_i}U_v
=
P_0.$$
\end{lem}

\begin{proof}
This is the mean ergodic theorem for amenable groups
\cite[Theorem 4.3.10]{BratteliRobinson003}.
\end{proof}

\begin{prop}[factor states cluster in the mean]\label{prop:factor-clustering}
Let a norm-asymptotically abelian action $\eta$ of an amenable discrete
group $G$ commute with $\tau$.
If $\oastate[\psi]\in K_\beta$ is $G$-invariant and
$\oa{M}_\psi$ is a factor, then every Følner net
$\net{F_i}{i\in I}$ satisfies
$$\lim_{i\in I}
\frac{1}{\abscard{F_i}}
\sum_{v\in F_i}
\fun{\oastate[\psi]}{B\fun{\eta_v}{A}}
=
\fun{\oastate[\psi]}{B}
\fun{\oastate[\psi]}{A}$$
for all $A,B\in\oa{A}$.
Moreover,
$$P_0\fun{\oarepn_\psi}{A}\oagnsvector[\Psi]
=
\fun{\oastate[\psi]}{A}\oagnsvector[\Psi].$$
\end{prop}

\begin{proof}
Set
$$M_i
=
\frac{1}{\abscard{F_i}}
\sum_{v\in F_i}
\fun{\oarepn_\psi}{\fun{\eta_v}{A}}.$$
Every weak-operator limit point of $\net{M_i}{i\in I}$ belongs to
$\oa{M}_\psi$.
Norm asymptotic abelianness gives
$$\wlim_{i\in I}
\commutator{M_i}{\fun{\oarepn_\psi}{B}}
=
0$$
for $B\in\oa{A}_{\txtloc}$.
Density gives the same identity for every $B\in\oa{A}$.
Every limit point therefore belongs to
$\oa{M}_\psi\cap\oacommutant{\fun{\oarepn_\psi}{\oa{A}}}=\oa{Z}_\psi$.
Factoriality and invariance identify it with
$\fun{\oastate[\psi]}{A}$.
Thus
$$\wlim_{i\in I}M_i
=
\fun{\oastate[\psi]}{A}.$$
Lemma \ref{lem:mean-ergodic} gives
$$\slim_{i\in I}M_i\oagnsvector[\Psi]
=
P_0\fun{\oarepn_\psi}{A}\oagnsvector[\Psi].$$
For every $B\in\oa{A}$,
$$\bkt{\fun{\oarepn_\psi}{\faadj{B}}\oagnsvector[\Psi]}
{P_0\fun{\oarepn_\psi}{A}\oagnsvector[\Psi]}
=
\fun{\oastate[\psi]}{B}
\fun{\oastate[\psi]}{A}
=
\bkt{\fun{\oarepn_\psi}{\faadj{B}}\oagnsvector[\Psi]}
{\fun{\oastate[\psi]}{A}\oagnsvector[\Psi]}.$$
Cyclicity of $\oagnsvector[\Psi]$ proves the vector identity.
The scalar weak-operator limit gives the stated correlation limit.
\end{proof}

\bibliography{myref.bib}

\end{document}